\documentclass[12pt]{article}

\usepackage{latexsym, amssymb}
\usepackage{amsmath,amsfonts,amsthm,latexsym, amssymb, wasysym}
\usepackage{authblk}
\usepackage[notlof,notlot,notindex]{tocbibind}
\usepackage{graphicx}
\usepackage{float}
\usepackage{animate}
\usepackage[usenames,dvipsnames]{color}
\usepackage{pstricks}
\usepackage{chemfig}
\usepackage{subfig}

\definecolor{darkblue}{rgb}{0.0, 0.0, 0.55}
\usepackage{hyperref}
\hypersetup{
    colorlinks,
    citecolor=black,
    filecolor=black,
    linkcolor=darkblue,
    urlcolor=black
}
\usepackage[fleqn,tbtags]{mathtools}
\mathtoolsset{showonlyrefs,showmanualtags}
\usepackage{breqn}
\usepackage{tikz}
\usetikzlibrary{decorations.fractals}

\newtheorem*{definition}{Definition}

\begin{document}

\title{A Unified Mathematical Language \\ for Medicine and Science}
\author{Patrick St-Amant}
\affil{Department of Mathematics, \\ Coll\'egial International Sainte-Anne, \\ Montreal, Canada}

\date{}

\maketitle

\begin{abstract}A unified mathematical language for medicine and science will be presented. Using this language, models for DNA replication, protein synthesis, chemical reactions, neurons and a cardiac cycle of a heart have been built. Models for Turing machines, cellular automaton, fractals and physical systems are also represented with the use of this language. Interestingly, the language comes with a way to represent probability theory concepts and also programming statements. With this language, questions and processes in medicine can be represented as systems of equations; and solutions to these equations are viewed as treatments or previously unknown processes. This language can serve as the framework for the creation of a large interactive open-access scientific database that allows extensive mathematical medicine computations. It can also serve as a basis for exploring ideas related to what could be called metascience.
\end{abstract}
\pagebreak
\tableofcontents
\pagebreak

\section{A Vision of Medicine}

Imagine a world where treatments are computed and where knowledge of medicine is merged in a large database. New treatments are computed based on mathematical models built by researchers and each new model, like a piece of a puzzle, can be connected within a vast system of knowledge. Potential new treatments would be tested within a computational model and effects would be observed virtually, making animal experimentation obsolete due to lack of accuracy. New processes explaining functions of the body would be computationally deduced and hypothesized. Ultimately, treatments with minimal sides effects would be tailored to each patient based on their personal biological data.

It is difficult to keep up-to-date with new important research and publications and impossible to cover all the new daily literature in one's domain of study. We are at a point where even a group of experts with the best communication tools and habits cannot efficiently integrate all the new knowledge of their field. In contrast, we now have the technological means to gather and analyse large amount of data.

A key component needed for the realization of this future is a common language for science and medicine. In this paper, we present a unified mathematical language which can be used to represent biological processes and scientific concepts. Although the examples are geared towards medicine, we will also present how this language can be used in different domains with the aim of demonstrating that effort in developing such a language can lead to improved communication and exchange of ideas between fields of studies.

 Translating each research paper into a set of mathematical expressions and integrating it into a large database would ensure the findings of each new research study would be integrated into the existing body of knowledge. A feature of this language is that it could encompass the results of a vast number of papers, so that computations, discoveries and analysis could also be done on the field of medicine as a whole field. This could be a step into a new domain which could be called meta-medicine. This could be an opportunity to bridge the gap between publishing results and the integration of the results by the community. Competing theories and contradicting results could be tested against each other through computation and help to improve the general model. Interestingly, experimental results which are usually considered inconclusive or `negative' could also contribute to enrich the global database.

\section{Purpose of the Language}

The objective behind these investigations is to build a language in which all scientific concepts and notations can be represented, while keeping the way each discipline functions and by enhancing the computational nature of each discipline. Along with the ability to represent all scientific concepts and its computational power, we would also need a language that allows navigation between different levels of complexity or scales (from DNA replication to body mechanics), can be used in a large database, has a visual notation and compact notation and has a relative ease of use regardless of the domain of study. With this language, dynamic models in biology, chemistry, neuroscience, computer science and physics have been built. We hope that sufficient steps have been taken towards demonstrating that the language has the qualities mentioned above and that this will inspire others to explore it further.

In mathematics, set theory combined with mathematical logic is known to be able to represent most mathematical expressions and concepts, but in practice we rarely directly use set theoretical expressions to represent or solve mathematical problems. Similarly, in computer science, although all programs can be reduced to Turing machines of zeros and ones, we rarely think or interact with high-level programming languages in this way. Whatever the language, there seems to be a gap between what we do in the discipline and the fundamental language. We could say that English along with all mathematical symbols is a complete language which can communicate any scientific concept, but unfortunately, computations cannot be done efficiently on words and sentences. The mathematical languages known as rewriting systems \cite{Terese} are very powerful and have been shown to represent grammar, plants and fractals \cite{Chomsky, Lindenmayer}. However, the scope of these systems is sometimes restricted to certain types of objects and focuses on automatic theorem proving, normal forms (expressions that cannot be transformed further) and termination.

The language which will be presented can be viewed as a higher order rewriting language that has no restrictions on the type of objects and on the objects that can be named. Alternatively, the language could be viewed as a higher order universal Turing machine where the symbols, states and tape can be any type of object. Furthermore, we could say that the head of the Turing machine could affect its own states and the symbols on the tape could also affect the states and properties of the head. Example of types of objects are models of atoms, molecules, DNA strands, cells, neurons, concepts, mathematical equations, programs, 3D geometrical structures, people and also infinite collections of different types of objects.

The driving question underlying this study can be written as:
\begin{center}What are the atoms of scientific concepts and how can we combine them to represent complex systems?\end{center}

\section{Essence of the Language}

In this section, we introduce the elementary elements of the language. This is only meant to offer a quick glance at the language and one should read further in the subsequent sections for a more accessible and progressive description of the language along with key applications to biology, medicine and other scientific domains.

Everywhere we look in the world, we see change. We observe changes at every level, whether it be atoms, humans or the stars; they are changing all the time. As observers, we can take note of these changes and record them. Based on these records, we can infer and extrapolate the past and the future. In studying these records, we recognize interesting structures and name them. The principles behind the unified mathematical language that will presented can be reduced to the following two statements:
\begin{enumerate}
  \item Observe a change in the world. Record what it was and what it became.
  \item All objects and changes can be named.
\end{enumerate}

For example, let $A$ and $B$ be objects. We will denote the transformation of object $A$ into $B$ by the notation $$\fbox{$B\rhd A$}.$$
This notation should be understood as $B$ having the potential to slide down over the triangle $\rhd$ in the direction of $A$ so that $B$ will take the place of $A$. In itself, this object that we call a \textit{transformation} is interesting, but it only represents a potential. For the change to occur, we first need the object $A$ to be present. This will be denoted by $$\fbox{$B\rhd A$}\rightarrow(A).$$
The right arrow means that we apply the transformation $\fbox{$B\rhd A$}$ to the object $A$ in the ordered set $(A)$. Therefore, changing $A$ into $B$ will result in $(B)$. In other words, $A$ is changed into $B$ by applying the transformation $\fbox{$B\rhd A$}$. This notation should be visualized as though the transformation will slide down the arrow $\rightarrow$ until the $A$ on the right-hand side of the transformation superposes over the $A$ inside the parentheses. When the $A$'s are matched, the $B$ slides down over the directed triangle to finally replace $A$.

%The following animation explains how we can view the application of a transformation on the ordered set $(A)$.
%
%\begin{center}\animategraphics[controls, scale=0.7, loop]{3}{BasicAnimation}{0}{26}\end{center}

Mathematically, the application of a transformation can be written this way by using the double arrow as follows.
$$\fbox{$B\rhd A$}\rightarrow(A)\Rightarrow(B)$$
This reads as the system which is composed of the transformation $\fbox{$B\rhd A$}$ applied to the object $(A)$ which \textit{reduces} to the object $(B)$.

When we hear the name of friend, we can access a lot of knowledge concerning our friend. The short string of letters of the name points to an array of knowledge and permits us to avoid the lengthy description of all that is known about the friend. Naming objects is also very important in mathematics and science. Scientific terms permit us to quickly build new theories and explain insightful understanding about the world. The calculative power of mathematics can be said to rely on its ability to condense long expressions and operations into a string of symbols where the operations are viewed as names that indicate how the symbols should be acted upon. In our language, we will name objects and transformations by using the symbol `$:=$'. For example, we can give the name \textit{t} to our transformation $\fbox{$B\rhd A$}$ by writing $t:=\fbox{$B\rhd A$}.$

For this language to be powerful, a key is to allow $A$ and $B$ to be anything we want. For example, we can take $A$ and $B$ to be numbers, models of DNA strands, neurons, people, behaviours, 3 dimensional geometric objects or even groups containing different types of objects.

Based on this approach, this means that the action of folding a model of a protein $P$ into the structure $S$ is written as $FoldingProtein:=\fbox{$S\rhd P$}$ and representing a person closing his hand can also be written in the same language as $ClosingHand:=\fbox{$\text{closed hand}\rhd \text{open hand}$}$.  A physical therapist will usually be interested in body movements while the biologist will be concerned about objects of the size of proteins, but now both can use this language without losing their respective level of interest. Thus, everything can be brought back to this language and facilitate comparisons and knowledge transfer between disciplines.

This way of viewing things can seem overly simplistic when we think of all the numerous systems and concepts in science it needs to represent, but this is similar to explaining all the physical objects we observe with only the concept of atoms. We will see that the combinations of transformations leads to highly complex systems such as the functioning of the body. As mentioned, transformations can act on any type of object, but they can also be applied in sequence or in parallel to groups of objects. Moreover, transformations can be applied to groups of transformations so that we can describe and study higher order languages. Also, it is possible to have transformations which apply to themselves, thus opening a wealth of possibilities.

Importantly, the language comes with computational power because a programming statement can be viewed as a collection of transformations of different types of objects. Moreover, we will see that scientific questions can be written as systems of equations such that the solutions to these equations are the answers. This way, each domain can rely on computing tools to deepen their respective explorations. In medicine, this means that with some effort in translating, creating models and computing, we can build a large dynamic model of the body where all different levels are represented. We can also insert variables into our models and create equations that represent important questions in medicine. Solutions to these equations would represent treatments or unknown biological processes, thus allowing us to concentrate our resources on solving certain equations to unlock new types of treatments or understanding.

Using transformations, which can be seen as the atoms of scientific concepts, we will be able to represent many systems. For example, DNA replication can be seen as composed of transformations aimed at separating the strands and transformations replacing nucleotides by pairs of nucleotides. Turing machines can be viewed as a collection of transformations which change series of zeros and ones on a tape. A mass falling to the ground can be represented by a transformation which replaces the mass at a certain height by the same mass slightly closer to the ground. We will present more refined versions of these examples in the following text and also give models for messenger RNA, proteins, a simplified heart, neurons, cell division, chemical reactions, programming statements, mathematical functions and cellular automata.

Another important outcome of this language is that discoveries in one scientific domain can lead to similar observations in another domain as the domains share similar mathematical structures. This points to what could be called \textit{metascience}. Having a seamless language for computer science, mathematics and biology would permit us to transform abstract mathematical concepts into the field of  biology, and conversely, study abstract biological processes as mathematical structures. We could say that a large part of modern mathematics stems from the plane geometric figures studied by the ancient Greeks, fractions and numbers. In light of this, one might wonder which kind of mathematical structures will be studied in the future.

\section{The Mathematical Language}

In the following section we will describe the language, present important types of transformations, define the equations of this language and discuss some properties of the language.

\subsection{Visual Language}

When we visualize molecules; we usually imagine them as a drawing or a three dimensional structure. One of the most efficient ways to learn is by seeing someone do it, the next best thing for many people is to see a drawing or an illustration for which there is a textual description. The inline notation where all is written as text is valuable if we want a compact form of a concept, but can be more difficult to understand. For example, molecules are more accurately represented in three dimensions, but they have a two dimension representation and an inline notation. If we do not want to lose information while reducing the number of dimensions, we have to add new symbols and notations making the representation less easily understood.

For example, a D-glucose chain can be represented inline as $C_{6}H_{12}O_{6}$ while the 2D skeletal notation and the 3D representation give more information as seen below. Note that, while not easy to read, it is possible to encode inline the 3D representation of a molecule.
\begin{figure}[H]%
    \centering
    \subfloat[Skeletal notation]{{\includegraphics[width=5cm]{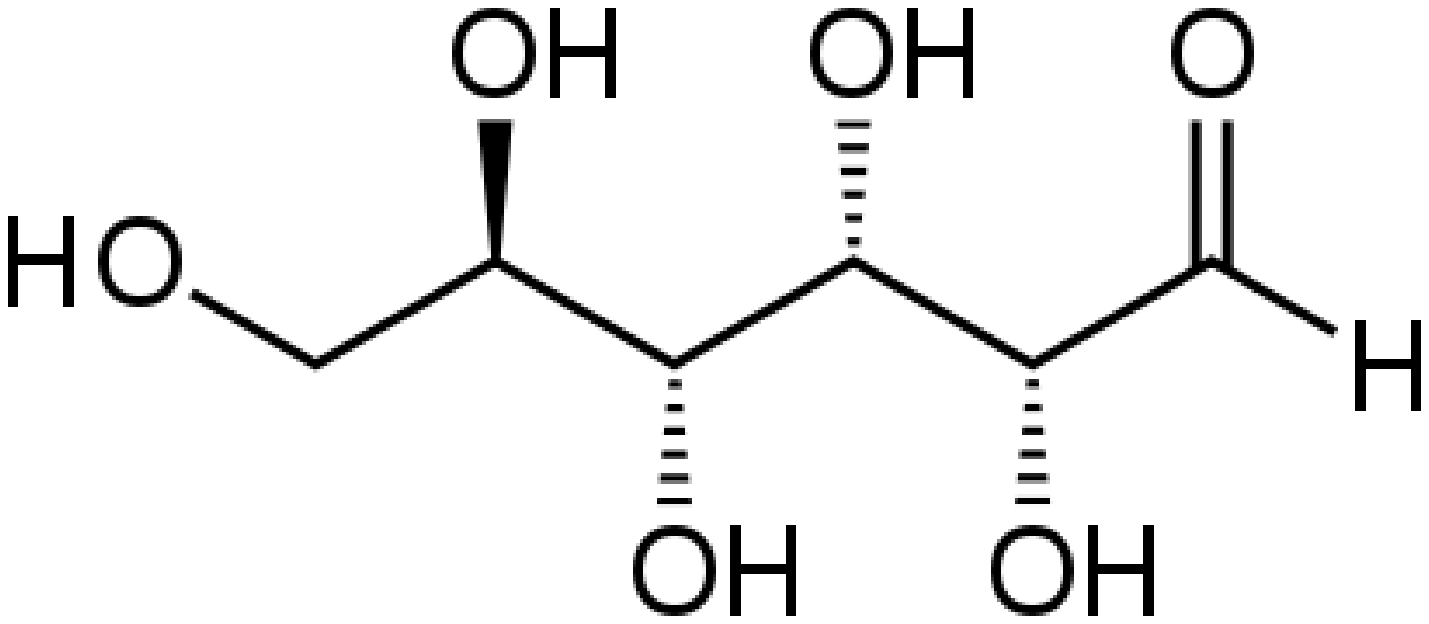} }}%
    \qquad
    \subfloat[3D representation]{{\includegraphics[width=5cm]{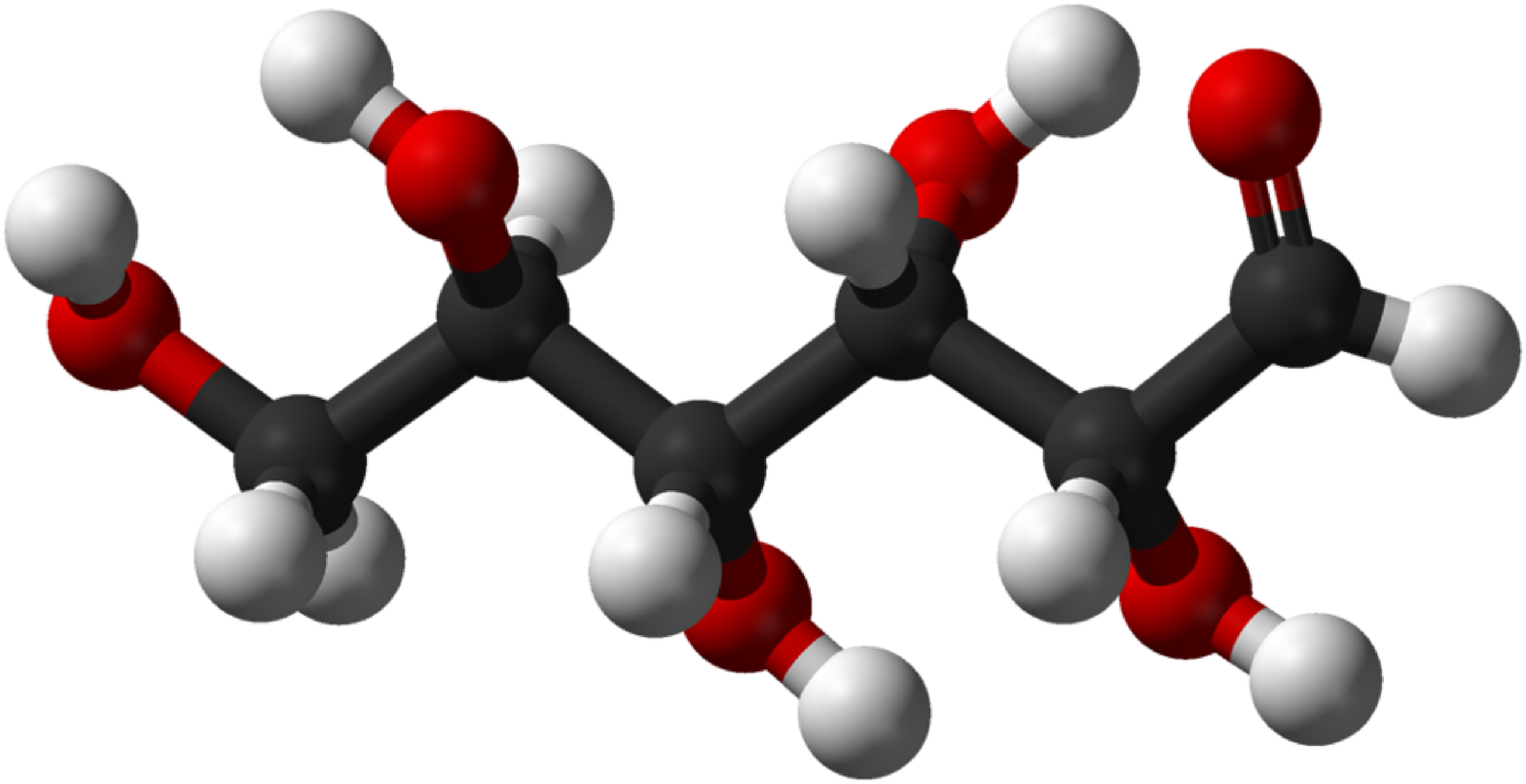} }}%
    \caption{D-glucose chain}%
    \label{glucose}%
\end{figure}

Usually, a mathematical language is restricted to certain types of symbols such as numbers, letters or operators. To have an all-encompassing language, we have to extend our notion of what a symbol is. For us, we will allow the symbols to be graphs, diagrams, geometrical shapes and even 3D objects (represented in 2D or not). This allows us to have a language which has a two dimensional notation, an inline notation and a three dimensional version. Diagrammatics notation is already being successfully used in physics, mathematics and biology. In Physics, Feynmann diagrams \cite{FeynmanA, FeynmanB} have proven to be a powerful tool and Penrose's tensor diagrammatic notation is proposing a new way to do tensor calculus \cite{Penrose}. In mathematics, diagrams are at the heart of graph theory and knot theory. As a notation, the language of commutative diagrams of category theory is now an essential tool in investigating highly abstract mathematical structures \cite{Mac Lane} and different diagrammatic notations have been used to further study category theory \cite{Selinger}. In Biology, textbooks use many types of diagrams to help the reader to understand. While these diagrams are not standardized, it could be fruitful to have a standard notation to enhance access to knowledge. With all the new capabilities of text editing and access 3D software, we are in a position to transition out of line-by-line mathematical notation and explore diagrams and 3D notations.

 In this paper, to help understand, we will often use the two dimensional visual language and make extensive use of colors to simplify the notation. Colors can be always be replaced by numbered subscript or superscript, but its use helps to make the notation lighter. The visual notation is good for communication and understanding, but the inline notation is very important since it is the basis of computation. It will quickly become clear that the inline version of a structure can be cumbersome, since even colors need to be replaced by a subscript or superscript notation. We will introduce the visual notation in parallel with the basic inline notation. In the construction of mathematical models, we will not restrict ourselves to a particular one, but will switch between them in a way which is beneficial for understanding.

\subsection{Basic Notation}

\subsubsection{Forms}
In our language, any collection of symbols will be called a \textit{form}. There are no restrictions on what a form can be. They can be letters, numbers, mathematical symbols, graphs, planar figures, 3D geometric shapes, people and collections of objects. We will only have to be careful when handling the special characters `$\rhd$' and `$=$'. This language could be seen as a calculus of forms where forms can be anything we decide to name or distinguish.

\subsubsection{Transformations}
As time passes, things or the forms we see are always changing. A seed becomes a sprout and the sprout becomes a tree. In our attempt to understand the world, we make observations such as ``this type of seed will give a certain type of tree with many different characteristics such as the shape of the leaves and cellular structure''. In the case of the seed-tree observations, we usually think of the seed as being the cause of the tree and not the other way around. As presented above, we will denote this are $\fbox{$Tree\rhd Seed$}$ and call such a structure a \textit{transformation}. Transformations are forms having potential to change other forms. The transformation $\fbox{$Tree\rhd Seed$}$ can be understood as representing the process where a Seed is replaced by a Tree, a seed transforming into a tree or a seed becoming a tree. The right-hand side of the symbol $\rhd$ will be called the \textit{cause} of the transformation and the left-hand side will be called the \textit{effect} of the transformation. It should be visualized as the potential for the word $Tree$ to slide down the directed triangle and replace the word $Seed$.

To observe the process of the becoming of a tree from a seed; there needs to be a seed present. This can be represented as $\fbox{$Tree\rhd Seed$}\rightarrow(Seed)$ and will be called a \textit{system}. The Seed at the right end of the arrow and between the parentheses `$(\,)$' indicates the presence of the Seed. This ordered set that is acted on by some transformations will be called the \textit{initial} form of the of these transformations. In the present case, we say that $(Seed)$ is the initial form of the transformation $\fbox{$Tree\rhd Seed$}$. The structure $\fbox{$Tree\rhd Seed$}$ can be seen as the potential of transformation of a seed into a tree and will only transform a seed in a tree when a seed is present. Eventually, we will have many elements in the initial form and unless said otherwise, we will consider that order in which the element appears as important. For example, $(A,B)$ is not considered to be the same as $(B,A)$.

We can represent the system $\fbox{$Tree\rhd Seed$}\rightarrow(Seed)$ in a diagram form as follows.

%Seed to Tree basic_a
\begin{equation}\scalebox{0.7} % Change this value to rescale the drawing.
{\begin{pspicture}(0,-0.87)(7.2,0.87)
\pscircle[linewidth=0.04,linestyle=dashed,dash=0.16cm 0.16cm,dimen=outer](6.33,0.0){0.87}
\usefont{T1}{ptm}{m}{n}
\rput(6.2767186,0.005){\large Seed}
\psline[linewidth=0.04cm,arrowsize=0.16cm 8.0,arrowlength=1.4,arrowinset=0.4]{->}(3.56,0.01)(5.12,0.01)
\psframe[linewidth=0.04,dimen=outer](3.32,0.79)(0.0,-0.67)
\usefont{T1}{ptm}{m}{n}
\rput(2.6367188,0.005){\large Seed}
\usefont{T1}{ptm}{m}{n}
\rput(0.7146875,0.005){\large Tree}
\rput{-90.0}(1.73,1.71){\pstriangle[linewidth=0.04,dimen=outer](1.72,-0.33)(0.72,0.64)}
\end{pspicture}
}
\end{equation}

\noindent Or it can be represented as the following diagram, since it is understood that the side to where the triangle is pointing is the form that will be replaced.

%%Seed to Tree basic_b
\begin{equation}\scalebox{0.7} % Change this value to rescale the drawing.
{\begin{pspicture}(0,-0.87)(7.12,0.87)
\pscircle[linewidth=0.04,linestyle=dashed,dash=0.16cm 0.16cm,dimen=outer](0.87,0.0){0.87}
\usefont{T1}{ptm}{m}{n}
\rput(0.81671876,0.005){\large Seed}
\psline[linewidth=0.04cm,arrowsize=0.16cm 8.0,arrowlength=1.4,arrowinset=0.4]{->}(3.66,-0.01)(2.1,-0.01)
\psframe[linewidth=0.04,dimen=outer](7.12,0.75)(3.8,-0.71)
\usefont{T1}{ptm}{m}{n}
\rput(6.436719,-0.035){\large Seed}
\usefont{T1}{ptm}{m}{n}
\rput(4.5146875,-0.035){\large Tree}
\rput{-90.0}(5.57,5.47){\pstriangle[linewidth=0.04,dimen=outer](5.52,-0.37)(0.72,0.64)}
\end{pspicture}
}
\end{equation}

In the diagram notation, the initial form will be denoted by a shape drawn with dash lines. We will use mostly shapes such as dashed circles, ellipses and rectangles.

\subsubsection{Reduction}
The act of applying the transformation $\fbox{$Tree\rhd Seed$}$ to the form $(Seed)$ is said to be a \textit{reduction} of the system $\fbox{$Tree\rhd Seed$}\rightarrow (Seed)$ to the form $(Tree)$. This reduction is indicated by a double arrow `$\Rightarrow$' and is written as
$$\fbox{$Tree\rhd Seed$}\rightarrow (Seed)\Rightarrow (Tree).$$

\noindent This expression is read as the transformation $\fbox{$Tree\rhd Seed$}$ applied to the form $(Seed)$ and reduces to the form $(Tree)$. It is important to note that the transformation is applied only once and then disappears after the reduction. When multiple transformations are used in a system, a reduction can refer to the final form where all transformations have been applied or all the steps leading to that final form are complete.

Seeds come in many types, so we can extend our language further and apply the transformation such as  $\fbox{$Pine\rhd Seed$}$ to multiple seeds. There are two interesting ways to do this. The first is by successively applying certain transformations to an ordered set containing multiple seeds. The second is by applying one transformation at a time without any definite order. The first is said to be applied in \textit{series} and the second in \textit{parallel}.

The number of transformations which can be applied on an initial form can be finite or infinite. In the theory of term rewriting systems, a main focus is to have systems which terminate after a finite number of reductions. For us, since we are mainly concerned about creating mathematical models, we are interested in all the steps of the reductions and this whether it terminates or not. An interesting question for us to ask is if a system has a period or a cycle. In biology, an example of periodic system is the cell cycle.

\subsubsection{Series of transformations}
A series of three transformation applied to an ordered set of three seeds is written as

$$\fbox{$Maple\rhd Seed$}\rightarrow\fbox{$Oak\rhd Seed$}\rightarrow\fbox{$Pine\rhd Seed$}\rightarrow (Seed, Seed, Seed)$$

The reduction sequence of this system is be done in three steps and is denoted as follows. The transformation closest to the initial form is applied first followed by the second closest and so on.

$$\begin{array}{c}
\fbox{$Maple\rhd Seed$}\rightarrow\fbox{$Oak\rhd Seed$}\rightarrow\fbox{$Pine\rhd Seed$}\rightarrow(Seed, Seed, Seed)\\
\Downarrow\\
\fbox{$Maple\rhd Seed$}\rightarrow\fbox{$Oak\rhd Seed$}\rightarrow(Seed, Pine, Seed)\\
\Downarrow\\
\fbox{$Maple\rhd Seed$}\rightarrow(Oak, Pine, Seed)\\
\Downarrow\\
(Oak, Pine, Maple)\\
\end{array}$$

Since all the elements of the initial form are the same, the transformation $\fbox{$Pine\rhd Seed$}$ can replace any choice of Seed. If one wants to differentiate between the Seeds, we could rename them as in the following example.

$$\fbox{$Maple\rhd Seed3$}\rightarrow\fbox{$Oak\rhd Seed2$}\rightarrow\fbox{$Pine\rhd Seed1$}\rightarrow(Seed3, Seed1, Seed2)$$

We now introduce superscript to the transformations. If we have a series of the same transformation being applied on an initial form, we will write this in a more compact notation by writing as a superscript an integer followed by the letter `$S$' which indicates that the transformation is applied in series. For example,
$$\fbox{$Tree\rhd Seed$}\rightarrow\fbox{$Tree\rhd Seed$}\rightarrow\fbox{$Tree\rhd Seed$}\rightarrow(Seed, Seed, Seed)$$
\noindent can be written as
$$\fbox{$Tree\rhd Seed$}^{\,3S}\rightarrow(Seed, Seed, Seed).$$
\noindent Another example with two different types of transformation is to write
$$\fbox{$Oak\rhd Seed$}^{\,7S}\rightarrow\fbox{$Pine\rhd Seed$}^{\,10S}\rightarrow(Seed, Seed, Seed, Seed, Seed)$$
\noindent to represent 10 successive applications of the transformation $\fbox{$Pine\rhd Seed$}$ followed by a series of 7 applications of $\fbox{$Oak\rhd Seed$}$.

It is important to note that when no superscript is written, it is assumed to mean a superscript of $1S$. For example, $\fbox{$Tree\rhd Seed$}^{\,1S}$ is equivalent to $\fbox{$Tree\rhd Seed$}$.

We can also use the infinity symbol `$\infty$' to denote that the transformation is continuously applied when needed. Thus the system
$$\fbox{$Oak\rhd Seed$}^{\,\infty S}\rightarrow(Seed, Seed, Seed, Seed, Seed),$$
\noindent reduces to
$$\fbox{$Oak\rhd Seed$}^{\,\infty S}\rightarrow(Oak, Oak, Oak, Oak, Oak).$$

If we want to replace all that can be replaced, we use the sharp symbol `$\sharp$' to denote that the transformation is applied until it reaches a step where it cannot replace another element of the initial form, then the transformation disappears. In other words, it replaces all it can and then disappears. For example, the system

$$\fbox{$Oak\rhd Seed$}^{\,\sharp S}\rightarrow(Seed, Seed, Seed, Seed, Seed),$$
\noindent reduces to
$$(Oak, Oak, Oak, Oak, Oak).$$

\subsubsection{Parallel transformations}
 When transformations are applied  in parallel, we will denote this by grouping them in box brackets `$[\,]$'. For example, two transformations applied on an initial form containing three seeds is written as

$$[\,\fbox{$Pine\rhd Seed$},\fbox{$Oak\rhd Seed$}\,]\rightarrow(Seed).$$

When parallel transformations are applied to an ordered set, any transformation that can be applied can be chosen to be applied first. After it has been applied, it disappears and another one is chosen. An example of reduction is
\begin{eqnarray*}
&[\,\fbox{$Pine\rhd Seed$},\fbox{$Oak\rhd Seed$}\,]\rightarrow(Seed)&\\
&\Downarrow&\\
&[\,\fbox{$Oak\rhd Seed$}\,]\rightarrow(Pine)&
\end{eqnarray*}

But since the transformations were in parallel, we could also have
\begin{eqnarray*}
&[\,\fbox{$Pine\rhd Seed$},\fbox{$Oak\rhd Seed$}\,]\rightarrow(Seed)&\\
&\Downarrow&\\
&[\,\fbox{$Pine\rhd Seed$}\,]\rightarrow(Oak)&
\end{eqnarray*}

We now introduce the diagram notation. The system $$[\,\fbox{$Pine\rhd Seed$},\fbox{$Oak\rhd Seed$},\fbox{$Maple\rhd Seed$}\,]\rightarrow(Seed, Seed, Seed),$$ which contains three parallel transformations can also be represented as the following diagram.

%seed to tree1 Final_a
\begin{equation}\label{seed1}\scalebox{0.7} % Change this value to rescale the drawing.
{\begin{pspicture}(0,-2.62)(9.08,2.62)
\pscircle[linewidth=0.04,linestyle=dashed,dash=0.16cm 0.16cm,dimen=outer](7.49,0.07){1.59}
\usefont{T1}{ptm}{m}{n}
\rput(7.2967186,1.155){\large Seed}
\usefont{T1}{ptm}{m}{n}
\rput(7.016719,-0.565){\large Seed}
\usefont{T1}{ptm}{m}{n}
\rput(8.316719,0.155){\large Seed}
\psline[linewidth=0.04cm,arrowsize=0.16cm 8.0,arrowlength=1.4,arrowinset=0.4]{->}(3.74,-0.02)(5.76,-0.02)
\psframe[linewidth=0.04,dimen=outer](3.56,2.62)(0.24,1.16)
\usefont{T1}{ptm}{m}{n}
\rput(2.8167188,1.835){\large Seed}
\usefont{T1}{ptm}{m}{n}
\rput(0.9278125,1.835){\large Pine}
\psframe[linewidth=0.04,dimen=outer](3.56,0.72)(0.24,-0.74)
\usefont{T1}{ptm}{m}{n}
\rput(2.7967188,-0.065){\large Seed}
\usefont{T1}{ptm}{m}{n}
\rput(0.9034375,-0.045){\large Oak}
\psframe[linewidth=0.04,dimen=outer](3.56,-1.16)(0.0,-2.62)
\usefont{T1}{ptm}{m}{n}
\rput(2.7767189,-1.865){\large Seed}
\usefont{T1}{ptm}{m}{n}
\rput(0.7678125,-1.885){\large Maple}
\rput{-90.0}(0.02,3.74){\pstriangle[linewidth=0.04,dimen=outer](1.88,1.54)(0.72,0.64)}
\rput{-90.0}(1.88,1.88){\pstriangle[linewidth=0.04,dimen=outer](1.88,-0.32)(0.72,0.64)}
\rput{-90.0}(3.72,0.0){\pstriangle[linewidth=0.04,dimen=outer](1.86,-2.18)(0.72,0.64)}
\psline[linewidth=0.04cm,arrowsize=0.16cm 8.0,arrowlength=1.4,arrowinset=0.4]{->}(3.8,-1.92)(5.84,-0.92)
\psline[linewidth=0.04cm,arrowsize=0.16cm 8.0,arrowlength=1.4,arrowinset=0.4]{->}(3.72,1.88)(5.76,0.88)
\end{pspicture}
}\end{equation}

\noindent It is also possible to reduce the number of arrows in the diagram by grouping the transformation under a bracket `$]$' as seen below.

%seed to tree1 Final_b
\begin{equation}\scalebox{0.7} % Change this value to rescale the drawing.
{
\begin{pspicture}(0,-3.02)(9.02,3.02)
\pscircle[linewidth=0.04,linestyle=dashed,dash=0.16cm 0.16cm,dimen=outer](7.43,0.07){1.59}
\usefont{T1}{ptm}{m}{n}
\rput(7.2367187,1.155){\large Seed}
\usefont{T1}{ptm}{m}{n}
\rput(6.956719,-0.565){\large Seed}
\usefont{T1}{ptm}{m}{n}
\rput(8.256719,0.155){\large Seed}
\psline[linewidth=0.04cm,arrowsize=0.16cm 8.0,arrowlength=1.4,arrowinset=0.4]{->}(4.14,-0.02)(5.7,-0.02)
\psframe[linewidth=0.04,dimen=outer](3.5,2.62)(0.18,1.16)
\usefont{T1}{ptm}{m}{n}
\rput(2.8167188,1.835){\large Seed}
\usefont{T1}{ptm}{m}{n}
\rput(0.8678125,1.835){\large Pine}
\psframe[linewidth=0.04,dimen=outer](3.5,0.72)(0.18,-0.74)
\usefont{T1}{ptm}{m}{n}
\rput(2.7767189,-0.085){\large Seed}
\usefont{T1}{ptm}{m}{n}
\rput(0.8234375,-0.065){\large Oak}
\psframe[linewidth=0.04,dimen=outer](3.5,-1.16)(0.0,-2.62)
\usefont{T1}{ptm}{m}{n}
\rput(2.7567186,-1.925){\large Seed}
\usefont{T1}{ptm}{m}{n}
\rput(0.7878125,-1.925){\large Maple}
\rput{-90.0}(0.08,3.72){\pstriangle[linewidth=0.04,dimen=outer](1.9,1.5)(0.72,0.64)}
\rput{-90.0}(1.98,1.82){\pstriangle[linewidth=0.04,dimen=outer](1.9,-0.4)(0.72,0.64)}
\rput{-90.0}(3.8,-0.04){\pstriangle[linewidth=0.04,dimen=outer](1.88,-2.24)(0.72,0.64)}
\psline[linewidth=0.04](3.3,3.0)(3.86,3.0)(3.9,-3.0)(3.32,-3.0)
\end{pspicture}
}\end{equation}

Similarly to series of transformations, there is a superscript notation if the same transformation is applied multiple times on an initial form. This is done by writing as a superscript an integer followed by the letter `$P$' which indicates that the transformation is applied in parallel. For example,
$$[\, \fbox{$Tree\rhd Seed$},\fbox{$Tree\rhd Seed$},\fbox{$Tree\rhd Seed$}\,] \rightarrow(Seed, Seed, Seed)$$
\noindent can be written as
$$[\,\fbox{$Tree\rhd Seed$}^{\,3P}\,]\rightarrow(Seed, Seed, Seed).$$
\noindent Another example with two different types of transformations, is to write
$$[\,\fbox{$Oak\rhd Seed$}^{\,2P},\fbox{$Pine\rhd Seed$}^{\,3P}\,]\rightarrow(Seed, Seed, Seed, Seed, Seed)$$
\noindent to represent the following diagram.

%Parallel superscripts
\begin{equation}\scalebox{0.7} % Change this value to rescale the drawing.
{
\begin{pspicture}(0,-4.38)(9.08,4.38)
\pscircle[linewidth=0.04,linestyle=dashed,dash=0.16cm 0.16cm,dimen=outer](7.49,-0.01){1.59}
\usefont{T1}{ptm}{m}{n}
\rput(7.6367188,0.955){\large Seed}
\usefont{T1}{ptm}{m}{n}
\rput(7.936719,-0.805){\large Seed}
\usefont{T1}{ptm}{m}{n}
\rput(8.296719,0.115){\large Seed}
\psline[linewidth=0.04cm,arrowsize=0.16cm 8.0,arrowlength=1.4,arrowinset=0.4]{->}(3.88,1.78)(5.62,1.28)
\psframe[linewidth=0.04,dimen=outer](4.56,4.38)(1.24,2.92)
\usefont{T1}{ptm}{m}{n}
\rput(3.8767188,3.595){\large Seed}
\usefont{T1}{ptm}{m}{n}
\rput(1.9278125,3.595){\large Pine}
\psframe[linewidth=0.04,dimen=outer](4.58,-2.92)(1.26,-4.38)
\usefont{T1}{ptm}{m}{n}
\rput(3.8567188,-3.725){\large Seed}
\usefont{T1}{ptm}{m}{n}
\rput(1.9034375,-3.705){\large Oak}
\rput{-90.0}(-0.62,6.54){\pstriangle[linewidth=0.04,dimen=outer](2.96,3.26)(0.72,0.64)}
\rput{-90.0}(6.7,-0.74){\pstriangle[linewidth=0.04,dimen=outer](2.98,-4.04)(0.72,0.64)}
\psframe[linewidth=0.04,dimen=outer](3.72,-1.1)(0.4,-2.56)
\usefont{T1}{ptm}{m}{n}
\rput(2.9967186,-1.905){\large Seed}
\usefont{T1}{ptm}{m}{n}
\rput(1.0434375,-1.885){\large Oak}
\rput{-90.0}(4.02,0.22){\pstriangle[linewidth=0.04,dimen=outer](2.12,-2.22)(0.72,0.64)}
\psframe[linewidth=0.04,dimen=outer](3.74,2.56)(0.42,1.1)
\usefont{T1}{ptm}{m}{n}
\rput(3.0567188,1.775){\large Seed}
\usefont{T1}{ptm}{m}{n}
\rput(1.1078125,1.775){\large Pine}
\rput{-90.0}(0.38,3.9){\pstriangle[linewidth=0.04,dimen=outer](2.14,1.44)(0.72,0.64)}
\psframe[linewidth=0.04,dimen=outer](3.32,0.7)(0.0,-0.76)
\usefont{T1}{ptm}{m}{n}
\rput(2.6367188,-0.085){\large Seed}
\usefont{T1}{ptm}{m}{n}
\rput(0.6878125,-0.085){\large Pine}
\rput{-90.0}(1.82,1.62){\pstriangle[linewidth=0.04,dimen=outer](1.72,-0.42)(0.72,0.64)}
\psline[linewidth=0.04cm,arrowsize=0.16cm 8.0,arrowlength=1.4,arrowinset=0.4]{->}(3.9,-1.86)(5.64,-1.36)
\psline[linewidth=0.04cm,arrowsize=0.16cm 8.0,arrowlength=1.4,arrowinset=0.4]{->}(4.72,3.56)(6.3,2.02)
\psline[linewidth=0.04cm,arrowsize=0.16cm 8.0,arrowlength=1.4,arrowinset=0.4]{->}(3.54,-0.02)(5.4,0.0)
\psline[linewidth=0.04cm,arrowsize=0.16cm 8.0,arrowlength=1.4,arrowinset=0.4]{->}(4.78,-3.68)(6.36,-2.14)
\usefont{T1}{ptm}{m}{n}
\rput(6.6967187,0.355){\large Seed}
\usefont{T1}{ptm}{m}{n}
\rput(6.7367187,-0.485){\large Seed}
\end{pspicture}
}\end{equation}

In a similar way to series of transformations, we can also use the infinity symbol `$\infty$' to denote that a transformation is continuously applied and the sharp symbol `$\sharp$' to indicate that the transformation replaces all it can from the initial form and then disappears.

\subsubsection{Arising and Dissolution}

This language permits creation of new objects in an ordered set by using a transformation in which the right or left hand-side of the transformation is empty. An example is given by the following transformation of making rabbits appear in a grass and flower field. This is an interesting property, since this gives us a mathematical way to add new objects to a set of forms.

$$\begin{array}{c}
\fbox{$rabbit\,\, \rhd \hspace{1cm}$}^{\,3S}\rightarrow  (grass, flowers) \\
\Downarrow\\
 \fbox{$rabbit\,\, \rhd \hspace{1cm}$}^{\,2S}(grass, flowers, rabbit)\\
 \Downarrow\\
 \fbox{$rabbit\,\, \rhd \hspace{1cm}$}(grass, rabbit, flowers, rabbit)\\
  \Downarrow\\
 (grass, rabbit, flowers, rabbit, rabbit)
\end{array}$$

Similarly, as seen in the following example we can dissolve a form.
$$\begin{array}{c}
\fbox{$\hspace{1cm}  \rhd \,\, cup$}^{\,2S}\rightarrow  (cup, spoon, cup) \\
\Downarrow\\
 \fbox{$\hspace{1cm}  \rhd \,\, cup$}\rightarrow (cup, spoon)\\
 \Downarrow\\
 (spoon)
\end{array}$$

\subsubsection{Position Pairing in Transformations}

Until now in our transformations, there was only one term on each side of the directed triangle. We now allow multiple terms on each side of the triangle. Let's revisit our seed example.

A seed turns into a tree if there is a soil, water and sun. After the appearance of the tree the soil is still there and the sun is still present.  We could view the sun and the soil to be the supporting conditions. After a transformation is applied, the supporting conditions will stay there and will not change. In practice, we can say that a very small amount of soil and energy from the sun was used, but in our present model, we will consider that it is a negligible amount. To do this we will identify together elements on each side of the triangle. This is done by respecting the position of the terms on both sides of the directed triangle. To indicate that a form $A$ in a certain position has been replace by another form $B$, we write the $A$ and $B$ at the same position on each side of the directed triangle. If a form is unaffected by the transformation, the form will appear at the same place on both sides. Another way to view this is to think that $B$ has been paired with $A$. This is similar to functions where each element of the domain is paired with an element of the range of the function.

Here is an example of our seed turning into a tree with the conditions of sun and soil.

%conditions_no water
\begin{equation}\scalebox{0.7} % Change this value to rescale the drawing.
{
\begin{pspicture}(0,-1.8)(13.22,1.8)
\pscircle[linewidth=0.04,linestyle=dashed,dash=0.16cm 0.16cm,dimen=outer](11.63,0.09){1.59}
\psline[linewidth=0.04cm,arrowsize=0.16cm 8.0,arrowlength=1.4,arrowinset=0.4]{->}(8.2,0.14)(9.76,0.14)
\psframe[linewidth=0.04,dimen=outer](7.8,1.8)(0.0,-1.8)
\usefont{T1}{ptm}{m}{n}
\rput(6.1728125,-0.925){\large Soil}
\usefont{T1}{ptm}{m}{n}
\rput(5.2167187,-0.245){\large  Seed}
\usefont{T1}{ptm}{m}{n}
\rput(5.8653126,1.035){\large Sun}
\rput{-90.0}(3.62,3.82){\pstriangle[linewidth=0.04,dimen=outer](3.72,-0.36)(1.0,0.92)}
\usefont{T1}{ptm}{m}{n}
\rput(2.0728126,-0.905){\large Soil}
\usefont{T1}{ptm}{m}{n}
\rput(1.1146874,-0.225){\large Tree}
\usefont{T1}{ptm}{m}{n}
\rput(1.7653126,1.055){\large Sun}
\usefont{T1}{ptm}{m}{n}
\rput(12.452812,0.595){\large Soil}
\usefont{T1}{ptm}{m}{n}
\rput(10.716719,0.575){\large  Seed}
\usefont{T1}{ptm}{m}{n}
\rput(11.685312,-0.625){\large Sun}
\end{pspicture}
}\end{equation}

\noindent This will reduce to

\begin{equation}\scalebox{0.7} % Change this value to rescale the drawing.
{
\begin{pspicture}(0,-1.59)(3.18,1.59)
\pscircle[linewidth=0.04,linestyle=dashed,dash=0.16cm 0.16cm,dimen=outer](1.59,0.0){1.59}
\usefont{T1}{ptm}{m}{n}
\rput(2.4128125,0.505){\large Soil}
\usefont{T1}{ptm}{m}{n}
\rput(0.6546875,0.485){\large  Tree}
\usefont{T1}{ptm}{m}{n}
\rput(1.6453125,-0.715){\large Sun}
\end{pspicture}
}
\end{equation}

\noindent Notice that the terms of initial form are not in the same configuration as seen in the transformation. The position is only important in the transformation and is meant to indicate which forms are paired together. When a transformation is applied on the initial form, the elements are replaced at the place they are found in the initial form as seen above.

We now look at another example where a seed and water becomes a tree based on the conditions of the sun and soil. Notice that the water disappeared and that the seed was replaced by a tree. Note that this system makes an implicit use of the dissolution transformations when the water is removed.

%conditions_with water
\begin{equation}\scalebox{0.7} % Change this value to rescale the drawing.
{
\begin{pspicture}(0,-1.8)(13.54,1.8)
\pscircle[linewidth=0.04,linestyle=dashed,dash=0.16cm 0.16cm,dimen=outer](11.95,0.09){1.59}
\psline[linewidth=0.04cm,arrowsize=0.16cm 8.0,arrowlength=1.4,arrowinset=0.4]{->}(8.52,0.14)(10.08,0.14)
\psframe[linewidth=0.04,dimen=outer](8.12,1.8)(0.0,-1.8)
\usefont{T1}{ptm}{m}{n}
\rput(6.1528125,-1.165){\large Soil}
\usefont{T1}{ptm}{m}{n}
\rput(5.2967186,-0.105){\large  Seed}
\usefont{T1}{ptm}{m}{n}
\rput(6.1653123,1.235){\large Sun}
\rput{-90.0}(4.04,4.28){\pstriangle[linewidth=0.04,dimen=outer](4.16,-0.34)(1.0,0.92)}
\usefont{T1}{ptm}{m}{n}
\rput(11.232813,-0.145){\large Soil}
\usefont{T1}{ptm}{m}{n}
\rput(11.816719,0.855){\large  Seed}
\usefont{T1}{ptm}{m}{n}
\rput(12.765312,0.155){\large Sun}
\usefont{T1}{ptm}{m}{n}
\rput(7.1917186,-0.465){\large Water}
\usefont{T1}{ptm}{m}{n}
\rput(12.151719,-0.785){\large Water}
\usefont{T1}{ptm}{m}{n}
\rput(1.6928124,-1.165){\large Soil}
\usefont{T1}{ptm}{m}{n}
\rput(0.9546875,-0.085){\large Tree}
\usefont{T1}{ptm}{m}{n}
\rput(1.7053125,1.235){\large Sun}
\end{pspicture}
}\end{equation}

In the next example, the water disappeared, the seed was replaced by a tree and an apple appears in some position.

\begin{equation}\scalebox{0.7} % Change this value to rescale the drawing.
{
\begin{pspicture}(0,-1.8)(13.54,1.8)
\pscircle[linewidth=0.04,linestyle=dashed,dash=0.16cm 0.16cm,dimen=outer](11.95,0.09){1.59}
\psline[linewidth=0.04cm,arrowsize=0.16cm 8.0,arrowlength=1.4,arrowinset=0.4]{->}(8.52,0.14)(10.08,0.14)
\psframe[linewidth=0.04,dimen=outer](8.12,1.8)(0.0,-1.8)
\usefont{T1}{ptm}{m}{n}
\rput(6.1528125,-1.165){\large Soil}
\usefont{T1}{ptm}{m}{n}
\rput(5.2967186,-0.105){\large  Seed}
\usefont{T1}{ptm}{m}{n}
\rput(6.1653123,1.235){\large Sun}
\rput{-90.0}(4.04,4.28){\pstriangle[linewidth=0.04,dimen=outer](4.16,-0.34)(1.0,0.92)}
\usefont{T1}{ptm}{m}{n}
\rput(11.232813,-0.145){\large Soil}
\usefont{T1}{ptm}{m}{n}
\rput(11.816719,0.855){\large  Seed}
\usefont{T1}{ptm}{m}{n}
\rput(12.765312,0.155){\large Sun}
\usefont{T1}{ptm}{m}{n}
\rput(7.1917186,-0.465){\large Water}
\usefont{T1}{ptm}{m}{n}
\rput(12.151719,-0.785){\large Water}
\usefont{T1}{ptm}{m}{n}
\rput(1.6928124,-1.165){\large Soil}
\usefont{T1}{ptm}{m}{n}
\rput(0.9546875,-0.085){\large Tree}
\usefont{T1}{ptm}{m}{n}
\rput(1.7053125,1.235){\large Sun}
\usefont{T1}{ptm}{m}{n}
\rput(2.4896874,0.495){\large Apple}
\end{pspicture}
}\end{equation}

The next example shows that the replacing object can be positioned at another position than the replaced object. Here the triangle contracts to a point in the center of the triangle.

%Contracting triangle
\begin{equation}\scalebox{0.5} % Change this value to rescale the drawing.
{
\begin{pspicture}(0,-5.17)(14.72,5.17)
\pscircle[linewidth=0.04,linestyle=dashed,dash=0.16cm 0.16cm,dimen=outer](12.81,3.18){1.91}
\psline[linewidth=0.04cm,arrowsize=0.16cm 8.0,arrowlength=1.4,arrowinset=0.4]{->}(8.4,3.37)(9.96,3.37)
\psframe[linewidth=0.04,dimen=outer](8.12,5.17)(0.0,1.57)
\rput{-90.0}(0.57,7.47){\pstriangle[linewidth=0.04,dimen=outer](4.02,2.99)(1.0,0.92)}
\pstriangle[linewidth=0.04,dimen=outer](12.85,2.65)(1.7,1.44)
\psdots[dotsize=0.4](12.0,2.65)
\psdots[dotsize=0.4](13.68,2.65)
\psdots[dotsize=0.4](12.84,4.03)
\pstriangle[linewidth=0.04,dimen=outer](6.17,2.65)(1.7,1.44)
\psdots[dotsize=0.4](5.32,2.65)
\psdots[dotsize=0.4](7.0,2.65)
\psdots[dotsize=0.4](6.16,4.03)
\psdots[dotsize=0.4](1.76,3.19)
\psline[linewidth=0.04cm,arrowsize=0.16cm 4.0,arrowlength=1.4,arrowinset=0.4,doubleline=true,doublesep=0.12,doublecolor=white]{->}(9.14,0.99)(9.14,-0.93)
\psdots[dotsize=0.4](9.16,-3.25)
\pscircle[linewidth=0.04,linestyle=dashed,dash=0.16cm 0.16cm,dimen=outer](9.17,-3.26){1.91}
\end{pspicture}
}
\end{equation}

\subsubsection{Preserving Connections}\label{Preserving Connections}

In science we are often interested in objects which have connections with one another. For example, atoms are connected by bonds to create molecules and networks of cities are sometimes represented as graphs of nodes connected by edges. We will later see that this will be very useful when we will be representing biological process such a DNA replication and representing a peptide bond.

Let's now look at a simplified network of the Canadian postal service with the cities of Vancouver, Calgary, Winnipeg, Montreal, Halifax and Toronto. In the United States, the simplified postal network covers the cites of New York, San Francisco, Denver, Washington, Los Angeles and Miami. After an agreement between both postal services; it is decided that both networks should be connected between the cities of Toronto and New York. Using our notation, this can be represented in the following transformation.

%network
\begin{equation}\label{network}\scalebox{0.6} % Change this value to rescale the drawing.
{
\begin{pspicture}(0,-3.7027662)(21.728354,3.7027662)
\definecolor{color1570}{rgb}{0.2,0.2,0.2}
\usefont{T1}{ptm}{m}{n}
\rput(10.672812,2.1719077){\large Vancouver}
\usefont{T1}{ptm}{m}{n}
\rput(13.100781,2.0519075){\large Calgary}
\usefont{T1}{ptm}{m}{n}
\rput(17.98297,1.4919076){\large Montreal}
\usefont{T1}{ptm}{m}{n}
\rput(17.285625,0.67190754){\large Toronto}
\usefont{T1}{ptm}{m}{n}
\rput(19.923594,1.9119076){\large Halifax}
\usefont{T1}{ptm}{m}{n}
\rput(18.546093,-1.1280924){\large Washington}
\usefont{T1}{ptm}{m}{n}
\rput(18.824219,-0.46809244){\large New York}
\usefont{T1}{ptm}{m}{n}
\rput(15.100938,-1.1480925){\large Denver}
\usefont{T1}{ptm}{m}{n}
\rput(18.282188,-2.5880926){\large Miami}
\usefont{T1}{ptm}{m}{n}
\rput(12.593437,-1.9680924){\large Los Angeles}
\usefont{T1}{ptm}{m}{n}
\rput(12.268594,-1.0280925){\large San Francisco}
\usefont{T1}{ptm}{m}{n}
\rput(15.493125,2.0119076){\large Winnipeg}
\psline[linewidth=0.04cm](11.72,2.0969076)(12.16,2.0369074)
\psline[linewidth=0.04cm](16.52,1.8969076)(17.0,1.6569076)
\psline[linewidth=0.04cm](17.84,1.2569076)(17.34,0.91690755)
\psline[linewidth=0.04cm](16.98,0.93690753)(16.4,1.6569076)
\psline[linewidth=0.04cm](18.9,1.4969076)(19.6,1.6769075)
\psline[linewidth=0.04cm](18.58,-0.66309243)(18.42,-0.9230924)
\psline[linewidth=0.04cm](18.36,-1.3830924)(18.22,-2.2830925)
\psline[linewidth=0.04cm](18.06,-2.2230926)(15.94,-1.3830924)
\psline[linewidth=0.04cm](15.98,-1.1630925)(17.26,-1.1830925)
\psline[linewidth=0.04cm](13.72,-1.0830925)(14.18,-1.1030924)
\psline[linewidth=0.04cm](12.38,-1.3230925)(12.52,-1.7230924)
\psline[linewidth=0.04cm](12.62,-1.7230924)(14.2,-1.4230925)
\psbezier[linewidth=0.04,linestyle=dashed,dash=0.16cm 0.16cm](9.66,2.9369075)(8.956538,2.226174)(8.985538,-2.2834187)(9.7,-2.9830925)(10.414462,-3.6827662)(20.297916,-3.5351872)(21.0,-2.8230925)(21.702084,-2.1109977)(21.708353,2.271049)(21.0,2.9769075)(20.291647,3.6827662)(10.3634615,3.647641)(9.66,2.9369075)
\usefont{T1}{ptm}{m}{n}
\rput(5.045625,0.5719075){\large  Toronto}
\usefont{T1}{ptm}{m}{n}
\rput(5.9842186,-0.42809245){\large  New York}
\psline[linewidth=0.04cm](13.88,1.9569075)(14.44,1.9569075)
\usefont{T1}{ptm}{m}{n}
\rput(0.965625,0.61190754){\large  Toronto}
\usefont{T1}{ptm}{m}{n}
\rput(1.9442188,-0.42809245){\large  New York}
\psline[linewidth=0.04cm,arrowsize=0.32cm 8.0,arrowlength=1.4,arrowinset=0.4]{->}(7.46,0.076907545)(8.92,0.076907545)
\psframe[linewidth=0.04,dimen=outer](7.28,1.2969075)(0.0,-1.2430924)
\psline[linewidth=0.04cm,linecolor=color1570](1.24,0.37690756)(2.22,-0.16309245)
\rput{-90.0}(3.4830925,3.8469076){\pstriangle[linewidth=0.04,dimen=outer](3.665,-0.21309246)(0.93,0.79)}
\end{pspicture}
}\end{equation}

Most of the time, to represent connections between forms, we will use lines or directed arrows. When we need to modify a form connected to other forms, we will consider that transformations will preserve connections. More precisely, we say that:

\textit{If form $A$ is replaced by form $B$, all the connections that node $A$ has are transferred to $B$ and $B$ keeps its previous connections.}

To illustrate this, we give the following two simple examples where $A-C$ is interpreted as node $A$ is connected to node $B$.

$$\begin{array}{c}
\fbox{$B \rhd A$}\rightarrow  (A-C) \\
\Downarrow\\
(B-C)
\end{array}$$
and
$$\begin{array}{c}
\fbox{$B \rhd A$}\rightarrow  (D-A-C) \\
\Downarrow\\
(D-B-C)
\end{array}$$

%This means that when a form $A$ is replaced by another form $B$ the connection of form $A$ are transferred to the form $B$.

If we add connections to the contracting triangle we have seen above, we have the following reduction where the connections of the black circle nodes are transferred to the empty square and empty circle nodes.

%Contracting triangle connections new
\begin{equation}\scalebox{0.5} % Change this value to rescale the drawing.
{
\begin{pspicture}(0,-6.46)(15.62,6.46)
\pscircle[linewidth=0.04,linestyle=dashed,dash=0.16cm 0.16cm,dimen=outer](13.09,3.93){2.53}
\psline[linewidth=0.04cm,arrowsize=0.16cm 8.0,arrowlength=1.4,arrowinset=0.4]{->}(8.66,4.1)(10.22,4.1)
\psframe[linewidth=0.04,dimen=outer](8.38,5.9)(0.0,2.3)
\rput{-90.0}(0.02,8.34){\pstriangle[linewidth=0.04,dimen=outer](4.18,3.7)(1.0,0.92)}
\pstriangle[linewidth=0.04,dimen=outer](13.11,3.38)(1.7,1.44)
\psdots[dotsize=0.4](12.26,3.38)
\psdots[dotsize=0.4](13.94,3.38)
\psdots[dotsize=0.4](13.1,4.76)
\psline[linewidth=0.04cm](13.1,4.76)(13.08,5.46)
\psline[linewidth=0.04cm](13.94,3.36)(14.5,2.84)
\psline[linewidth=0.04cm](12.24,3.34)(11.74,2.82)
\pstriangle[linewidth=0.04,dimen=outer](6.43,3.38)(1.7,1.44)
\psdots[dotsize=0.4](5.58,3.38)
\psdots[dotsize=0.4](7.26,3.38)
\psdots[dotsize=0.4](6.42,4.76)
\psline[linewidth=0.04cm,arrowsize=0.16cm 4.0,arrowlength=1.4,arrowinset=0.4,doubleline=true,doublesep=0.12,doublecolor=white]{->}(9.1,0.98)(9.1,-0.94)
\pscircle[linewidth=0.04,linestyle=dashed,dash=0.16cm 0.16cm,dimen=outer](9.11,-3.93){2.53}
\pstriangle[linewidth=0.04,dimen=outer](1.87,3.38)(1.7,1.44)
\psdots[dotsize=0.4](1.02,3.38)
\psdots[dotsize=0.4,fillstyle=solid,dotstyle=o](2.7,3.38)
\psdots[dotsize=0.4,fillstyle=solid,dotstyle=square](1.86,4.76)
\pstriangle[linewidth=0.04,dimen=outer](9.15,-4.54)(1.7,1.44)
\psdots[dotsize=0.4](8.3,-4.54)
\psline[linewidth=0.04cm](9.14,-3.16)(9.12,-2.46)
\psline[linewidth=0.04cm](9.98,-4.56)(10.54,-5.08)
\psline[linewidth=0.04cm](8.28,-4.58)(7.78,-5.1)
\psdots[dotsize=0.4,fillstyle=solid,dotstyle=square](9.14,-3.16)
\psdots[dotsize=0.4,fillstyle=solid,dotstyle=o](9.98,-4.54)
\end{pspicture}
}\end{equation}

\subsubsection{Colors for Pairing}\label{Colors for Pairing}

%This also applies when the form $A$ is a set containing many forms.

Returning to our example of the Canadian and US postal services, another way to connect Toronto and New York is to write the following.

%network2
\begin{equation}\scalebox{0.65} % Change this value to rescale the drawing.
{
\begin{pspicture}(0,-6.62)(12.18,6.62)
\usefont{T1}{ptm}{m}{n}
\rput(1.6528125,5.575){\large Vancouver}
\usefont{T1}{ptm}{m}{n}
\rput(3.8807812,5.415){\large Calgary}
\usefont{T1}{ptm}{m}{n}
\rput(8.762969,4.895){\large Montreal}
\usefont{T1}{ptm}{m}{n}
\rput(7.985625,4.075){\large Toronto}
\usefont{T1}{ptm}{m}{n}
\rput(10.623593,5.315){\large Halifax}
\usefont{T1}{ptm}{m}{n}
\rput(8.746094,-4.425){\large Washington}
\usefont{T1}{ptm}{m}{n}
\rput(9.224218,-3.705){\large New York}
\usefont{T1}{ptm}{m}{n}
\rput(5.5609374,-4.485){\large Denver}
\usefont{T1}{ptm}{m}{n}
\rput(8.682187,-5.825){\large Miami}
\usefont{T1}{ptm}{m}{n}
\rput(2.9934375,-5.205){\large Los Angeles}
\usefont{T1}{ptm}{m}{n}
\rput(2.8685937,-4.345){\large San Francisco}
\usefont{T1}{ptm}{m}{n}
\rput(6.193125,5.355){\large Winnipeg}
\psline[linewidth=0.04cm](2.64,5.54)(3.08,5.48)
\psline[linewidth=0.04cm](7.16,5.26)(7.84,5.02)
\psline[linewidth=0.04cm](8.54,4.66)(8.04,4.32)
\psline[linewidth=0.04cm](7.8,4.32)(7.08,5.04)
\psline[linewidth=0.04cm](9.6,4.9)(10.64,5.1)
\psline[linewidth=0.04cm](8.98,-3.9)(8.82,-4.16)
\psline[linewidth=0.04cm](8.74,-4.72)(8.66,-5.46)
\psline[linewidth=0.04cm](8.46,-5.46)(6.34,-4.64)
\psline[linewidth=0.04cm](6.44,-4.38)(7.66,-4.42)
\psline[linewidth=0.04cm](4.22,-4.32)(4.82,-4.36)
\psline[linewidth=0.04cm](2.76,-4.64)(2.92,-4.96)
\psline[linewidth=0.04cm](3.02,-4.96)(4.76,-4.58)
\psline[linewidth=0.04cm](4.58,5.36)(5.32,5.32)
\usefont{T1}{ptm}{m}{n}
\rput(8.864219,3.195){\large  New York}
\psline[linewidth=0.04cm](8.08,3.82)(8.66,3.42)
\psline[linewidth=0.04cm,arrowsize=0.32cm 8.0,arrowlength=1.4,arrowinset=0.4]{->}(6.02,-1.14)(6.02,-2.38)
\psellipse[linewidth=0.04,linestyle=dashed,dash=0.16cm 0.16cm,dimen=outer](5.91,-4.66)(5.41,1.96)
\usefont{T1}{ptm}{m}{n}
\rput(8.884219,-0.225){\large  New York}
\psframe[linewidth=0.04,dimen=outer](12.18,6.62)(0.0,-0.92)
\rput{-180.0}(12.0,4.32){\pstriangle[linewidth=0.04,dimen=outer](6.0,1.72)(1.04,0.88)}
\end{pspicture}
}
\end{equation}

From this, it is not clear if the position of New York on the small side of the directed triangle corresponds to New York or Montreal on the other side. Also, it is not a very compact notation. A solution to this problem is to introduce colors to help pairing forms. Forms of the same color on both sides of the directed triangle are paired together, meaning that the form being replaced is substituted by the form of the same color. Here we consider that black is not a color.

We can now rewrite our network while making sure that the connection is made with New York and not Montreal. The two postal networks are now linked together through Toronto and New York.

%network3
\begin{equation}\label{network3}\scalebox{0.65} % Change this value to rescale the drawing.
{
\begin{pspicture}(0,-6.03)(12.18,6.03)
\definecolor{color1624}{rgb}{0.2,1.0,0.0}
\usefont{T1}{ptm}{m}{n}
\rput(1.6528125,4.985){\large Vancouver}
\usefont{T1}{ptm}{m}{n}
\rput(3.8807812,4.825){\large Calgary}
\usefont{T1}{ptm}{m}{n}
\rput(8.762969,4.305){\large Montreal}
\usefont{T1}{ptm}{m}{n}
\rput(7.985625,3.485){\large Toronto}
\usefont{T1}{ptm}{m}{n}
\rput(10.623593,4.725){\large Halifax}
\usefont{T1}{ptm}{m}{n}
\rput(8.726093,-3.835){\large Washington}
\usefont{T1}{ptm}{m}{n}
\rput(9.204219,-3.115){\large New York}
\usefont{T1}{ptm}{m}{n}
\rput(5.5409374,-3.895){\large Denver}
\usefont{T1}{ptm}{m}{n}
\rput(8.662188,-5.235){\large Miami}
\usefont{T1}{ptm}{m}{n}
\rput(2.9734375,-4.615){\large Los Angeles}
\usefont{T1}{ptm}{m}{n}
\rput(2.8485937,-3.755){\large San Francisco}
\usefont{T1}{ptm}{m}{n}
\rput(6.193125,4.765){\large Winnipeg}
\psline[linewidth=0.04cm](2.64,4.95)(3.08,4.89)
\psline[linewidth=0.04cm](7.16,4.67)(7.84,4.43)
\psline[linewidth=0.04cm](8.54,4.07)(8.04,3.73)
\psline[linewidth=0.04cm](7.8,3.73)(7.08,4.45)
\psline[linewidth=0.04cm](9.6,4.31)(10.64,4.51)
\psline[linewidth=0.04cm](8.96,-3.31)(8.8,-3.57)
\psline[linewidth=0.04cm](8.72,-4.13)(8.64,-4.87)
\psline[linewidth=0.04cm](8.44,-4.87)(6.32,-4.05)
\psline[linewidth=0.04cm](6.42,-3.79)(7.64,-3.83)
\psline[linewidth=0.04cm](4.2,-3.73)(4.8,-3.77)
\psline[linewidth=0.04cm](2.74,-4.05)(2.9,-4.37)
\psline[linewidth=0.04cm](3.0,-4.37)(4.74,-3.99)
\psline[linewidth=0.04cm](4.58,4.77)(5.32,4.73)
\usefont{T1}{ptm}{m}{n}
\rput(8.864219,2.605){\large \color{color1624}
 New York}
\psline[linewidth=0.04cm](8.08,3.23)(8.66,2.83)
\psline[linewidth=0.04cm,arrowsize=0.32cm 8.0,arrowlength=1.4,arrowinset=0.4]{->}(6.0,-0.55)(6.0,-1.79)
\psellipse[linewidth=0.04,linestyle=dashed,dash=0.16cm 0.16cm,dimen=outer](5.89,-4.07)(5.41,1.96)
\usefont{T1}{ptm}{m}{n}
\rput(8.884219,0.645){\large \color{color1624}
 New York}
\psframe[linewidth=0.04,dimen=outer](12.18,6.03)(0.0,-0.15)
\rput{-180.0}(12.0,3.14){\pstriangle[linewidth=0.04,dimen=outer](6.0,1.13)(1.04,0.88)}
\end{pspicture}
}
\end{equation}

We can also use two colors to indicate the pairing. The blue color in the following example can be interpreted as connecting San Francisco with Vancouver and New York with Toronto.

%network4
\begin{equation}\scalebox{0.65} % Change this value to rescale the drawing.
{
\begin{pspicture}(0,-6.03)(12.18,6.03)
\definecolor{color5707}{rgb}{0.2,1.0,0.0}
\definecolor{color5763}{rgb}{0.0,0.2,0.8}
\usefont{T1}{ptm}{m}{n}
\rput(1.5128125,5.305){\large  Vancouver}
\usefont{T1}{ptm}{m}{n}
\rput(3.9207811,5.145){\large Calgary}
\usefont{T1}{ptm}{m}{n}
\rput(8.802969,4.625){\large Montreal}
\usefont{T1}{ptm}{m}{n}
\rput(8.025625,3.805){\large Toronto}
\usefont{T1}{ptm}{m}{n}
\rput(10.663593,5.045){\large Halifax}
\usefont{T1}{ptm}{m}{n}
\rput(8.726093,-3.835){\large Washington}
\usefont{T1}{ptm}{m}{n}
\rput(9.204219,-3.115){\large New York}
\usefont{T1}{ptm}{m}{n}
\rput(5.5409374,-3.895){\large Denver}
\usefont{T1}{ptm}{m}{n}
\rput(8.662188,-5.235){\large Miami}
\usefont{T1}{ptm}{m}{n}
\rput(2.9734375,-4.615){\large Los Angeles}
\usefont{T1}{ptm}{m}{n}
\rput(2.8485937,-3.755){\large San Francisco}
\usefont{T1}{ptm}{m}{n}
\rput(6.233125,5.085){\large Winnipeg}
\psline[linewidth=0.04cm](2.68,5.27)(3.12,5.21)
\psline[linewidth=0.04cm](7.2,4.99)(7.88,4.75)
\psline[linewidth=0.04cm](8.58,4.39)(8.08,4.05)
\psline[linewidth=0.04cm](7.84,4.05)(7.12,4.77)
\psline[linewidth=0.04cm](9.64,4.63)(10.68,4.83)
\psline[linewidth=0.04cm](8.96,-3.31)(8.8,-3.57)
\psline[linewidth=0.04cm](8.72,-4.13)(8.64,-4.87)
\psline[linewidth=0.04cm](8.44,-4.87)(6.32,-4.05)
\psline[linewidth=0.04cm](6.42,-3.79)(7.64,-3.83)
\psline[linewidth=0.04cm](4.2,-3.73)(4.8,-3.77)
\psline[linewidth=0.04cm](2.74,-4.05)(2.9,-4.37)
\psline[linewidth=0.04cm](3.0,-4.37)(4.74,-3.99)
\psline[linewidth=0.04cm](4.62,5.09)(5.36,5.05)
\usefont{T1}{ptm}{m}{n}
\rput(8.904219,2.925){\large \color{color5707}New York}
\psline[linewidth=0.04cm](8.12,3.55)(8.7,3.15)
\psline[linewidth=0.04cm,arrowsize=0.32cm 8.0,arrowlength=1.4,arrowinset=0.4]{->}(6.0,-0.55)(6.0,-1.79)
\psellipse[linewidth=0.04,linestyle=dashed,dash=0.16cm 0.16cm,dimen=outer](5.89,-4.07)(5.41,1.96)
\psframe[linewidth=0.04,dimen=outer](12.18,6.03)(0.0,-0.35)
\rput{-180.0}(12.12,2.86){\pstriangle[linewidth=0.04,dimen=outer](6.06,0.99)(1.04,0.88)}
\usefont{T1}{ptm}{m}{n}
\rput(2.8685937,2.305){\large \color{color5763} San Francisco}
\psline[linewidth=0.04cm](1.5,5.07)(2.76,2.61)
\usefont{T1}{ptm}{m}{n}
\rput(8.904219,0.745){\large \color{color5707}New York}
\usefont{T1}{ptm}{m}{n}
\rput(2.8685937,0.125){\large \color{color5763} San Francisco}
\end{pspicture}
}\end{equation}

This systems reduces to

%Network Full
\begin{equation}\scalebox{0.65} % Change this value to rescale the drawing.
{
\begin{pspicture}(0,-3.6)(13.26,3.6)
\usefont{T1}{ptm}{m}{n}
\rput(2.3128126,1.855){\large   Vancouver}
\usefont{T1}{ptm}{m}{n}
\rput(4.7207813,1.695){\large Calgary}
\usefont{T1}{ptm}{m}{n}
\rput(9.602969,1.175){\large Montreal}
\usefont{T1}{ptm}{m}{n}
\rput(8.825625,0.355){\large Toronto}
\usefont{T1}{ptm}{m}{n}
\rput(11.4635935,1.595){\large Halifax}
\usefont{T1}{ptm}{m}{n}
\rput(9.546094,-1.265){\large Washington}
\usefont{T1}{ptm}{m}{n}
\rput(10.024219,-0.545){\large New York}
\usefont{T1}{ptm}{m}{n}
\rput(6.3609376,-1.325){\large Denver}
\usefont{T1}{ptm}{m}{n}
\rput(9.482187,-2.665){\large Miami}
\usefont{T1}{ptm}{m}{n}
\rput(3.7934375,-2.045){\large Los Angeles}
\usefont{T1}{ptm}{m}{n}
\rput(3.6685936,-1.185){\large San Francisco}
\usefont{T1}{ptm}{m}{n}
\rput(7.033125,1.635){\large Winnipeg}
\psline[linewidth=0.04cm](3.48,1.82)(3.92,1.76)
\psline[linewidth=0.04cm](8.0,1.54)(8.68,1.3)
\psline[linewidth=0.04cm](9.38,0.94)(8.88,0.6)
\psline[linewidth=0.04cm](8.64,0.6)(7.92,1.32)
\psline[linewidth=0.04cm](10.44,1.18)(11.48,1.38)
\psline[linewidth=0.04cm](9.78,-0.74)(9.62,-1.0)
\psline[linewidth=0.04cm](9.54,-1.56)(9.46,-2.3)
\psline[linewidth=0.04cm](9.26,-2.3)(7.14,-1.48)
\psline[linewidth=0.04cm](7.24,-1.22)(8.46,-1.26)
\psline[linewidth=0.04cm](5.02,-1.16)(5.62,-1.2)
\psline[linewidth=0.04cm](3.56,-1.48)(3.72,-1.8)
\psline[linewidth=0.04cm](3.82,-1.8)(5.56,-1.42)
\psline[linewidth=0.04cm](5.42,1.64)(6.16,1.6)
\psline[linewidth=0.04cm](8.92,0.1)(9.5,-0.3)
\psellipse[linewidth=0.04,linestyle=dashed,dash=0.16cm 0.16cm,dimen=outer](6.63,0.0)(6.63,3.6)
\psline[linewidth=0.04cm](2.3,1.62)(3.56,-0.84)
\end{pspicture}
}\end{equation}

Take the system $\fbox{$D-B \rhd A$}\rightarrow  (A-C)$ written inline where the connections between the nodes identified by letters, it is not clear how we should interpret it. There are a few possibilities. If we replace $A$ by $D-B$, what will happen with the connection between $A$ and $C$? We could replace the form $A$ by $B$ and keep the connection between $B$ and $D$ or replace $A$ by $D$ and keep the connection between $D$ and $B$. For this example, we can also use a color to indicate what should be paired together in the transformation.

By identifying $A$ with $B$ using the color green, we are saying that $A$ is replaced by $B$ and not by $D-B$. Thus, we have the following reduction.

$$\begin{array}{c}
\fbox{$D-{\color{green}B} \rhd {\color{green}A}$}\rightarrow  (A-C) \\
\Downarrow\\
(D-B-C)
\end{array}$$

If we choose another color pairing, we find a different reduction.

$$\begin{array}{c}
\fbox{${\color{green}D}-B \rhd {\color{green}A}$}\rightarrow  (A-C) \\
\Downarrow\\
(B-D-C)
\end{array}$$

Colors indicate which form should be replaced by which other form, but they also ensure that the connections are preserved. We have stated that if a form is in the same position, then the connections are preserved. But if the position of objects in a transformation does not correspond, how can we preserve the connections? For example, the contracting triangle with connections must be reduced as follows.

%Contracting triangle no color
\begin{equation}\scalebox{0.5} % Change this value to rescale the drawing.
{
\begin{pspicture}(0,-6.37)(15.36,6.37)
\pscircle[linewidth=0.04,linestyle=dashed,dash=0.16cm 0.16cm,dimen=outer](12.83,3.84){2.53}
\psline[linewidth=0.04cm,arrowsize=0.16cm 8.0,arrowlength=1.4,arrowinset=0.4]{->}(8.4,4.01)(9.96,4.01)
\psframe[linewidth=0.04,dimen=outer](8.12,5.81)(0.0,2.21)
\rput{-90.0}(-0.15,7.99){\pstriangle[linewidth=0.04,dimen=outer](3.92,3.61)(1.0,0.92)}
\pstriangle[linewidth=0.04,dimen=outer](12.85,3.29)(1.7,1.44)
\psdots[dotsize=0.4](12.0,3.29)
\psdots[dotsize=0.4](13.68,3.29)
\psdots[dotsize=0.4](12.84,4.67)
\psline[linewidth=0.04cm](12.84,4.67)(12.82,5.37)
\psline[linewidth=0.04cm](13.68,3.27)(14.24,2.75)
\psline[linewidth=0.04cm](11.98,3.25)(11.48,2.73)
\pstriangle[linewidth=0.04,dimen=outer](6.17,3.29)(1.7,1.44)
\psdots[dotsize=0.4](5.32,3.29)
\psdots[dotsize=0.4](7.0,3.29)
\psdots[dotsize=0.4](6.16,4.67)
\psdots[dotsize=0.4](1.76,3.83)
\psline[linewidth=0.04cm,arrowsize=0.16cm 4.0,arrowlength=1.4,arrowinset=0.4,doubleline=true,doublesep=0.12,doublecolor=white]{->}(8.84,0.89)(8.84,-1.03)
\pscircle[linewidth=0.04,linestyle=dashed,dash=0.16cm 0.16cm,dimen=outer](8.87,-3.84){2.53}
\psline[linewidth=0.04cm](8.88,-3.01)(8.86,-2.31)
\psline[linewidth=0.04cm](9.72,-4.41)(10.28,-4.93)
\psline[linewidth=0.04cm](8.02,-4.43)(7.52,-4.95)
\psdots[dotsize=0.4](8.86,-3.85)
\end{pspicture}
}\end{equation}

By using a color, we ensure that the connections are preserved since we are clearly indicating which objects are corresponding.

%Contracting triangle color
\begin{equation}\scalebox{0.5} % Change this value to rescale the drawing.
{
\begin{pspicture}(0,-6.37)(15.36,6.37)
\definecolor{color1031}{rgb}{0.0,0.8,0.0}
\definecolor{color1035}{rgb}{0.2,0.8,0.0}
\pscircle[linewidth=0.04,linestyle=dashed,dash=0.16cm 0.16cm,dimen=outer](12.83,3.84){2.53}
\psline[linewidth=0.04cm,arrowsize=0.16cm 8.0,arrowlength=1.4,arrowinset=0.4]{->}(8.4,4.01)(9.96,4.01)
\psframe[linewidth=0.04,dimen=outer](8.12,5.81)(0.0,2.21)
\rput{-90.0}(-0.15,7.99){\pstriangle[linewidth=0.04,dimen=outer](3.92,3.61)(1.0,0.92)}
\pstriangle[linewidth=0.04,dimen=outer](12.85,3.29)(1.7,1.44)
\psdots[dotsize=0.4](12.0,3.29)
\psdots[dotsize=0.4](13.68,3.29)
\psdots[dotsize=0.4](12.84,4.67)
\psline[linewidth=0.04cm](12.84,4.67)(12.82,5.37)
\psline[linewidth=0.04cm](13.68,3.27)(14.24,2.75)
\psline[linewidth=0.04cm](11.98,3.25)(11.48,2.73)
\pstriangle[linewidth=0.04,linecolor=color1031,dimen=outer](6.17,3.29)(1.7,1.44)
\psdots[dotsize=0.4,linecolor=color1031](5.32,3.29)
\psdots[dotsize=0.4,linecolor=color1031](7.0,3.29)
\psdots[dotsize=0.4,linecolor=color1031](6.16,4.67)
\psdots[dotsize=0.4,linecolor=color1035](1.76,3.83)
\psline[linewidth=0.04cm,arrowsize=0.16cm 4.0,arrowlength=1.4,arrowinset=0.4,doubleline=true,doublesep=0.12,doublecolor=white]{->}(8.84,0.89)(8.84,-1.03)
\pscircle[linewidth=0.04,linestyle=dashed,dash=0.16cm 0.16cm,dimen=outer](8.87,-3.84){2.53}
\psline[linewidth=0.04cm](8.86,-3.75)(8.84,-3.05)
\psline[linewidth=0.04cm](8.92,-3.89)(9.48,-4.41)
\psline[linewidth=0.04cm](8.74,-3.91)(8.24,-4.43)
\psdots[dotsize=0.4](8.86,-3.85)
\end{pspicture}
}
\end{equation}

\subsubsection{Distance Preservation}\label{Distance Preservation}

 The distances between the elements in the initial form are important when we use a dashed line or the inline parentheses notation `$(\,\,)$'. This is useful for ordered sets, for models of maps and objects in space like molecules. This means that terms of the same color can appear in different positions in a transformation and will move the corresponding element in the initial form.

 If we represent the initial form with a full line or the inline parentheses notation `$\{\,\,\}$', this will mean that the position is not important. Even if the position is not important; the connections stay connected even if the nodes are moved around. This is used for sets where the order is not important and for graphs where the length of edges or position of the vertices are not relevant.

 It is important to note that by default a transformation with disjoint elements on the right-hand side of a transformation does not require that the order or the space between them to be the same as in the initial form to be applied. For example,
  $$\fbox{$X, Y \rhd C, A$}\rightarrow  (A, B, C)$$
 \noindent reduces to $(Y, B, X)$.

When reading this transformation, we should view $X, Y$ to be the effect of the transformation and $C, A$ to be the cause.  The commas and the position of these letters is important. It means that the transformation will search the initial form for a $C$ with the potential to replace it by $X$ and will search the initial form for a $A$ with the potential to replace it by $Y$, but the transformation will only be applied when $C$ and $A$ are present in the initial form.

 If we want the order or the space to be respected, we will write $rigid$ on the subscript of $\rhd$. Thus,
 $$\fbox{$X, Y \rhd_{rigid} C, A$}\rightarrow  (A, B, C)$$
  \noindent does not reduce further, but the following system does.
   $$\fbox{$X, Y \rhd_{rigid} B, C$}\rightarrow  (A, B, C)\Rightarrow (A, X, Y).$$

\subsubsection{Transformation exceptions}

A transformation which replaces a form $A$ by a form $B$ will replace $A$ by $B$ regardless of what is around $A$. If we want this transformation to not replace $B$ when $C$ is present with $A$ in the initial form, we can write $C$ in red to mean that the transformation will not replace $A$ when $C$ is also present in the initial form. Thus, the transformation $\fbox{$B\rhd A,{\color{red}C}$}$ will reduce the initial form $(D, A, F)$ to $(D, B, F)$, but will leave $(C, B, F)$ unaffected because $C$ is present in the initial form.

We will consider the color red to be a reserved color and will only be used to indicate that a transformation cannot be applied if the form in red appears in the initial form.

\subsection{Naming}\label{naming}

We will now introduce three type of naming: the simple identifier or label, a type of naming which uses variables and a way to name based on abstraction.

\subsubsection{Identifiers}

We have already briefly introduced the \textit{naming} symbol `:=' to name transformations. In what follow, we will explain more concerning this notation and we will introduce a construction similar but more flexible than the objects and classes in object-oriented programming.

The reason for the `:' next to the equality symbols indicates that the symbols on the right is the name or label associated to the transformation written on the left. For example, $Growth :=\fbox{$Tree \rhd Seed$}$ or $\fbox{$Tree \rhd Seed$}=: Growth$ means that we give the label $Growth$ to the transformation $Tree \rightarrow Seed$. If we want to change or shorten the name of a transformation we do this by using the symbol $:=:$. For example, we can make $G_1$ a perfect synonym of $Growth$ by writing $G_1 :=: Growth$. We can also give names to initial sets. For example, $Field:=(Seed, Seed, Seed, Soil)$. Such a name or label given to a form will be know as an \textit{identifier}.

The understanding of this notation is similar to the usual understanding of mathematical equations. For example, if we have the equation $2x+y=1$ and we know that $y=4$ we can replace $y$ by $4$ in the equation to get $2x+4=1$. We will treat the naming symbols in the same way. For example, if we have the system $\fbox{$Tree \rhd Seed$}\rightarrow Field$ where $Field:=(Seed, Seed, Seed, Soil)$, then we can write
 $$\fbox{$Tree \rhd Seed$}^{\,3S}\rightarrow (Seed, Seed, Seed, Soil).$$
 \noindent Therefore, naming a form means that we can replace each occurrence of this form by this name and each occurrence of this name in a form can be replaced by the form it represents.

 When reducing a system, it is important that we have objects of the same type in the initial form and in the transformations. For example, $\fbox{$Tree \rhd Seed$}\rightarrow \text{Field}$ could not be reduced unless we replace the name $Field$ so that we have the system
  $$\fbox{$Tree \rhd Seed$}^{\,3S}\rightarrow (Seed, Seed, Seed, Soil).$$

\subsubsection{Function Forms}\label{function forms}

In object-oriented programming, classes are powerful tools that allow encapsulation of data, help making a program more transparent and encourage code reusability. One way to understand classes and objects is given in the following general example.

If we are writing a program that refers to different horses, we would need to access all these different horses. Every time we need to refer to a horse in our program, we could write the code describing this horse. If we have to refer to many different horses this could be quite long and repetitive. Another way to do this is to build a class or template for all types of horses. Suppose we are only concerned with characteristics such as color, age, running speed, height and tameness. We can define a class called `Horses' that will group the attributes of color (C), age (A), running speed (S), and the behaviours such as walk, run and jump. In the present case, the behaviour jump could be a very complex description of the mechanics of a horse jumping. Now when we want refer to a particular horse, we only need to insert the values $C=white, A=12 \,\, years, S=80 \,\, km/hour$ in our Horses class. Our horse would also come with the behaviours of the class and thus make the horse more than a list of attributes. In this setting, we don't have to redefine each new horse we are referring to, we only need to set values to the characteristics contained in the class. In programming, a horse would be said to be an instance or an objet of the Horses class.

In the language of forms and transformations, we can include the powerful property of object-oriented programming by naming a form or transformation with a label followed by a list of values written under parentheses as  $Label(V_1=value, V_2=value,..., V_n= value)$. We will refer to these values as the \textit{variables}. This is related to what is done when defining a function by writing $f(x)$ to refer to an algebraic expression with a variable $x$. Although we will often use the notation $Label(V_1=value, V_2=value,..., V_n= value)$, we will not restrict the language to allow only this form of label. We will see below examples where there is no restriction on the form of label. This will be convenient to represent arithmetic operations such as addition and multiplication.

 An example with one variable is $Growth(Tree) :=\fbox{$Tree \rhd Seed$}$. This is interpreted as turning $Tree$ into a variable. Now, if we want to refer to the growth of an oak tree or a maple tree we would respectively write $Growth(Oak) :=$ and $Growth(Maple)$. An example with two variables is $Growth(Tree, Sun) :=\fbox{$Tree \rhd Seed, Sun$}$. Thus, if we want to refer to the growth of an oak under the midday sun, we would write $Growth(Oak, Midday sun)$. It is important to note that these names such as $Growth(Oak, Midday sun)$ are also considered to be forms, since we could also apply transformations to these names. This name associated to a form containing variables will be called a \textit{function form}.

 For a more complex example, let's define a form for the Canadian postal network. Note that we will use dotted lines to surround the collection of forms to be named.

%Canada network
\begin{equation}\scalebox{.65}
{
\begin{pspicture}(0,-1.6728125)(11.804063,1.7128125)
\usefont{T1}{ptm}{m}{n}
\rput(1.736875,0.4621875){\large Vancouver}
\usefont{T1}{ptm}{m}{n}
\rput(3.9648438,0.3021875){\large Calgary}
\usefont{T1}{ptm}{m}{n}
\rput(8.847032,-0.2178125){\large Montreal}
\usefont{T1}{ptm}{m}{n}
\rput(8.069688,-1.0378125){\large Toronto}
\usefont{T1}{ptm}{m}{n}
\rput(11.057656,0.1621875){\large Y}
\usefont{T1}{ptm}{m}{n}
\rput(5.83875,0.1021875){\large X}
\psline[linewidth=0.04cm](2.7240624,0.4271875)(3.1640625,0.3671875)
\psline[linewidth=0.04cm](6.3240623,0.1271875)(7.8440623,-0.0728125)
\psline[linewidth=0.04cm](8.624063,-0.4528125)(8.124063,-0.7928125)
\psline[linewidth=0.04cm](7.8840623,-0.7928125)(6.3040624,-0.1328125)
\psline[linewidth=0.04cm](9.684063,-0.2128125)(10.724063,-0.0128125)
\psline[linewidth=0.04cm](4.6640625,0.2471875)(5.4040623,0.2071875)
\psframe[linewidth=0.04,linestyle=dotted,dotsep=0.16cm,dimen=outer](11.804063,1.0471874)(0.5040625,-1.6728125)
\usefont{T1}{ptm}{m}{n}
\rput(1.0059375,1.5021875){\large Can(X,Y):=}
\end{pspicture}
}
\end{equation}

%American network
\begin{equation}\scalebox{.65} % Change this value to rescale the drawing.
{
\begin{pspicture}(0,-1.9728125)(10.6375,2.0128126)
\usefont{T1}{ptm}{m}{n}
\rput(8.303594,0.0021875){\large Washington}
\usefont{T1}{ptm}{m}{n}
\rput(8.631875,0.8821875){\large V}
\usefont{T1}{ptm}{m}{n}
\rput(5.1184373,-0.0578125){\large Denver}
\usefont{T1}{ptm}{m}{n}
\rput(8.200469,-1.4178125){\large W}
\usefont{T1}{ptm}{m}{n}
\rput(2.5509374,-0.7778125){\large Los Angeles}
\usefont{T1}{ptm}{m}{n}
\rput(2.4260938,0.0821875){\large San Francisco}
\psline[linewidth=0.04cm](8.5375,0.5271875)(8.3775,0.2671875)
\psline[linewidth=0.04cm](8.2975,-0.2928125)(8.2175,-1.0328125)
\psline[linewidth=0.04cm](8.0175,-1.0328125)(5.8975,-0.2128125)
\psline[linewidth=0.04cm](5.9975,0.0471875)(7.2175,0.0071875)
\psline[linewidth=0.04cm](3.7775,0.1071875)(4.3775,0.0671875)
\psline[linewidth=0.04cm](2.3175,-0.2128125)(2.4775,-0.5328125)
\psline[linewidth=0.04cm](2.5775,-0.5328125)(4.3175,-0.1528125)
\psframe[linewidth=0.04,linestyle=dotted,dotsep=0.16cm,dimen=outer](10.6375,1.3671875)(0.3975,-1.9728125)
\usefont{T1}{ptm}{m}{n}
\rput(1.0460937,1.8021874){\large US(V, W):=}
\end{pspicture}
}
\end{equation}

We can now rewrite the transformation \refeq{network} of section \ref{Preserving Connections} as

%Network names
\begin{equation}\scalebox{0.7} % Change this value to rescale the drawing.
{
\begin{pspicture}(0,-1.72)(15.18,1.72)
\definecolor{color510}{rgb}{0.2,0.2,0.2}
\psbezier[linewidth=0.04,linestyle=dashed,dash=0.16cm 0.16cm](9.62327,1.3557045)(9.3,1.0276233)(9.313326,-1.054048)(9.641652,-1.3770239)(9.969978,-1.7)(14.511845,-1.6318761)(14.834482,-1.3031664)(15.157119,-0.97445667)(15.16,1.0483379)(14.834482,1.374169)(14.508964,1.7)(9.946541,1.6837859)(9.62327,1.3557045)
\usefont{T1}{ptm}{m}{n}
\rput(5.045625,0.535){\large  Toronto}
\usefont{T1}{ptm}{m}{n}
\rput(5.9842186,-0.465){\large  New York}
\usefont{T1}{ptm}{m}{n}
\rput(0.965625,0.575){\large  Toronto}
\usefont{T1}{ptm}{m}{n}
\rput(1.9442188,-0.465){\large  New York}
\psline[linewidth=0.04cm,arrowsize=0.32cm 8.0,arrowlength=1.4,arrowinset=0.4]{->}(7.46,0.04)(8.92,0.04)
\psframe[linewidth=0.04,dimen=outer](7.28,1.26)(0.0,-1.28)
\psline[linewidth=0.04cm,linecolor=color510](1.24,0.34)(2.22,-0.2)
\rput{-90.0}(3.52,3.81){\pstriangle[linewidth=0.04,dimen=outer](3.665,-0.25)(0.93,0.79)}
\usefont{T1}{ptm}{m}{n}
\rput(12.19625,0.815){\large Can(Winnipeg, Halifax)}
\usefont{T1}{ptm}{m}{n}
\rput(12.202969,-0.785){\large US(New York, Miami)}
\end{pspicture}
}
\end{equation}

A change to these networks can be done only by changing the names under the parentheses. For example, we could change New York to Boston just by writing $US(Boston, Miami)$ instead of $US(New York)$.

\subsubsection{Abstraction}\label{abstraction}

We have constructed transformations that resemble object programming classes and mathematical functions. We can start with a form and turn any part of that form into a variable. Such a process will be called form \textit{abstraction}. When abstracting a symbol, all occurrences of that symbol in the form become a variable. For example, the initial form $(A, B, B, C, A, A)$ can be abstracted by writing $f(A)=(A, B, B, C, A, A)$. This means that $A$ is now considered to be a variable and that $A$ can be replaced by any form. Example of instances of $f(A)$ are $f(E)=(E, B, B, C, E, E)$ and $f(\fbox{$M\rhd N$}):=(\fbox{$M\rhd N$}, B, B, C, \fbox{$M\rhd N$}, \fbox{$M\rhd N$})$. It is important to note that when giving instances, we will not write the symbol $:=$ but will write only $=$.

We can also use colors if there is more than one element which has the same name and we want to abstract only one of these. For example, $g({\color{blue}A}):=(A, B, B, C, {\color{blue}A}, A)$ will be evaluated as $g(E)=(A, B, B, C, E, A)$. If we want to abstract to have two variables we write $h({\color{blue}A},{\color{green}B}):=(A, B, {\color{green}B}, C, A, {\color{blue}A})$ and evaluate as $h(E, F)=(A, B, F, C, A, E)$. If the variables have the same name, this is done by using different colors so that $k({\color{blue}A},{\color{green}A}):=({\color{blue}A}, B, B, C, A, {\color{green}A})$ will be evaluated as $k(E, F)=(E, B, F, C, A, F)$.

If we want to abstract a group of elements we can use a single color for each element. For example,  $j({\color{green}A, B, B,}):=({\color{green}A, B, B,} C, A, A)$ will be evaluated as $j(E)=(E, C, A, A)$ or $j(E, F)=(E, F, C, A, A)$.

Abstraction can also be made on any symbol. Here are three examples.

 $$Arrow({\color{blue}\rightarrow}):=\fbox{$Tree \rhd Seed$}^{\,3S}{\color{blue}\rightarrow} (Seed, Seed, Seed, Soil).$$
 \noindent This can be useful to stop the application of a transformation by removing the arrow by writing $Arrow(\hspace{1cm})$.

 $$Superscript({\color{green}3S}):=\fbox{$Tree \rhd Seed$}^{\,{\color{green}3S}}\rightarrow (Seed, Seed, Seed, Soil).$$
 \noindent This can be useful to change the number of applications and series to parallel transformations by writing $Superscript(2P)$.

\begin{equation*}LettersAI({\color{green}a}, {\color{blue}i}):=C{\color{green}a}n(W{\color{blue}i}nn{\color{blue}i}peg, H{\color{green}a}lifax)\end{equation*}

This means that $LettersAI(A,I)$ gives $$[CAn(WInnIpeg, HAlifax)$$
However, in this case there is no meaning for this abstraction, we are allowing this because we want to have a very flexible language. Restrictions can always be applied when needed depending on the context. We now have a lot of flexibility with our notation and can modify almost all symbols in a transformation. The goal behind such a flexible notation is to be able to bridge the gaps between different domains of science by making sure that all can be represented with the use of transformations and useful naming process.

Until now, the variables which were abstracted were written under parentheses like is usually done with functions, but we do not need to follow this way of naming.  We can abstract forms which are similar to the operation of addition and multiplication, respectively $a+b$ and $a\cdot b$.

Integers can be represented by an amount of black dots. Thus, $1:=\bullet$, $2:=\bullet, \bullet$, $3:=\bullet, \bullet, \bullet$ and so on.
We can represent $1+2$ as a system of transformations by writing
 $$1+2:=\fbox{$\bullet \rhd \hspace{1cm}$}\rightarrow \{\bullet, \bullet\}.$$
 \noindent We can abstract this to define the addition operation by writing
 $${\color{green}\bullet}+{\color{blue}\bullet, \bullet}:=\fbox{${\color{green}\bullet} \rhd \hspace{1cm}$}\rightarrow \{{\color{blue}\bullet, \bullet}\}.$$
 \noindent This way, if we want to find the result of $2+3$, we only have that
 $$2+3:=\fbox{$2 \rhd \hspace{1cm}$}\rightarrow \{3\}.$$
 By replacing $2$ and $3$ with dots on the right hand-side, we find
 $$2+3:=\fbox{$\bullet, \bullet \rhd \hspace{1cm}$}\rightarrow \{\bullet, \bullet, \bullet\},$$
 \noindent which will reduce to $\{\bullet, \bullet, \bullet, \bullet, \bullet\}$ or the name $5$.

Similarly, we can define multiplication by

$$a\times b:=\fbox{$a\rhd \bullet$}^{\, \sharp S}\rightarrow \{b\}.$$

\noindent Recall that the superscript $\sharp S$ indicates that the transformation disappears after it is applied until each element of the initial form has been replaced. Note that only the elements of the initial forms are replaces not the elements that are introduced during the reduction. In this case, this is not viewed as an abstraction, but only a definition. For $2\times 3=\fbox{$2\rhd \bullet$}^{\, \sharp S}\rightarrow \{3\}$, we find the following reduction.

$$\begin{array}{c}
\fbox{$2\rhd \bullet$}^{\, \sharp S}\rightarrow \{3\}\\
=\\
\fbox{$\bullet, \bullet\rhd \bullet$}^{\, \sharp S}\rightarrow \{\bullet, \bullet, \bullet\} \\
\Downarrow\\
\{\bullet, \bullet, \bullet, \bullet, \bullet,\bullet\}
\end{array}$$

\subsection{Reduction view}

When we are more interested in the result of each step of a reduction, it is possible to use a convenient notation. When a form $X$ is reduced to a form $Y$ by a transformation, we can write the name of the transformation above the reduction double arrow to indicate the transformation which was used in that reduction.

\begin{definition}Let $\text{T}:=\fbox{$X\rhd Y$}$, then the following three notations are equivalent.
$$\fbox{$X\rhd Y$} \rightarrow Z \Rightarrow W,$$

$$T \rightarrow Z \Rightarrow W,$$

$$Z \xRightarrow[]{\text{T}} W.$$\end{definition}

We can also name steps of a reduction by using the naming symbol $:=$ written over the reduction double arrow. For example, if we have the reduction $A\Rightarrow B\Rightarrow C\Rightarrow D$, we can name the step between $C$ and $D$ as the transformation $t$ by writing
$$A\Rightarrow B\Rightarrow C\xRightarrow[]{\text{t}:=\,\,} D.$$

\subsection{Three dimensional language}

The language is not restricted to a 2 dimensional notation, but can also use a 3 dimensional notation.  Their are many advantages in using a 3d representation. One advantage is that instead of using subscript and superscript, we can use different shapes to represent different types of transformations. In some cases, we could quickly identify the transformation to be applied. Examples of this, from an inline notation to a 2d notation is the Penrose tensor diagram notation and Feynman diagrams. For example, a transformation which rotates an object by $\pi/2$ radians and doubles the height of an object could be named as follows.

\begin{figure}[H]\label{figure}
\centering
\fbox{\setlength{\fboxsep}{0pt}{\includegraphics[scale=0.2]{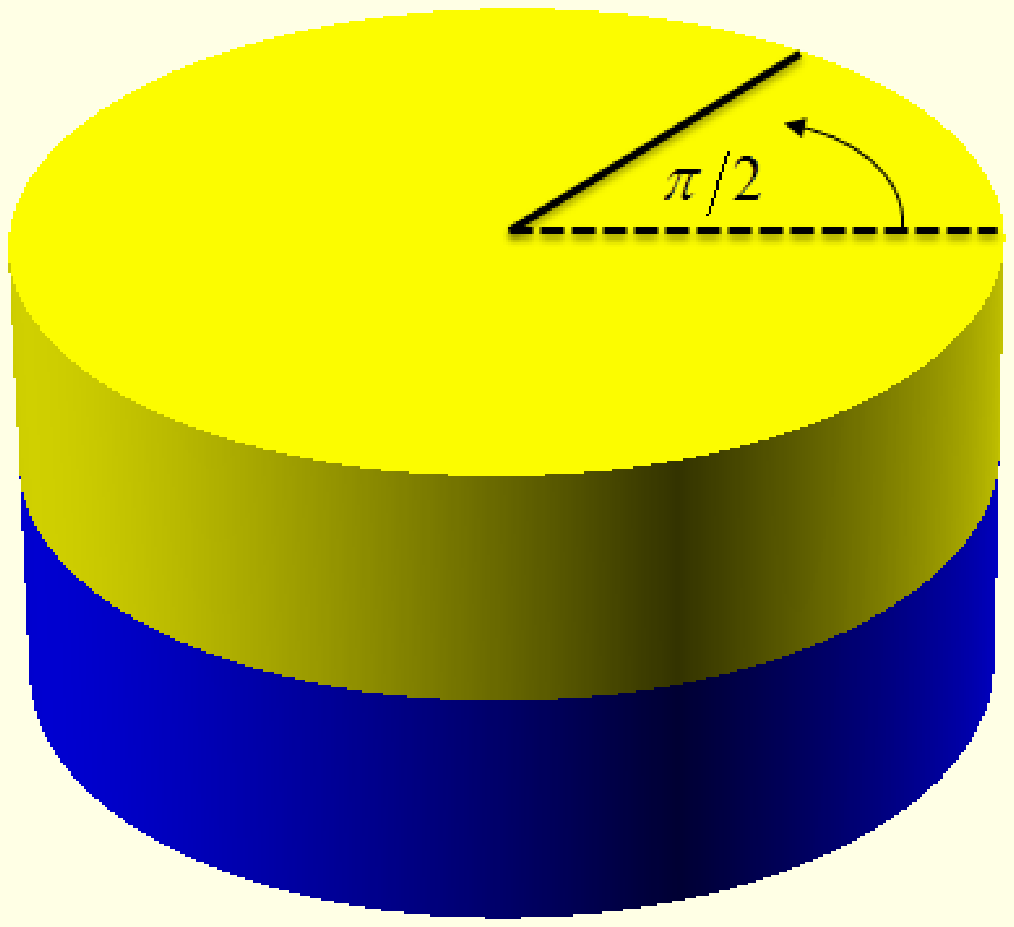}}}
%\caption{3D notation for a rotation of $45^\circ$ and scaling by $2$.}
\end{figure}

A transformation which rotates an object by $\pi/2$ radians, but triples the height of an object could be named as follows.

\begin{figure}[H]\label{figure}
\centering
\fbox{\setlength{\fboxsep}{0pt}{\includegraphics[scale=0.2]{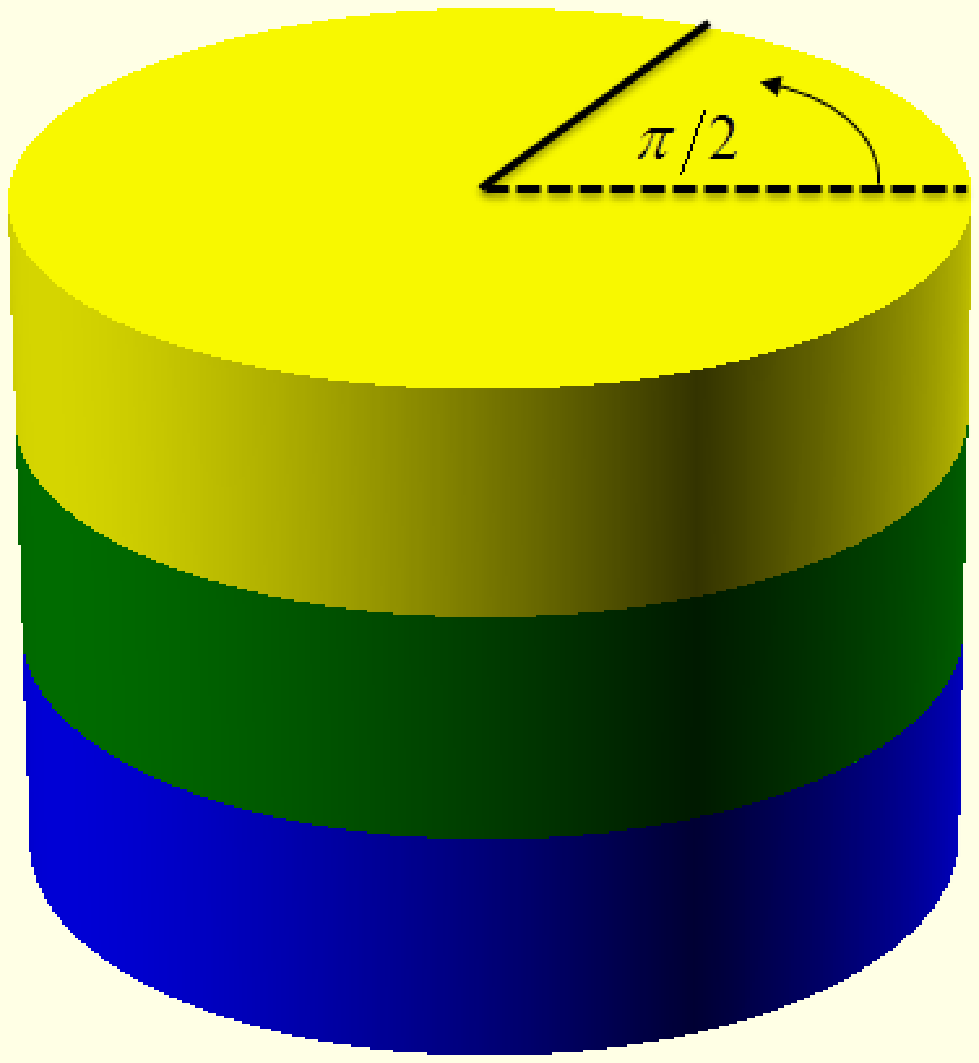}}}
%\caption{3D notation for a rotation of $45^\circ$ and scaling by $2$.}
\end{figure}

Naming a transformation with a 3D shape can be interesting if we also define a way to combine the names of similar transformations together in a geometrical way. For example, applying the tripling transformation to the geometrical shape representing the doubling transformation, we will find the following transformation.

\begin{figure}[H]\label{figure}
\centering
\fbox{\setlength{\fboxsep}{0pt}{\includegraphics[scale=0.2]{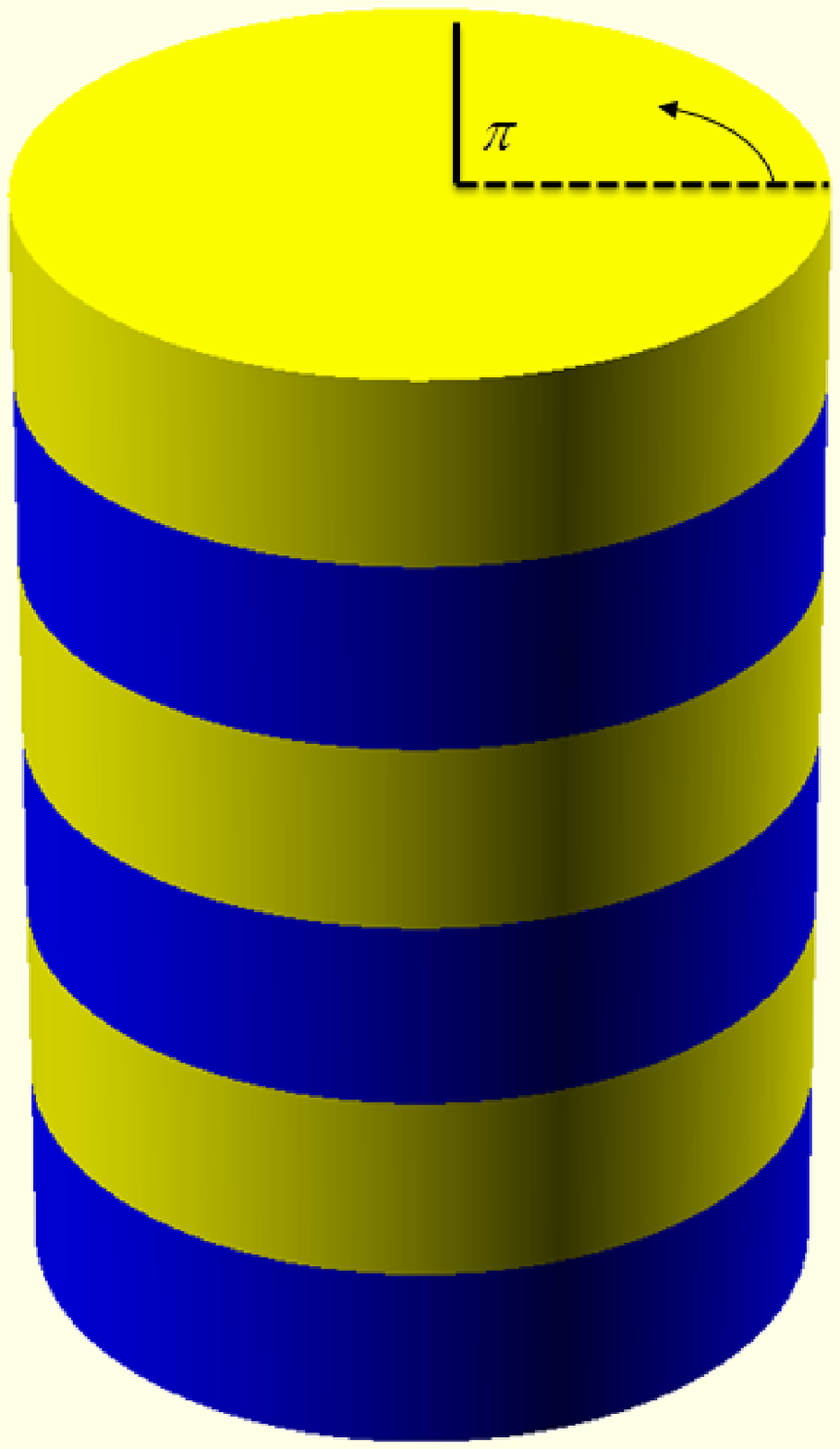}}}
%\caption{3D notation for a rotation of $45^\circ$ and scaling by $2$.}
\end{figure}

This transformation is interpreted as scaling the figure by a factor of $6$ and rotating by $\pi$ radians. Applied on a geometrical shape such as a tetrahedron, this is equivalent to first applying the doubling transformation and then applying the tripling transformation.

Another advantage is that we can now have a mathematical notation to study and modify geometrical objects. In mathematics, this is related to the field of topology. We will now give an example where an empty sphere with identified poles is reduced to a torus through a sequence of transformations.

We can apply the transformation
 \begin{figure}[H]\label{figure2}
\centering
\fbox{\setlength{\fboxsep}{0pt}{\includegraphics[scale=0.2]{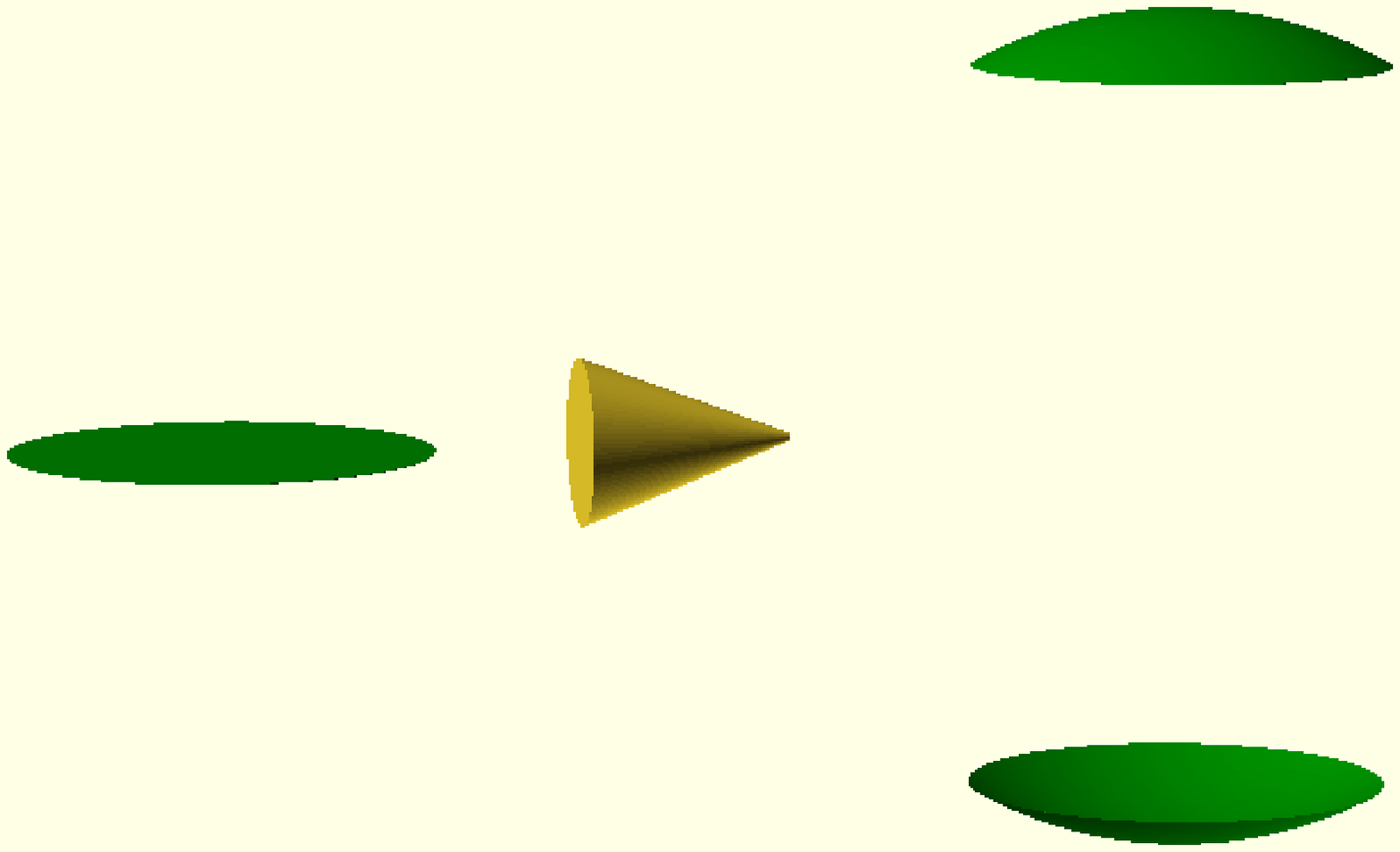}}}
%\caption{3D notation for a rotation of $45^\circ$ and scaling by $2$.}
\end{figure}
\noindent on the empty sphere with identified blue poles given by
\begin{figure}[H]\label{figure3}
\centering
\fbox{\setlength{\fboxsep}{0pt}{\includegraphics[scale=0.2]{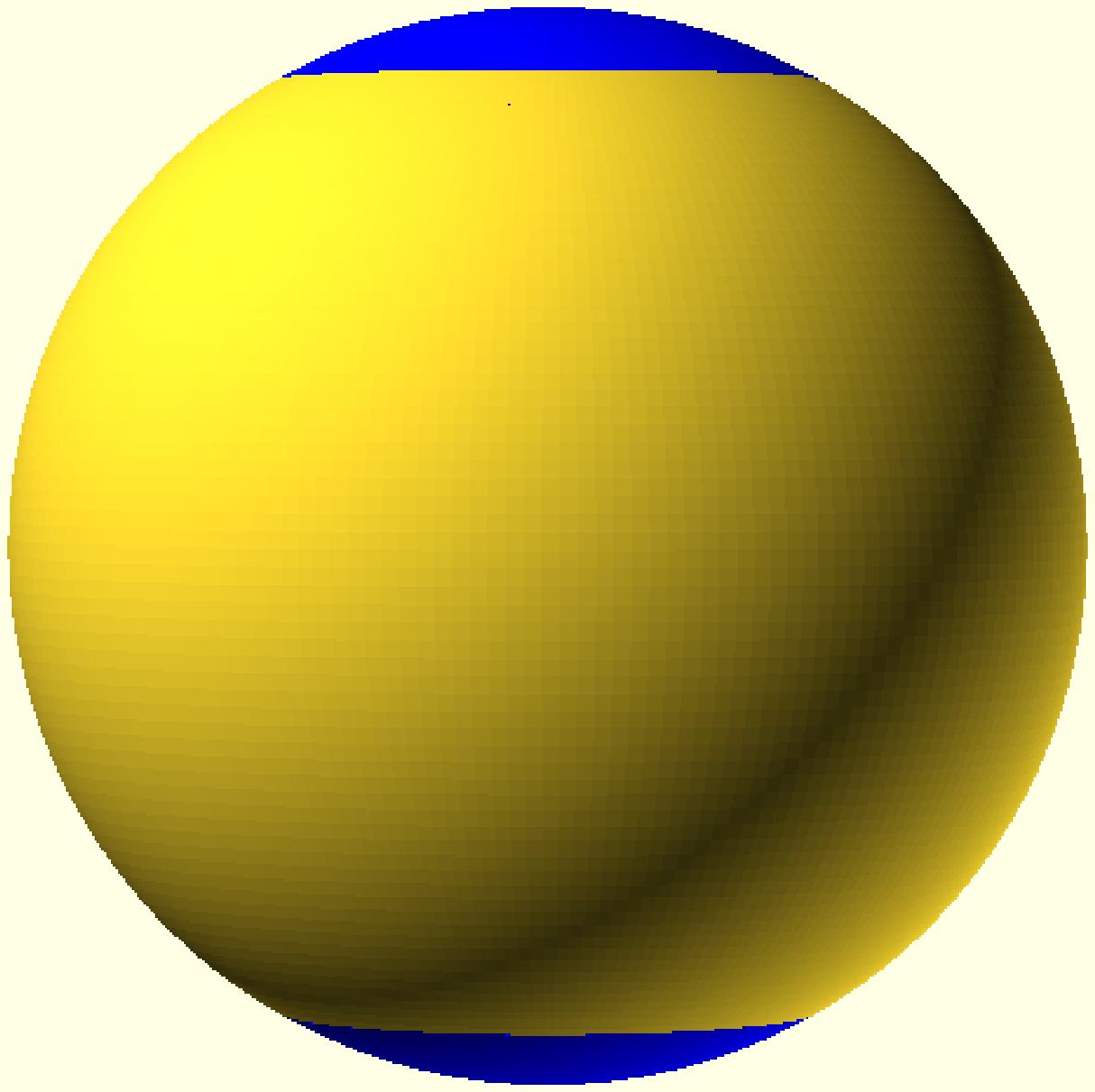}}}
.
%\caption{3D notation for a rotation of $45^\circ$ and scaling by $2$.}
\end{figure}

\noindent This gives the geometric object
\begin{figure}[H]\label{figure4}
\centering
\fbox{\setlength{\fboxsep}{0pt}{\includegraphics[scale=0.2]{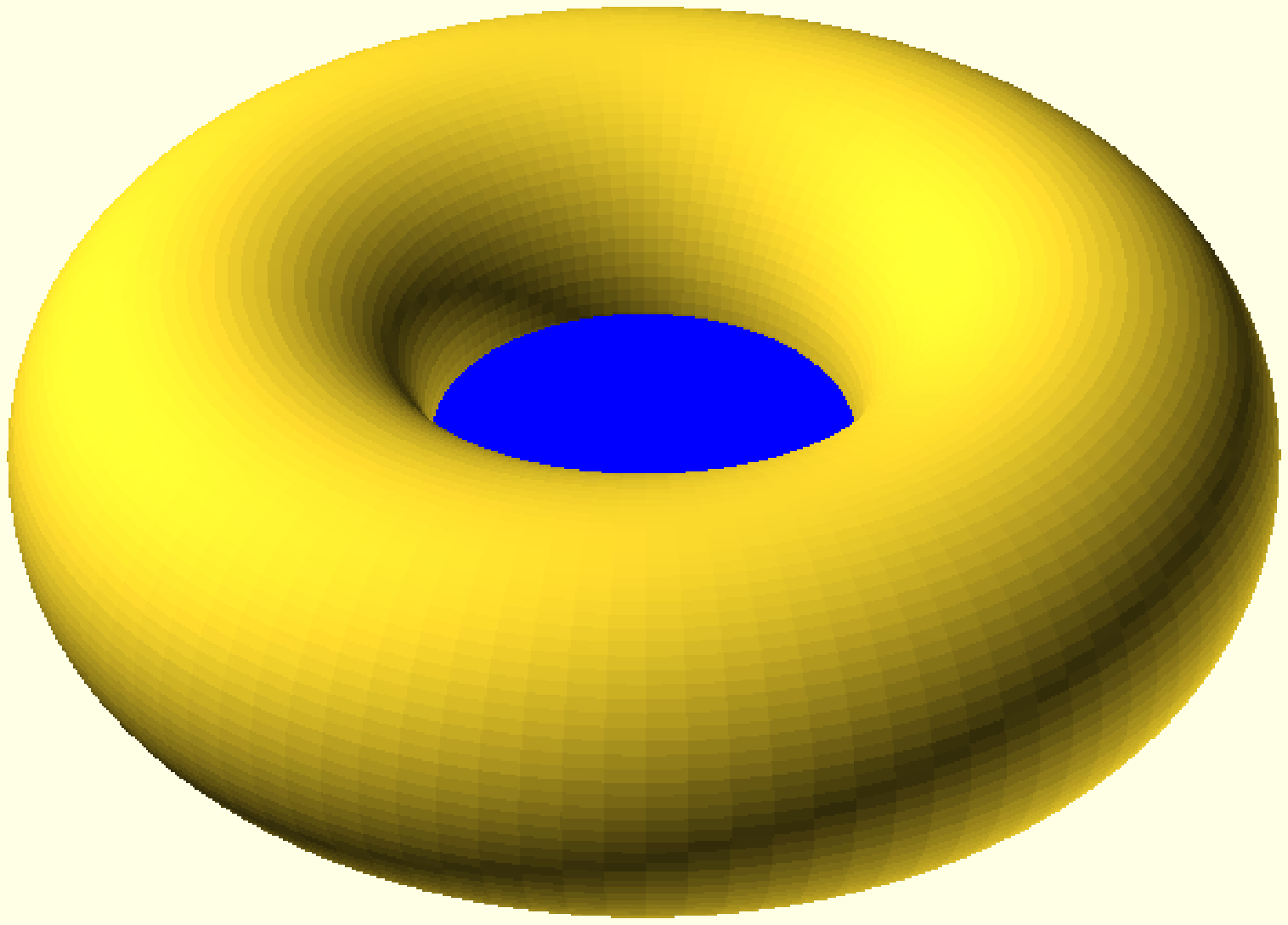}}}
%\caption{3D notation for a rotation of $45^\circ$ and scaling by $2$.}
\end{figure}
\noindent Note that since the connections need to be kept, the surface of the sphere must follow the circle in the middle of the sphere.

Then, by applying the transformation
\begin{figure}[H]\label{figure5}
\centering
\fbox{\setlength{\fboxsep}{0pt}{\includegraphics[scale=0.2]{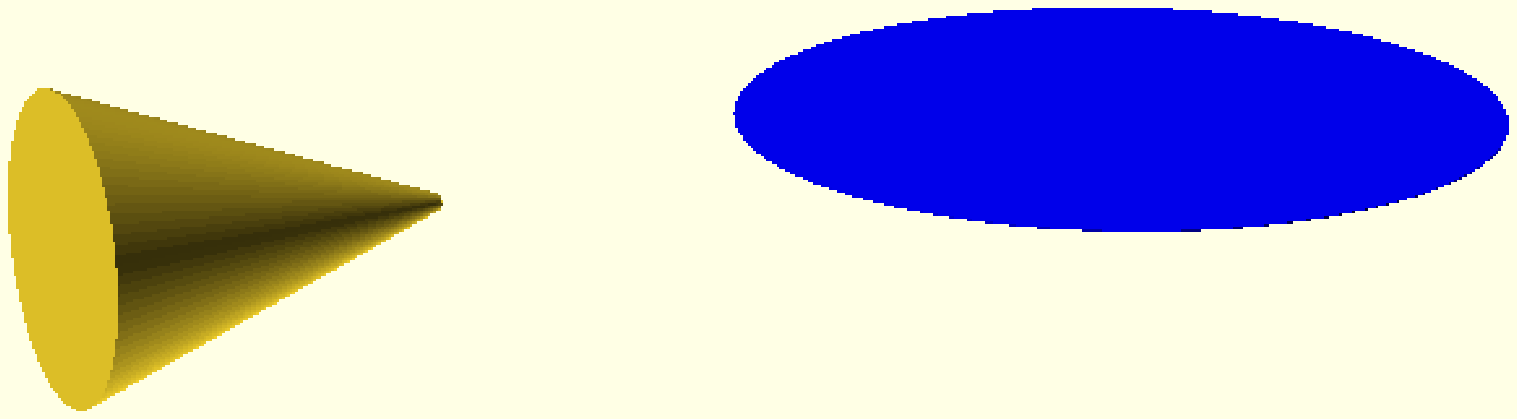}}}
,
%\caption{3D notation for a rotation of $45^\circ$ and scaling by $2$.}
\end{figure}
\noindent which dissolves the blue circle, we get the following torus.
\begin{figure}[H]\label{figure6}
\centering
\fbox{\setlength{\fboxsep}{0pt}{\includegraphics[scale=0.2]{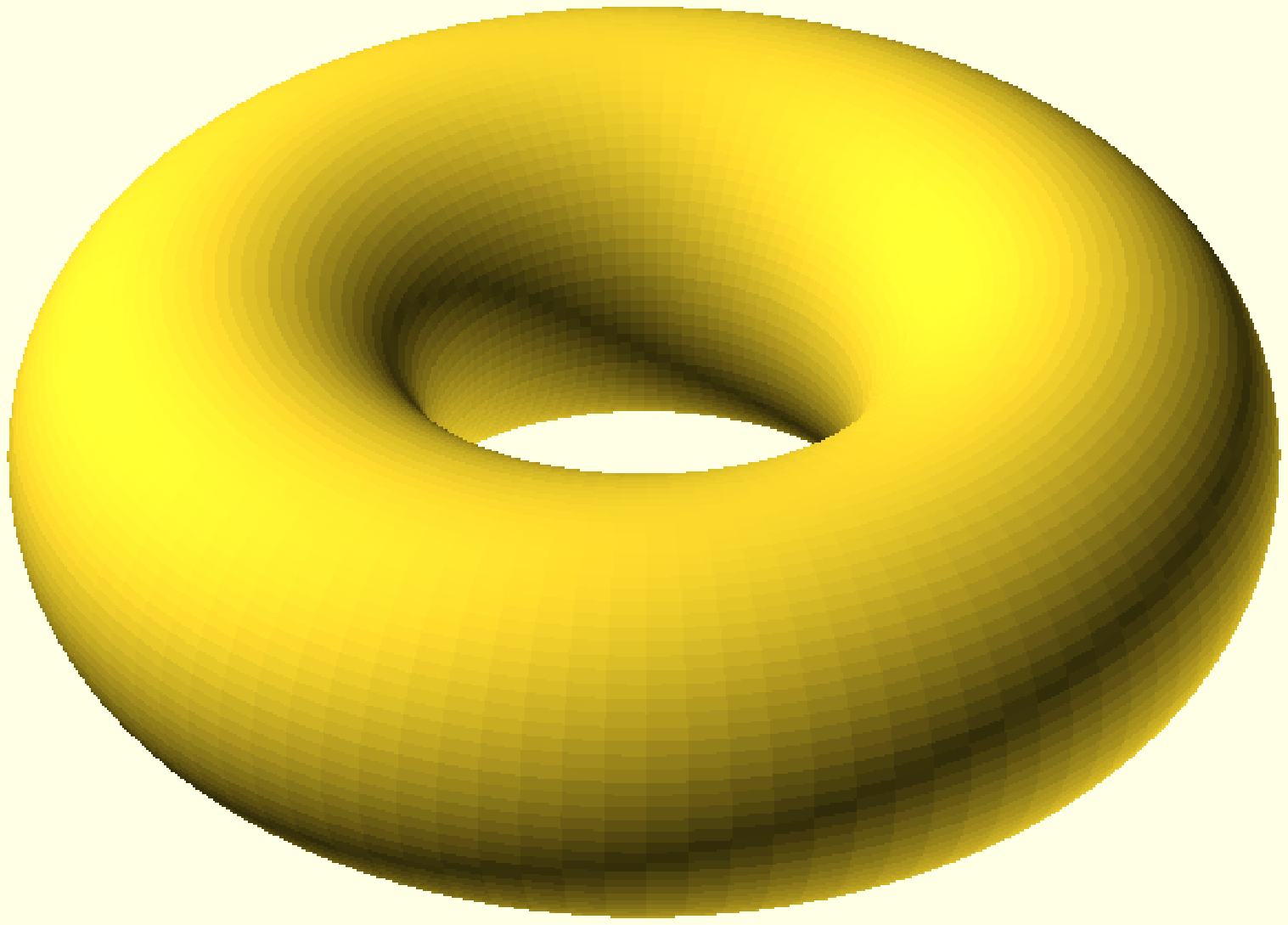}}}
.
%\caption{3D notation for a rotation of $45^\circ$ and scaling by $2$.}
\end{figure}

A three dimensional notation could be the new way to present scientific and mathematical ideas. It is possible that scientific papers of the future won't take the form of a written paper, but will be presented in a 3D virtual environment. The preparations and computations involved in such papers might be better accomplished in an interactive dynamic 3D virtual environment.

\subsection{Diagram versus Inline Notation}

Until now we have used a combination of two dimensional notations and inline notations (which is a notation which can be written was a line of text). The advantage of this is that it is sometimes faster to write and can be the basis of a computing language. The disadvantage is that it is difficult to refer to geometric properties such as angles and shapes.

Eventually, all 2D or 3D diagrams should be able to be encoded inline; although the inline notation might be hard to read; that language is closer to classical mathematics and programming. The best would be to become fluent in diagram and inline notation in a way which would allow the use of the strengths of each notation.

We now introduce the symbol ``$!$'' which when placed immediately before similar symbols will indicate that we are referring to the same symbol. For example, in the set $\{x,x,x\}$ we have three occurrences of the symbol $x$, but in the set $\{!x,!x,x\}$ there are exactly two occurrences of the symbol $x$. This is a useful notation to write geometric shapes inline.

For example, the triangular graph $\scalebox{0.3} % Change this value to rescale the drawing.
{
\begin{pspicture}(0,-0.75)(1.78,0.75)
\psline[linewidth=0.04cm](0.96,0.33)(1.38,-0.29)
\psline[linewidth=0.04cm](0.46,-0.51)(1.28,-0.51)
\psdots[dotsize=0.4](1.56,-0.51)
\psdots[dotsize=0.4](0.2,-0.53)
\psdots[dotsize=0.4](0.82,0.53)
\psline[linewidth=0.04cm](0.62,0.33)(0.3,-0.31)
\end{pspicture}
}
$ can be write inline as $!\bullet-\bullet-\bullet-!\bullet$.

\subsection{Equations and solutions}

One of the first mathematical equations we can think of is $2+2=4$. This is an equation since we agree that the symbols $2+2$ mean exactly $4$ or that the statement $2+2=4$ is true. One key thing about equations is that we can also put a variable $x$ in an equation and then ask to find the value of $x$ which will make the equation true. In this case, the value of $x$ would be known as a solution of the equation. For example, if we want to solve the equation $2+2=x$, we can say that $4$ is a solution of this equation. A solution of the equation $2+x=10$ would be $8$. There is not always a unique solution, for example $x^2=4$ has $2$ and $-2$ as solutions.

Equations are not only restricted to mathematics, but are in fact quite common. Day to day questions such as `What is the time?' and `How old are you?' can also be considered to be equations. For example, we can view the question `What is the time?' as being asked to solve the equation $\text{Present time} = x$ and `How old are you?' as solving for $x$ the equation $x= \text{(Today's date)} - \text{(Date of your birth)}$.

\subsubsection*{Equations}

Equations and solutions to equations are an essential part of a language. Finding a remedy or discovering a more precise biological process can be viewed as finding the solution of equations involving transformations. Because systems of transformations can also be reduced, our equations will take a different form. For now we will only define three types of equations, one involving the naming symbols and the other involving the reduction symbol. The equations $a:=b$, $a=:b$, $a:=:b$ and $a=b$ will mean that  the label $a$ can be used instead of $b$ in any circumstance. Note that after naming a transformation by using $a:=b$, we can stop using the colon symbol and write $a=b$. The second type of equation is $x\Rightarrow y$ which can be read as `the system of transformations x reduces to y'.  If the system $y$ also reduces to $x$ we will write $x\Leftrightarrow y$. An example of this is when the two systems are the same but with different labels and names. Note that $x\Leftrightarrow x$ is always true.

If both $x$ and $y$ reduces to the same form $c$ we will write $x\Rightarrow c \Leftarrow y$. This is also considered to be an equation. For example, the following two systems $S_1$ and $S_2$ reduce to the same form $(abab)$.
$$S_1:=\fbox{$ab\rhd c$}^{\,2S}\rightarrow (cc)$$
$$S_2:=\fbox{$abab\rhd b$}\rightarrow(b)$$
Thus we can write the equation
$$S_1\Rightarrow (abab) \Leftarrow S_2.$$

When we have such an equation, we will say that the system $S_1$ and $S_2$ are \textit{reduction equivalent} and write this as $S_1\asymp S_2$. This is our third type of equation.

Usually, equations will be expressions involving variables, forms, transformations, systems, naming symbols, the double arrow reduction symbols and the symbol `$\asymp$'. Since reductions do not always terminate, it would eventually be interesting to define new types of relations to help compare different systems.

\subsubsection*{Solutions}

We now present two short examples of solutions of equations. We will look at solutions to equations in more detail in section \ref{Computing remedies}.
Now that we have equations (or reductions) involving forms, we can include variables in an equation and ask which forms are solutions to this equation. For example, with $X$ as a variable, we can wonder which form will satisfy the equation
$$S_1:=\fbox{$ab\rhd X$}^{\,2S}\rightarrow (cc)\Rightarrow (abab).$$
The solution to this equation is $X=c$. In the case of the equation
$$S_1:=\fbox{$ab\rhd X$}^{\,2S}\rightarrow (efef)\Rightarrow (abab),$$
the solution would be $X=ef$.

\subsection{Finer and coarser models}
We now give an example of finer and coarser model. Take the transformation $T(C,A):=\fbox{$C\rhd A$}$. Applied to the form $(A)$ we get $(C)$. Applying the transformation $T(B,A):=\fbox{$B\rhd A$}$ followed by $T(C,B):=\fbox{$C\rhd B$}$ to $(A)$ will reduce also to $(C)$. This means that $T(C,A)\Rightarrow (A)$ is reduction equivalent to $T(C,B)\rightarrow T(B,A)\rightarrow (A)$. Since the two systems are initially both applied on the same initial form, we will also say that $T(C,B)\rightarrow T(B,A)$ is a \textit{finer} model than $T(C,A)$ or that $T(C,A)$ is a \textit{coarser} model than $T(C,B)\rightarrow T(B,A)$.

This can be seen as having refined the process of changing $A$ into $B$ by writing the subprocess involved in this coarser change.

\subsection{Higher Order}\label{higher order}

We have seen that each initial form is affected by a collection of transformations and how reductions are accomplished when there is one initial form. We now extend our language to include multiple initial forms in a system such that transformations which are applied to an initial form can themselves be affected by other transformations, thus making the language a higher order language. Moreover, a transformation will also be able to affect itself, thus allowing recurrence relations and transformations.

Let's look at a first example.

%Higher order change seed
\begin{equation}\scalebox{0.7} % Change this value to rescale the drawing.
{
\begin{pspicture}(0,-2.59)(13.22,2.59)
\pscircle[linewidth=0.04,linestyle=dashed,dash=0.16cm 0.16cm,dimen=outer](11.63,-1.0){1.59}
\usefont{T1}{ptm}{m}{n}
\rput(11.572657,-0.455){\large OakSeed}
\usefont{T1}{ptm}{m}{n}
\rput(11.586875,-1.595){\large MapleSeed}
\psline[linewidth=0.04cm,arrowsize=0.16cm 8.0,arrowlength=1.4,arrowinset=0.4]{->}(7.94,-0.99)(10.04,-1.01)
\psframe[linewidth=0.04,dimen=outer](7.82,-0.25)(2.42,-1.71)
\usefont{T1}{ptm}{m}{n}
\rput(6.206875,-0.995){\large MapleSeed}
\usefont{T1}{ptm}{m}{n}
\rput(3.2676563,-0.995){\large Sprout}
\psframe[linewidth=0.04,dimen=outer](5.42,2.59)(0.0,1.13)
\usefont{T1}{ptm}{m}{n}
\rput(4.046875,1.865){\large MapleSeed}
\usefont{T1}{ptm}{m}{n}
\rput(1.1126562,1.865){\large OakSeed}
\rput{-90.0}(5.45,3.47){\pstriangle[linewidth=0.04,dimen=outer](4.46,-1.31)(0.72,0.64)}
\rput{-90.0}(0.65,4.47){\pstriangle[linewidth=0.04,dimen=outer](2.56,1.59)(0.72,0.64)}
\psline[linewidth=0.04cm,arrowsize=0.16cm 8.0,arrowlength=1.4,arrowinset=0.4]{->}(5.48,1.89)(6.22,-0.51)
\psellipse[linewidth=0.04,linestyle=dashed,dash=0.16cm 0.16cm,dimen=outer](6.17,-1.01)(1.39,0.58)
\end{pspicture}
}
\end{equation}

We notice that depending on which transformation is applied first, we will have a different type of seed sprouting. We thus have to introduce a new notation to indicate the order in which the arrows are applied.

Numbers over the arrows indicate the order in which each arrow is applied. The arrow with the smallest number is applied first. It is understood that there is no number for series of transformations since there is only one initial form and the closest transformation to the initial form is applied first. Similarly to parallel transformations, if there are no numbers on the arrows or the same number on multiple arrows, then the choice of the applied transformation is arbitrary.

Here are the reductions for different numberings. Note that if both arrows are not numbered or numbered with the same number, the reduction will be an arbitrary reduction between these two.

%Higher order compare reductions
\begin{equation}\scalebox{0.5} % Change this value to rescale the drawing.
{
\begin{pspicture}(0,-5.44)(27.72,5.44)
\pscircle[linewidth=0.04,linestyle=dashed,dash=0.16cm 0.16cm,dimen=outer](11.63,1.83){1.59}
\usefont{T1}{ptm}{m}{n}
\rput(11.572657,2.375){\large OakSeed}
\usefont{T1}{ptm}{m}{n}
\rput(11.586875,1.235){\large MapleSeed}
\psline[linewidth=0.04cm,arrowsize=0.16cm 8.0,arrowlength=1.4,arrowinset=0.4]{->}(7.94,1.84)(10.04,1.82)
\psframe[linewidth=0.04,dimen=outer](7.82,2.58)(2.42,1.12)
\usefont{T1}{ptm}{m}{n}
\rput(6.206875,1.835){\large MapleSeed}
\usefont{T1}{ptm}{m}{n}
\rput(3.2676563,1.835){\large Sprout}
\psframe[linewidth=0.04,dimen=outer](5.42,5.42)(0.0,3.96)
\usefont{T1}{ptm}{m}{n}
\rput(4.046875,4.695){\large MapleSeed}
\usefont{T1}{ptm}{m}{n}
\rput(1.1126562,4.695){\large OakSeed}
\rput{-90.0}(2.54,6.26){\pstriangle[linewidth=0.04,dimen=outer](4.4,1.54)(0.72,0.64)}
\rput{-90.0}(-2.18,7.3){\pstriangle[linewidth=0.04,dimen=outer](2.56,4.42)(0.72,0.64)}
\psline[linewidth=0.04cm,arrowsize=0.16cm 8.0,arrowlength=1.4,arrowinset=0.4]{->}(5.48,4.72)(6.22,2.32)
\psellipse[linewidth=0.04,linestyle=dashed,dash=0.16cm 0.16cm,dimen=outer](6.17,1.82)(1.39,0.58)
\usefont{T1}{ptm}{m}{n}
\rput(6.037344,3.735){\large 1}
\usefont{T1}{ptm}{m}{n}
\rput(8.79625,2.175){\large 2}
\psline[linewidth=0.04cm,arrowsize=0.16cm 4.0,arrowlength=1.4,arrowinset=0.4,doubleline=true,doublesep=0.12]{->}(7.58,-0.14)(7.6,-1.88)
\pscircle[linewidth=0.04,linestyle=dashed,dash=0.16cm 0.16cm,dimen=outer](7.53,-3.85){1.59}
\usefont{T1}{ptm}{m}{n}
\rput(7.467656,-3.305){\large Sprout}
\usefont{T1}{ptm}{m}{n}
\rput(7.486875,-4.445){\large MapleSeed}
\pscircle[linewidth=0.04,linestyle=dashed,dash=0.16cm 0.16cm,dimen=outer](26.07,1.85){1.59}
\usefont{T1}{ptm}{m}{n}
\rput(26.012657,2.395){\large OakSeed}
\usefont{T1}{ptm}{m}{n}
\rput(26.026875,1.255){\large MapleSeed}
\psline[linewidth=0.04cm,arrowsize=0.16cm 8.0,arrowlength=1.4,arrowinset=0.4]{->}(22.38,1.86)(24.48,1.84)
\psframe[linewidth=0.04,dimen=outer](22.26,2.6)(16.86,1.14)
\usefont{T1}{ptm}{m}{n}
\rput(20.646875,1.855){\large MapleSeed}
\usefont{T1}{ptm}{m}{n}
\rput(17.707657,1.855){\large Sprout}
\psframe[linewidth=0.04,dimen=outer](19.86,5.44)(14.44,3.98)
\usefont{T1}{ptm}{m}{n}
\rput(18.486876,4.715){\large MapleSeed}
\usefont{T1}{ptm}{m}{n}
\rput(15.552656,4.715){\large OakSeed}
\rput{-90.0}(16.98,20.7){\pstriangle[linewidth=0.04,dimen=outer](18.84,1.54)(0.72,0.64)}
\rput{-90.0}(12.24,21.76){\pstriangle[linewidth=0.04,dimen=outer](17.0,4.44)(0.72,0.64)}
\psline[linewidth=0.04cm,arrowsize=0.16cm 8.0,arrowlength=1.4,arrowinset=0.4]{->}(19.92,4.74)(20.66,2.34)
\psellipse[linewidth=0.04,linestyle=dashed,dash=0.16cm 0.16cm,dimen=outer](20.61,1.84)(1.39,0.58)
\usefont{T1}{ptm}{m}{n}
\rput(20.51625,3.755){\large 2}
\usefont{T1}{ptm}{m}{n}
\rput(23.197344,2.195){\large 1}
\psline[linewidth=0.04cm,arrowsize=0.16cm 4.0,arrowlength=1.4,arrowinset=0.4,doubleline=true,doublesep=0.12]{->}(22.02,-0.12)(22.04,-1.86)
\pscircle[linewidth=0.04,linestyle=dashed,dash=0.16cm 0.16cm,dimen=outer](26.13,-3.59){1.59}
\usefont{T1}{ptm}{m}{n}
\rput(26.072657,-3.045){\large OakSeed}
\usefont{T1}{ptm}{m}{n}
\rput(26.047657,-4.165){\large Sprout}
\psline[linewidth=0.04cm,arrowsize=0.16cm 8.0,arrowlength=1.4,arrowinset=0.4]{->}(22.44,-3.58)(24.54,-3.6)
\usefont{T1}{ptm}{m}{n}
\rput(23.29625,-3.245){\large 2}
\psframe[linewidth=0.04,dimen=outer](22.38,-2.86)(16.96,-4.32)
\usefont{T1}{ptm}{m}{n}
\rput(21.006874,-3.585){\large MapleSeed}
\usefont{T1}{ptm}{m}{n}
\rput(18.072657,-3.585){\large OakSeed}
\rput{-90.0}(23.06,15.98){\pstriangle[linewidth=0.04,dimen=outer](19.52,-3.86)(0.72,0.64)}
\end{pspicture}
}
\end{equation}

Having access to higher order transformations can be interesting to model complex systems and also for testing of systems. When building a first order model, collections of transformations applied to the whole model can help refine the model.

We can also have a transformation point to two different initial forms. Note that when a transformation points to two forms, it will disappear after it was applied unless there is a number at the superscript indicating the presence of multiple copies of the transformation.

%Higher order transformation to 2 initial.
\begin{equation}\scalebox{0.65} % Change this value to rescale the drawing.
{
\begin{pspicture}(0,-4.72)(20.34,4.72)
\psframe[linewidth=0.04,dimen=outer](6.14,2.62)(2.06,0.4)
\usefont{T1}{ptm}{m}{n}
\rput(5.58875,1.52){\Large A}
\rput{-90.0}(3.4,6.44){\pstriangle[linewidth=0.04,dimen=outer](4.92,1.2)(0.72,0.64)}
\psline[linewidth=0.04cm,arrowsize=0.16cm 8.0,arrowlength=1.4,arrowinset=0.4]{->}(6.26,1.56)(7.84,1.52)
\usefont{T1}{ptm}{m}{n}
\rput(8.72875,1.52){\Large A}
\psframe[linewidth=0.04,dimen=outer](2.92,4.72)(0.0,3.58)
\usefont{T1}{ptm}{m}{n}
\rput(2.28875,4.12){\Large A}
\rput{-90.0}(-2.6,5.68){\pstriangle[linewidth=0.04,dimen=outer](1.54,3.82)(0.72,0.64)}
\usefont{T1}{ptm}{m}{n}
\rput(0.62296873,4.12){\Large B}
\usefont{T1}{ptm}{m}{n}
\rput(3.48875,2.04){\Large A}
\usefont{T1}{ptm}{m}{n}
\rput(3.8829687,1.12){\Large C}
\usefont{T1}{ptm}{m}{n}
\rput(2.9829688,1.12){\Large C}
\psline[linewidth=0.04cm,arrowsize=0.16cm 8.0,arrowlength=1.4,arrowinset=0.4]{->}(3.0,4.06)(3.5,2.4)
\pscircle[linewidth=0.04,linestyle=dashed,dash=0.16cm 0.16cm,dimen=outer](3.47,1.51){0.95}
\psline[linewidth=0.04cm,arrowsize=0.16cm 8.0,arrowlength=1.4,arrowinset=0.4]{->}(2.98,4.16)(8.0,2.04)
\psline[linewidth=0.04cm,arrowsize=0.16cm 4.0,arrowlength=1.4,arrowinset=0.4,doubleline=true,doublesep=0.12,doublecolor=white]{->}(10.16,2.88)(12.32,2.88)
\usefont{T1}{ptm}{m}{n}
\rput(5.384844,3.5){\Large 2}
\usefont{T1}{ptm}{m}{n}
\rput(6.8104687,1.82){\Large 3}
\usefont{T1}{ptm}{m}{n}
\rput(3.3826563,3.38){\Large 1}
\pscircle[linewidth=0.04,linestyle=dashed,dash=0.16cm 0.16cm,dimen=outer](8.75,1.51){0.95}
\usefont{T1}{ptm}{m}{n}
\rput(14.102969,3.28){\Large B}
\usefont{T1}{ptm}{m}{n}
\rput(14.522968,2.36){\Large C}
\usefont{T1}{ptm}{m}{n}
\rput(13.622969,2.36){\Large C}
\pscircle[linewidth=0.04,linestyle=dashed,dash=0.16cm 0.16cm,dimen=outer](14.11,2.75){0.95}
\psframe[linewidth=0.04,dimen=outer](6.14,-2.5)(2.06,-4.72)
\usefont{T1}{ptm}{m}{n}
\rput(5.58875,-3.6){\Large A}
\rput{-90.0}(8.52,1.32){\pstriangle[linewidth=0.04,dimen=outer](4.92,-3.92)(0.72,0.64)}
\psline[linewidth=0.04cm,arrowsize=0.16cm 8.0,arrowlength=1.4,arrowinset=0.4]{->}(6.26,-3.56)(7.84,-3.6)
\usefont{T1}{ptm}{m}{n}
\rput(8.72875,-3.6){\Large A}
\psframe[linewidth=0.04,dimen=outer](2.92,-0.4)(0.0,-1.54)
\usefont{T1}{ptm}{m}{n}
\rput(2.28875,-1.0){\Large A}
\rput{-90.0}(2.52,0.56){\pstriangle[linewidth=0.04,dimen=outer](1.54,-1.3)(0.72,0.64)}
\usefont{T1}{ptm}{m}{n}
\rput(0.62296873,-1.0){\Large B}
\usefont{T1}{ptm}{m}{n}
\rput(3.48875,-3.08){\Large A}
\usefont{T1}{ptm}{m}{n}
\rput(3.8829687,-4.0){\Large C}
\usefont{T1}{ptm}{m}{n}
\rput(2.9829688,-4.0){\Large C}
\psline[linewidth=0.04cm,arrowsize=0.16cm 8.0,arrowlength=1.4,arrowinset=0.4]{->}(3.0,-1.06)(3.5,-2.72)
\pscircle[linewidth=0.04,linestyle=dashed,dash=0.16cm 0.16cm,dimen=outer](3.47,-3.61){0.95}
\psline[linewidth=0.04cm,arrowsize=0.16cm 8.0,arrowlength=1.4,arrowinset=0.4]{->}(2.98,-0.96)(8.0,-3.08)
\psline[linewidth=0.04cm,arrowsize=0.16cm 4.0,arrowlength=1.4,arrowinset=0.4,doubleline=true,doublesep=0.12,doublecolor=white]{->}(10.16,-2.24)(12.32,-2.24)
\usefont{T1}{ptm}{m}{n}
\rput(5.342656,-1.62){\Large 1}
\usefont{T1}{ptm}{m}{n}
\rput(6.8104687,-3.3){\Large 3}
\usefont{T1}{ptm}{m}{n}
\rput(3.4248438,-1.74){\Large 2}
\pscircle[linewidth=0.04,linestyle=dashed,dash=0.16cm 0.16cm,dimen=outer](8.75,-3.61){0.95}
\psframe[linewidth=0.04,dimen=outer](16.78,-1.2)(12.7,-3.42)
\usefont{T1}{ptm}{m}{n}
\rput(16.22875,-2.3){\Large A}
\rput{-90.0}(17.86,13.26){\pstriangle[linewidth=0.04,dimen=outer](15.56,-2.62)(0.72,0.64)}
\psline[linewidth=0.04cm,arrowsize=0.16cm 8.0,arrowlength=1.4,arrowinset=0.4]{->}(16.9,-2.26)(18.48,-2.3)
\usefont{T1}{ptm}{m}{n}
\rput(19.342968,-2.3){\Large B}
\usefont{T1}{ptm}{m}{n}
\rput(14.12875,-1.78){\Large A}
\usefont{T1}{ptm}{m}{n}
\rput(14.522968,-2.7){\Large C}
\usefont{T1}{ptm}{m}{n}
\rput(13.622969,-2.7){\Large C}
\pscircle[linewidth=0.04,linestyle=dashed,dash=0.16cm 0.16cm,dimen=outer](14.11,-2.31){0.95}
\usefont{T1}{ptm}{m}{n}
\rput(17.450468,-2.0){\Large 3}
\pscircle[linewidth=0.04,linestyle=dashed,dash=0.16cm 0.16cm,dimen=outer](19.39,-2.31){0.95}
\end{pspicture}
}\end{equation}

We now show an example of a transformation affecting itself.

%Higher order self affecting
\begin{equation}\scalebox{0.7} % Change this value to rescale the drawing.
{
\begin{pspicture}(0,-1.57)(15.52,1.55)
\psline[linewidth=0.04cm,arrowsize=0.16cm 8.0,arrowlength=1.4,arrowinset=0.4]{->}(2.18,-1.55)(1.4,0.01)
\psframe[linewidth=0.04,dimen=outer](4.8,1.55)(0.0,-0.39)
\usefont{T1}{ptm}{m}{n}
\rput(3.28875,0.59){\Large A}
\rput{-90.0}(1.93,3.19){\pstriangle[linewidth=0.04,dimen=outer](2.56,0.31)(0.72,0.64)}
\usefont{T1}{ptm}{m}{n}
\rput(0.7029688,0.59){\Large B}
\usefont{T1}{ptm}{m}{n}
\rput(1.64875,0.59){\Large A}
\psellipse[linewidth=0.04,linestyle=dashed,dash=0.16cm 0.16cm,dimen=outer](1.17,0.62)(0.91,0.63)
\rput{-90.0}(5.41,4.45){\psarc[linewidth=0.04](4.93,-0.48){1.07}{0.0}{180.0}}
\psline[linewidth=0.04cm](4.96,-1.55)(2.16,-1.55)
\psline[linewidth=0.04cm,arrowsize=0.16cm 4.0,arrowlength=1.4,arrowinset=0.4,doubleline=true,doublesep=0.12]{->}(6.6,0.57)(8.18,0.57)
\psframe[linewidth=0.04,dimen=outer](15.52,1.51)(8.92,-0.43)
\usefont{T1}{ptm}{m}{n}
\rput(13.42875,0.55){\Large A}
\rput{-90.0}(12.11,13.29){\pstriangle[linewidth=0.04,dimen=outer](12.7,0.27)(0.72,0.64)}
\usefont{T1}{ptm}{m}{n}
\rput(10.842969,0.55){\Large B}
\usefont{T1}{ptm}{m}{n}
\rput(11.78875,0.55){\Large A}
\psellipse[linewidth=0.04,linestyle=dashed,dash=0.16cm 0.16cm,dimen=outer](10.76,0.58)(1.46,0.63)
\usefont{T1}{ptm}{m}{n}
\rput(9.862968,0.55){\Large B}
\end{pspicture}
}
\end{equation}

In the following example, we have a transformation which is applied 3 times to itself. To denote this, we will use a superscript with a number followed by the letter $R$ indicating a recurrence.

%Higher order self affecting three
\begin{equation}\scalebox{0.7} % Change this value to rescale the drawing.
{
\begin{pspicture}(0,-3.61875)(11.44,3.65875)
\psline[linewidth=0.04cm,arrowsize=0.16cm 8.0,arrowlength=1.4,arrowinset=0.4]{->}(2.22,0.42125)(1.44,1.98125)
\psframe[linewidth=0.04,dimen=outer](4.84,3.52125)(0.04,1.58125)
\usefont{T1}{ptm}{m}{n}
\rput(3.32875,2.56125){\Large A}
\rput{-90.0}(-0.00125,5.20125){\pstriangle[linewidth=0.04,dimen=outer](2.6,2.28125)(0.72,0.64)}
\usefont{T1}{ptm}{m}{n}
\rput(0.74296874,2.56125){\Large B}
\usefont{T1}{ptm}{m}{n}
\rput(1.68875,2.56125){\Large A}
\psellipse[linewidth=0.04,linestyle=dashed,dash=0.16cm 0.16cm,dimen=outer](1.21,2.59125)(0.91,0.63)
\rput{-90.0}(3.47875,6.46125){\psarc[linewidth=0.04](4.97,1.49125){1.07}{0.0}{180.0}}
\psline[linewidth=0.04cm](5.0,0.42125)(2.2,0.42125)
\psline[linewidth=0.04cm,arrowsize=0.16cm 4.0,arrowlength=1.4,arrowinset=0.4,doubleline=true,doublesep=0.12,doublecolor=white]{->}(3.04,0.16125)(3.04,-1.35875)
\psframe[linewidth=0.04,dimen=outer](11.44,-1.67875)(0.0,-3.61875)
\usefont{T1}{ptm}{m}{n}
\rput(6.60875,-2.65875){\Large A}
\rput{-90.0}(8.39875,3.24125){\pstriangle[linewidth=0.04,dimen=outer](5.82,-2.89875)(0.72,0.64)}
\usefont{T1}{ptm}{m}{n}
\rput(1.9229687,-2.63875){\Large B}
\usefont{T1}{ptm}{m}{n}
\rput(2.8429687,-2.63875){\Large B}
\psellipse[linewidth=0.04,linestyle=dashed,dash=0.16cm 0.16cm,dimen=outer](2.81,-2.64875)(2.43,0.73)
\usefont{T1}{ptm}{m}{n}
\rput(0.9429687,-2.63875){\Large B}
\usefont{T1}{ptm}{m}{n}
\rput(5.2298436,3.45625){\large 3R}
\usefont{T1}{ptm}{m}{n}
\rput(3.8629687,-2.65875){\Large B}
\usefont{T1}{ptm}{m}{n}
\rput(4.76875,-2.65875){\Large A}
\end{pspicture}
}
\end{equation}

\section{Probabilities}

When we have two transformations applied in parallel to an initial set, we are naturally faced with the question of which transformation should be applied. We can take advantage of this by introducing the concept of probability in our language. Take the transformation

$$[\:\fbox{$heads\rhd coin$},\fbox{$tails\rhd coin$}\,]\rightarrow \{coin\}$$

As we already mentioned, we will interpret the transformation as saying that there is a $1$ chance out of $2$ that the initial set containing a single coin is reduced to heads and a $1$ chance out of $2$ (or $0.5$ chance) that the initial set is reduced to tails. The chances of being applied can also be indicated by using numbers on the subscript of each transformation as follows.

$$[\:\fbox{$heads\rhd coin$}_{\,0.5},\fbox{$tails\rhd coin$}_{\,0.5}\,]\rightarrow \{coin\}$$

For more occurrences of each transformation, we have to differentiate between series and parallel transformations. For example,
$$[\:\fbox{$heads\rhd coin$}^{\,2P},\fbox{$tails\rhd coin$}\,]\rightarrow \{coin\}$$
is understood as each of the three transformation having $1/3$ chance of being applied.

For series of transformations, the example
$$[\:\fbox{$heads\rhd coin$}^{\,2S},\fbox{$tails\rhd coin$}^{\,2S}\,]\rightarrow \{coin, coin\}$$
is understood as a $1/2$ chance for each transformation to be applied first, but before the same transformation is applied a second time, all other transformations need to be applied once if they can be applied. This system will automatically result
$$[\:\fbox{$heads\rhd coin$}^{\,1S},\fbox{$tails\rhd coin$}^{\,1S}\,]\rightarrow \{heads, tails\}$$
The following system
$$[\:\fbox{$heads\rhd coin$}^{\,2S},\fbox{$tails\rhd coin$}^{\,2S}\,]\rightarrow \{coin, coin\}$$
has $1/2$ chance of reducing to $$[\:\fbox{$tails\rhd coin$}^{\,2S}\,]\rightarrow \{heads, tails, heads\}$$ and $1/2$ chance of reducing to $$[\:\fbox{$heads\rhd coin$}^{\,2S}\,]\rightarrow \{heads, tails, tails\}.$$

Another example is
$$[\:\fbox{$1\rhd dice$}_{\,\frac{1}{6}},\fbox{$2\rhd dice$}_{\,\frac{1}{6}},\fbox{$3\rhd dice$}_{\,\frac{1}{6}},\fbox{$4\rhd dice$}_{\,\frac{1}{6}},\fbox{$5\rhd dice$}_{\,\frac{1}{6}},\fbox{$6\rhd dice$}_{\,\frac{1}{6}}\,]\rightarrow \{dice\}.$$
We will interpret the transformation as saying that there is a $\frac{1}{6}$ chance of getting any of numbers 1 to 6 on one roll.

The transformation
$$[\:\fbox{$heads\rhd coin$}_{\,\frac{1}{3}},\fbox{$tails\rhd coin$}_{\,\frac{1}{3}},\fbox{$tails\rhd coin$}_{\,\frac{1}{3}}\,]\rightarrow \{coin\}$$
is interpreted as $1/3$ chance of getting heads and $1/3$ chance of getting tails for each tails, for a total of $2/3$ chance of getting tails. So in general, if $n$ is the number of parallel transformations, then there is $1/n$ chance of that transformation to be applied. Note that in this case also, we do not need to use subscripts since no subscript on parallel transformations indicates similar chances of being applied.

We now add a new subscript notation, which will give different probabilities to each of the parallel transformations. The transformation
$$[\:\fbox{$heads\rhd coin$}_{\,0.1},\fbox{$tails\rhd coin$}_{\,0.9}\,]\rightarrow \{coin\}$$
is interpreted as saying that there is a $10\%$ chance that the initial set containing a single coin is reduced to heads and a $90\%$ chance that the initial set is reduced to tails. Note that if there is no subscript on the parallel transformations, then we consider they have the same chances of being applied and as mentioned above; have a probability of $1/n$.

The sum of the probabilities does not necessarily need to be $1$ as in our example $0.1+.90$. If the sum is $S$ and is smaller than $1$, we will consider that the remainder $1-S$ is the probability that no transformation is applied. For example, $$[\:\fbox{$heads\rhd coin$}_{\,0.25},\fbox{$tails\rhd coin$}_{\,0.15}\,]\rightarrow \{coin\}$$
means that there is a $0.6=1-(0.25+0.15)$ chance that a flipped coin does not reach the ground or that the coin lands on its side, thus not giving a heads or tails value.

Each subscript does not need to be smaller than $1$. If we have
$$[\:\fbox{$heads\rhd coin$}_{\,1},\fbox{$tails\rhd coin$}_{\,1}\,]\rightarrow \{coin, coin\},$$
this means that the two transformations will be applied simultaneously and will give $\{heads, tails\}$. When applying transformations simultaneously, there there is a possibility of conflict. For example, if we have
$$[\:\fbox{$heads\rhd coin$}_{\,1},\fbox{$tails\rhd coin$}_{\,1}\,]\rightarrow \{coin\},$$
we should apply both transformations simultaneously, but there is only one coin. This will be reduced to $\{heads\wedge tails\}$, which is read `both heads and tails'. Importantly, this can only happen if we cannot avoid the conflict, meaning that $[\:\fbox{$heads\rhd coin$}_{\,1},\fbox{$tails\rhd coin$}_{\,1}\,]\rightarrow \{coin, coin\},$ will not reduce to $\{heads\wedge tails, coin\}$ because there are enough coins in the initial set.

If we have
$$[\:\fbox{$heads\rhd coin$}_{\,1.1},\fbox{$tails\rhd coin$}_{\,1.9}\,]\rightarrow \{coin, coin, coin\},$$
we will interpret this as applying simultaneously the two transformations and then applying once one of the two transformations with respective probabilities of $0.1$ and $0.9$. Thus, this expression will reduce after the first step to $$[\:\fbox{$heads\rhd coin$}_{\,1.1},\fbox{$tails\rhd coin$}_{\,1.9}\,]\rightarrow \{heads, tails, coin\},$$
which will reduce after the second step to $\{heads, tails, heads\}$ with a $0.1$ chance and  $\{heads, tails, tails\}$ with a $0.9$ chance.

We can also use the probability notation in a series of transformations.

$$\fbox{$heads\rhd coin$}_{\,0.85}\rightarrow\fbox{$tails\rhd coin$}_{\,0.2}\rightarrow \{coin\}$$
means that there is $0.2$ chance of the coin becoming tails after the first step and $0.8$ chance of staying a coin. At the second step, there is $0.85$ chance of the coin becoming tails after the first step and $0.15$ chance of staying a coin.

\section{Summary of the notations}

We now give a summary of the notations which were defined above.

\subsubsection*{Basic notation}
\begin{itemize}
  \item Any symbol or collection of symbols is called a form. There are no restrictions on what a form can be. Example of forms are strings of letters, diagrams and three dimensional objects.
  \item $\fbox{$Y \rhd X$}$ is a transformation and represents the potential to replace $X$ by $Y$.
  \item The right-hand side of the symbol $\rhd$ is called the cause of the transformation and the left-hand side is called effect of the transformation.
  \item The initial form is the form to which the transformations are applied.
  \item $\fbox{$Y \rhd X$}\rightarrow (Z)$ is said to be a system.
  \item A system can be reduced when the right-hand side of the $\rhd$ symbol corresponds to an element in the initial form. Reduction is indicated with the symbol $\Rightarrow$. For example,
  $\fbox{$Y \rhd X$}\rightarrow (A, X, B)\Rightarrow (A, Y, B)$.
\end{itemize}

\subsubsection*{Naming}

\begin{itemize}
  \item Names are given by using the notation `$:=$', `$=:$' and `$:=:$' where the position of the colon indicates the side where the name appears.
  \item Identifier is usually only a word to designate a form. For example, $\text{XYtransformation}:=\fbox{$Y \rhd X$}$.
  \item Function forms are composed of a label and variables where we can substitute the variable by forms. For example $\text{transform(X,Y)}:=\fbox{$Y \rhd X$}$ is evaluated at $a$ and $b$ by writing $\text{transform(a,b)}:=\fbox{$a \rhd b$}$.
  \item Abstraction of a form is done by taking a form and changing some elements into variables. For example, the ordered set $\{a, b, c\}$ can be abstracted in $b$ by writing $Abstraction(X):=\{a, X, c\}$.
\end{itemize}

\subsubsection*{Series and parallel transformations}

\begin{itemize}
  \item Transformations applied in series are written as $\fbox{$Y \rhd X$}\rightarrow\fbox{$W \rhd V$}\rightarrow (Z)$ where the closest to the initial form is applied first.
  \item When $n$ copies of a transformation is applied in series, we can denote this by the superscript $nS$ as in $\fbox{$Y \rhd X$}^{\,nS}\rightarrow (Z)$. We can have an infinite number of copies by writing $\infty$ instead of $n$.
  \item Transformations applied in parallel are written under square brackets as $[\fbox{$Y \rhd X$},\fbox{$W \rhd V$}]\rightarrow (Z)$.
  \item When $n$ copies of a transformation are applied in parallel, we can denote this by the superscript $nP$ as in $[\fbox{$Y \rhd X$}^{\,nP}]\rightarrow (Z)$.
  \item For series and parallel transformations, we can have an infinite number of copies of transformations by writing $\infty$ instead of $n$.
  \item For series and parallel transformations, the $\sharp$ symbol on the superscript indicates that the transformations replace all they can until they cannot be applied, then the transformations disappear.
\end{itemize}

\subsubsection*{Distance and order}

  \begin{itemize}
    \item If the distances or order of the elements of the initial form are important, we use ellipses with dashed lines in diagram view; and parentheses $(\,\,)$ in the inline view. When the distances and order are not important, we use ellipses with full lines in diagram view and braces $\{\,\,\}$ in the inline view.
    \item A transformation with disjoint elements on the cause of a transformation does not require that the order or the distance between them to be the same as in the initial form to be applied. For example, we have that $\fbox{$X, Y \rhd C, A$}\rightarrow  (A, B, C)\Rightarrow(Y, B, X)$
    \item If we write the subscript $rigid$ on $\rhd$, then order and distance need to be respected. For example, $\fbox{$X, Y \rhd_{rigid} B, C$}\rightarrow  (A, B, C)\Rightarrow (A, X, Y).$
  \end{itemize}

\subsubsection*{Colors}

\begin{itemize}
  \item Black forms at the same position in the cause and effect of a transformation indicates that the form in the cause will replace the one from the effect.
  \item Forms of the same color in the cause and effect of a transformation indicates that the form in the cause will replace the one from the effect.
  \item The forms in red in the cause of a transformation indicates that the transformation cannot be applied if the form in red
appears in the initial form. Red is reserved for this interpretation.
\end{itemize}

\subsubsection*{Connections}
\begin{itemize}
  \item Connections between nodes or vertices are usually denoted by lines.
  \item If form A is replaced by form B, all the connections that node A has are transferred to B and B keeps its previous connections.
\end{itemize}

\section{Mathematical models in medicine}

We now present the main application of our language which is mathematical medicine. To enhance our motivation, have objectives and have an idea of the important points regarding the construction for an efficient mathematical language of medicine, we can keep in mind the following list of milestones to be attained.
\bigskip

\textbf{\large{Milestones}}
\begin{itemize}
  \item A large open database incorporating all known biological processes (this is mostly a translation phase, where all papers are translated into a common mathematical framework).
  \item Computing treatments with positive results.
  \item Discovering new processes or refining a process.
  \item Having a global model which makes animal testing irrelevant.
  \item Having individually tailored medicine.
  \item Widely accessible medicinal compounds which can be printed, sent or created in local laboratories.
  \item Curing, understanding or controlling all known diseases and conditions.
  \item Prediction of potential future diseases and study of theoretical biological systems in different types of environments.
\end{itemize}

\bigskip

We will present in the following sections different biological models ranging from DNA replication up to a beating heart.

\subsection{Cell-division cycle}\label{Cell-division cycle}

The division cycle of a cell with a membrane-bound nucleus can be divided into three periods: interphase, mitosis and cytokinesis. In the interphase, the DNA is duplicated, during the mitosis, the duplicated DNA strands are separated and at the end of cytokinesis, we observe two daughters cells.

At the coarsest level, cell division can be modelled as follows. Note that in the following reduction, each transformation is named by using the ``$:=$'' symbol.

%Cell division
\begin{equation*}
\scalebox{0.6} % Change this value to rescale the drawing.
{
\begin{pspicture}(0,-1.37)(21.88,1.37)
\pscircle[linewidth=0.04,dimen=outer](1.17,0.02){1.17}
\pscustom[linewidth=0.04,linecolor=red]
{
\newpath
\moveto(0.9,0.67)
\lineto(0.89,0.59)
\curveto(0.885,0.55)(0.88,0.44)(0.88,0.37)
\curveto(0.88,0.3)(0.885,0.215)(0.9,0.17)
}
\pscustom[linewidth=0.04,linecolor=blue]
{
\newpath
\moveto(1.34,-0.01)
\lineto(1.33,-0.09)
\curveto(1.325,-0.13)(1.32,-0.245)(1.32,-0.32)
\curveto(1.32,-0.395)(1.325,-0.51)(1.33,-0.55)
\curveto(1.335,-0.59)(1.34,-0.635)(1.34,-0.65)
}
\pscustom[linewidth=0.04,linecolor=blue]
{
\newpath
\moveto(0.64,0.09)
\lineto(0.63,0.01)
\curveto(0.625,-0.03)(0.62,-0.14)(0.62,-0.21)
\curveto(0.62,-0.28)(0.625,-0.365)(0.64,-0.41)
}
\pscustom[linewidth=0.04,linecolor=red]
{
\newpath
\moveto(1.64,0.59)
\lineto(1.63,0.51)
\curveto(1.625,0.47)(1.62,0.355)(1.62,0.28)
\curveto(1.62,0.205)(1.625,0.09)(1.63,0.05)
\curveto(1.635,0.01)(1.64,-0.035)(1.64,-0.05)
}
\pscircle[linewidth=0.04,dimen=outer](5.93,0.0){1.37}
\pscustom[linewidth=0.04,linecolor=red]
{
\newpath
\moveto(5.62,0.67)
\lineto(5.61,0.59)
\curveto(5.605,0.55)(5.6,0.44)(5.6,0.37)
\curveto(5.6,0.3)(5.605,0.215)(5.62,0.17)
}
\pscustom[linewidth=0.04,linecolor=blue]
{
\newpath
\moveto(6.06,-0.01)
\lineto(6.05,-0.09)
\curveto(6.045,-0.13)(6.04,-0.245)(6.04,-0.32)
\curveto(6.04,-0.395)(6.045,-0.51)(6.05,-0.55)
\curveto(6.055,-0.59)(6.06,-0.635)(6.06,-0.65)
}
\pscustom[linewidth=0.04,linecolor=blue]
{
\newpath
\moveto(5.36,0.09)
\lineto(5.35,0.01)
\curveto(5.345,-0.03)(5.34,-0.14)(5.34,-0.21)
\curveto(5.34,-0.28)(5.345,-0.365)(5.36,-0.41)
}
\pscustom[linewidth=0.04,linecolor=red]
{
\newpath
\moveto(6.36,0.59)
\lineto(6.35,0.51)
\curveto(6.345,0.47)(6.34,0.355)(6.34,0.28)
\curveto(6.34,0.205)(6.345,0.09)(6.35,0.05)
\curveto(6.355,0.01)(6.36,-0.035)(6.36,-0.05)
}
\pscustom[linewidth=0.04,linecolor=red]
{
\newpath
\moveto(6.28,0.59)
\lineto(6.29,0.51)
\curveto(6.295,0.47)(6.3,0.355)(6.3,0.28)
\curveto(6.3,0.205)(6.295,0.09)(6.29,0.05)
\curveto(6.285,0.01)(6.28,-0.035)(6.28,-0.05)
}
\pscustom[linewidth=0.04,linecolor=red]
{
\newpath
\moveto(5.54,0.67)
\lineto(5.55,0.59)
\curveto(5.555,0.55)(5.56,0.44)(5.56,0.37)
\curveto(5.56,0.3)(5.555,0.215)(5.54,0.17)
}
\pscustom[linewidth=0.04,linecolor=blue]
{
\newpath
\moveto(5.98,-0.01)
\lineto(5.99,-0.09)
\curveto(5.995,-0.13)(6.0,-0.245)(6.0,-0.32)
\curveto(6.0,-0.395)(5.995,-0.51)(5.99,-0.55)
\curveto(5.985,-0.59)(5.98,-0.635)(5.98,-0.65)
}
\pscustom[linewidth=0.04,linecolor=blue]
{
\newpath
\moveto(5.28,0.09)
\lineto(5.29,0.01)
\curveto(5.295,-0.03)(5.3,-0.14)(5.3,-0.21)
\curveto(5.3,-0.28)(5.295,-0.365)(5.28,-0.41)
}
\pscircle[linewidth=0.04,dimen=outer](11.2,0.05){1.24}
\pscircle[linewidth=0.04,dimen=outer](13.04,0.05){1.26}
\pscustom[linewidth=0.04,linecolor=red]
{
\newpath
\moveto(12.98,0.75)
\lineto(12.97,0.67)
\curveto(12.965,0.63)(12.96,0.52)(12.96,0.45)
\curveto(12.96,0.38)(12.965,0.295)(12.98,0.25)
}
\pscustom[linewidth=0.04,linecolor=blue]
{
\newpath
\moveto(13.42,0.07)
\lineto(13.41,-0.01)
\curveto(13.405,-0.05)(13.4,-0.165)(13.4,-0.24)
\curveto(13.4,-0.315)(13.405,-0.43)(13.41,-0.47)
\curveto(13.415,-0.51)(13.42,-0.555)(13.42,-0.57)
}
\pscustom[linewidth=0.04,linecolor=blue]
{
\newpath
\moveto(12.72,0.17)
\lineto(12.71,0.09)
\curveto(12.705,0.05)(12.7,-0.06)(12.7,-0.13)
\curveto(12.7,-0.2)(12.705,-0.285)(12.72,-0.33)
}
\pscustom[linewidth=0.04,linecolor=red]
{
\newpath
\moveto(13.72,0.67)
\lineto(13.71,0.59)
\curveto(13.705,0.55)(13.7,0.435)(13.7,0.36)
\curveto(13.7,0.285)(13.705,0.17)(13.71,0.13)
\curveto(13.715,0.09)(13.72,0.045)(13.72,0.03)
}
\psframe[linewidth=0.04,linecolor=white,dimen=outer,fillstyle=solid](12.52,0.87)(11.74,-0.75)
\psline[linewidth=0.06cm,linestyle=dashed,dash=0.16cm 0.16cm](12.1,0.91)(12.1,-0.81)
\pscustom[linewidth=0.04,linecolor=red]
{
\newpath
\moveto(11.22,0.75)
\lineto(11.23,0.67)
\curveto(11.235,0.63)(11.24,0.52)(11.24,0.45)
\curveto(11.24,0.38)(11.235,0.295)(11.22,0.25)
}
\pscustom[linewidth=0.04,linecolor=blue]
{
\newpath
\moveto(10.78,0.07)
\lineto(10.79,-0.01)
\curveto(10.795,-0.05)(10.8,-0.165)(10.8,-0.24)
\curveto(10.8,-0.315)(10.795,-0.43)(10.79,-0.47)
\curveto(10.785,-0.51)(10.78,-0.555)(10.78,-0.57)
}
\pscustom[linewidth=0.04,linecolor=blue]
{
\newpath
\moveto(11.48,0.17)
\lineto(11.49,0.09)
\curveto(11.495,0.05)(11.5,-0.06)(11.5,-0.13)
\curveto(11.5,-0.2)(11.495,-0.285)(11.48,-0.33)
}
\pscustom[linewidth=0.04,linecolor=red]
{
\newpath
\moveto(10.48,0.67)
\lineto(10.49,0.59)
\curveto(10.495,0.55)(10.5,0.435)(10.5,0.36)
\curveto(10.5,0.285)(10.495,0.17)(10.49,0.13)
\curveto(10.485,0.09)(10.48,0.045)(10.48,0.03)
}
\pscircle[linewidth=0.04,dimen=outer](20.71,0.02){1.17}
\pscustom[linewidth=0.04,linecolor=red]
{
\newpath
\moveto(20.44,0.67)
\lineto(20.43,0.59)
\curveto(20.425,0.55)(20.42,0.44)(20.42,0.37)
\curveto(20.42,0.3)(20.425,0.215)(20.44,0.17)
}
\pscustom[linewidth=0.04,linecolor=blue]
{
\newpath
\moveto(20.88,-0.01)
\lineto(20.87,-0.09)
\curveto(20.865,-0.13)(20.86,-0.245)(20.86,-0.32)
\curveto(20.86,-0.395)(20.865,-0.51)(20.87,-0.55)
\curveto(20.875,-0.59)(20.88,-0.635)(20.88,-0.65)
}
\pscustom[linewidth=0.04,linecolor=blue]
{
\newpath
\moveto(20.18,0.09)
\lineto(20.17,0.01)
\curveto(20.165,-0.03)(20.16,-0.14)(20.16,-0.21)
\curveto(20.16,-0.28)(20.165,-0.365)(20.18,-0.41)
}
\pscustom[linewidth=0.04,linecolor=red]
{
\newpath
\moveto(21.18,0.59)
\lineto(21.17,0.51)
\curveto(21.165,0.47)(21.16,0.355)(21.16,0.28)
\curveto(21.16,0.205)(21.165,0.09)(21.17,0.05)
\curveto(21.175,0.01)(21.18,-0.035)(21.18,-0.05)
}
\pscircle[linewidth=0.04,dimen=outer](18.13,0.04){1.17}
\pscustom[linewidth=0.04,linecolor=red]
{
\newpath
\moveto(18.4,0.69)
\lineto(18.41,0.61)
\curveto(18.415,0.57)(18.42,0.46)(18.42,0.39)
\curveto(18.42,0.32)(18.415,0.235)(18.4,0.19)
}
\pscustom[linewidth=0.04,linecolor=blue]
{
\newpath
\moveto(17.96,0.01)
\lineto(17.97,-0.07)
\curveto(17.975,-0.11)(17.98,-0.225)(17.98,-0.3)
\curveto(17.98,-0.375)(17.975,-0.49)(17.97,-0.53)
\curveto(17.965,-0.57)(17.96,-0.615)(17.96,-0.63)
}
\pscustom[linewidth=0.04,linecolor=blue]
{
\newpath
\moveto(18.66,0.11)
\lineto(18.67,0.03)
\curveto(18.675,-0.01)(18.68,-0.12)(18.68,-0.19)
\curveto(18.68,-0.26)(18.675,-0.345)(18.66,-0.39)
}
\pscustom[linewidth=0.04,linecolor=red]
{
\newpath
\moveto(17.66,0.61)
\lineto(17.67,0.53)
\curveto(17.675,0.49)(17.68,0.375)(17.68,0.3)
\curveto(17.68,0.225)(17.675,0.11)(17.67,0.07)
\curveto(17.665,0.03)(17.66,-0.015)(17.66,-0.03)
}
\psline[linewidth=0.06cm,arrowsize=0.02cm 2.0,arrowlength=1.4,arrowinset=0.4,doubleline=true,doublesep=0.12,doublecolor=white]{->}(14.94,0.11)(16.38,0.13)
\usefont{T1}{ptm}{m}{n}
\rput(3.3314064,1.16){$Interphase$}
\usefont{T1}{ptm}{m}{n}
\rput(8.441406,1.16){$Mitosis$}
\usefont{T1}{ptm}{m}{n}
\rput(15.551406,1.16){$Cytokinesis$}
\psline[linewidth=0.06cm,arrowsize=0.02cm 2.0,arrowlength=1.4,arrowinset=0.4,doubleline=true,doublesep=0.12,doublecolor=white]{->}(7.8,0.11)(9.24,0.13)
\psline[linewidth=0.06cm,arrowsize=0.02cm 2.0,arrowlength=1.4,arrowinset=0.4,doubleline=true,doublesep=0.12,doublecolor=white]{->}(2.72,0.09)(4.16,0.11)
\usefont{T1}{ptm}{m}{n}
\rput{-90.0}(7.775,9.105938){\rput(8.421406,0.66){$:=$}}
\usefont{T1}{ptm}{m}{n}
\rput{-90.0}(2.635,3.8859375){\rput(3.2414062,0.62){$:=$}}
\usefont{T1}{ptm}{m}{n}
\rput{-90.0}(14.855,16.185938){\rput(15.501407,0.66){$:=$}}
\end{pspicture}
}
\end{equation*}

\subsection{Interphase}\label{Interphase}

The transformation $Interphase$ can be further refined by defining the following model named $Interphase3$. Both transformations applied on an initial cell will reduce to the same form, but since $Interphase3$ is composed of three transformations instead of one, we write the number $3$ next to the name interphase. As a general rule, the higher the number, the more refined the model, but this might not always be the case.

%interphase3
\begin{equation*}\scalebox{0.7} % Change this value to rescale the drawing.
{
\begin{pspicture}(0,-1.9092188)(17.58,1.9492188)
\definecolor{color69}{rgb}{0.0,0.8,0.0}
\definecolor{color73}{rgb}{0.2,0.8,0.0}
\pscustom[linewidth=0.04,linecolor=color69]
{
\newpath
\moveto(9.930552,0.012394104)
\lineto(9.915749,-0.110186465)
\curveto(9.908347,-0.17147675)(9.900945,-0.34768617)(9.900945,-0.4626056)
\curveto(9.900945,-0.577525)(9.908347,-0.75373477)(9.915749,-0.81502503)
\curveto(9.923149,-0.8763153)(9.930552,-0.9452673)(9.930552,-0.968251)
}
\pscustom[linewidth=0.04,linecolor=color69]
{
\newpath
\moveto(7.738268,0.012394104)
\lineto(7.7234645,-0.110186465)
\curveto(7.716063,-0.17147675)(7.7086616,-0.34768617)(7.7086616,-0.4626056)
\curveto(7.7086616,-0.577525)(7.716063,-0.75373477)(7.7234645,-0.81502503)
\curveto(7.730866,-0.8763153)(7.738268,-0.9452673)(7.738268,-0.968251)
}
\pscustom[linewidth=0.04,linecolor=color69]
{
\newpath
\moveto(7.6198425,0.012394104)
\lineto(7.634646,-0.110186465)
\curveto(7.642047,-0.17147675)(7.649449,-0.34768617)(7.649449,-0.4626056)
\curveto(7.649449,-0.577525)(7.642047,-0.75373477)(7.634646,-0.81502503)
\curveto(7.6272445,-0.8763153)(7.6198425,-0.9452673)(7.6198425,-0.968251)
}
\psframe[linewidth=0.04,dimen=outer](10.72,0.41078126)(6.96,-1.4892187)
\pscircle[linewidth=0.04,linecolor=color73,dimen=outer](15.95,-0.51921874){0.59}
\psline[linewidth=0.02cm,linestyle=dashed,dash=0.16cm 0.16cm](15.94,-0.5492188)(16.5,-0.5492188)
\usefont{T1}{ptm}{m}{n}
\rput(15.98,-0.35921875){r}
\pscircle[linewidth=0.04,linecolor=color73,dimen=outer](13.5,-0.50921875){0.68}
\psline[linewidth=0.02cm,linestyle=dashed,dash=0.16cm 0.16cm](13.54,-0.52921873)(14.16,-0.52921873)
\usefont{T1}{ptm}{m}{n}
\rput(13.561406,-0.25921875){$R_1$}
\psframe[linewidth=0.04,dimen=outer](17.04,0.41078126)(12.34,-1.4892187)
\rput{-90.0}(15.325309,14.392398){\pstriangle[linewidth=0.04,dimen=outer](14.858853,-0.747602)(0.64552706,0.56229365)}
\rput{-90.0}(9.305308,8.332398){\pstriangle[linewidth=0.04,dimen=outer](8.818853,-0.767602)(0.64552706,0.56229365)}
\psline[linewidth=0.04cm,arrowsize=0.16cm 8.0,arrowlength=1.4,arrowinset=0.4]{->}(10.94,-0.5492188)(12.12,-0.5492188)
\usefont{T1}{ptm}{m}{n}
\rput(14.151406,0.74078125){$CellGrowth(R_1, r):=$}
\usefont{T1}{ptm}{m}{n}
\rput(8.401406,0.72078127){$DnaReplication:=$}
\psline[linewidth=0.04cm,arrowsize=0.16cm 8.0,arrowlength=1.4,arrowinset=0.4]{->}(5.52,-0.50921875)(6.7,-0.50921875)
\psframe[linewidth=0.04,linestyle=dotted,dotsep=0.16cm,dimen=outer](17.58,1.3107812)(0.0,-1.9092188)
\usefont{T1}{ptm}{m}{n}
\rput(1.4514062,1.7607813){$Interphase_3=$}
\pscircle[linewidth=0.04,linecolor=color73,dimen=outer](4.2,-0.5492188){0.68}
\psline[linewidth=0.02cm,linestyle=dashed,dash=0.16cm 0.16cm](4.24,-0.56921875)(4.86,-0.56921875)
\usefont{T1}{ptm}{m}{n}
\rput(4.2614064,-0.29921874){$R_1$}
\psframe[linewidth=0.04,dimen=outer](5.22,0.39078125)(0.52,-1.5092187)
\rput{-90.0}(3.5253084,2.552398){\pstriangle[linewidth=0.04,dimen=outer](3.0388532,-0.767602)(0.64552706,0.56229365)}
\usefont{T1}{ptm}{m}{n}
\rput(2.4814062,0.72078127){$CellGrowth(R_2,R_1)=$}
\pscircle[linewidth=0.04,linecolor=color73,dimen=outer](1.64,-0.5492188){0.8}
\psline[linewidth=0.02cm,linestyle=dashed,dash=0.16cm 0.16cm](1.74,-0.56921875)(2.38,-0.56921875)
\usefont{T1}{ptm}{m}{n}
\rput(1.7014062,-0.31921875){$R_2$}
\usefont{T1}{ptm}{m}{n}
\rput(11.031406,0.46078125){$\sharp S$}
\end{pspicture}
}
\end{equation*}

$CellGrowth(r,R_1)$ is usually called the $G_1$-phase, $CellGrowth(R_1,R_2)$ is called the $G_2$-phase and $DnaReplication$ is called $S$-phase.

\subsection{DNA replication}\label{DNA replication}

We will now give a system that replicates DNA. The collection of transformations used in the process can be considered to be a refined model of the transformation $DnaReplication$.

\subsubsection{Diagram for the replication of the right leading strand}

%DNA replication leading_final
\begin{equation*}\scalebox{0.7} % Change this value to rescale the drawing.
{
\begin{pspicture}(0,-7.2475)(21.522812,7.2875)
\definecolor{color952}{rgb}{0.2,0.8,0.0}
\definecolor{color957}{rgb}{0.6,0.0,0.6}
\definecolor{color967b}{rgb}{1.0,1.0,0.6}
\definecolor{color1001b}{rgb}{1.0,0.6,0.0}
\definecolor{color1010b}{rgb}{0.6,0.8,1.0}
\definecolor{color1140}{rgb}{0.0,0.8,0.0}
\rput{-90.0}(13.618438,2.1234374){\pstriangle[linewidth=0.04,dimen=outer](7.8709373,-6.0275)(0.66,0.56)}
\psframe[linewidth=0.04,dimen=outer,fillstyle=solid,fillcolor=color967b](9.200937,-5.1075)(8.800938,-5.8475)
\psframe[linewidth=0.04,dimen=outer,fillstyle=solid,fillcolor=color967b](10.320937,-5.1075)(9.920938,-5.8475)
\pswedge[linewidth=0.04](10.120937,-5.8275){0.18}{0.0}{180.0}
\psframe[linewidth=0.04,dimen=outer](9.200937,-5.8075)(8.800938,-6.5475)
\psframe[linewidth=0.04,dimen=outer,fillstyle=solid,fillcolor=color967b](5.9809375,-5.0875)(5.5809374,-5.8275)
\psframe[linewidth=0.04,dimen=outer,fillstyle=solid,fillcolor=color967b](7.1009374,-5.0875)(6.7009373,-5.8275)
\psframe[linewidth=0.04,dimen=outer](5.9809375,-5.7875)(5.5809374,-6.5275)
\psline[linewidth=0.04cm](6.3009377,-6.5275)(5.2209377,-6.5275)
\psframe[linewidth=0.04,dimen=outer](7.1009374,-5.7875)(6.7009373,-6.5275)
\psline[linewidth=0.04cm](7.4209375,-6.5275)(6.3009377,-6.5275)
\psline[linewidth=0.04cm](9.540937,-6.5275)(8.420938,-6.5275)
\usefont{T1}{ptm}{m}{n}
\rput(6.912344,-4.0375){$\text{Polymerase III leading(C,G)}=$}
\psframe[linewidth=0.04,dimen=outer](11.120937,-4.3675)(4.7009373,-7.2475)
\usefont{T1}{ptm}{m}{n}
\rput(10.108125,-5.3375){C}
\usefont{T1}{ptm}{m}{n}
\rput(6.9020314,-6.2575){G}
\usefont{T1}{ptm}{m}{n}
\rput(6.888125,-5.3375){C}
\usefont{T1}{ptm}{m}{n}
\rput(3.3023438,2.7825){$\text{Initiate Polymerase III leading(T,A)}=$}
\psframe[linewidth=0.04,dimen=outer](6.6409373,2.5325)(2.0809374,-0.3075)
\psframe[linewidth=0.04,dimen=outer,fillstyle=solid,fillcolor=color967b](3.3809376,1.9125)(2.9809375,1.1725)
\psframe[linewidth=0.04,dimen=outer](3.3809376,1.2125)(2.9809375,0.4725)
\usefont{T1}{ptm}{m}{n}
\rput(3.1875,0.7225){A}
\usefont{T1}{ptm}{m}{n}
\rput(3.1609375,1.7025){T}
\psline[linewidth=0.04cm](3.7009375,0.4925)(2.6209376,0.4925)
\rput{-90.0}(3.2584374,5.7234373){\pstriangle[linewidth=0.04,dimen=outer](4.4909377,0.9525)(0.66,0.56)}
\psframe[linewidth=0.04,dimen=outer,fillstyle=solid,fillcolor=color1001b](5.8409376,1.9325)(5.4409375,1.1925)
\pswedge[linewidth=0.04](5.6409373,1.2325){0.18}{0.0}{180.0}
\usefont{T1}{ptm}{m}{n}
\rput(5.6209373,1.7225){T}
\psline[linewidth=0.04cm](11.780937,-0.1675)(9.4609375,-0.1675)
\psframe[linewidth=0.04,dimen=outer,fillstyle=solid,fillcolor=color1001b](10.180938,-0.1475)(9.780937,-0.8875)
\psframe[linewidth=0.04,dimen=outer,fillstyle=solid,fillcolor=color1010b](10.180938,-0.8475)(9.780937,-1.5875)
\usefont{T1}{ptm}{m}{n}
\rput(9.9609375,-0.4375){T}
\usefont{T1}{ptm}{m}{n}
\rput(9.9675,-1.2775){A}
\psframe[linewidth=0.04,dimen=outer,fillstyle=solid,fillcolor=color967b](11.260938,-0.1475)(10.860937,-0.8875)
\psline[linewidth=0.04cm](11.680938,-1.5675)(9.4609375,-1.5675)
\psframe[linewidth=0.04,dimen=outer,fillstyle=solid,fillcolor=color1010b](11.260938,-0.8475)(10.860937,-1.5875)
\usefont{T1}{ptm}{m}{n}
\rput(11.028125,-0.4575){C}
\usefont{T1}{ptm}{m}{n}
\rput(11.042031,-1.2775){G}
\psline[linewidth=0.04cm](13.780937,-0.1675)(11.660937,-0.1675)
\psline[linewidth=0.04cm](13.780937,-1.5675)(11.660937,-1.5675)
\psframe[linewidth=0.04,dimen=outer,fillstyle=solid,fillcolor=color967b](12.380938,-0.1475)(11.980938,-0.8875)
\psframe[linewidth=0.04,dimen=outer,fillstyle=solid,fillcolor=color1010b](12.380938,-0.8475)(11.980938,-1.5875)
\usefont{T1}{ptm}{m}{n}
\rput(12.1675,-0.4575){A}
\usefont{T1}{ptm}{m}{n}
\rput(12.160937,-1.2775){T}
\psframe[linewidth=0.04,dimen=outer,fillstyle=solid,fillcolor=color967b](13.4609375,-0.1475)(13.060938,-0.8875)
\psframe[linewidth=0.04,dimen=outer,fillstyle=solid,fillcolor=color1010b](13.4609375,-0.8475)(13.060938,-1.5875)
\usefont{T1}{ptm}{m}{n}
\rput(13.2475,-0.4375){A}
\usefont{T1}{ptm}{m}{n}
\rput(13.260938,-1.2575){T}
\usefont{T1}{ptm}{m}{n}
\rput(3.5123436,-1.1575){$\text{Initiate Polymerase III leading(A,T)}^1$}
\usefont{T1}{ptm}{m}{n}
\rput(3.5623438,-2.5175){$\text{Initiate Polymerase III leading(C,G)}^1$}
\usefont{T1}{ptm}{m}{n}
\rput(3.5223436,-1.8575){$\text{Initiate Polymerase III leading(G,C)}^1$}
\usefont{T1}{ptm}{m}{n}
\rput(16.812344,-4.6775){$\text{Polymerase III leading(T,A)}^{\infty S}$}
\usefont{T1}{ptm}{m}{n}
\rput(15.792344,-5.2775){$\text{Polymerase III leading(A,T)}^{\infty S}$}
\usefont{T1}{ptm}{m}{n}
\rput(17.982344,-4.0175){$\text{Polymerase III leading(G,C)}^{\infty S}$}
\psline[linewidth=0.04cm,arrowsize=0.16cm 8.0,arrowlength=1.4,arrowinset=0.4]{->}(11.360937,2.7325)(11.360937,1.6875)
\psline[linewidth=0.04cm,arrowsize=0.16cm 8.0,arrowlength=1.4,arrowinset=0.4]{->}(7.1609373,0.9525)(8.500937,0.2325)
\psline[linewidth=0.04cm,arrowsize=0.16cm 8.0,arrowlength=1.4,arrowinset=0.4]{->}(7.0409374,-1.1475)(8.300938,-0.9875)
\psline[linewidth=0.04cm,arrowsize=0.16cm 8.0,arrowlength=1.4,arrowinset=0.4]{->}(7.0809374,-1.8075)(8.440937,-1.5075)
\psline[linewidth=0.04cm,arrowsize=0.16cm 8.0,arrowlength=1.4,arrowinset=0.4]{->}(7.2009373,-2.4875)(8.680938,-2.1275)
\psline[linewidth=0.04cm,arrowsize=0.16cm 8.0,arrowlength=1.4,arrowinset=0.4]{->}(9.540937,-3.6675)(9.980938,-2.4475)
\psline[linewidth=0.04cm,arrowsize=0.16cm 8.0,arrowlength=1.4,arrowinset=0.4]{->}(12.560938,-4.8275)(12.320937,-2.8675)
\psline[linewidth=0.04cm,arrowsize=0.16cm 8.0,arrowlength=1.4,arrowinset=0.4]{->}(13.720938,-4.2675)(13.060938,-2.5675)
\psline[linewidth=0.04cm,arrowsize=0.16cm 8.0,arrowlength=1.4,arrowinset=0.4]{->}(14.980938,-3.7075)(13.540937,-2.2075)
\usefont{T1}{ptm}{m}{n}
\rput(6.8678126,2.1225){1}
\usefont{T1}{ptm}{m}{n}
\rput(11.592343,-4.6375){$\infty S$}
\usefont{T1}{ptm}{m}{n}
\rput(9.964063,-0.8775){H}
\psline[linewidth=0.04cm](15.940937,-0.1675)(13.600938,-0.1675)
\psline[linewidth=0.04cm](15.940937,-1.5675)(13.620937,-1.5675)
\psframe[linewidth=0.04,dimen=outer,fillstyle=solid,fillcolor=color967b](14.540937,-0.1475)(14.140938,-0.8875)
\psframe[linewidth=0.04,dimen=outer,fillstyle=solid,fillcolor=color1010b](14.540937,-0.8475)(14.140938,-1.5875)
\usefont{T1}{ptm}{m}{n}
\rput(14.322031,-0.4575){G}
\usefont{T1}{ptm}{m}{n}
\rput(14.328125,-1.2775){C}
\psframe[linewidth=0.04,dimen=outer,fillstyle=solid,fillcolor=color967b](15.620937,-0.1475)(15.220938,-0.8875)
\psframe[linewidth=0.04,dimen=outer,fillstyle=solid,fillcolor=color1010b](15.620937,-0.8475)(15.220938,-1.5875)
\usefont{T1}{ptm}{m}{n}
\rput(15.388125,-0.4375){C}
\usefont{T1}{ptm}{m}{n}
\rput(15.442031,-1.2575){G}
\usefont{T1}{ptm}{m}{n}
\rput(12.317187,4.8475){H}
\psframe[linewidth=0.04,dimen=outer](14.000937,6.8375)(8.4609375,2.9375)
\usefont{T1}{ptm}{m}{n}
\rput(9.350938,7.1075){Helicase:=}
\psframe[linewidth=0.04,linecolor=color952,dimen=outer](9.500937,6.4975)(9.100938,5.7575)
\pswedge[linewidth=0.04,linecolor=color952](9.300938,5.7775){0.18}{0.0}{180.0}
\rput{-180.0}(18.601875,7.955){\pswedge[linewidth=0.04,linecolor=color957](9.300938,3.9775){0.18}{0.0}{180.0}}
\rput{-90.0}(6.5334377,16.328438){\pstriangle[linewidth=0.04,dimen=outer](11.430938,4.6175)(0.66,0.56)}
\psframe[linewidth=0.04,linecolor=color957,dimen=outer](9.500937,3.9975)(9.100938,3.2575)
\psframe[linewidth=0.04,linecolor=color952,dimen=outer](12.540937,5.5975)(12.140938,4.8575)
\psframe[linewidth=0.04,linecolor=color957,dimen=outer](12.540937,4.8975)(12.140938,4.1575)
\psline[linewidth=0.04cm,linecolor=color952](9.480938,6.4775)(10.260938,5.5775)
\psline[linewidth=0.04cm,linecolor=color957](9.480938,3.2775)(10.280937,4.1975)
\psline[linewidth=0.04cm,linecolor=color1140](12.500937,5.5775)(13.240937,5.5775)
\psline[linewidth=0.04cm,linecolor=color957](12.500937,4.1775)(13.240937,4.1775)
\psframe[linewidth=0.04,dimen=outer](13.620937,5.5975)(13.220938,4.8575)
\psframe[linewidth=0.04,dimen=outer](13.620937,4.8975)(13.220938,4.1575)
\psframe[linewidth=0.04,dimen=outer](10.640938,5.5975)(10.240937,4.8575)
\psframe[linewidth=0.04,dimen=outer](10.640938,4.8975)(10.240937,4.1575)
\usefont{T1}{ptm}{m}{n}
\rput(10.417188,4.8675){H}
\usefont{T1}{ptm}{m}{n}
\rput(14.425938,6.4775){\small $\infty P$}
\end{pspicture}
}
\end{equation*}

To avoid rewriting each type of $\textit{\text{Initiate Polymerase III leading}}$, we used abstraction of the form $\textit{\text{Initiate Polymerase III leading(A,T)}}$. Since transformations preserve connections; the bonds between nucleotides are preserved. The letter written on the nucleotides is considered to be connected to the nucleotide than contains it, thus when nothing is written on the nucleotide, it is assumed that the letter is still on the nucleotide after the application of a transformation.

The system modeling the replication of the right lagging strand is now given. Instead of writing all combinations of pairs of nucleotides, we introduce the notation \textit{For all }$(X,Y)\in\{(A,T),(T,A),(G,C),(C,G)\}$ which we write at the bottom of the diagram. This notation means that each transformation over $(X,Y)$ will come in four copies, that is one for each ordered pair. Note that we have given another way to write the transformation for helicase.

\subsubsection{Diagram for the replication of the right lagging strand}

%DNA replication lagging_final
\begin{equation*}\scalebox{0.7} % Change this value to rescale the drawing.
{
\begin{pspicture}(0,-10.043906)(15.812813,10.043906)
\definecolor{color1493b}{rgb}{1.0,1.0,0.6}
\definecolor{color1496b}{rgb}{0.6,0.8,1.0}
\definecolor{color1504b}{rgb}{1.0,0.4,0.0}
\definecolor{color1523}{rgb}{0.2,0.8,0.0}
\definecolor{color1528}{rgb}{0.6,0.0,0.6}
\definecolor{color1665}{rgb}{0.0,0.8,0.0}
\usefont{T1}{ptm}{m}{n}
\rput(12.963125,7.603906){H}
\psline[linewidth=0.04cm](14.98,-0.06609375)(12.86,-0.06609375)
\psframe[linewidth=0.04,dimen=outer,fillstyle=solid,fillcolor=color1493b](14.66,-0.04609375)(14.26,-0.7860938)
\psframe[linewidth=0.04,dimen=outer,fillstyle=solid,fillcolor=color1493b](13.58,-0.04609375)(13.18,-0.7860938)
\psline[linewidth=0.04cm](14.98,-1.4660938)(12.86,-1.4660938)
\psframe[linewidth=0.04,dimen=outer,fillstyle=solid,fillcolor=color1496b](14.66,-0.74609375)(14.26,-1.4860938)
\psframe[linewidth=0.04,dimen=outer,fillstyle=solid,fillcolor=color1496b](13.58,-0.74609375)(13.18,-1.4860938)
\usefont{T1}{ptm}{m}{n}
\rput(13.366563,-0.35609376){A}
\usefont{T1}{ptm}{m}{n}
\rput(13.36,-1.1760937){T}
\psline[linewidth=0.04cm](12.98,-0.06609375)(10.66,-0.06609375)
\psframe[linewidth=0.04,dimen=outer,fillstyle=solid,fillcolor=color1493b](12.46,-0.04609375)(12.06,-0.7860938)
\psframe[linewidth=0.04,dimen=outer,fillstyle=solid,fillcolor=color1504b](11.38,-0.04609375)(10.98,-0.7860938)
\psline[linewidth=0.04cm](12.88,-1.4660938)(10.66,-1.4660938)
\psframe[linewidth=0.04,dimen=outer,fillstyle=solid,fillcolor=color1496b](12.46,-0.74609375)(12.06,-1.4860938)
\psframe[linewidth=0.04,dimen=outer,fillstyle=solid,fillcolor=color1496b](11.38,-0.74609375)(10.98,-1.4860938)
\usefont{T1}{ptm}{m}{n}
\rput(11.166562,-1.1760937){A}
\usefont{T1}{ptm}{m}{n}
\rput(11.16,-0.33609375){T}
\usefont{T1}{ptm}{m}{n}
\rput(12.227187,-0.35609376){C}
\usefont{T1}{ptm}{m}{n}
\rput(12.241094,-1.1760937){G}
\usefont{T1}{ptm}{m}{n}
\rput(14.46,-1.1560937){T}
\usefont{T1}{ptm}{m}{n}
\rput(14.446563,-0.33609375){A}
\psframe[linewidth=0.04,dimen=outer](14.64,9.593906)(9.1,5.6939063)
\usefont{T1}{ptm}{m}{n}
\rput(9.995,9.863906){Helicase:=}
\psframe[linewidth=0.04,linecolor=color1523,dimen=outer](10.14,9.253906)(9.74,8.5139065)
\pswedge[linewidth=0.04,linecolor=color1523](9.94,8.533906){0.18}{0.0}{180.0}
\rput{-180.0}(19.88,13.467813){\pswedge[linewidth=0.04,linecolor=color1528](9.94,6.7339063){0.18}{0.0}{180.0}}
\rput{-90.0}(4.416094,19.723906){\pstriangle[linewidth=0.04,dimen=outer](12.07,7.373906)(0.66,0.56)}
\psframe[linewidth=0.04,linecolor=color1528,dimen=outer](10.14,6.7539062)(9.74,6.0139065)
\psframe[linewidth=0.04,linecolor=color1523,dimen=outer](13.18,8.353907)(12.78,7.6139064)
\psframe[linewidth=0.04,linecolor=color1528,dimen=outer](13.18,7.6539063)(12.78,6.913906)
\psframe[linewidth=0.04,dimen=outer](6.42,1.1339062)(0.0,-1.7460938)
\psline[linewidth=0.04cm](0.54,0.33390626)(1.66,0.33390626)
\usefont{T1}{ptm}{m}{n}
\rput(2.4751563,1.5039062){Polymerase III lagging(X,Y):=}
\rput{-180.0}(2.12,-0.0321875){\psframe[linewidth=0.04,linecolor=color1523,dimen=outer](1.26,0.35390624)(0.86,-0.38609374)}
\rput{-180.0}(2.12,-0.0321875){\psframe[linewidth=0.04,dimen=outer](1.26,0.35390624)(0.86,-0.38609374)}
\psline[linewidth=0.04cm](1.66,0.33390626)(2.74,0.33390626)
\rput{-180.0}(4.36,-1.4321876){\psframe[linewidth=0.04,dimen=outer,fillstyle=solid,fillcolor=color1496b](2.38,-0.34609374)(1.98,-1.0860938)}
\rput{-180.0}(4.36,-0.0321875){\psframe[linewidth=0.04,dimen=outer](2.38,0.35390624)(1.98,-0.38609374)}
\rput{-180.0}(4.36,-1.4321876){\psframe[linewidth=0.04,dimen=outer,fillstyle=solid,fillcolor=color1496b](2.38,-0.34609374)(1.98,-1.0860938)}
\rput{-180.0}(4.36,-0.0321875){\psframe[linewidth=0.04,dimen=outer](2.38,0.35390624)(1.98,-0.38609374)}
\rput{-90.0}(3.6360939,2.9439063){\pstriangle[linewidth=0.04,dimen=outer](3.29,-0.62609375)(0.66,0.56)}
\rput{-90.0}(3.6360939,2.9439063){\pstriangle[linewidth=0.04,dimen=outer](3.29,-0.62609375)(0.66,0.56)}
\rput{-180.0}(8.44,-1.4321876){\psframe[linewidth=0.04,dimen=outer,fillstyle=solid,fillcolor=color1496b](4.42,-0.34609374)(4.02,-1.0860938)}
\rput{-180.0}(8.44,-0.7321875){\pswedge[linewidth=0.04](4.22,-0.36609375){0.18}{0.0}{180.0}}
\rput{-180.0}(8.44,-1.4321876){\psframe[linewidth=0.04,dimen=outer,fillstyle=solid,fillcolor=color1496b](4.42,-0.34609374)(4.02,-1.0860938)}
\rput{-180.0}(8.44,-0.7321875){\pswedge[linewidth=0.04](4.22,-0.36609375){0.18}{0.0}{180.0}}
\psline[linewidth=0.04cm](4.8,0.33390626)(5.92,0.33390626)
\psline[linewidth=0.04cm](4.8,0.33390626)(5.92,0.33390626)
\rput{-180.0}(10.68,-1.4321876){\psframe[linewidth=0.04,dimen=outer,fillstyle=solid,fillcolor=color1496b](5.54,-0.34609374)(5.14,-1.0860938)}
\rput{-180.0}(10.68,-0.0321875){\psframe[linewidth=0.04,dimen=outer](5.54,0.35390624)(5.14,-0.38609374)}
\rput{-180.0}(10.68,-1.4321876){\psframe[linewidth=0.04,dimen=outer,fillstyle=solid,fillcolor=color1496b](5.54,-0.34609374)(5.14,-1.0860938)}
\rput{-180.0}(10.68,-0.0321875){\psframe[linewidth=0.04,dimen=outer](5.54,0.35390624)(5.14,-0.38609374)}
\psframe[linewidth=0.04,dimen=outer](7.96,7.433906)(3.92,5.6339064)
\usefont{T1}{ptm}{m}{n}
\rput(4.8751564,7.8039064){Primase:=}
\rput{-180.0}(9.72,12.607813){\psframe[linewidth=0.04,dimen=outer,fillstyle=solid,fillcolor=color1496b](5.06,6.6739063)(4.66,5.933906)}
\rput{-180.0}(9.72,13.667812){\psframe[linewidth=0.04,dimen=outer,fillstyle=solid](5.06,7.0339065)(4.66,6.6339064)}
\rput{-90.0}(-0.46390626,12.443906){\pstriangle[linewidth=0.04,dimen=outer](5.99,6.1739063)(0.66,0.56)}
\rput{-90.0}(-0.46390626,12.443906){\pstriangle[linewidth=0.04,dimen=outer](5.99,6.1739063)(0.66,0.56)}
\rput{-180.0}(14.04,12.607813){\psframe[linewidth=0.04,dimen=outer,fillstyle=solid,fillcolor=color1496b](7.22,6.6739063)(6.82,5.933906)}
\rput{-180.0}(14.04,13.307813){\pswedge[linewidth=0.04](7.02,6.6539063){0.18}{0.0}{180.0}}
\psframe[linewidth=0.04,dimen=outer](6.2,4.773906)(0.1,2.1339064)
\usefont{T1}{ptm}{m}{n}
\rput(2.9735937,5.163906){Initiate Polymerase III lagging(X,Y):=}
\psline[linewidth=0.04cm](0.6,4.1739063)(1.68,4.1739063)
\rput{-180.0}(2.24,7.6478124){\psframe[linewidth=0.04,dimen=outer](1.32,4.1939063)(0.92,3.4539063)}
\rput{-180.0}(4.48,6.2878127){\psframe[linewidth=0.04,dimen=outer,fillstyle=solid,fillcolor=color1496b](2.44,3.5139062)(2.04,2.7739062)}
\rput{-180.0}(4.48,7.3478127){\psframe[linewidth=0.04,dimen=outer,fillstyle=solid](2.44,3.8739061)(2.04,3.4739063)}
\rput{-180.0}(4.48,6.9878125){\pswedge[linewidth=0.04](2.24,3.4939063){0.18}{0.0}{180.0}}
\rput{-90.0}(-0.16390625,6.8639064){\pstriangle[linewidth=0.04,dimen=outer](3.35,3.2339063)(0.66,0.56)}
\rput{-180.0}(8.36,6.2478123){\psframe[linewidth=0.04,dimen=outer,fillstyle=solid,fillcolor=color1496b](4.38,3.4939063)(3.98,2.7539062)}
\rput{-180.0}(8.36,6.9478126){\pswedge[linewidth=0.04](4.18,3.4739063){0.18}{0.0}{180.0}}
\rput{-180.0}(10.64,6.2878127){\psframe[linewidth=0.04,dimen=outer,fillstyle=solid,fillcolor=color1496b](5.52,3.5139062)(5.12,2.7739062)}
\rput{-180.0}(10.64,7.3478127){\psframe[linewidth=0.04,dimen=outer,fillstyle=solid](5.52,3.8739061)(5.12,3.4739063)}
\rput{-180.0}(10.64,6.9878125){\pswedge[linewidth=0.04](5.32,3.4939063){0.18}{0.0}{180.0}}
\psframe[linewidth=0.04,dimen=outer](7.26,-6.086094)(1.12,-8.726093)
\rput{-180.0}(13.16,-14.332188){\psframe[linewidth=0.04,linecolor=color1523,dimen=outer,fillstyle=solid](6.78,-6.9660935)(6.38,-7.3660936)}
\psline[linewidth=0.04cm](6.0,-6.646094)(4.88,-6.646094)
\rput{-180.0}(10.92,-15.392187){\psframe[linewidth=0.04,dimen=outer,fillstyle=solid,fillcolor=color1496b](5.66,-7.3260937)(5.26,-8.066093)}
\rput{-180.0}(10.92,-13.9921875){\psframe[linewidth=0.04,dimen=outer](5.66,-6.626094)(5.26,-7.3660936)}
\rput{-90.0}(11.496094,-3.1960938){\pstriangle[linewidth=0.04,dimen=outer](4.15,-7.626094)(0.66,0.56)}
\usefont{T1}{ptm}{m}{n}
\rput(3.4651563,-5.7360935){Polymerase I right(X,Y):=}
\rput{-180.0}(6.16,-15.432187){\psframe[linewidth=0.04,dimen=outer,fillstyle=solid,fillcolor=color1496b](3.28,-7.3460937)(2.88,-8.086094)}
\psline[linewidth=0.04cm](2.48,-6.666094)(1.4,-6.666094)
\rput{-180.0}(6.16,-14.032187){\psframe[linewidth=0.04,linecolor=color1523,dimen=outer](3.28,-6.646094)(2.88,-7.3860936)}
\rput{-180.0}(3.92,-15.432187){\psframe[linewidth=0.04,dimen=outer,fillstyle=solid,fillcolor=color1496b](2.16,-7.3460937)(1.76,-8.086094)}
\rput{-180.0}(3.92,-14.032187){\psframe[linewidth=0.04,dimen=outer](2.16,-6.646094)(1.76,-7.3860936)}
\psline[linewidth=0.04cm](5.72,-3.1660938)(6.84,-3.1660938)
\rput{-180.0}(12.52,-8.432187){\psframe[linewidth=0.04,dimen=outer,fillstyle=solid,fillcolor=color1496b](6.46,-3.8460937)(6.06,-4.586094)}
\rput{-180.0}(12.52,-7.0321875){\psframe[linewidth=0.04,dimen=outer](6.46,-3.1460938)(6.06,-3.8860939)}
\rput{-180.0}(10.28,-8.432187){\psframe[linewidth=0.04,dimen=outer,fillstyle=solid,fillcolor=color1496b](5.34,-3.8460937)(4.94,-4.586094)}
\rput{-180.0}(10.28,-7.3721876){\psframe[linewidth=0.04,linecolor=color1523,dimen=outer,fillstyle=solid](5.34,-3.4860938)(4.94,-3.8860939)}
\rput{-90.0}(8.076094,0.34390625){\pstriangle[linewidth=0.04,dimen=outer](4.21,-4.146094)(0.66,0.56)}
\usefont{T1}{ptm}{m}{n}
\rput(3.3751562,-2.1960938){Polymerase I left(X,Y):=}
\psline[linewidth=0.04cm](2.58,-3.1860938)(3.66,-3.1860938)
\rput{-180.0}(6.2,-8.472187){\psframe[linewidth=0.04,dimen=outer,fillstyle=solid,fillcolor=color1496b](3.3,-3.8660936)(2.9,-4.606094)}
\rput{-180.0}(6.2,-7.0721874){\psframe[linewidth=0.04,dimen=outer](3.3,-3.1660938)(2.9,-3.9060938)}
\psline[linewidth=0.04cm,linecolor=color1523,fillcolor=color1523,dotsize=0.07055555cm 2.0]{*-*}(1.44,-3.1860938)(2.56,-3.1860938)
\rput{-180.0}(3.96,-7.0721874){\psframe[linewidth=0.04,linecolor=color1523,dimen=outer](2.18,-3.1660938)(1.78,-3.9060938)}
\psframe[linewidth=0.04,dimen=outer](13.56,-6.546094)(9.52,-8.346094)
\psline[linewidth=0.04cm,linecolor=color1523](9.76,-7.4260936)(10.88,-7.4260936)
\usefont{T1}{ptm}{m}{n}
\rput(10.454532,-6.1960936){Ligase:=}
\rput{-90.0}(18.976093,4.123906){\pstriangle[linewidth=0.04,dimen=outer](11.55,-7.706094)(0.66,0.56)}
\rput{-90.0}(18.976093,4.123906){\pstriangle[linewidth=0.04,dimen=outer](11.55,-7.706094)(0.66,0.56)}
\psline[linewidth=0.04cm,linecolor=color1523,fillcolor=color1523,dotsize=0.07055555cm 2.0]{*-*}(12.18,-7.4060936)(13.3,-7.4060936)
\psframe[linewidth=0.04,dimen=outer](7.24,-2.5660937)(1.1,-5.206094)
\psline[linewidth=0.04cm,arrowsize=0.16cm 8.0,arrowlength=1.4,arrowinset=0.4]{->}(10.94,5.1339064)(11.38,3.0139062)
\psline[linewidth=0.04cm,arrowsize=0.16cm 8.0,arrowlength=1.4,arrowinset=0.4]{->}(7.16,5.473906)(9.62,2.7339063)
\psline[linewidth=0.04cm,arrowsize=0.16cm 8.0,arrowlength=1.4,arrowinset=0.4]{->}(6.62,3.0339062)(8.62,1.9339062)
\psline[linewidth=0.04cm,arrowsize=0.16cm 8.0,arrowlength=1.4,arrowinset=0.4]{->}(6.9,-0.40609375)(8.74,-0.46609375)
\psline[linewidth=0.04cm,arrowsize=0.16cm 8.0,arrowlength=1.4,arrowinset=0.4]{->}(7.66,-3.9460938)(9.4,-2.8860939)
\psline[linewidth=0.04cm,arrowsize=0.16cm 8.0,arrowlength=1.4,arrowinset=0.4]{->}(7.68,-7.4860935)(9.9,-3.6260939)
\psline[linewidth=0.04cm,arrowsize=0.16cm 8.0,arrowlength=1.4,arrowinset=0.4]{->}(11.72,-5.546094)(11.84,-3.8060937)
\usefont{T1}{ptm}{m}{n}
\rput(1.125625,3.9039063){Y}
\usefont{T1}{ptm}{m}{n}
\rput(4.1865625,2.9439063){X}
\usefont{T1}{ptm}{m}{n}
\rput(4.2265625,-0.85609376){X}
\usefont{T1}{ptm}{m}{n}
\rput(1.065625,0.08390625){Y}
\usefont{T1}{ptm}{m}{n}
\rput(5.1465626,-4.336094){X}
\usefont{T1}{ptm}{m}{n}
\rput(1.985625,-3.4360938){Y}
\usefont{T1}{ptm}{m}{n}
\rput(6.5865626,-7.836094){X}
\usefont{T1}{ptm}{m}{n}
\rput(3.0865624,-7.836094){X}
\usefont{T1}{ptm}{m}{n}
\rput(3.085625,-6.936094){Y}
\usefont{T1}{ptm}{m}{n}
\rput(4.320625,-9.816093){\textit{For all }$(X,Y)\in\{(A,T),(T,A),(G,C),(C,G)\}$}
\psline[linewidth=0.04cm,linestyle=dotted,dotsep=0.16cm](15.1,-0.06609375)(15.6,-0.06609375)
\psline[linewidth=0.04cm,linestyle=dotted,dotsep=0.16cm](15.14,-1.4460938)(15.6,-1.4460938)
\psline[linewidth=0.04cm,linecolor=color1523](10.12,9.233906)(10.9,8.333906)
\psline[linewidth=0.04cm,linecolor=color1528](10.12,6.0339065)(10.92,6.953906)
\psline[linewidth=0.04cm,linecolor=color1665](13.14,8.333906)(13.88,8.333906)
\psline[linewidth=0.04cm,linecolor=color1528](13.14,6.933906)(13.88,6.933906)
\psframe[linewidth=0.04,dimen=outer](14.26,8.353907)(13.86,7.6139064)
\psframe[linewidth=0.04,dimen=outer](14.26,7.6539063)(13.86,6.913906)
\psframe[linewidth=0.04,dimen=outer](11.28,8.353907)(10.88,7.6139064)
\psframe[linewidth=0.04,dimen=outer](11.28,7.6539063)(10.88,6.913906)
\rput{-180.0}(2.24,6.3278127){\psframe[linewidth=0.04,dimen=outer,fillstyle=solid,fillcolor=color1496b](1.32,3.5339062)(0.92,2.7939062)}
\usefont{T1}{ptm}{m}{n}
\rput(1.1265625,2.9839063){X}
\rput{-180.0}(2.12,-1.4321876){\psframe[linewidth=0.04,dimen=outer,fillstyle=solid,fillcolor=color1496b](1.26,-0.34609374)(0.86,-1.0860938)}
\rput{-180.0}(2.12,-1.4321876){\psframe[linewidth=0.04,dimen=outer,fillstyle=solid,fillcolor=color1496b](1.26,-0.34609374)(0.86,-1.0860938)}
\usefont{T1}{ptm}{m}{n}
\rput(1.0665625,-0.87609375){X}
\rput{-180.0}(3.96,-8.472187){\psframe[linewidth=0.04,dimen=outer,fillstyle=solid,fillcolor=color1496b](2.18,-3.8660936)(1.78,-4.606094)}
\usefont{T1}{ptm}{m}{n}
\rput(1.9865625,-4.336094){X}
\rput{-180.0}(13.16,-15.392187){\psframe[linewidth=0.04,dimen=outer,fillstyle=solid,fillcolor=color1496b](6.78,-7.3260937)(6.38,-8.066093)}
\pswedge[linewidth=0.04](6.58,-7.3460937){0.18}{-180.0}{0.0}
\usefont{T1}{ptm}{m}{n}
\rput(6.5665627,-7.836094){X}
\rput{-180.0}(10.28,-7.7321873){\pswedge[linewidth=0.04](5.14,-3.8660936){0.18}{0.0}{180.0}}
\rput{-180.0}(9.72,13.307813){\pswedge[linewidth=0.04](4.86,6.6539063){0.18}{0.0}{180.0}}
\psline[linewidth=0.04cm,linecolor=color1523,fillcolor=color1523,dotsize=0.07055555cm 2.0]{*-*}(3.64,-6.666094)(2.52,-6.666094)
\usefont{T1}{ptm}{m}{n}
\rput(11.163125,-0.7760937){H}
\usefont{T1}{ptm}{m}{n}
\rput(11.063125,7.623906){H}
\usefont{T1}{ptm}{m}{n}
\rput(6.6275,4.313906){\small $\infty S$}
\usefont{T1}{ptm}{m}{n}
\rput(15.0875,9.233906){\small $\infty S$}
\usefont{T1}{ptm}{m}{n}
\rput(8.3875,7.0339065){\small $\infty S$}
\usefont{T1}{ptm}{m}{n}
\rput(6.8675,0.6539062){\small $\infty S$}
\usefont{T1}{ptm}{m}{n}
\rput(7.6875,-2.9860938){\small $\infty S$}
\usefont{T1}{ptm}{m}{n}
\rput(14.0075,-7.046094){\small $\infty S$}
\usefont{T1}{ptm}{m}{n}
\rput(7.6675,-6.626094){\small $\infty S$}
\end{pspicture}
}
\end{equation*}

We will now define the same right DNA replication process but written in inline notation. We will also define the left replication inline.

DNA will be encoded in an array $\langle a_1,a_2,...,a_n\rangle$ and in each component $a_i$ of the array there will be a pair of nucleotides $(X,Y)$. The kth component is underlined to indicate where helicase will start.

$$dna:=\langle(x_1,y_1)\mid(x_2,y_2)\mid ... \mid\underline{(x_{k-1},\overleftarrow{y}_{k-1})}\mid\underline{(\overrightarrow{x}_k,y_k)}\mid(x_{k+1},y_{k+1})\mid ... \mid(x_n,y_n)\rangle$$

We now define helicase and polymerase. Note that in our language, if the name starts with a lowercase, it will refer to the inline notation and if the name starts by an uppercase, it will refer to the diagram notation.

\subsubsection{Inline replication of the right leading strand}

Note that in the following transformations, we use an asterisk `$\ast$' to indicate an empty component of a pair.

\begin{flalign*}&\text{helicase right}:=\fbox{$(x_i,\ast)(\ast,y_i)\mid\underline{(x_{i+1},y_{i+1})}\rhd \underline{(x_i,y_i)}\mid(x_{i+1},y_{i+1})$}&\end{flalign*}

\begin{flalign*}&\text{helicase final right}:=\fbox{$(x_i,\ast)(\ast,y_i)\rangle\rhd \underline{(x_i,y_i)}\rangle$}&\end{flalign*}

\begin{flalign*}&\text{initiate polymerase leading right(X,Y)}:=\fbox{$(X,Y)\rhd (\overrightarrow{X},\ast)$}&\end{flalign*}

\begin{flalign*}&\text{polymerase leading right(X,Y)}:=\fbox{$(K,L)(M,N)\mid(X,Y)\rhd (K,{\color{red}\ast})(M,N)\mid(X,\ast)$}&\end{flalign*}

%\begin{flalign*}&polyR(X,Y):=:\text{polymerase leading right(X,Y)}:=\fbox{$(V,W),(X,Y)\rhd (V,W),(X,\ast)$}&\end{flalign*}
Where $K, L, M$ and $N$ can be any nucleotide or the symbol $\ast$, but where ${\color{red}\ast}$ means that it cannot be the asterisk symbol.
We have a transformation \textit{for all }$(X,Y)\in\{(A,T),(T,A),(G,C),(C,G)\}$, \textit{for all } $(V,W)\in\{(A,T),(T,A),(G,C),(C,G)\}$.

\subsubsection{Inline replication of the left leading strand}

\begin{flalign*}&\text{helicase left}:=\fbox{$\underline{(x_{j-1},y_{j-1})}\mid(x_j,\ast)(\ast,y_j)\rhd (x_{j-1},y_{j-1})\mid\underline{(x_j,y_j)}$}&\end{flalign*}

\begin{flalign*}&\text{helicase final left}:=\fbox{$\langle(x_j,\ast)(\ast,y_j)\rhd \langle\underline{(x_j,y_j)}$}&\end{flalign*}

\begin{flalign*}&\text{initiate polymerase leading left(X,Y)}:=\fbox{$(X,Y)\rhd (\ast,\overleftarrow{Y}) $}&\end{flalign*}

\begin{flalign*}&\text{polymerase leading left(X,Y)}:=\fbox{$(X,Y)\mid(K,L)(M,N)\rhd (\ast,Y)\mid(K,L)({\color{red}\ast},N)$}&\end{flalign*}

After applying all the previous transformations on the leading strand, we find the following structure which is the replicated leading strand.

\begin{equation*}\begin{split}repLeading:=\langle(x_1,\ast)(x_1,y_1)\mid(x_2,\ast)(x_2,y_2)\mid ... \mid(x_{k-1},\ast)(x_{k-1},y_{k-1}), \\
\,\,\,\,\,\, (x_k,y_k)(\ast,y_k)\mid(x_{k+1},y_{k+1})(\ast,y_{k+1})\mid ... \mid(x_n,y_n)(\ast,y_n)\rangle\end{split}\end{equation*}

\subsubsection{Inline replication of the right lagging strand}

\begin{flalign*}&\text{primaseR}:=\fbox{$(\Box,Y)\rhd (\ast,Y)$}&\end{flalign*}

\begin{flalign*}&\text{initiate poly III lagging right(X,Y)}:=\fbox{$(X,Y)\mid(\Box,B)\rhd (\ast,Y)\mid(\Box,B)$}&\end{flalign*}

\begin{flalign*}&\text{polymerase III lagging right(X,Y)}:=\fbox{$(X,Y)\mid(A,B)\rhd (\ast,Y)\mid(A,B)$}&\end{flalign*}

\begin{flalign*}&\text{polymerase Ia lagging right(X,Y)}:=\fbox{$(\tilde{X},Y)\mid(A,B)({\color{red}\Box},E)\rhd (\Box,Y)\mid(A,B)({\color{red}\Box},E)$}&\end{flalign*}

\begin{flalign*}&\text{polymerase Ib lagging right(X,Y)}:=\fbox{$(A,B)(D,E)\mid(\tilde{X},Y)\rhd ({\color{red}\Box},B)(D,E)\mid(\Box,Y)$}\end{flalign*}

\begin{flalign*}&ligase:=\fbox{$X\rhd \tilde{X}$}&\end{flalign*}

\subsubsection{Inline replication of the left lagging strand}

\begin{flalign*}&\text{primaseL}:=\fbox{$(X,\Box)\rhd (X¸,\ast)$}&\end{flalign*}

\begin{flalign*}&\text{initiate poly III lagging left(X,Y)}:=\fbox{$(A,\Box)\mid(X,Y)\rhd (A,\Box)\mid(X,\ast)$}&\end{flalign*}

\begin{flalign*}&\text{polymerase III lagging left(X,Y)}:=\fbox{$(A,B)\mid(X,Y)\rhd (A,B)\mid(X,\ast)$}&\end{flalign*}

\begin{flalign*}&\text{polymerase Ia lagging left(X,Y)}:=\fbox{$(A,B)(D,E)\mid(X,\tilde{Y})\rhd (A,{\color{red}\Box})(D,E)\mid(X,\Box)$}&\end{flalign*}

\begin{flalign*}&\text{polymerase Ib lagging left(X,Y)}:=\fbox{$(X,\tilde{Y})\mid(A,B)(D,E)\rhd (Y,\Box)\mid(A,B)(D,{\color{red}\Box})$}\end{flalign*}

\begin{flalign*}&ligase:=\fbox{$Y\rhd \tilde{Y}$}&\end{flalign*}

\subsection{Messenger RNA}

The transcription of DNA into pre-messenger RNA is very similar to DNA replication of the right leading strand.

%RNA transcription
\begin{equation*}\scalebox{0.7} % Change this value to rescale the drawing.
{
\begin{pspicture}(0,-6.979219)(20.966875,7.019219)
\definecolor{color50b}{rgb}{1.0,1.0,0.6}
\definecolor{color50}{rgb}{0.2,0.8,0.0}
\definecolor{color67b}{rgb}{1.0,0.6,1.0}
\definecolor{color104b}{rgb}{0.6,0.8,1.0}
\definecolor{color195}{rgb}{0.2,1.0,0.0}
\definecolor{color196}{rgb}{0.0,0.8,0.0}
\psframe[linewidth=0.04,linecolor=color50,dimen=outer,fillstyle=solid,fillcolor=color50b](8.660937,6.2207813)(8.260938,5.480781)
\pswedge[linewidth=0.04,linecolor=color50](8.4609375,5.500781){0.18}{0.0}{180.0}
\psframe[linewidth=0.04,linecolor=color50,dimen=outer](8.660937,3.7207813)(8.260938,2.9807813)
\psframe[linewidth=0.04,dimen=outer](13.320937,6.540781)(7.8609376,2.6407812)
\rput{-90.0}(6.050156,15.171719){\pstriangle[linewidth=0.04,dimen=outer](10.610937,4.2807813)(0.66,0.56)}
\usefont{T1}{ptm}{m}{n}
\rput(10.212344,6.8107815){$\text{RNA Helicase open:=}$}
\rput{-90.0}(13.350156,2.3917189){\pstriangle[linewidth=0.04,dimen=outer](7.8709373,-5.7592187)(0.66,0.56)}
\psframe[linewidth=0.04,dimen=outer,fillstyle=solid,fillcolor=color50b](9.200937,-4.8392186)(8.800938,-5.579219)
\psframe[linewidth=0.04,dimen=outer,fillstyle=solid,fillcolor=color50b](10.320937,-4.8392186)(9.920938,-5.579219)
\pswedge[linewidth=0.04](10.120937,-5.559219){0.18}{0.0}{180.0}
\psframe[linewidth=0.04,dimen=outer,fillstyle=solid,fillcolor=color67b](9.200937,-5.7392187)(8.800938,-6.479219)
\psframe[linewidth=0.04,dimen=outer,fillstyle=solid,fillcolor=color50b](5.9809375,-4.8192186)(5.5809374,-5.559219)
\psframe[linewidth=0.04,dimen=outer,fillstyle=solid,fillcolor=color50b](7.1009374,-4.8192186)(6.7009373,-5.559219)
\psframe[linewidth=0.04,dimen=outer,fillstyle=solid,fillcolor=color67b](5.9809375,-5.7392187)(5.5809374,-6.479219)
\psline[linewidth=0.04cm](6.3009377,-6.459219)(5.2209377,-6.459219)
\psframe[linewidth=0.04,linecolor=color50,dimen=outer,fillstyle=solid,fillcolor=color67b](7.1009374,-5.7392187)(6.7009373,-6.479219)
\psline[linewidth=0.04cm,linecolor=color50](7.4209375,-6.459219)(6.3009377,-6.459219)
\psline[linewidth=0.04cm](9.540937,-6.459219)(8.420938,-6.459219)
\usefont{T1}{ptm}{m}{n}
\rput(6.592344,-3.7692187){$\text{RNA Polymerase(C,G)}:=$}
\psframe[linewidth=0.04,dimen=outer](11.120937,-4.099219)(4.7009373,-6.979219)
\usefont{T1}{ptm}{m}{n}
\rput(10.108125,-5.0692186){C}
\usefont{T1}{ptm}{m}{n}
\rput(6.9020314,-6.209219){G}
\usefont{T1}{ptm}{m}{n}
\rput(6.888125,-5.0692186){C}
\usefont{T1}{ptm}{m}{n}
\rput(2.9423437,3.0507812){$\text{Initiate RNA Polymerase(T,A)}=$}
\psframe[linewidth=0.04,dimen=outer](6.6409373,2.8007812)(2.0809374,-0.03921875)
\psframe[linewidth=0.04,dimen=outer,fillstyle=solid,fillcolor=color50b](3.3809376,2.1807814)(2.9809375,1.4407812)
\psframe[linewidth=0.04,linecolor=color50,dimen=outer,fillstyle=solid,fillcolor=color67b](3.3809376,1.2607813)(2.9809375,0.5207813)
\usefont{T1}{ptm}{m}{n}
\rput(3.1875,0.7707813){A}
\usefont{T1}{ptm}{m}{n}
\rput(3.1609375,1.9307812){T}
\psline[linewidth=0.04cm,linecolor=color50](3.7009375,0.54078126)(2.6209376,0.54078126)
\rput{-90.0}(2.9901562,5.991719){\pstriangle[linewidth=0.04,dimen=outer](4.4909377,1.2207812)(0.66,0.56)}
\psframe[linewidth=0.04,dimen=outer,fillstyle=solid,fillcolor=color50b](5.8409376,2.2007813)(5.4409375,1.4607812)
\pswedge[linewidth=0.04](5.6409373,1.5007813){0.18}{0.0}{180.0}
\usefont{T1}{ptm}{m}{n}
\rput(5.6209373,1.9507812){T}
\psline[linewidth=0.04cm](11.740937,0.5207813)(9.420938,0.5207813)
\psframe[linewidth=0.04,dimen=outer,fillstyle=solid,fillcolor=color50b](10.140938,0.54078126)(9.740937,-0.19921875)
\psframe[linewidth=0.04,dimen=outer,fillstyle=solid,fillcolor=color104b](10.140938,-0.15921874)(9.740937,-0.89921874)
\usefont{T1}{ptm}{m}{n}
\rput(9.920938,0.25078124){T}
\usefont{T1}{ptm}{m}{n}
\rput(9.9275,-0.58921874){A}
\psframe[linewidth=0.04,dimen=outer,fillstyle=solid,fillcolor=color50b](11.220938,0.54078126)(10.820937,-0.19921875)
\psline[linewidth=0.04cm](11.640938,-0.87921876)(9.420938,-0.87921876)
\psframe[linewidth=0.04,dimen=outer,fillstyle=solid,fillcolor=color104b](11.220938,-0.15921874)(10.820937,-0.89921874)
\usefont{T1}{ptm}{m}{n}
\rput(10.988125,0.23078126){C}
\usefont{T1}{ptm}{m}{n}
\rput(11.002031,-0.58921874){G}
\psline[linewidth=0.04cm](13.740937,0.5207813)(11.620937,0.5207813)
\psline[linewidth=0.04cm](13.740937,-0.87921876)(11.620937,-0.87921876)
\psframe[linewidth=0.04,dimen=outer,fillstyle=solid,fillcolor=color50b](12.340938,0.54078126)(11.940937,-0.19921875)
\psframe[linewidth=0.04,dimen=outer,fillstyle=solid,fillcolor=color104b](12.340938,-0.15921874)(11.940937,-0.89921874)
\usefont{T1}{ptm}{m}{n}
\rput(12.1275,0.23078126){A}
\usefont{T1}{ptm}{m}{n}
\rput(12.120937,-0.58921874){T}
\psframe[linewidth=0.04,dimen=outer,fillstyle=solid,fillcolor=color50b](13.420938,0.54078126)(13.020938,-0.19921875)
\psframe[linewidth=0.04,dimen=outer,fillstyle=solid,fillcolor=color104b](13.420938,-0.15921874)(13.020938,-0.89921874)
\usefont{T1}{ptm}{m}{n}
\rput(13.2075,0.25078124){A}
\usefont{T1}{ptm}{m}{n}
\rput(13.220938,-0.56921875){T}
\usefont{T1}{ptm}{m}{n}
\rput(3.1623437,-0.88921875){$\text{Initiate RNA Polymerase(A,U)}^1$}
\usefont{T1}{ptm}{m}{n}
\rput(3.2023437,-2.2492187){$\text{Initiate RNA Polymerase(C,G)}^1$}
\usefont{T1}{ptm}{m}{n}
\rput(3.1623437,-1.5892187){$\text{Initiate RNA Polymerase(G,C)}^1$}
\usefont{T1}{ptm}{m}{n}
\rput(16.462343,-4.409219){$\text{RNA Polymerase(T,A)}^{\infty P}$}
\usefont{T1}{ptm}{m}{n}
\rput(15.452344,-5.0092187){$\text{RNA Polymerase(A,U)}^{\infty P}$}
\usefont{T1}{ptm}{m}{n}
\rput(17.632343,-3.7492187){$\text{RNA Polymerase(G,C)}^{\infty P}$}
\psline[linewidth=0.04cm,arrowsize=0.16cm 8.0,arrowlength=1.4,arrowinset=0.4]{->}(9.820937,2.3607812)(10.100938,1.3607812)
\psline[linewidth=0.04cm,arrowsize=0.16cm 8.0,arrowlength=1.4,arrowinset=0.4]{->}(7.1609373,1.2207812)(8.500937,0.50078124)
\psline[linewidth=0.04cm,arrowsize=0.16cm 8.0,arrowlength=1.4,arrowinset=0.4]{->}(7.0409374,-0.87921876)(8.300938,-0.71921873)
\psline[linewidth=0.04cm,arrowsize=0.16cm 8.0,arrowlength=1.4,arrowinset=0.4]{->}(7.0809374,-1.5392188)(8.440937,-1.2392187)
\psline[linewidth=0.04cm,arrowsize=0.16cm 8.0,arrowlength=1.4,arrowinset=0.4]{->}(7.2009373,-2.2192187)(8.680938,-1.8592187)
\psline[linewidth=0.04cm,arrowsize=0.16cm 8.0,arrowlength=1.4,arrowinset=0.4]{->}(9.540937,-3.3992188)(9.980938,-2.1792188)
\psline[linewidth=0.04cm,arrowsize=0.16cm 8.0,arrowlength=1.4,arrowinset=0.4]{->}(12.560938,-4.559219)(12.320937,-2.5992188)
\psline[linewidth=0.04cm,arrowsize=0.16cm 8.0,arrowlength=1.4,arrowinset=0.4]{->}(13.720938,-3.9992187)(13.060938,-2.2992187)
\psline[linewidth=0.04cm,arrowsize=0.16cm 8.0,arrowlength=1.4,arrowinset=0.4]{->}(14.980938,-3.4392188)(13.540937,-1.9392188)
\usefont{T1}{ptm}{m}{n}
\rput(6.8478127,2.5307813){1}
\psline[linewidth=0.04cm](16.020937,0.5207813)(13.700937,0.5207813)
\psframe[linewidth=0.04,dimen=outer,fillstyle=solid,fillcolor=color50b](14.420938,0.54078126)(14.020938,-0.19921875)
\psframe[linewidth=0.04,dimen=outer,fillstyle=solid,fillcolor=color104b](14.420938,-0.15921874)(14.020938,-0.89921874)
\usefont{T1}{ptm}{m}{n}
\rput(14.200937,0.25078124){T}
\usefont{T1}{ptm}{m}{n}
\rput(14.2075,-0.58921874){A}
\psframe[linewidth=0.04,dimen=outer,fillstyle=solid,fillcolor=color50b](15.500937,0.54078126)(15.100938,-0.19921875)
\psline[linewidth=0.04cm](15.920938,-0.87921876)(13.700937,-0.87921876)
\psframe[linewidth=0.04,dimen=outer,fillstyle=solid,fillcolor=color104b](15.500937,-0.15921874)(15.100938,-0.89921874)
\usefont{T1}{ptm}{m}{n}
\rput(15.282031,0.23078126){G}
\usefont{T1}{ptm}{m}{n}
\rput(15.268125,-0.58921874){C}
\psline[linewidth=0.04cm](18.020937,0.5207813)(15.900937,0.5207813)
\psline[linewidth=0.04cm](18.020937,-0.87921876)(15.900937,-0.87921876)
\psframe[linewidth=0.04,dimen=outer,fillstyle=solid,fillcolor=color50b](16.620937,0.54078126)(16.220938,-0.19921875)
\psframe[linewidth=0.04,dimen=outer,fillstyle=solid,fillcolor=color104b](16.620937,-0.15921874)(16.220938,-0.89921874)
\usefont{T1}{ptm}{m}{n}
\rput(16.388124,0.23078126){C}
\usefont{T1}{ptm}{m}{n}
\rput(16.422031,-0.58921874){G}
\psframe[linewidth=0.04,dimen=outer,fillstyle=solid,fillcolor=color50b](17.700937,0.54078126)(17.300938,-0.19921875)
\psframe[linewidth=0.04,dimen=outer,fillstyle=solid,fillcolor=color104b](17.700937,-0.15921874)(17.300938,-0.89921874)
\usefont{T1}{ptm}{m}{n}
\rput(17.4875,0.25078124){A}
\usefont{T1}{ptm}{m}{n}
\rput(17.500938,-0.56921875){T}
\usefont{T1}{ptm}{m}{n}
\rput(11.005938,0.7157813){\footnotesize Gi}
\usefont{T1}{ptm}{m}{n}
\rput(16.391874,0.73578125){\footnotesize Ge}
\usefont{T1}{ptm}{m}{n}
\rput(5.6059375,2.3757813){\footnotesize Gi}
\psframe[linewidth=0.04,dimen=outer,fillstyle=solid,fillcolor=color50b](9.720938,5.320781)(9.320937,4.5807815)
\psframe[linewidth=0.04,dimen=outer](9.720938,4.6207814)(9.320937,3.8807812)
\psframe[linewidth=0.04,linecolor=color50,dimen=outer,fillstyle=solid,fillcolor=color50b](11.780937,5.3007812)(11.380938,4.5607815)
\psframe[linewidth=0.04,linecolor=color50,dimen=outer](11.780937,4.6007814)(11.380938,3.8607812)
\psline[linewidth=0.04cm,linecolor=color50](12.640938,3.8807812)(11.560938,3.8807812)
\psframe[linewidth=0.04,dimen=outer](12.900937,4.6007814)(12.500937,3.8607812)
\psline[linewidth=0.04cm,linecolor=color50](12.620937,5.2807813)(11.4609375,5.2807813)
\psline[linewidth=0.04cm,linecolor=color50](8.620937,6.2007813)(9.340938,5.3007812)
\psline[linewidth=0.04cm,linecolor=color195](8.620937,3.0007813)(9.340938,3.9007812)
\psline[linewidth=0.04cm,linecolor=color196](9.520938,4.7807813)(9.520938,4.420781)
\psline[linewidth=0.04cm,linecolor=color50](11.580937,4.7607813)(11.580937,4.400781)
\usefont{T1}{ptm}{m}{n}
\rput(11.551875,5.4757814){\footnotesize \color{red}Ge}
\psline[linewidth=0.04cm](11.020938,0.0)(11.020938,-0.35921875)
\psframe[linewidth=0.04,dimen=outer,fillstyle=solid,fillcolor=color50b](12.900937,5.3007812)(12.500937,4.5607815)
\usefont{T1}{ptm}{m}{n}
\rput(3.1659374,2.3557813){\footnotesize Gi}
\psframe[linewidth=0.04,linecolor=color50,dimen=outer,fillstyle=solid,fillcolor=color50b](19.620937,6.2607813)(19.220938,5.520781)
\psframe[linewidth=0.04,linecolor=color50,dimen=outer](19.620937,3.7607813)(19.220938,3.0207813)
\psframe[linewidth=0.04,dimen=outer](19.980938,6.540781)(14.520938,2.6407812)
\rput{-90.0}(12.710156,21.831718){\pstriangle[linewidth=0.04,dimen=outer](17.270937,4.2807813)(0.66,0.56)}
\usefont{T1}{ptm}{m}{n}
\rput(16.982344,6.8307815){$\text{RNA Helicase close:=}$}
\psframe[linewidth=0.04,dimen=outer,fillstyle=solid,fillcolor=color50b](18.560938,5.360781)(18.160938,4.6207814)
\psframe[linewidth=0.04,dimen=outer](18.560938,4.6607814)(18.160938,3.9207811)
\psframe[linewidth=0.04,linecolor=color50,dimen=outer,fillstyle=solid,fillcolor=color50b](16.540937,5.320781)(16.140938,4.5807815)
\psframe[linewidth=0.04,linecolor=color50,dimen=outer](16.540937,4.6207814)(16.140938,3.8807812)
\psline[linewidth=0.04cm,linecolor=color50](15.280937,3.9007812)(16.360937,3.9007812)
\psframe[linewidth=0.04,dimen=outer](15.420938,4.6207814)(15.020938,3.8807812)
\psline[linewidth=0.04cm,linecolor=color50](15.300938,5.3007812)(16.460938,5.3007812)
\psline[linewidth=0.04cm,linecolor=color50](19.260937,6.2407813)(18.540937,5.340781)
\psline[linewidth=0.04cm,linecolor=color195](19.260937,3.0407813)(18.540937,3.9407814)
\psframe[linewidth=0.04,dimen=outer,fillstyle=solid,fillcolor=color50b](15.420938,5.320781)(15.020938,4.5807815)
\psline[linewidth=0.04cm,arrowsize=0.16cm 8.0,arrowlength=1.4,arrowinset=0.4]{->}(15.060938,2.3407812)(14.800938,1.4607812)
\psdots[dotsize=0.12](5.7809377,-5.4192185)
\psdots[dotsize=0.12](3.1809375,1.5607812)
\psdots[dotsize=0.12](19.420937,5.6407814)
\psdots[dotsize=0.12](9.000937,-5.4392185)
\psdots[dotsize=0.12](6.9009376,-5.4392185)
\usefont{T1}{ptm}{m}{n}
\rput(13.706562,6.045781){\scriptsize $\infty P$}
\usefont{T1}{ptm}{m}{n}
\rput(20.366562,6.065781){\scriptsize $\infty P$}
\usefont{T1}{ptm}{m}{n}
\rput(11.506562,-4.534219){\scriptsize $\infty P$}
\end{pspicture}
}\end{equation*}

\subsection{RNA Splicing}

After the pre-messenger RNA has been coded, the introns must be removed by the spliceosome. We will simplify the model by assigning to the pre-mRNA three points: branch site \textit{B}, 5' splice site \textit{S5} and the 3' splice site \textit{S3}. These points are determined in practice by the spliceosome, pairs of nucleotides for the splice sites (\textit{GU} for the 5'site and \textit{AG} for the 3' site) and sequences of nucleotides for the branch site. The sites will be indicated with the use of superscripts written on the nucleotides. Note that the left-hand side of the pre-mRNA is the 5' and right-hand side is 3'.

%\begin{equation*}pre-mRNA:=\langle X_1-X_2-X_3-...X_i-X_{i+1}-...-X_j-X_{j+1}-...-X_k-X_{k+1}-...-X_n-1-X_n\rangle\end{equation*}

\begin{flalign*}&\text{pre-mRNA}:=A-...-D-E^{S5}-F-...-G-H^{B}-I-...-K-L^{S3}-M-...-Q&\end{flalign*}

\begin{flalign*}&\text{fold}:=\fbox{$\begin{array}{ccc}
    {\color{green}F} & {\color{green}\textbf{\text{---}}} & {\color{green}E^{S5}} \\
    |&   & | \\
    {\color{blue}G} & {\color{blue}\textbf{\text{---}}} & {\color{blue}H^{B}} \\
  \end{array}\rhd {\color{green}E^{S5}-F}, {\color{blue}G-H^{B}}$}&\end{flalign*}

\begin{flalign*}&\text{spliceS5}:=\fbox{$D^{*}\,\,\,\,\,\,\, E^{S5}\rhd D-E^{S5}$}&\end{flalign*}

\begin{flalign*}&\text{join}:=\fbox{$\begin{array}{ccc}
    &   & {\color{green}D^{*}} \\
    &   & | \\
    {\color{blue}L^{S3}} & {\color{blue}\textbf{\text{---}}} & {\color{blue}M} \\
  \end{array}\rhd {\color{green}D^{*}},{\color{blue}L^{S3}-M}$}&\end{flalign*}

\begin{flalign*}&\text{spliceS3}:=\fbox{$L^{S3} \,\,\,\,\,\,\, M\rhd L^{S3}-M$}&\end{flalign*}

Thus the splicing of an intron is given by
\begin{equation*}\text{spliceS3}\rightarrow\text{join}\rightarrow\text{spliceS5}\rightarrow\text{fold}\rightarrow\{\text{pre-mRNA}\}\end{equation*}
\noindent and gives two pieces of RNA. The piece with the loop is the intron and the other is a strand composed of two exons.

\subsection{Proteins}

After splicing, we have messenger RNA (mRNA). To differentiate between pre-mRNA and mRNA, we will identify the first nucleotide with a tilde. Here is an example:

\begin{equation*}\text{mRNA}:=\widetilde{A}-B-C-D-M-N-O-P-...-Z.\end{equation*}

Initiation of the protein encoding is done with the transformation
\begin{equation*}\text{initiate protein}:=\fbox{$ \begin{array}{c}
   init\\
   \|\\
   \widehat{U} \\
     \end{array}\rhd
    \begin{array}{c}
     \\
     \\
    \widetilde{U}\\
     \end{array}$}\end{equation*}

To synthesize a protein, the ribosome reads a sequence of triplets composed of different combinations of the nucleotides C, A, G and U. We will now give the general form of codons needed to synthesize a protein.

\begin{equation*}codon(L,UVW):=\fbox{$ \begin{array}{r}
    \text{K}\sim L\\
   \|\,\\
   U-V-W-\widehat{X} \\
     \end{array}\rhd
    \begin{array}{l}
     K\\
     \,\|\\
    \widehat{U}-V-W-X\\
     \end{array}$}\end{equation*}

The top part with connection symbols `$\sim$' is the protein being assembled and the bottom part is the RNA-messenger being decoded. The cause of the transformation is equivalent to reading the code $VWX$ and in the effect, $VWX$ is skipped to be in position for the next reading. Also in the effect, after reading the code $VWX$, the amino acid $L$ is added on the right end of the protein next to $K$.

 Examples of specific codons are $codon(Leucine,CCG)$, $codon(Leucine,CCC)$, $codon(Phenylalanine,UUU)$, $codon(Lycine,AAA)$ and the stop codon $codon(stop,UAG)$ which will stop the synthesis of the protein.

\subsection{Heart model}

We now present a simplified model of a beating heart. The heart is composed of the right ventricle, the left ventricle, right atrium and left atrium along with the heart conduction system. The conduction system is composed of the sinuatrial node ($SA$), at the top in blue, we have the Bachmann's bundle and in blue near the ventricles we have the bundle of His with left and right bundle branches. An edge connected to two nodes represents a cell (cardiac myocyte) which has the ability to contract. The cells denoted by $R$ are markers to initiate the relaxation process. Note that in this model the distances matter in the initial forms and the transformations.

%Heart4
\begin{equation*}\scalebox{1.3}  % Change this value to rescale the drawing.
{
\begin{pspicture}(0,-4.5804687)(8.781525,4.610469)
\definecolor{color919}{rgb}{0.6,0.0,0.6}
\definecolor{color1060}{rgb}{0.0,0.2,1.0}
\psline[linewidth=0.03cm,linecolor=blue](6.9585495,-0.66885614)(7.4029064,-1.4094958)
\psline[linewidth=0.03cm,linecolor=blue](6.9585495,-0.66885614)(7.2319846,-1.4693768)
\psline[linewidth=0.03cm,linecolor=blue,fillcolor=blue,dotsize=0.07055555cm 2.0]{-*}(6.2510405,-0.7201449)(6.949893,-0.6508267)
\psline[linewidth=0.03cm,linecolor=blue](6.1059594,0.27511963)(6.6288137,0.5261645)
\psline[linewidth=0.04cm,fillcolor=black,dotsize=0.07055555cm 2.0]{-*}(6.196107,0.31840324)(5.628537,0.06807439)
\psline[linewidth=0.04cm,fillcolor=black,dotsize=0.07055555cm 2.0]{-*}(5.629253,0.11279002)(5.5745134,-0.51251274)
\psline[linewidth=0.04cm,fillcolor=black,dotsize=0.07055555cm 2.0]{-*}(5.5745134,-0.51251274)(5.854392,-0.9105928)
\psline[linewidth=0.04cm,fillcolor=black,dotsize=0.07055555cm 2.0]{-*}(5.8724217,-0.90193605)(6.2510405,-0.7201449)
\psline[linewidth=0.04cm,fillcolor=black,dotsize=0.07055555cm 2.0]{-*}(6.2777267,-0.7295177)(6.6837473,-0.5123837)
\psline[linewidth=0.04cm,fillcolor=black,dotsize=0.07055555cm 2.0]{-*}(6.494015,-0.071020074)(6.160048,0.3010898)
\psline[linewidth=0.04cm,fillcolor=black,dotsize=0.07055555cm 2.0]{-*}(5.9056168,-1.0634851)(6.0016212,-1.7717105)
\psline[linewidth=0.04cm,fillcolor=black,dotsize=0.07055555cm 2.0]{-*}(6.0016212,-1.7717105)(6.4675875,-2.2801154)
\psline[linewidth=0.04cm,fillcolor=black,dotsize=0.07055555cm 2.0]{-*}(6.4675875,-2.2801154)(7.0972505,-2.6211784)
\psline[linewidth=0.04cm,fillcolor=black,dotsize=0.07055555cm 2.0]{-*}(7.0972505,-2.6211784)(7.6309743,-2.808633)
\psline[linewidth=0.04cm,fillcolor=black,dotsize=0.07055555cm 2.0]{-*}(7.6497197,-2.7552607)(7.524881,-2.1718094)
\psline[linewidth=0.04cm,fillcolor=black,dotsize=0.07055555cm 2.0]{-*}(7.524881,-2.1718094)(7.213955,-1.4780335)
\psline[linewidth=0.04cm,fillcolor=black,dotsize=0.07055555cm 2.0]{-*}(7.2319846,-1.4693768)(6.8922243,-0.90037495)
\psline[linewidth=0.04cm,fillcolor=black,dotsize=0.07055555cm 2.0]{-*}(6.865538,-0.89100224)(6.4104414,-0.82109696)
\psline[linewidth=0.04cm,fillcolor=black,dotsize=0.07055555cm 2.0]{-*}(6.4645295,-0.79512686)(5.9229302,-1.099544)
\psline[linewidth=0.04cm,fillcolor=black,dotsize=0.07055555cm 2.0]{-*}(6.6750903,-0.49435425)(6.5207014,-0.08039281)
\psline[linewidth=0.04cm,fillcolor=black,dotsize=0.07055555cm 2.0]{-*}(6.6288137,0.5261645)(7.17907,0.8125522)
\psline[linewidth=0.04cm,fillcolor=black,dotsize=0.07055555cm 2.0]{-*}(7.1437273,0.83995444)(7.66593,0.49166667)
\psline[linewidth=0.04cm,fillcolor=black,dotsize=0.07055555cm 2.0]{-*}(7.66593,0.49166667)(7.8015733,0.024332892)
\psline[linewidth=0.04cm,fillcolor=black,dotsize=0.07055555cm 2.0]{-*}(7.783544,0.015676172)(7.4049253,-0.16611493)
\psline[linewidth=0.04cm,fillcolor=black,dotsize=0.07055555cm 2.0]{-*}(7.3955526,-0.1928011)(6.9722185,-0.37387618)
\psline[linewidth=0.04cm,fillcolor=black,dotsize=0.07055555cm 2.0]{-*}(6.7464275,0.050173994)(6.6648726,0.54347795)
\psline[linewidth=0.04cm,fillcolor=black,dotsize=0.07055555cm 2.0]{-*}(7.8888564,-0.11124602)(8.381509,-0.62902355)
\psline[linewidth=0.04cm,fillcolor=black,dotsize=0.07055555cm 2.0]{-*}(8.381509,-0.62902355)(8.486886,-1.3105627)
\psline[linewidth=0.04cm,fillcolor=black,dotsize=0.07055555cm 2.0]{-*}(8.486886,-1.3105627)(8.359312,-2.015208)
\psline[linewidth=0.04cm,fillcolor=black,dotsize=0.07055555cm 2.0]{-*}(8.359312,-2.015208)(8.171858,-2.5489314)
\psline[linewidth=0.04cm,fillcolor=black,dotsize=0.07055555cm 2.0]{-*}(8.118485,-2.530186)(7.7412343,-2.0679288)
\psline[linewidth=0.04cm,fillcolor=black,dotsize=0.07055555cm 2.0]{-*}(7.7412343,-2.0679288)(7.3942494,-1.3914663)
\psline[linewidth=0.04cm,fillcolor=black,dotsize=0.07055555cm 2.0]{-*}(7.37622,-1.400123)(7.1446366,-0.7791809)
\psline[linewidth=0.04cm,fillcolor=black,dotsize=0.07055555cm 2.0]{-*}(7.1540093,-0.7524947)(7.384032,-0.35363415)
\psline[linewidth=0.04cm,fillcolor=black,dotsize=0.07055555cm 2.0]{-*}(7.3299437,-0.3796043)(7.90617,-0.14730494)
\psline[linewidth=0.04cm,fillcolor=black,dotsize=0.07055555cm 2.0]{-*}(6.9635615,-0.35584673)(6.737055,0.02348782)
\usefont{T1}{ptm}{m}{n}
\rput{25.647673}(0.8098903,-2.581766){\rput(6.058367,0.48040754){\scriptsize SA}}
\psline[linewidth=0.04cm,fillcolor=black,dotsize=0.07055555cm 2.0]{-*}(3.7653124,3.9495313)(3.2053125,3.9495313)
\psline[linewidth=0.04cm,linecolor=orange,fillcolor=orange,dotsize=0.07055555cm 2.0]{*-*}(2.2453125,3.9495313)(1.8853126,3.9495313)
\rput{-90.0}(-1.2042187,6.7148438){\pstriangle[linewidth=0.02,dimen=outer](2.7553124,3.7895312)(0.38,0.34)}
\rput{-90.0}(0.11578125,5.154844){\pstriangle[linewidth=0.02,dimen=outer](2.6353126,2.3495312)(0.38,0.34)}
\rput{-90.0}(3.4757812,2.3548439){\pstriangle[linewidth=0.02,dimen=outer](2.9153125,-0.73046875)(0.38,0.34)}
\psline[linewidth=0.03cm,linecolor=blue](3.1453125,2.5495312)(3.6853125,2.5495312)
\pscircle[linewidth=0.04,dimen=outer,fillstyle=solid,fillcolor=black](3.7453125,2.5495312){0.08}
\rput{25.647673}(0.40305296,-3.0722878){\pscircle[linewidth=0.04,linecolor=blue,dimen=outer,fillstyle=solid,fillcolor=blue](6.949893,-0.6508267){0.08}}
\pscircle[linewidth=0.04,linecolor=orange,dimen=outer,fillstyle=solid,fillcolor=orange](3.7653124,3.9495313){0.08}
\pscircle[linewidth=0.04,linecolor=orange,dimen=outer,fillstyle=solid,fillcolor=orange](3.0853126,2.5495312){0.08}
\rput{-42.31023}(-0.7404762,3.1898315){\pscircle[linewidth=0.04,dimen=outer,fillstyle=solid,fillcolor=black](3.7512739,2.551669){0.08}}
\psline[linewidth=0.03cm,linecolor=color919](1.4453125,2.5495312)(1.9853125,2.5495312)
\pscircle[linewidth=0.04,linecolor=orange,dimen=outer,fillstyle=solid,fillcolor=orange](1.3853126,2.5495312){0.08}
\rput{-42.31023}(-1.1937271,2.01856){\pscircle[linewidth=0.04,linecolor=orange,dimen=outer,fillstyle=solid,fillcolor=orange](2.0112739,2.551669){0.08}}
\rput{-90.0}(0.79578125,4.154844){\pstriangle[linewidth=0.02,dimen=outer](2.4753125,1.5095313)(0.38,0.34)}
\rput{-90.0}(3.3357813,9.214844){\pstriangle[linewidth=0.04,dimen=outer](6.2753124,2.7695312)(0.38,0.34)}
\rput{-42.31023}(-0.74289995,5.543568){\pscircle[linewidth=0.04,dimen=outer,fillstyle=solid,fillcolor=black](6.791274,3.731669){0.08}}
\usefont{T1}{ptm}{m}{n}
\rput(6.776094,3.9745312){\scriptsize SA}
\rput{-42.31023}(-0.4909223,4.594718){\pscircle[linewidth=0.04,linecolor=orange,dimen=outer,fillstyle=solid,fillcolor=orange](5.6912737,2.931669){0.08}}
\usefont{T1}{ptm}{m}{n}
\rput(5.6760936,3.1745312){\scriptsize SA}
\psline[linewidth=0.03cm,linecolor=blue](3.6253126,1.6695312)(3.7053125,0.8095313)
\psline[linewidth=0.03cm,linecolor=blue](3.6253126,1.6695312)(3.5253124,0.82953125)
\psline[linewidth=0.03cm,linecolor=blue,fillcolor=blue,dotsize=0.07055555cm 2.0]{-*}(2.9653125,1.9295312)(3.6253126,1.6895312)
\pscircle[linewidth=0.04,linecolor=blue,dimen=outer,fillstyle=solid,fillcolor=blue](3.6253126,1.6895312){0.08}
\pscircle[linewidth=0.04,linecolor=orange,dimen=outer,fillstyle=solid,fillcolor=orange](2.9853125,1.9095312){0.08}
\rput{-103.10508}(3.5082123,4.430878){\pscircle[linewidth=0.04,dimen=outer,fillstyle=solid,fillcolor=black](3.5130315,0.8227844){0.08}}
\rput{-123.863205}(5.1128535,4.368619){\pscircle[linewidth=0.04,dimen=outer,fillstyle=solid,fillcolor=black](3.7211916,0.82111615){0.08}}
\psline[linewidth=0.03cm,linecolor=color919](1.8653125,1.6695312)(1.9453125,0.8095313)
\psline[linewidth=0.03cm,linecolor=color919](1.8653125,1.6695312)(1.7653126,0.82953125)
\psline[linewidth=0.03cm,linecolor=color919,fillcolor=color919,dotsize=0.07055555cm 2.0]{-*}(1.2053125,1.9295312)(1.8653125,1.6895312)
\pscircle[linewidth=0.04,linecolor=color919,dimen=outer,fillstyle=solid,fillcolor=color919](1.8653125,1.6895312){0.08}
\pscircle[linewidth=0.04,linecolor=orange,dimen=outer,fillstyle=solid,fillcolor=orange](1.2253125,1.9095312){0.08}
\rput{-103.10508}(1.349154,2.716716){\pscircle[linewidth=0.04,linecolor=orange,dimen=outer,fillstyle=solid,fillcolor=orange](1.7530315,0.8227844){0.08}}
\rput{-123.863205}(2.3721604,2.9071672){\pscircle[linewidth=0.04,linecolor=orange,dimen=outer,fillstyle=solid,fillcolor=orange](1.9611915,0.82111615){0.08}}
\psline[linewidth=0.04cm](3.7653124,3.2495313)(3.2053125,3.2495313)
\psline[linewidth=0.04cm,linecolor=orange,fillcolor=orange,dotsize=0.07055555cm 2.0]{*-*}(2.2453125,3.2495313)(1.8853126,3.2495313)
\rput{-90.0}(-0.50421876,6.014844){\pstriangle[linewidth=0.02,dimen=outer](2.7553124,3.0895312)(0.38,0.34)}
\pscircle[linewidth=0.04,linecolor=orange,dimen=outer,fillstyle=solid,fillcolor=orange](3.7653124,3.2495313){0.08}
\pscircle[linewidth=0.04,linecolor=orange,dimen=outer,fillstyle=solid,fillcolor=orange](3.2053125,3.2495313){0.08}
\psline[linewidth=0.04cm,linecolor=orange,fillcolor=orange,dotsize=0.07055555cm 2.0]{*-*}(3.7053125,-0.55046874)(3.3453126,-0.55046874)
\usefont{T1}{ptm}{m}{n}
\rput(3.5376563,-0.40546876){\scriptsize R}
\usefont{T1}{ptm}{m}{n}
\rput(2.0976562,-0.40546876){\scriptsize R}
\psline[linewidth=0.04cm,fillcolor=black,dotsize=0.07055555cm 2.0]{-*}(2.3853126,-0.5704687)(1.8253125,-0.5704687)
\pscircle[linewidth=0.04,dimen=outer,fillstyle=solid,fillcolor=black](2.3853126,-0.5704687){0.08}
\psline[linewidth=0.04cm,linecolor=orange](3.7053125,-1.2304688)(3.3453126,-1.2304688)
\rput{-90.0}(4.1157813,1.6348437){\pstriangle[linewidth=0.02,dimen=outer](2.8753126,-1.4104687)(0.38,0.34)}
\psline[linewidth=0.04cm](2.3653126,-1.2304688)(1.8053125,-1.2304688)
\pscircle[linewidth=0.04,dimen=outer,fillstyle=solid,fillcolor=black](2.3653126,-1.2304688){0.08}
\pscircle[linewidth=0.04,dimen=outer,fillstyle=solid,fillcolor=black](1.8053125,-1.2304688){0.08}
\pscircle[linewidth=0.04,dimen=outer,fillstyle=solid,fillcolor=black](3.7053125,-1.2304688){0.08}
\pscircle[linewidth=0.04,linecolor=orange,dimen=outer,fillstyle=solid,fillcolor=orange](3.3453126,-1.2304688){0.08}
\psline[linewidth=0.04cm,linecolor=orange](3.7053125,-1.8504688)(3.3453126,-1.8504688)
\rput{-90.0}(4.735781,1.0148437){\pstriangle[linewidth=0.02,dimen=outer](2.8753126,-2.0304687)(0.38,0.34)}
\psline[linewidth=0.04cm](2.3653126,-1.8504688)(1.8053125,-1.8504688)
\pscircle[linewidth=0.04,dimen=outer,fillstyle=solid,fillcolor=black](2.3653126,-1.8504688){0.08}
\pscircle[linewidth=0.04,dimen=outer,fillstyle=solid,fillcolor=black](1.8053125,-1.8504688){0.08}
\pscircle[linewidth=0.04,dimen=outer,fillstyle=solid,fillcolor=black](3.7053125,-1.8504688){0.08}
\pscircle[linewidth=0.04,dimen=outer,fillstyle=solid,fillcolor=black](3.3453126,-1.8504688){0.08}
\usefont{T1}{ptm}{m}{n}
\rput{25.647673}(-0.32068443,-2.860772){\rput(6.118736,-2.1424344){\scriptsize \psframebox*[framesep=0, boxsep=false,fillcolor=white] {R}}}
\usefont{T1}{ptm}{m}{n}
\rput{25.647673}(0.44153553,-3.8199923){\rput(8.606801,-0.9478072){\scriptsize R}}
\usefont{T1}{ptm}{m}{n}
\rput{25.647673}(0.29700914,-2.6534336){\rput(5.972159,-0.68198407){\scriptsize R}}
\usefont{T1}{ptm}{m}{n}
\rput{25.647673}(0.7664924,-3.244258){\rput(7.504662,0.05383703){\scriptsize R}}
\usefont{T1}{ptm}{m}{n}
\rput{25.647673}(0.0013150312,-3.2081223){\rput(7.040513,-1.6088288){\scriptsize I}}
\psline[linewidth=0.04cm,fillcolor=black,dotsize=0.07055555cm 2.0]{*-*}(7.3853126,2.1095312)(7.4053125,2.8695312)
\psline[linewidth=0.04cm,fillcolor=black,dotsize=0.07055555cm 2.0]{-*}(7.4053125,2.8695312)(7.3453126,3.5295312)
\usefont{T1}{ptm}{m}{n}
\rput(7.1651564,2.8145313){\scriptsize I}
\psframe[linewidth=0.02,dimen=outer](3.9453125,4.2295313)(1.7053125,3.6695313)
\psframe[linewidth=0.01,dimen=outer](3.9453125,3.4895313)(1.7053125,2.9295313)
\psframe[linewidth=0.01,dimen=outer](3.9453125,2.8095312)(1.2053125,2.2495313)
\psframe[linewidth=0.01,dimen=outer](3.9053125,-0.23046875)(1.6653125,-0.79046875)
\psframe[linewidth=0.01,dimen=outer](3.9053125,-0.9504688)(1.6653125,-1.5104687)
\psframe[linewidth=0.01,dimen=outer](3.8853126,-1.5704688)(1.6453125,-2.1304688)
\psframe[linewidth=0.01,dimen=outer](3.9453125,2.1295311)(1.0053124,0.6095312)
\psframe[linewidth=0.01,dimen=outer](8.025312,4.2495313)(5.4253125,1.9095312)
\psline[linewidth=0.04cm,fillcolor=black,dotsize=0.07055555cm 2.0]{*-*}(7.614375,2.1095312)(7.594375,2.8695312)
\psline[linewidth=0.04cm,fillcolor=black,dotsize=0.07055555cm 2.0]{-*}(7.594375,2.8695312)(7.654375,3.5295312)
\usefont{T1}{ptm}{m}{n}
\rput(7.793906,2.8145313){\scriptsize I}
\usefont{T1}{ptm}{m}{n}
\rput{25.647673}(0.15838481,-3.4057877){\rput(7.553226,-1.3626534){\scriptsize I}}
\rput{-90.0}(5.1557813,0.03484375){\pstriangle[linewidth=0.02,dimen=outer](2.5953126,-2.7304688)(0.38,0.34)}
\psline[linewidth=0.03cm,linecolor=color919](3.1053126,-2.5304687)(3.6453125,-2.5304687)
\pscircle[linewidth=0.04,dimen=outer,fillstyle=solid,fillcolor=black](3.7053125,-2.5304687){0.08}
\pscircle[linewidth=0.04,dimen=outer,fillstyle=solid,fillcolor=black](3.0453124,-2.5304687){0.08}
\rput{-42.31023}(2.6686785,1.8396212){\pscircle[linewidth=0.04,dimen=outer,fillstyle=solid,fillcolor=black](3.711274,-2.528331){0.08}}
\psline[linewidth=0.03cm,linecolor=blue](1.4053125,-2.5304687)(1.9453125,-2.5304687)
\pscircle[linewidth=0.04,dimen=outer,fillstyle=solid,fillcolor=black](1.3453125,-2.5304687){0.08}
\rput{-42.31023}(2.2154276,0.6683497){\pscircle[linewidth=0.04,dimen=outer,fillstyle=solid,fillcolor=black](1.9712739,-2.528331){0.08}}
\rput{-90.0}(5.835781,-0.92515624){\pstriangle[linewidth=0.02,dimen=outer](2.4553125,-3.5504687)(0.38,0.34)}
\psline[linewidth=0.03cm,linecolor=color919](3.6053126,-3.3904688)(3.6853125,-4.2504687)
\psline[linewidth=0.03cm,linecolor=color919](3.6053126,-3.3904688)(3.5053124,-4.2304688)
\psline[linewidth=0.03cm,linecolor=color919,fillcolor=color919,dotsize=0.07055555cm 2.0]{-*}(2.9453125,-3.1304688)(3.6053126,-3.3704689)
\pscircle[linewidth=0.04,linecolor=color919,dimen=outer,fillstyle=solid,fillcolor=color919](3.6053126,-3.3704689){0.08}
\pscircle[linewidth=0.04,dimen=outer,fillstyle=solid,fillcolor=black](2.9653125,-3.1504688){0.08}
\rput{-103.10508}(8.411894,-1.7958933){\pscircle[linewidth=0.04,dimen=outer,fillstyle=solid,fillcolor=black](3.4930315,-4.2372155){0.08}}
\rput{-123.863205}(9.283383,-3.5274813){\pscircle[linewidth=0.04,dimen=outer,fillstyle=solid,fillcolor=black](3.7011917,-4.238884){0.08}}
\psline[linewidth=0.03cm,linecolor=blue](1.8453125,-3.3904688)(1.9253125,-4.2504687)
\psline[linewidth=0.03cm,linecolor=blue](1.8453125,-3.3904688)(1.7453125,-4.2304688)
\psline[linewidth=0.03cm,linecolor=color1060,fillcolor=color1060,dotsize=0.07055555cm 2.0]{-*}(1.1853125,-3.1304688)(1.8453125,-3.3704689)
\pscircle[linewidth=0.04,linecolor=blue,dimen=outer,fillstyle=solid,fillcolor=blue](1.8453125,-3.3704689){0.08}
\pscircle[linewidth=0.04,dimen=outer,fillstyle=solid,fillcolor=black](1.2053125,-3.1504688){0.08}
\rput{-103.10508}(6.2528358,-3.5100558){\pscircle[linewidth=0.04,dimen=outer,fillstyle=solid,fillcolor=black](1.7330315,-4.2372155){0.08}}
\rput{-123.863205}(6.54269,-4.988933){\pscircle[linewidth=0.04,dimen=outer,fillstyle=solid,fillcolor=black](1.9411916,-4.238884){0.08}}
\psframe[linewidth=0.01,dimen=outer](3.9053125,-2.2704687)(1.1653125,-2.8304687)
\psframe[linewidth=0.01,dimen=outer](3.9253125,-2.9304688)(0.9853125,-4.4504685)
\psline[linewidth=0.04cm,arrowsize=0.16cm 4.0,arrowlength=1.4,arrowinset=0.4]{->}(6.7053127,1.8495313)(6.7253127,1.2295313)
\psline[linewidth=0.04cm,arrowsize=0.16cm 4.0,arrowlength=1.4,arrowinset=0.4]{->}(4.2053127,0.7295312)(4.8453126,0.7295312)
\psline[linewidth=0.02cm](4.1853123,0.44953126)(4.2053127,4.329531)
\psline[linewidth=0.02cm]{cc-}(4.2053127,4.329531)(3.9853125,4.329531)
\psline[linewidth=0.02cm](4.1853123,0.44953126)(3.9653125,0.44953126)
\psline[linewidth=0.04cm,arrowsize=0.16cm 4.0,arrowlength=1.4,arrowinset=0.4]{->}(4.1653123,-1.2704687)(4.8053126,-1.2704687)
\psline[linewidth=0.02cm]{cc-cc}(4.1453123,-4.570469)(4.1853123,-0.13046876)
\psline[linewidth=0.02cm]{cc-}(4.1853123,-0.13046876)(3.9653125,-0.13046876)
\psline[linewidth=0.02cm](4.1453123,-4.570469)(3.9253125,-4.570469)
\usefont{T1}{ptm}{m}{n}
\rput(1.8309375,4.454531){\scriptsize $\text{Contraction sequence}:=$}
\usefont{T1}{ptm}{m}{n}
\rput(1.9309375,0.05453125){\scriptsize $\text{Relaxation sequence}:=$}
\usefont{T1}{ptm}{m}{n}
\rput(6.6309376,4.454531){\scriptsize $\text{Initiate contraction}:=$}
\usefont{T1}{ptm}{m}{n}
\rput(4.4890623,4.034531){\tiny $\infty P$}
\usefont{T1}{ptm}{m}{n}
\rput(4.4690623,-0.38546875){\tiny $\infty P$}
\usefont{T1}{ptm}{m}{n}
\rput(8.153906,3.9145312){\tiny 1}
\end{pspicture}
}
\end{equation*}

Based on the same principles, it would be possible to construct a 3D model of a beating heart. Interestingly, it would also be possible to simulate conditions like cardiac arrhythmia by putting different restrictions in the model or applying different types of transformations to the model. For example, by randomly contracting pacemaker cells of the model we could study the effects on the heart's functions.

\subsection{Skin Healing}

The skin's epidermis, in particular the stratum spinosum, can modeled very coarsely as seen below.

%Skin
\begin{equation*}\scalebox{0.18} % Change this value to rescale the drawing.
{
\begin{pspicture}(0,-14.95)(36.3,14.95)
\definecolor{color2944b}{rgb}{0.8,0.2,0.0}
\psframe[linewidth=0.04,dimen=outer,fillstyle=solid,fillcolor=color2944b](8.88,-2.35)(6.36,-4.87)
\psframe[linewidth=0.04,dimen=outer,fillstyle=solid,fillcolor=color2944b](11.4,-2.35)(8.88,-4.87)
\psframe[linewidth=0.04,dimen=outer,fillstyle=solid,fillcolor=color2944b](13.92,-2.35)(11.4,-4.87)
\psframe[linewidth=0.04,dimen=outer,fillstyle=solid,fillcolor=color2944b](16.44,-2.35)(13.92,-4.87)
\psframe[linewidth=0.04,dimen=outer,fillstyle=solid,fillcolor=color2944b](18.96,-2.35)(16.44,-4.87)
\psframe[linewidth=0.04,dimen=outer,fillstyle=solid,fillcolor=color2944b](21.48,-2.35)(18.96,-4.87)
\psframe[linewidth=0.04,dimen=outer,fillstyle=solid,fillcolor=color2944b](31.52,-2.35)(29.0,-4.87)
\psframe[linewidth=0.04,dimen=outer,fillstyle=solid,fillcolor=color2944b](34.04,-2.35)(31.52,-4.87)
\psframe[linewidth=0.04,dimen=outer,fillstyle=solid,fillcolor=color2944b](8.88,-4.87)(6.36,-7.39)
\psframe[linewidth=0.04,dimen=outer,fillstyle=solid,fillcolor=color2944b](11.4,-4.87)(8.88,-7.39)
\psframe[linewidth=0.04,dimen=outer,fillstyle=solid,fillcolor=color2944b](13.92,-4.87)(11.4,-7.39)
\psframe[linewidth=0.04,dimen=outer,fillstyle=solid,fillcolor=color2944b](16.44,-4.87)(13.92,-7.39)
\psframe[linewidth=0.04,dimen=outer,fillstyle=solid,fillcolor=color2944b](18.96,-4.87)(16.44,-7.39)
\psframe[linewidth=0.04,dimen=outer,fillstyle=solid,fillcolor=color2944b](21.48,-4.87)(18.96,-7.39)
\psframe[linewidth=0.04,dimen=outer,fillstyle=solid,fillcolor=color2944b](31.52,-4.87)(29.0,-7.39)
\psframe[linewidth=0.04,dimen=outer,fillstyle=solid,fillcolor=color2944b](34.04,-4.87)(31.52,-7.39)
\psframe[linewidth=0.04,dimen=outer,fillstyle=solid,fillcolor=color2944b](8.88,-7.37)(6.36,-9.89)
\psframe[linewidth=0.04,dimen=outer,fillstyle=solid,fillcolor=color2944b](11.4,-7.37)(8.88,-9.89)
\psframe[linewidth=0.04,dimen=outer,fillstyle=solid,fillcolor=color2944b](13.92,-7.37)(11.4,-9.89)
\psframe[linewidth=0.04,dimen=outer,fillstyle=solid,fillcolor=color2944b](16.44,-7.37)(13.92,-9.89)
\psframe[linewidth=0.04,dimen=outer,fillstyle=solid,fillcolor=color2944b](18.96,-7.37)(16.44,-9.89)
\psframe[linewidth=0.04,dimen=outer,fillstyle=solid,fillcolor=color2944b](21.48,-7.37)(18.96,-9.89)
\psframe[linewidth=0.04,dimen=outer,fillstyle=solid,fillcolor=color2944b](31.52,-7.37)(29.0,-9.89)
\psframe[linewidth=0.04,dimen=outer,fillstyle=solid,fillcolor=color2944b](34.04,-7.37)(31.52,-9.89)
\psframe[linewidth=0.04,dimen=outer,fillstyle=solid,fillcolor=color2944b](8.88,-9.89)(6.36,-12.41)
\psframe[linewidth=0.04,dimen=outer,fillstyle=solid,fillcolor=color2944b](11.4,-9.89)(8.88,-12.41)
\psframe[linewidth=0.04,dimen=outer,fillstyle=solid,fillcolor=color2944b](13.92,-9.89)(11.4,-12.41)
\psframe[linewidth=0.04,dimen=outer,fillstyle=solid,fillcolor=color2944b](16.44,-9.89)(13.92,-12.41)
\psframe[linewidth=0.04,dimen=outer,fillstyle=solid,fillcolor=color2944b](18.96,-9.89)(16.44,-12.41)
\psframe[linewidth=0.04,dimen=outer,fillstyle=solid,fillcolor=color2944b](21.48,-9.89)(18.96,-12.41)
\psframe[linewidth=0.04,dimen=outer,fillstyle=solid,fillcolor=color2944b](31.52,-9.89)(29.0,-12.41)
\psframe[linewidth=0.04,dimen=outer,fillstyle=solid,fillcolor=color2944b](34.04,-9.89)(31.52,-12.41)
\psline[linewidth=0.04cm,linestyle=dashed,dash=0.16cm 0.16cm](34.06,-2.37)(36.28,-2.37)
\psline[linewidth=0.04cm,linestyle=dashed,dash=0.16cm 0.16cm](34.02,-4.89)(36.24,-4.89)
\psline[linewidth=0.04cm,linestyle=dashed,dash=0.16cm 0.16cm](34.06,-7.39)(36.28,-7.39)
\psline[linewidth=0.04cm,linestyle=dashed,dash=0.16cm 0.16cm](34.02,-9.91)(36.24,-9.91)
\psline[linewidth=0.04cm,linestyle=dashed,dash=0.16cm 0.16cm](34.0,-12.43)(36.22,-12.43)
\psline[linewidth=0.04cm,linestyle=dashed,dash=0.16cm 0.16cm](4.18,-2.37)(6.4,-2.37)
\psline[linewidth=0.04cm,linestyle=dashed,dash=0.16cm 0.16cm](4.14,-4.89)(6.36,-4.89)
\psline[linewidth=0.04cm,linestyle=dashed,dash=0.16cm 0.16cm](4.18,-7.39)(6.4,-7.39)
\psline[linewidth=0.04cm,linestyle=dashed,dash=0.16cm 0.16cm](4.14,-9.91)(6.36,-9.91)
\psline[linewidth=0.04cm,linestyle=dashed,dash=0.16cm 0.16cm](4.12,-12.43)(6.34,-12.43)
\psframe[linewidth=0.04,dimen=outer,fillstyle=solid,fillcolor=color2944b](13.6,14.17)(11.08,11.65)
\psframe[linewidth=0.04,dimen=outer,fillstyle=solid,fillcolor=color2944b](13.6,11.65)(11.08,9.13)
\psframe[linewidth=0.04,dimen=outer,fillstyle=solid,fillcolor=color2944b](16.12,11.65)(13.6,9.13)
\rput{-90.0}(-2.65,20.05){\pstriangle[linewidth=0.04,dimen=outer](8.7,10.49)(2.08,1.72)}
\psframe[linewidth=0.04,dimen=outer,fillstyle=solid,fillcolor=color2944b](3.64,14.23)(1.12,11.71)
\psframe[linewidth=0.04,dimen=outer,fillstyle=solid,fillcolor=color2944b](3.64,11.71)(1.12,9.19)
\psframe[linewidth=0.04,dimen=outer,fillstyle=solid,fillcolor=color2944b](6.16,11.71)(3.64,9.19)
\psframe[linewidth=0.04,dimen=outer](17.08,14.95)(0.0,8.45)
\psframe[linewidth=0.04,dimen=outer,fillstyle=solid,fillcolor=color2944b](16.14,6.47)(13.62,3.95)
\psframe[linewidth=0.04,dimen=outer,fillstyle=solid,fillcolor=color2944b](16.14,3.95)(13.62,1.43)
\psframe[linewidth=0.04,dimen=outer,fillstyle=solid,fillcolor=color2944b](13.62,3.95)(11.1,1.43)
\rput{-90.0}(5.07,12.37){\pstriangle[linewidth=0.04,dimen=outer](8.72,2.79)(2.08,1.72)}
\psframe[linewidth=0.04,dimen=outer,fillstyle=solid,fillcolor=color2944b](6.18,6.53)(3.66,4.01)
\psframe[linewidth=0.04,dimen=outer,fillstyle=solid,fillcolor=color2944b](6.18,4.01)(3.66,1.49)
\psframe[linewidth=0.04,dimen=outer,fillstyle=solid,fillcolor=color2944b](3.66,4.01)(1.14,1.49)
\psframe[linewidth=0.04,dimen=outer](17.1,7.25)(0.02,0.75)
\psframe[linewidth=0.04,dimen=outer,fillstyle=solid,fillcolor=color2944b](8.88,-12.39)(6.36,-14.91)
\psframe[linewidth=0.04,dimen=outer,fillstyle=solid,fillcolor=color2944b](11.4,-12.39)(8.88,-14.91)
\psframe[linewidth=0.04,dimen=outer,fillstyle=solid,fillcolor=color2944b](13.92,-12.39)(11.4,-14.91)
\psframe[linewidth=0.04,dimen=outer,fillstyle=solid,fillcolor=color2944b](16.44,-12.39)(13.92,-14.91)
\psframe[linewidth=0.04,dimen=outer,fillstyle=solid,fillcolor=color2944b](18.96,-12.39)(16.44,-14.91)
\psframe[linewidth=0.04,dimen=outer,fillstyle=solid,fillcolor=color2944b](21.48,-12.39)(18.96,-14.91)
\psframe[linewidth=0.04,dimen=outer,fillstyle=solid,fillcolor=color2944b](31.52,-12.39)(29.0,-14.91)
\psframe[linewidth=0.04,dimen=outer,fillstyle=solid,fillcolor=color2944b](34.04,-12.39)(31.52,-14.91)
\psline[linewidth=0.04cm,linestyle=dashed,dash=0.16cm 0.16cm](34.02,-14.93)(36.24,-14.93)
\psline[linewidth=0.04cm,linestyle=dashed,dash=0.16cm 0.16cm](4.14,-14.93)(6.36,-14.93)
\psframe[linewidth=0.04,dimen=outer,fillstyle=solid,fillcolor=color2944b](23.96,-2.35)(21.44,-4.87)
\psframe[linewidth=0.04,dimen=outer,fillstyle=solid,fillcolor=color2944b](26.48,-2.35)(23.96,-4.87)
\psframe[linewidth=0.04,dimen=outer,fillstyle=solid,fillcolor=color2944b](29.0,-2.35)(26.48,-4.87)
\psframe[linewidth=0.04,dimen=outer,fillstyle=solid,fillcolor=color2944b](23.96,-4.87)(21.44,-7.39)
\psframe[linewidth=0.04,dimen=outer,fillstyle=solid,fillcolor=color2944b](26.48,-4.87)(23.96,-7.39)
\psframe[linewidth=0.04,dimen=outer,fillstyle=solid,fillcolor=color2944b](29.0,-4.87)(26.48,-7.39)
\psframe[linewidth=0.04,dimen=outer,fillstyle=solid,fillcolor=color2944b](23.96,-7.37)(21.44,-9.89)
\psframe[linewidth=0.04,dimen=outer,fillstyle=solid,fillcolor=color2944b](26.48,-7.37)(23.96,-9.89)
\psframe[linewidth=0.04,dimen=outer,fillstyle=solid,fillcolor=color2944b](29.0,-7.37)(26.48,-9.89)
\psframe[linewidth=0.04,dimen=outer,fillstyle=solid,fillcolor=color2944b](23.96,-9.89)(21.44,-12.41)
\psframe[linewidth=0.04,dimen=outer,fillstyle=solid,fillcolor=color2944b](26.48,-9.89)(23.96,-12.41)
\psframe[linewidth=0.04,dimen=outer,fillstyle=solid,fillcolor=color2944b](29.0,-9.89)(26.48,-12.41)
\psframe[linewidth=0.04,dimen=outer,fillstyle=solid,fillcolor=color2944b](23.96,-12.39)(21.44,-14.91)
\psframe[linewidth=0.04,dimen=outer,fillstyle=solid,fillcolor=color2944b](26.48,-12.39)(23.96,-14.91)
\psframe[linewidth=0.04,dimen=outer,fillstyle=solid,fillcolor=color2944b](29.0,-12.39)(26.48,-14.91)
\psline[linewidth=0.04cm,arrowsize=0.64cm 16.0,arrowlength=1.4,arrowinset=0.4]{->}(17.4,11.39)(22.5,0.49)
\psline[linewidth=0.04cm,arrowsize=0.64cm 16.0,arrowlength=1.4,arrowinset=0.4]{->}(17.4,3.75)(18.82,0.33)
\psframe[linewidth=0.04,dimen=outer,fillstyle=solid,fillcolor=color2944b](3.66,6.53)(1.14,4.01)
\psframe[linewidth=0.04,dimen=outer,fillstyle=solid,fillcolor=color2944b](6.16,14.23)(3.64,11.71)
\usefont{T1}{ptm}{m}{n}
\rput(18.12453,13.57){\Huge $\infty P$}
\usefont{T1}{ptm}{m}{n}
\rput(18.10453,5.49){\Huge $\infty P$}
\end{pspicture}
}\end{equation*}

When we remove cells from our model to represent a skin injury, the two transformations will replace the cells from the bottom up. This can be interpreted as an upward migration of the keratinocytes.

The following figure is an example of a skin injury to which the two transformations can be applied.

%\begin{center}\animategraphics[controls, scale=0.18, loop]{3}{Healing_}{1}{21}\end{center}

\begin{figure}[H]
\centering
\includegraphics[scale=0.15]{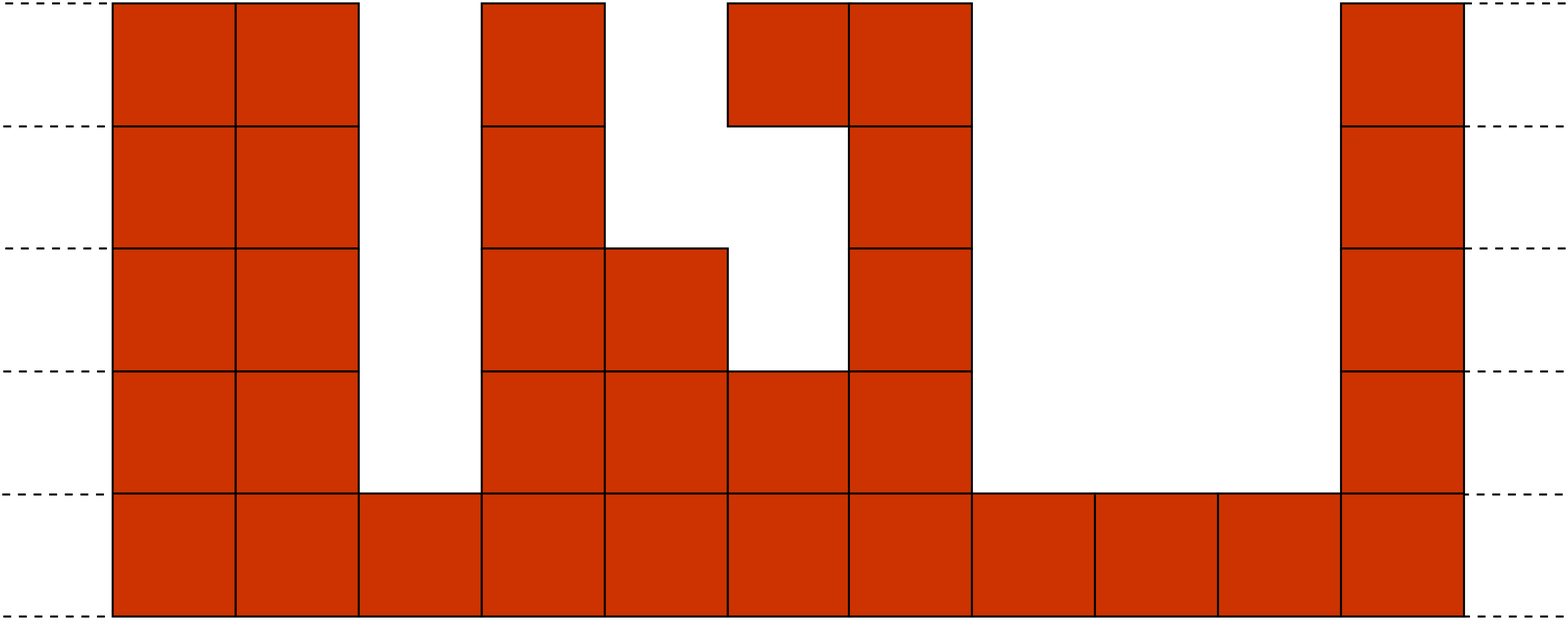}
%\caption{3D notation for a rotation of $45^\circ$ and scaling by $2$.}
\end{figure}

\bigskip

Note that the two transformations can be simplified into  \includegraphics[scale=0.05]{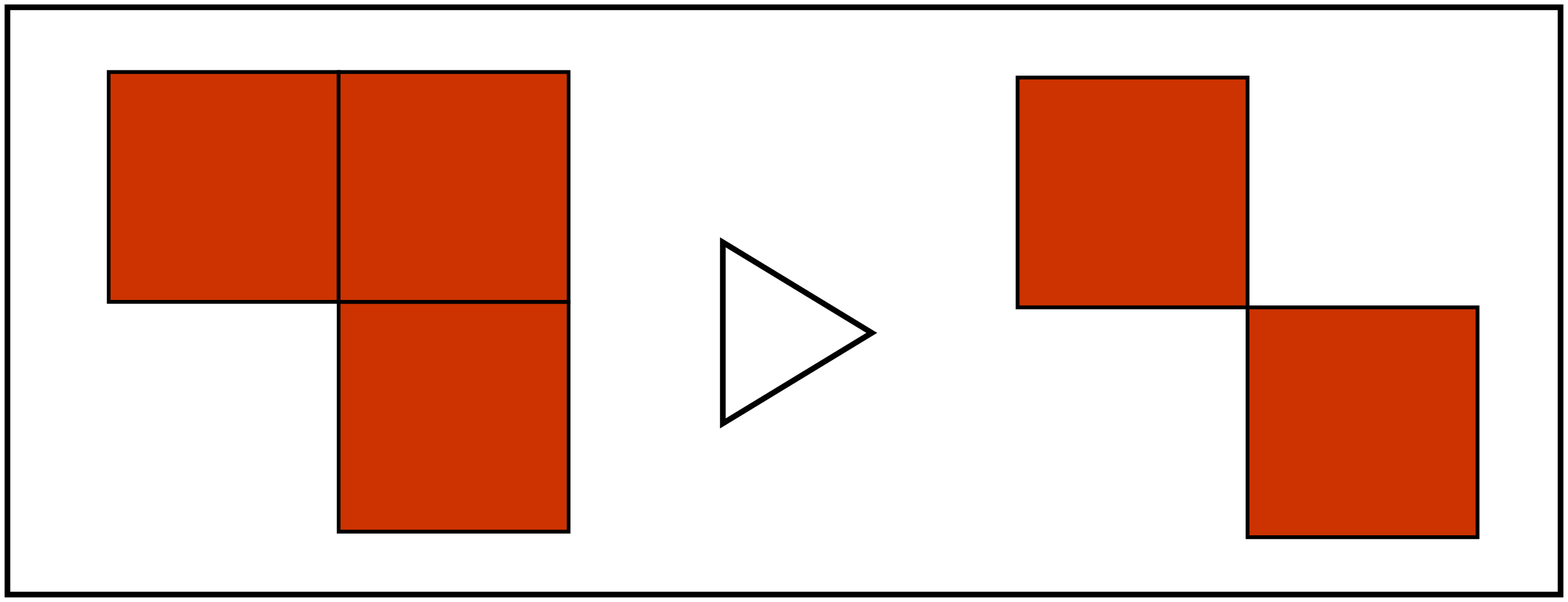} and \includegraphics[scale=0.05]{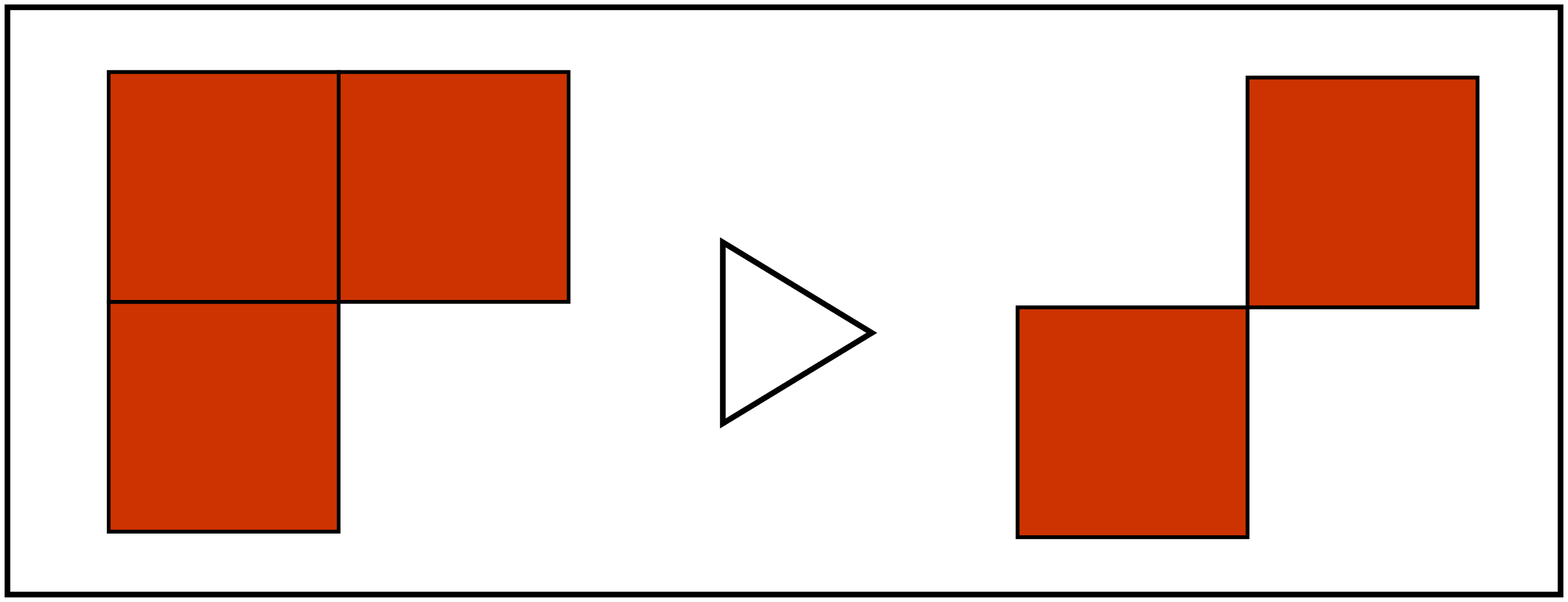}. Moreover, only one of these would be enough for our model skin healing.

\subsection{Neurons}

We will now give a model of a network of neurons including the transformations associated with it.

\subsubsection{Diagram model}\label{Neuron_Diagram model}
Here is a model of a single neuron.
%Neuron
\begin{equation*}\scalebox{0.8} % Change this value to rescale the drawing.
{
\begin{pspicture}(0,-1.52)(7.6075,1.52)
\pscircle[linewidth=0.06,dimen=outer](1.9,0.0){0.82}
\psline[linewidth=0.04cm,tbarsize=0.07055555cm 5.0,rbracketlength=0.15]{)-}(3.0,1.5)(2.36,0.68)
\psline[linewidth=0.04cm,tbarsize=0.07055555cm 5.0,rbracketlength=0.15]{)-}(0.92,1.5)(1.5,0.7)
\psline[linewidth=0.04cm,tbarsize=0.07055555cm 5.0,rbracketlength=0.15]{)-}(0.0,0.0)(1.1,-0.02)
\psline[linewidth=0.04cm,tbarsize=0.07055555cm 5.0,rbracketlength=0.15]{)-}(2.92,-1.44)(2.34,-0.68)
\psline[linewidth=0.04cm,tbarsize=0.07055555cm 5.0,rbracketlength=0.15]{)-}(0.92,-1.5)(1.5,-0.7)
\psframe[linewidth=0.04,dimen=outer,fillstyle=solid,fillcolor=black](6.42,0.12)(2.66,-0.12)
\psline[linewidth=0.04cm,dotsize=0.07055555cm 10.0]{-o}(6.4,0.1)(7.08,0.92)
\psline[linewidth=0.04cm,dotsize=0.07055555cm 10.0]{-o}(6.42,0.04)(7.46,0.38)
\usefont{T1}{ptm}{m}{n}
\rput(7.4525,0.375){\footnotesize X}
\usefont{T1}{ptm}{m}{n}
\rput(7.087656,0.895){\footnotesize W}
\psline[linewidth=0.04cm,dotsize=0.07055555cm 10.0]{-o}(6.4,-0.088125)(7.08,-0.908125)
\psline[linewidth=0.04cm,dotsize=0.07055555cm 10.0]{-o}(6.42,-0.028125)(7.46,-0.368125)
\usefont{T1}{ptm}{m}{n}
\rput(7.451719,-0.385){\footnotesize Y}
\usefont{T1}{ptm}{m}{n}
\rput(7.07125,-0.925){\footnotesize Z}
\end{pspicture}
}\end{equation*}

Neurons are connected as follows. Note that we have also included the threshold transformation of each neuron.

%Neurons
\begin{equation*}\scalebox{0.8} % Change this value to rescale the drawing.
{
\begin{pspicture}(0,-3.29)(15.179063,3.29)
\pscircle[linewidth=0.06,dimen=outer](9.28,-0.87){0.82}
\psline[linewidth=0.04cm,tbarsize=0.07055555cm 5.0,rbracketlength=0.15]{)-}(10.38,0.63)(9.74,-0.19)
\psline[linewidth=0.04cm,tbarsize=0.07055555cm 5.0,rbracketlength=0.15]{)-}(8.3,0.63)(8.88,-0.17)
\psline[linewidth=0.04cm,tbarsize=0.07055555cm 5.0,rbracketlength=0.15]{)-}(7.38,-0.87)(8.48,-0.89)
\psline[linewidth=0.04cm,tbarsize=0.07055555cm 5.0,rbracketlength=0.15]{)-}(10.3,-2.31)(9.72,-1.55)
\psline[linewidth=0.04cm,tbarsize=0.07055555cm 5.0,rbracketlength=0.15]{)-}(8.3,-2.37)(8.88,-1.57)
\psframe[linewidth=0.04,dimen=outer,fillstyle=solid,fillcolor=black](13.8,-0.75)(10.04,-0.99)
\psline[linewidth=0.04cm,dotsize=0.07055555cm 10.0]{-o}(13.78,-0.77)(14.46,0.05)
\usefont{T1}{ptm}{m}{n}
\rput(11.325625,-1.385){\small $\quad\quad\text{Threshold}_3^{\infty S}$}
\psline[linewidth=0.04cm,dotsize=0.07055555cm 10.0]{-o}(13.8,-0.83)(14.84,-0.49)
\usefont{T1}{ptm}{m}{n}
\rput(14.82375,-0.485){\tiny $X_3$}
\usefont{T1}{ptm}{m}{n}
\rput(14.45375,0.055){\tiny $W_3$}
\psline[linewidth=0.04cm,dotsize=0.07055555cm 10.0]{-o}(13.78,-0.958125)(14.46,-1.778125)
\psline[linewidth=0.04cm,dotsize=0.07055555cm 10.0]{-o}(13.8,-0.898125)(14.84,-1.238125)
\usefont{T1}{ptm}{m}{n}
\rput(14.82375,-1.225){\tiny $Y_3$}
\usefont{T1}{ptm}{m}{n}
\rput(14.45375,-1.765){\tiny $Z_3$}
\pscircle[linewidth=0.06,dimen=outer](2.92,1.77){0.82}
\psline[linewidth=0.04cm,tbarsize=0.07055555cm 5.0,rbracketlength=0.15]{)-}(4.02,3.27)(3.38,2.45)
\psline[linewidth=0.04cm,tbarsize=0.07055555cm 5.0,rbracketlength=0.15]{)-}(1.94,3.27)(2.52,2.47)
\psline[linewidth=0.04cm,tbarsize=0.07055555cm 5.0,rbracketlength=0.15]{)-}(1.02,1.77)(2.12,1.75)
\psline[linewidth=0.04cm,tbarsize=0.07055555cm 5.0,rbracketlength=0.15]{)-}(3.94,0.33)(3.36,1.09)
\psline[linewidth=0.04cm,tbarsize=0.07055555cm 5.0,rbracketlength=0.15]{)-}(1.94,0.27)(2.52,1.07)
\psframe[linewidth=0.04,dimen=outer,fillstyle=solid,fillcolor=black](7.44,1.89)(3.68,1.65)
\psline[linewidth=0.04cm,linestyle=dashed,dash=0.16cm 0.16cm](7.42,1.87)(8.1,2.69)
\usefont{T1}{ptm}{m}{n}
\rput(5.005625,1.215){\small $\quad\quad\text{Threshold}_2^{\infty S}$}
\psline[linewidth=0.04cm,linestyle=dashed,dash=0.16cm 0.16cm](7.44,1.81)(8.48,2.15)
\psline[linewidth=0.04cm,dotsize=0.07055555cm 10.0]{-o}(7.42,1.681875)(8.1,0.861875)
\psline[linewidth=0.04cm,linestyle=dashed,dash=0.16cm 0.16cm](7.44,1.741875)(8.48,1.401875)
\usefont{T1}{ptm}{m}{n}
\rput(8.09375,0.875){\tiny $Z_2$}
\pscircle[linewidth=0.06,dimen=outer](1.9,-1.77){0.82}
\psline[linewidth=0.04cm,tbarsize=0.07055555cm 5.0,rbracketlength=0.15]{)-}(3.0,-0.27)(2.36,-1.09)
\psline[linewidth=0.04cm,tbarsize=0.07055555cm 5.0,rbracketlength=0.15]{)-}(0.92,-0.27)(1.5,-1.07)
\psline[linewidth=0.04cm,tbarsize=0.07055555cm 5.0,rbracketlength=0.15]{)-}(0.0,-1.77)(1.1,-1.79)
\psline[linewidth=0.04cm,tbarsize=0.07055555cm 5.0,rbracketlength=0.15]{)-}(2.92,-3.21)(2.34,-2.45)
\psline[linewidth=0.04cm,tbarsize=0.07055555cm 5.0,rbracketlength=0.15]{)-}(0.92,-3.27)(1.5,-2.47)
\psframe[linewidth=0.04,dimen=outer,fillstyle=solid,fillcolor=black](6.42,-1.65)(2.66,-1.89)
\psline[linewidth=0.04cm,dotsize=0.07055555cm 10.0]{-o}(6.4,-1.67)(7.08,-0.85)
\usefont{T1}{ptm}{m}{n}
\rput(3.925625,-2.345){\small $\quad\quad\text{Threshold}_1^{\infty S}$}
\psline[linewidth=0.04cm,linestyle=dashed,dash=0.16cm 0.16cm](6.42,-1.73)(7.46,-1.39)
\usefont{T1}{ptm}{m}{n}
\rput(7.07375,-0.845){\tiny $W_1$}
\psline[linewidth=0.04cm,linestyle=dashed,dash=0.16cm 0.16cm](6.4,-1.858125)(7.08,-2.678125)
\psline[linewidth=0.04cm,dotsize=0.07055555cm 10.0]{-o}(6.42,-1.798125)(8.1,-2.59)
\usefont{T1}{ptm}{m}{n}
\rput(8.08375,-2.585){\tiny $Y_1$}
\psline[linewidth=0.04cm,arrowsize=0.16cm 4.0,arrowlength=1.4,arrowinset=0.4]{->}(10.7,-1.39)(9.98,-1.29)
\psline[linewidth=0.04cm,arrowsize=0.16cm 4.0,arrowlength=1.4,arrowinset=0.4]{->}(4.32,1.23)(3.6,1.33)
\psline[linewidth=0.04cm,arrowsize=0.16cm 4.0,arrowlength=1.4,arrowinset=0.4]{->}(3.28,-2.35)(2.56,-2.25)
\end{pspicture}
}\end{equation*}

In our simple model, we will be concerned with two actions of the neurotransmitters, excitatory ($E$) and inhibitory ($I$). Thus, The variables such as $W, X, Y$ and $Z$ will only take the values $E$ or $I$.

Each neuron can have a different threshold.  An example of the transformation which signals that the threshold has been attained is
$$Threshold:=\fbox{$Ready \rhd E,E,E$}.$$
\noindent Inhibition is represented by the transformation
$$Inhibition:=\fbox{$\quad \rhd E,I$}$$
 \noindent which cancels out $E$ and $I$ together. The initial set of these transformations is the content of the main large circle of the model neuron. There are two possibilities concerning the inhibition transformation. It can be applied in parallel to the threshold transformation or it can be applied in series with a `$\sharp$' on the superscript followed by the threshold transformation. If it is parallel, then even if there are still letters $I$, the threshold can be reached. If it is in series, the threshold can be reached only if there are no more letters $I$. Here is a list of the transformations needed.

%Neuron FiringWXYZ
\begin{equation*}\scalebox{0.7} % Change this value to rescale the drawing.
{
\begin{pspicture}(0,-3.5379686)(14.9,3.5779688)
\rput{-90.0}(5.8879685,8.972032){\pstriangle[linewidth=0.04,dimen=outer](7.43,1.1420312)(0.82,0.8)}
\pscircle[linewidth=0.06,dimen=outer](8.94,1.5220313){0.82}
\psframe[linewidth=0.04,dimen=outer,fillstyle=solid,fillcolor=black](13.46,1.6420312)(9.7,1.4020313)
\psline[linewidth=0.04cm,dotsize=0.07055555cm 10.0]{-o}(13.44,1.6220312)(14.12,2.4420311)
\usefont{T1}{ptm}{m}{n}
\rput(8.91625,1.5370313){\footnotesize Ready}
\psline[linewidth=0.04cm,dotsize=0.07055555cm 10.0]{-o}(13.46,1.5620313)(14.5,1.9020313)
\usefont{T1}{ptm}{m}{n}
\rput(14.5125,1.8770312){\footnotesize X}
\usefont{T1}{ptm}{m}{n}
\rput(14.1076565,2.4370313){\footnotesize W}
\psline[linewidth=0.04cm,dotsize=0.07055555cm 10.0]{-o}(13.44,1.4339062)(14.12,0.61390626)
\psline[linewidth=0.04cm,dotsize=0.07055555cm 10.0]{-o}(13.46,1.4939063)(14.5,1.1539062)
\usefont{T1}{ptm}{m}{n}
\rput(14.491718,1.1370312){\footnotesize Y}
\usefont{T1}{ptm}{m}{n}
\rput(14.13125,0.59703124){\footnotesize Z}
\pscircle[linewidth=0.06,linecolor=yellow,dimen=outer](0.94,1.5820312){0.82}
\psframe[linewidth=0.04,linecolor=yellow,dimen=outer,fillstyle=solid,fillcolor=yellow](5.46,1.7020313)(1.7,1.4620312)
\psline[linewidth=0.04cm,linecolor=yellow,dotsize=0.07055555cm 10.0]{-o}(5.44,1.6820313)(6.12,2.5020313)
\psline[linewidth=0.04cm,linecolor=yellow,dotsize=0.07055555cm 10.0]{-o}(5.46,1.6220312)(6.5,1.9620312)
\usefont{T1}{ptm}{m}{n}
\rput(6.4725,1.9370313){\footnotesize X}
\usefont{T1}{ptm}{m}{n}
\rput(6.1276565,2.4970312){\footnotesize W}
\psline[linewidth=0.04cm,linecolor=yellow,dotsize=0.07055555cm 10.0]{-o}(5.44,1.4939063)(6.12,0.67390627)
\psline[linewidth=0.04cm,linecolor=yellow,dotsize=0.07055555cm 10.0]{-o}(5.46,1.5539062)(6.5,1.2139063)
\usefont{T1}{ptm}{m}{n}
\rput(6.5117188,1.1970313){\footnotesize Y}
\usefont{T1}{ptm}{m}{n}
\rput(6.15125,0.65703124){\footnotesize Z}
\pscircle[linewidth=0.06,linecolor=yellow,dimen=outer](8.96,-2.2379687){0.82}
\psframe[linewidth=0.04,linecolor=yellow,dimen=outer,fillstyle=solid,fillcolor=yellow](13.48,-2.1179688)(9.72,-2.3579688)
\psline[linewidth=0.04cm,linecolor=yellow,dotsize=0.07055555cm 10.0]{-o}(13.46,-2.1379688)(14.14,-1.3179687)
\psline[linewidth=0.04cm,linecolor=yellow,dotsize=0.07055555cm 10.0]{-o}(13.48,-2.1979687)(14.52,-1.8579688)
\usefont{T1}{ptm}{m}{n}
\rput(14.495937,-1.8429687){\footnotesize x}
\usefont{T1}{ptm}{m}{n}
\rput(14.132969,-1.3029687){\footnotesize w}
\psline[linewidth=0.04cm,linecolor=yellow,dotsize=0.07055555cm 10.0]{-o}(13.46,-2.3260937)(14.14,-3.1460938)
\psline[linewidth=0.04cm,linecolor=yellow,dotsize=0.07055555cm 10.0]{-o}(13.48,-2.2660937)(14.52,-2.6060936)
\usefont{T1}{ptm}{m}{n}
\rput(14.518125,-2.5829687){\footnotesize y}
\usefont{T1}{ptm}{m}{n}
\rput(14.145156,-3.1229687){\footnotesize z}
\pscircle[linewidth=0.06,dimen=outer](0.98,-2.0379686){0.82}
\psframe[linewidth=0.04,dimen=outer,fillstyle=solid,fillcolor=black](5.5,-1.9179688)(1.74,-2.1579688)
\psline[linewidth=0.04cm,dotsize=0.07055555cm 10.0]{-o}(5.48,-1.9379687)(6.16,-1.1179688)
\psline[linewidth=0.04cm,dotsize=0.07055555cm 10.0]{-o}(5.5,-1.9979688)(6.54,-1.6579688)
\psline[linewidth=0.04cm,dotsize=0.07055555cm 10.0]{-o}(5.48,-2.1260939)(6.16,-2.9460938)
\psline[linewidth=0.04cm,dotsize=0.07055555cm 10.0]{-o}(5.5,-2.0660937)(6.54,-2.4060938)
\rput{-90.0}(9.687969,5.2920313){\pstriangle[linewidth=0.04,dimen=outer](7.49,-2.5979688)(0.82,0.8)}
\psframe[linewidth=0.02,dimen=outer](14.9,-0.67796874)(0.0,-3.5379686)
\psframe[linewidth=0.02,dimen=outer](14.9,3.0020313)(0.0,0.14203125)
\usefont{T1}{ptm}{m}{n}
\rput(1.2673438,3.3570313){\large $Firing:=$}
\usefont{T1}{ptm}{m}{n}
\rput(1.9973438,-0.32296875){\large $Reinitialization:=$}
\usefont{T1}{ptm}{m}{n}
\rput(6.5325,-1.6829687){\footnotesize X}
\usefont{T1}{ptm}{m}{n}
\rput(6.1676564,-1.1429688){\footnotesize W}
\usefont{T1}{ptm}{m}{n}
\rput(6.5317187,-2.4229689){\footnotesize Y}
\usefont{T1}{ptm}{m}{n}
\rput(6.15125,-2.9629688){\footnotesize Z}
\end{pspicture}
}\end{equation*}

%Neurons Contributions
\begin{equation*}\scalebox{0.8} % Change this value to rescale the drawing.
{
\begin{pspicture}(0,-1.9679687)(14.64,2.0079687)
\pscircle[linewidth=0.06,dimen=outer](5.9,-0.86796874){0.82}
\psline[linewidth=0.04cm,tbarsize=0.07055555cm 5.0,rbracketlength=0.15]{)-}(4.92,0.63203126)(5.5,-0.16796875)
\pscircle[linewidth=0.04,linecolor=yellow,dimen=outer,fillstyle=solid](4.74,0.85203123){0.24}
\rput{-90.0}(4.257969,3.0420313){\pstriangle[linewidth=0.04,dimen=outer](3.65,-1.0079688)(0.82,0.8)}
\pscircle[linewidth=0.06,dimen=outer](1.78,-0.84796876){0.82}
\psline[linewidth=0.04cm,tbarsize=0.07055555cm 5.0,rbracketlength=0.15]{)-}(0.8,0.65203124)(1.38,-0.14796875)
\pscircle[linewidth=0.04,dimen=outer,linecolor=yellow,fillstyle=solid](0.62,0.8720313){0.24}
\usefont{T1}{ptm}{m}{n}
\rput(1.5507812,-0.39296874){\footnotesize E}
\usefont{T1}{ptm}{m}{n}
\rput(4.730781,0.84703124){\footnotesize E}
\pscircle[linewidth=0.06,dimen=outer](13.52,-0.8879688){0.82}
\psline[linewidth=0.04cm,tbarsize=0.07055555cm 5.0,rbracketlength=0.15]{)-}(12.54,0.6120312)(13.12,-0.18796875)
\pscircle[linewidth=0.04,linecolor=yellow,dimen=outer,fillstyle=solid](12.36,0.83203125){0.24}
\rput{-90.0}(11.897968,10.642032){\pstriangle[linewidth=0.04,dimen=outer](11.27,-1.0279688)(0.82,0.8)}
\pscircle[linewidth=0.06,dimen=outer](9.4,-0.86796874){0.82}
\psline[linewidth=0.04cm,tbarsize=0.07055555cm 5.0,rbracketlength=0.15]{)-}(8.42,0.63203126)(9.0,-0.16796875)
\pscircle[linewidth=0.04,dimen=outer,linecolor=yellow,fillstyle=solid](8.24,0.85203123){0.24}
\usefont{T1}{ptm}{m}{n}
\rput(12.365313,0.82703125){\footnotesize I}
\usefont{T1}{ptm}{m}{n}
\rput(9.125313,-0.41296875){\footnotesize I}
\usefont{T1}{ptm}{m}{n}
\rput(0.600625,0.8670313){\footnotesize e}
\usefont{T1}{ptm}{m}{n}
\rput(8.215938,0.84703124){\footnotesize i}
\psframe[linewidth=0.02,dimen=outer](7.02,1.4120313)(0.0,-1.9479687)
\psframe[linewidth=0.02,dimen=outer](14.64,1.3920312)(7.62,-1.9679687)
\usefont{T1}{ptm}{m}{n}
\rput(9.417344,1.7870313){\large $I\text{-}contribution:=$}
\usefont{T1}{ptm}{m}{n}
\rput(1.9873438,1.7870313){\large $E\text{-}contribution:=$}
\end{pspicture}
}\end{equation*}

The process of a neuron firing is as follows. It receives the $E$ and $I$ contributions when the input neurons are firing. Inhibition is applied in parallel or in series to the threshold transformation. When the threshold is reached, the word `Ready' in written in the neuron. Then the neuron fires with the firing transformation. After this, contributions are added to other neurons and when all contributions have been added to the other neurons, we only have lowercase letters on the terminals of the axon. Then the neuron is reinitialized and will fire again when the threshold is reached.

\subsubsection{Inline model}

We now give an inline model of a few interconnected neurons. Each of these neurons will also have 5 input ($A,B,C,D,E$) and 4 output terminals ($W,X,Y,Z$). Each neuron will be assigned a number $n$ written as a subscript.

\begin{equation*}neuron_n:=[(\;\;\;,A_n)(\;\;\;,B_n)(\;\;\;,C_n)(\;\;\;,D_n)(\;\;\;,E_n)\ll \square \gg W_n,X_n,Y_n,Z_n]\end{equation*}

The letter $A_n, B_n, C_n, D_n$ and $E_n$ in the pairs such as $(\;\;\;,A_n)$ will be replaced with the terminals of other neurons when we create a network. Examples of such terminals are $W_5, X_2, W_{16}$ and $Z_{6002}$. The empty left component of the pairs will be replaced by the type of connection, that is $E$ for excitatory and $I$ for inhibitory. Note that the order in which the pairs appear is important, that is $(\;\;\;,A_n)(\;\;\;,B_n)$ is not the same as $(\;\;\;,B_n)(\;\;\;,A_n)$. The main circle of our diagram model is written inline as $\ll \square \gg$ where a white box means that the neuron is not ready to fire, while a black box means that the threshold has been reached.

\begin{equation*}\text{e-contribution}_n:=\fbox{$(e,D_n),\ll E, \quad\quad \rhd \quad\quad({\color{yellow}E},D_n),\ll \quad$} \end{equation*}

\begin{equation*}\text{i-contribution}_n:=\fbox{$(i,D_n),\ll I, \quad\quad \rhd \quad\quad({\color{yellow}I},D_n),\ll \quad$} \end{equation*}

\begin{equation*}threshold_n:= \fbox{$\quad\quad\quad\quad\quad \blacksquare \quad\quad\rhd \quad\quad\text{set containing letters E and I},\square$}\end{equation*}

The $\text{e-contribution}_n$ acts by transferring neurotransmitters represented by `$E,$' into the center part by writing $\ll E, \square \gg$. The $\text{i-contribution}_n$ works in the same way. An example of threshold for a neuron $k$ is
\begin{equation*}threshold_k:= \fbox{$\quad\quad\;\;\;\; \blacksquare \quad\quad\rhd \quad\quad\text{E,E,E,I},\square$}\end{equation*}

\begin{equation*}firing_n:=\fbox{$\square \gg {\color{yellow}W_n},{\color{yellow}X_n},{\color{yellow}Y_n},{\color{yellow}Z_n}\quad\quad \rhd \quad\quad\blacksquare \gg W_n, X_n, Y_n, Z_n$}\end{equation*}

\begin{equation*}reinitialization_n:=\fbox{$\ll \square \gg W_n,X_n,Y_n,Z_n\quad\quad\rhd\quad\quad\ll \square \gg w_n,x_n,y_n,z_n$}\end{equation*}

An active neuron is the collection of the following transformations: $neuron_n$, $threshold_n$, $\text{e-contribution}_n$, $\text{i-contribution}_n$, $firing_n$ and $reinitialization_n$. For the transformations $\text{e-contribution}_n$, $\text{i-contribution}_n$,  $firing_n$ and $reinitialization_n$, only the subscript will change between neurons. So in general, to define an active neuron, we will only mention the transformations $neuron_n$ and $threshold_n$.

We now define a network of three neurons which is the same as our network of the three neurons diagram above. Note that the question mark means that we are not showing what connection it is and if we have a space as in  $(\;\;\; , A_3)$, we understand that there is no connection.

\begin{equation*}neuron_1:=[(?,?)(?,?)(?,?)(?,?)(?,?)\ll \square \gg W_1,?,Y_1,?]\end{equation*}
\begin{equation*}threshold_1:= \fbox{$\;\;\;\;\;\;\;\blacksquare \quad\quad\rhd \quad\quad\text{E,E},\square$}\end{equation*}

\begin{equation*}neuron_2:=[(?,?)(?,?)(?,?)(?,?)(?,?)\ll \square \gg \; ?,?,?,Z_2]\end{equation*}
\begin{equation*}threshold_2:= \fbox{$\;\;\;\;\;\;\;\blacksquare \quad\quad\rhd \quad\quad\text{E,E},\square$}\end{equation*}

The third neuron is the one on the left that $neuron_1$ and $neuron_2$ connect to.

\begin{equation*}neuron_3:=[(\;\;\; ,\;\;)(E,Y_1)(E,W_1)(I,Z_2)( \;\;\; ,\;\;)\ll \square \gg W_3,X_3,Y_3,Z_3]\end{equation*}
\begin{equation*}threshold_1:= \fbox{$\;\;\;\;\;\;\;\blacksquare \quad\quad\rhd \quad\quad\text{E,E,I},\square$}\end{equation*}

It is possible to generalize our model by allowing more input and output on each neuron. We can also augment the number of different types of neurotransmitters and even consider neurons that have more than one type of neurotransmitter. For a neuron, we can associate a series of threshold transformations that are different from each other to model changes in firing frequency. From a more general view, we can define transformations that change the connections, grow connections and change the threshold of some neurons.

\subsection{Biology diagrams}

Because of its flexibility, the language of transformations can create helpful biology diagrams to understand processes. This is useful for textbooks and academic papers. As we have seen in the case of DNA, we can have a visual diagram and an inline notation (which is more computational). This reduces the gap between what is presented in diagrams and the formal mathematical process of the inline notation.

In the future we can think that there will be an automatic way to switch between 2D notation and inline notation. This will help understanding while keeping the computational power of the mathematical notation. It is possible that mathematicians and biologists will be encoding models and trying to refine them by switching between 2D and inline notations, expand a single form into its basic constituents and using powerful algorithms to expand our knowledge of biology.

\subsection{3D Mesh transformations}

 3D computer graphics are built with a set of interconnected polygons. The polygons are connected with different angles in 3 dimensions and create the surface of an object. We can think of these polygons as the equivalent of digital pixels. With a large 3D database containing the mesh of 3D objects representing organs and body parts, we could transform the shape of 3D representation of body parts by using \textit{mesh transformations}. Examples of mesh transformations on a triangular mesh are dividing triangles into structures composed of smaller triangles, changing the angle between triangles and dissolving triangles.

 The reason for introducing mesh transformations is to reduce the gap between the way biological processes work and how we represent them as 3D computer graphics. For example, having a 3D model of a heart along with a set of transformations similar to the 2D heart presented above, we could modify the shape of the heart to adapt to different individuals. We could also define surgical mesh transformations to represent surgical incisions. Another way to think about some parts of our 2D heart model is to think that the contractions and relaxations of the cell were in fact mesh transformations.
 Interacting with 3D digital objects with mesh transformations allows us to not to leave our system of transformations that is used to represent biological processes.

\subsection{Meta-medicine}

In a large medicine database written in our language, we could recognize and abstract important biological structures and apply these to the understanding problems in other fields of medicine. Investigations such as these have been very fruitful in mathematics, where abstracting structures of the number systems gave rise to abstract algebra.

Also, the knowledge and abstract framework at different scale levels can be used at other scale levels. By extracting abstract principles observed in the human body we could use the same abstract structure to help understand the entire population. For example, if we consider RNA-messenger as information, we could try to understand based on biological processes how to improve knowledge dissemination during epidemics. One wonders if we can find new ways to deal with world health concerns as a whole by understanding how our own body functions.

\section{Chemistry}

We will now see that we can also use systems of transformations to represent chemical reactions. Note that we will use the arrow $\rightarrowtail$ instead of the classical notation $\rightarrow$ for chemical equations.

\subsection{Chemical reactions}
 Examples of reactions are the combustion of carbon in oxygen $C+O_2\rightarrowtail CO_2$ and the combustion of hydrogen as used in space shuttles $2H_2+O_2\rightarrowtail 2H_2O$. Water ($H_2O$), hydrogen gas ($H_2$), oxygen gas ($O_2$) and carbon dioxyde ($CO_2$) are written as Lewis structures as follows.

\begin{center}
\chemname{\chemfig{\lewis{1:3:,O}(-[:217.775]H)(-[:-37.775]H)}}{Water}\hspace{1.5cm}
\chemname{\chemfig{H(-[:0]H)}}{Hydrogen gas}\hspace{1.5cm}
\chemname{\chemfig{\lewis{3:5:,O}(=[:0]\lewis{7:1:,O})}}{Oxygen gas}\hspace{1.5cm}
\chemname{\chemfig{C(=[:180]\lewis{3:5:,O})(=[:0]\lewis{7:1:,O})}}{Carbon dioxide}\end{center}

We can write the carbon combustion reaction in the form of a Lewis structure as follows.

\begin{center}\lewis{0.2.4.6.,C}\chemsign+\chemfig{\lewis{3:5:,O}(=[:0]\lewis{7:1:,O})}
\chemrel{->}\chemfig{C(=[:180]\lewis{3:5:,O})(=[:0]\lewis{7:1:,O})}\end{center}

As a coarse transformation, we can write carbon combustion as \begin{equation*}\fbox{$CO_2\rhd C,O_2$},\end{equation*}
\noindent and we can write hydrogen combustion as \begin{equation*}\fbox{$H_2O, H_2O\rhd H_2,H_2,O_2$}. \end{equation*}

These can also be written in Lewis notation as
\begin{equation*}\setlength{\fboxsep}{10pt}
\fbox{$\chemfig{C(=[:180]\lewis{3:5:,O})(=[:0]\lewis{7:1:,O})}\quad\rhd\quad \lewis{0.2.4.6.,C}\quad,\quad\chemfig{\lewis{3:5:,O}(=[:0]\lewis{7:1:,O})}$}\end{equation*}
and
\begin{equation*}\setlength{\fboxsep}{10pt}
\fbox{$\chemfig{\lewis{1:3:,O}(-[:217.775]H)(-[:-37.775]H)}\quad,\quad \chemfig{\lewis{1:3:,O}(-[:217.775]H)(-[:-37.775]H)}\quad\rhd\quad \chemfig{H(-[:0]H)} \quad,\quad \chemfig{H(-[:0]H)} \quad,\quad \chemfig{\lewis{3:5:,O}(=[:0]\lewis{7:1:,O})}$}\end{equation*}

\subsection{Studying hydrogen combustion}

One way to write a more precise model of the reaction $$2H_2+O_2\rightarrowtail 2H_2O$$ which represent hydrogen combustion in oxygen is by using the following transformations. When hydrogen is burned in the presence of oxygen the product is water. From here we can assume that one of the paths goes through hydrogen peroxide.

\begin{equation*}\setlength{\fboxsep}{10pt}
\text{H-oxidation}:=\fbox{$H^+\quad ,\quad H^+\quad ,\quad e^- \quad ,\quad e^-\quad\rhd\quad \chemfig{H(-[:0]H)}$}\end{equation*}

\begin{equation*}\setlength{\fboxsep}{10pt}
\text{O-reduction}:=\fbox{$O^-\quad\rhd\quad O^{\color{red}-}\quad ,\quad e^-$}\end{equation*}

\begin{equation*}\setlength{\fboxsep}{10pt}
\text{Bond}:=\fbox{$\chemfig{H(-[:0]O)}\quad\rhd\quad O^-\quad ,\quad H^+$}\end{equation*}

%\begin{equation*}\setlength{\fboxsep}{10pt}
%\text{Bond}:=\fbox{$\chemfig{H(-[:0]O)}\quad\rhd\quad \lewis{4.,O^-}\quad ,\quad H^+$}\end{equation*}

By applying these three transformations in parallel to $2H_2+O_2$, the two $H_2$ will become two $H^+$ and two free electrons because of $\text{H-oxidation}$. Then $\text{O-reduction}$ will give an extra electron to the two oxygens. This will result in the temporary molecule $$\chemfig{H>:[:45]O=[:0]O<[:-45]H}$$ Now, when each oxygen shares a covalent bond with a hydrogen atom, a bond is removed between the oxygens to give a hydrogen peroxide molecule ($H_2O_2$). This is represented by
\begin{equation*}\setlength{\fboxsep}{10pt}
\fbox{$\chemfig{H>:[:45]O-[:0]O<[:-45]H}\quad\rhd\quad \chemfig{H>:[:45]O=[:0]O<[:-45]H}$}.\end{equation*}

Note that the black triangle is a classical notation to indicate that the $HO$ bond is oriented towards the viewer and the dashed triangle means that the bond is oriented away from the viewer.

Now, since the oxygens do not have a negative superscript anymore, thus the transformation O-reduction can be applied again to each O. This means that the transformation $Bond$ will be applied again to one of the oxygens. We now only need to dissolve the second bond between the oxygen atoms with the following transformation.

\begin{equation*}\setlength{\fboxsep}{10pt}
\fbox{$\chemfig{O(<:[:135]H)(<[:225]H)([:0])}\quad ,\quad O\quad\rhd\quad \chemfig{O(<:[:135]H)(<[:225]H)(-[:0]O)}$}\end{equation*}

In the present case, we dissolved the bond when there were three attached $H$ atoms instead of four, this is probably more likely. Finally, $OH^-$ hydroxide ion will react with $H$ through the transformation $Bond$ to form the second $H_2O$ molecule. This is one of the many paths to go from $H_2+O_2$ to $H_2O$ and when hydrogen is burned, it is likely that all the paths are followed in different proportions.

All these possible pathways could be computationally studied to understand the inner working of a reaction.

\subsection{Reaction mechanisms}

We will now give an example of a representation of a reaction mechanism.

The decomposition of nitrogen dioxide into nitric oxide and oxygen is an example of a reaction mechanism. The net balanced equation is $2 NO_2\rightarrowtail 2NO+O_2$.

The mechanism of this reaction is believed to follow these steps.
\begin{enumerate}
  \item $2 NO_2\rightarrowtail NO_3+NO$
  \item $NO_3\rightarrowtail NO+O_2$
\end{enumerate}

This can be represented as a system of transformations as
$$[\;\fbox{$NO, O_2\rhd NO_3$},\fbox{$NO_3,NO\rhd NO_2, NO_2$}\;]\rightarrow\{NO_2, NO_2\}.$$

\section{Absolute versus relative transformations}

 Until now we were mostly interested in transformations from a global perspective. To model the evolution of a system, an absolute observer defines transformations which explain the changes observed in the system. We mostly defined transformations that somewhat stand out of the system and are then applied in the system when the conditions are right. The transformations defined as such are only based on our evaluation of the system and are used to describe what we see. From the point of view of a molecule, there is no global law that comes to affect the molecule. It is the same for biological processes, there is nowhere in the body where there is a set of transformations waiting to modify parts of the body.

As we have seen, the absolute transformations can model many things and follow the way we understand and describe the world. This leads us to the concept of relative transformations. It is not a really big change in the notation, but mostly a change in perspective. This perspective is closely related to the idea that until a transformation is applied we cannot detect it. We will now give two examples of systems that can be said to be relative.

\subsection{Relative molecules}

Instead of defining an initial form containing different elements and a collection of transformations applied to it, we will define elements that come with their transformation. We have already seen in section \ref{Neuron_Diagram model} that we can define neurons that come with their respective threshold transformation. This is an example of a form which is composed of a certain structure and a transformation where we are closer to the relative view.

We will now look at another example which is related to how molecules behave in a laboratory jar. To do this we take an initial form and put an atom $A$ in it. Assume that if atom $A$ has no bond and meets an atom $B$ it will create a bond with the atom $B$. This is represented as follows.

%Molecules relative
\begin{equation}\scalebox{0.8} % Change this value to rescale the drawing.
{
\begin{pspicture}(0,-0.8)(6.74,0.8)
\pscircle[linewidth=0.04,linestyle=dashed,dash=0.16cm 0.16cm,dimen=outer](5.94,0.0){0.8}
\psframe[linewidth=0.04,dimen=outer](4.1,0.76)(0.0,-0.7)
\usefont{T1}{ptm}{m}{n}
\rput(2.7946875,0.055){\large A}
\rput{-90.0}(2.02,2.14){\pstriangle[linewidth=0.04,dimen=outer](2.08,-0.26)(0.72,0.64)}
\psline[linewidth=0.04cm,arrowsize=0.16cm 8.0,arrowlength=1.4,arrowinset=0.4]{->}(4.12,-0.02)(5.2,-0.04)
\usefont{T1}{ptm}{m}{n}
\rput(5.6546874,-0.025){\large A}
\usefont{T1}{ptm}{m}{n}
\rput(3.5510938,0.055){\large B}
\usefont{T1}{ptm}{m}{n}
\rput(0.5546875,0.015){\large A}
\usefont{T1}{ptm}{m}{n}
\rput(1.3110938,0.015){\large B}
\psline[linewidth=0.04cm](0.7,0.02)(1.14,0.02)
\usefont{T1}{ptm}{m}{n}
\rput(4.465625,0.525){\scriptsize $\infty S$}
\psline[linewidth=0.04cm,linecolor=red](2.96,0.04)(3.4,0.04)
\end{pspicture}
}
\end{equation}

This defines an atom $A$ along with its property to create a bond with an atom $B$. In a solution, this atom will move and when an atom $B$ enters in its initial form, then a bond will be created with the atom $B$. As indicated by the bond in red in the cause of the transformation, the bond will be created only if the atom $A$ does not already have a bond. The initial set is understood to have a size and is moving along with the atom $A$. The atom $A$ meeting an atom $B$ will result in $A-B$ in the initial form with the transformation still pointing to the initial form. Note that by writing $\sharp$ instead of $\infty$ will make the transformation disappear. This can be understood as a relative model where each form has as a certain range of interactions given by the size of the initial form. Eventually, each atom or molecule could be defined as a collection of forms and transformations. We could also extend approach in a way that the transformations themselves also react to each other.

\subsection{Chains of transformations}

With the use of higher order transformation as defined in section \ref{higher order}, we can define the following chain of transformations. The $N$'s in the initial forms can be viewed as the initial state of the cell $\fbox{$\,\,\,\rhd N$}$. Changing the state of the first cell to $Active$ will induce a domino effect and change the state of the other cells to active. Note that since it is not needed in the following model, we have not written a number over the arrow to indicate the order in which they should be applied.

\begin{equation}\scalebox{0.6} % Change this value to rescale the drawing.
{
\begin{pspicture}(0,-2.35)(22.96,2.35)
\usefont{T1}{ptm}{m}{n}
\rput(20.95875,-1.575){\large N}
\psframe[linewidth=0.04,dimen=outer](7.16,-0.89)(3.54,-2.35)
\usefont{T1}{ptm}{m}{n}
\rput(6.29875,-1.635){\large N}
\usefont{T1}{ptm}{m}{n}
\rput(4.33875,-1.655){\large N}
\rput{-90.0}(7.15,3.81){\pstriangle[linewidth=0.04,dimen=outer](5.48,-1.99)(0.72,0.64)}
\psframe[linewidth=0.04,dimen=outer](3.62,2.35)(0.0,0.89)
\usefont{T1}{ptm}{m}{n}
\rput(2.81875,1.605){\large N}
\usefont{T1}{ptm}{m}{n}
\rput(0.8196875,1.605){\large Active}
\rput{-90.0}(0.33,3.51){\pstriangle[linewidth=0.04,dimen=outer](1.92,1.27)(0.72,0.64)}
\rput{-90.0}(22.31,21.35){\psarc[linewidth=0.04](21.83,-0.48){1.11}{0.0}{180.0}}
\psline[linewidth=0.04cm](21.82,0.63)(6.44,0.59)
\psline[linewidth=0.04cm,arrowsize=0.24cm 6.0,arrowlength=1.4,arrowinset=0.4]{->}(6.44,0.59)(4.84,-1.15)
\pscircle[linewidth=0.04,linestyle=dashed,dash=0.16cm 0.16cm,dimen=outer](4.33,-1.62){0.67}
\psframe[linewidth=0.04,dimen=outer](12.04,-0.87)(8.42,-2.33)
\usefont{T1}{ptm}{m}{n}
\rput(11.17875,-1.635){\large N}
\usefont{T1}{ptm}{m}{n}
\rput(9.21875,-1.635){\large N}
\rput{-90.0}(12.01,8.71){\pstriangle[linewidth=0.04,dimen=outer](10.36,-1.97)(0.72,0.64)}
\pscircle[linewidth=0.04,linestyle=dashed,dash=0.16cm 0.16cm,dimen=outer](9.21,-1.6){0.67}
\psframe[linewidth=0.04,dimen=outer](21.8,-0.85)(18.18,-2.31)
\usefont{T1}{ptm}{m}{n}
\rput(18.91875,-1.615){\large N}
\rput{-90.0}(21.75,18.49){\pstriangle[linewidth=0.04,dimen=outer](20.12,-1.95)(0.72,0.64)}
\pscircle[linewidth=0.04,linestyle=dashed,dash=0.16cm 0.16cm,dimen=outer](18.97,-1.58){0.67}
\psframe[linewidth=0.04,dimen=outer](17.0,-0.89)(13.38,-2.35)
\usefont{T1}{ptm}{m}{n}
\rput(16.13875,-1.635){\large N}
\usefont{T1}{ptm}{m}{n}
\rput(14.17875,-1.635){\large N}
\rput{-90.0}(16.99,13.65){\pstriangle[linewidth=0.04,dimen=outer](15.32,-1.99)(0.72,0.64)}
\pscircle[linewidth=0.04,linestyle=dashed,dash=0.16cm 0.16cm,dimen=outer](14.17,-1.62){0.67}
\psline[linewidth=0.04cm,arrowsize=0.24cm 6.0,arrowlength=1.4,arrowinset=0.4]{->}(7.16,-1.59)(8.66,-1.61)
\psline[linewidth=0.04cm,arrowsize=0.24cm 6.0,arrowlength=1.4,arrowinset=0.4]{->}(12.04,-1.57)(13.58,-1.59)
\psline[linewidth=0.04cm,arrowsize=0.24cm 6.0,arrowlength=1.4,arrowinset=0.4]{->}(16.98,-1.61)(18.52,-1.63)
\psline[linewidth=0.04cm,arrowsize=0.24cm 6.0,arrowlength=1.4,arrowinset=0.4]{->}(3.62,1.59)(4.24,-1.01)
\usefont{T1}{ptm}{m}{n}
\rput(3.9514062,2.1){$1$}
\usefont{T1}{ptm}{m}{n}
\rput(7.505625,-1.145){\scriptsize $\infty S$}
\usefont{T1}{ptm}{m}{n}
\rput(12.385625,-1.125){\scriptsize $\infty S$}
\usefont{T1}{ptm}{m}{n}
\rput(17.325624,-1.105){\scriptsize $\infty S$}
\usefont{T1}{ptm}{m}{n}
\rput(22.125626,-1.085){\scriptsize $\infty S$}
\end{pspicture}
}
\end{equation}

 Interestingly, changing the first cell back to $N$ with the transformation $\fbox{$N \rhd Active$}$ will change all the other cells to $N$. This abstract example can be interpreted as a solid-state drive that keeps the data $N$ or $Active$.

 The following example of a chain, will change the state of the three lower cells to $Active$ and when it is done, it will automatically change it back to $N$. This abstract model could be viewed as a neuron firing or muscle cells contracting and relaxing.

%ChainAlternating
\begin{equation}\scalebox{0.6} % Change this value to rescale the drawing.
{
\begin{pspicture}(0,-3.16)(22.565937,3.14)
\usefont{T1}{ptm}{m}{n}
\rput(20.87875,2.415){\large N}
\psframe[linewidth=0.04,dimen=outer](8.64,-0.24)(5.02,-1.7)
\usefont{T1}{ptm}{m}{n}
\rput(7.77875,-0.985){\large N}
\usefont{T1}{ptm}{m}{n}
\rput(5.81875,-1.005){\large N}
\rput{-90.0}(7.98,5.94){\pstriangle[linewidth=0.04,dimen=outer](6.96,-1.34)(0.72,0.64)}
\psframe[linewidth=0.04,dimen=outer](3.62,1.2)(0.0,-0.26)
\usefont{T1}{ptm}{m}{n}
\rput(2.91875,0.455){\large N}
\usefont{T1}{ptm}{m}{n}
\rput(0.8196875,0.455){\large Active}
\rput{-90.0}(1.48,2.36){\pstriangle[linewidth=0.04,dimen=outer](1.92,0.12)(0.72,0.64)}
\rput{-90.0}(20.62,16.32){\psarc[linewidth=0.04](18.47,-2.15){0.99}{0.0}{180.0}}
\psline[linewidth=0.04cm](18.02,2.36)(2.6,2.36)
\psline[linewidth=0.04cm,arrowsize=0.24cm 6.0,arrowlength=1.4,arrowinset=0.4]{->}(2.62,2.38)(1.18,0.96)
\pscircle[linewidth=0.04,linestyle=dashed,dash=0.16cm 0.16cm,dimen=outer](5.81,-0.97){0.67}
\psframe[linewidth=0.04,dimen=outer](13.52,-0.22)(9.9,-1.68)
\usefont{T1}{ptm}{m}{n}
\rput(12.65875,-0.985){\large N}
\usefont{T1}{ptm}{m}{n}
\rput(10.69875,-0.985){\large N}
\rput{-90.0}(12.84,10.84){\pstriangle[linewidth=0.04,dimen=outer](11.84,-1.32)(0.72,0.64)}
\pscircle[linewidth=0.04,linestyle=dashed,dash=0.16cm 0.16cm,dimen=outer](10.69,-0.95){0.67}
\psframe[linewidth=0.04,dimen=outer](21.64,3.14)(18.02,1.68)
\usefont{T1}{ptm}{m}{n}
\rput(18.75875,2.375){\large N}
\rput{-90.0}(17.48,22.24){\pstriangle[linewidth=0.04,dimen=outer](19.86,2.06)(0.72,0.64)}
\pscircle[linewidth=0.04,linestyle=dashed,dash=0.16cm 0.16cm,dimen=outer](20.91,2.41){0.67}
\psframe[linewidth=0.04,dimen=outer](18.48,-0.24)(14.86,-1.7)
\usefont{T1}{ptm}{m}{n}
\rput(17.61875,-0.985){\large N}
\usefont{T1}{ptm}{m}{n}
\rput(15.65875,-0.985){\large N}
\rput{-90.0}(17.82,15.78){\pstriangle[linewidth=0.04,dimen=outer](16.8,-1.34)(0.72,0.64)}
\pscircle[linewidth=0.04,linestyle=dashed,dash=0.16cm 0.16cm,dimen=outer](15.65,-0.97){0.67}
\psline[linewidth=0.04cm,arrowsize=0.24cm 6.0,arrowlength=1.4,arrowinset=0.4]{->}(8.64,-0.94)(10.14,-0.96)
\psline[linewidth=0.04cm,arrowsize=0.24cm 6.0,arrowlength=1.4,arrowinset=0.4]{->}(13.52,-0.92)(15.06,-0.94)
\psline[linewidth=0.04cm,arrowsize=0.24cm 6.0,arrowlength=1.4,arrowinset=0.4]{->}(19.52,-0.96)(20.58,1.84)
\psline[linewidth=0.04cm,arrowsize=0.24cm 6.0,arrowlength=1.4,arrowinset=0.4]{->}(3.62,0.44)(5.18,-1.04)
\usefont{T1}{ptm}{m}{n}
\rput(8.985625,-0.495){\scriptsize $\infty S$}
\usefont{T1}{ptm}{m}{n}
\rput(13.865625,-0.475){\scriptsize $\infty S$}
\usefont{T1}{ptm}{m}{n}
\rput(18.805626,-0.455){\scriptsize $\infty S$}
\usefont{T1}{ptm}{m}{n}
\rput(21.965626,2.905){\scriptsize $\infty S$}
\psline[linewidth=0.04cm](18.5,-0.94)(19.54,-0.94)
\usefont{T1}{ptm}{m}{n}
\rput(4.005625,0.945){\scriptsize $\infty S$}
\pscircle[linewidth=0.04,linestyle=dashed,dash=0.16cm 0.16cm,dimen=outer](0.83,0.47){0.67}
\psline[linewidth=0.04cm](18.48,-3.14)(3.56,-3.14)
\psline[linewidth=0.04cm,arrowsize=0.24cm 6.0,arrowlength=1.4,arrowinset=0.4]{->}(3.6,-3.12)(3.02,-0.14)
\pscircle[linewidth=0.04,linestyle=dashed,dash=0.16cm 0.16cm,dimen=outer](2.91,0.47){0.67}
\end{pspicture}
}\end{equation}

\section{Computing treatments and processes}\label{Computing remedies}

Using our language, we can mathematically and visually represent biological systems. This is useful to communicate and find new lines of investigations to finding remedies. We can write questions as equations and solve the equations. The ultimate goal is to compute treatments and unknown processes by solving equations.

\subsection{Solving for processes}
Earlier in section \ref{Cell-division cycle}, we coarsely defined interphase as follows.

%Interphase_coarse
\begin{equation*}\scalebox{0.7} % Change this value to rescale the drawing.
{
\begin{pspicture}(0,-1.37)(7.3,1.37)
\pscircle[linewidth=0.04,dimen=outer](1.17,0.02){1.17}
\pscustom[linewidth=0.04,linecolor=orange]
{
\newpath
\moveto(0.9,0.67)
\lineto(0.89,0.59)
\curveto(0.885,0.55)(0.88,0.44)(0.88,0.37)
\curveto(0.88,0.3)(0.885,0.215)(0.9,0.17)
}
\pscustom[linewidth=0.04,linecolor=blue]
{
\newpath
\moveto(1.34,-0.01)
\lineto(1.33,-0.09)
\curveto(1.325,-0.13)(1.32,-0.245)(1.32,-0.32)
\curveto(1.32,-0.395)(1.325,-0.51)(1.33,-0.55)
\curveto(1.335,-0.59)(1.34,-0.635)(1.34,-0.65)
}
\pscustom[linewidth=0.04,linecolor=blue]
{
\newpath
\moveto(0.64,0.09)
\lineto(0.63,0.01)
\curveto(0.625,-0.03)(0.62,-0.14)(0.62,-0.21)
\curveto(0.62,-0.28)(0.625,-0.365)(0.64,-0.41)
}
\pscustom[linewidth=0.04,linecolor=orange]
{
\newpath
\moveto(1.64,0.59)
\lineto(1.63,0.51)
\curveto(1.625,0.47)(1.62,0.355)(1.62,0.28)
\curveto(1.62,0.205)(1.625,0.09)(1.63,0.05)
\curveto(1.635,0.01)(1.64,-0.035)(1.64,-0.05)
}
\pscircle[linewidth=0.04,dimen=outer](5.93,0.0){1.37}
\pscustom[linewidth=0.04,linecolor=orange]
{
\newpath
\moveto(5.62,0.67)
\lineto(5.61,0.59)
\curveto(5.605,0.55)(5.6,0.44)(5.6,0.37)
\curveto(5.6,0.3)(5.605,0.215)(5.62,0.17)
}
\pscustom[linewidth=0.04,linecolor=blue]
{
\newpath
\moveto(6.06,-0.01)
\lineto(6.05,-0.09)
\curveto(6.045,-0.13)(6.04,-0.245)(6.04,-0.32)
\curveto(6.04,-0.395)(6.045,-0.51)(6.05,-0.55)
\curveto(6.055,-0.59)(6.06,-0.635)(6.06,-0.65)
}
\pscustom[linewidth=0.04,linecolor=blue]
{
\newpath
\moveto(5.36,0.09)
\lineto(5.35,0.01)
\curveto(5.345,-0.03)(5.34,-0.14)(5.34,-0.21)
\curveto(5.34,-0.28)(5.345,-0.365)(5.36,-0.41)
}
\pscustom[linewidth=0.04,linecolor=orange]
{
\newpath
\moveto(6.36,0.59)
\lineto(6.35,0.51)
\curveto(6.345,0.47)(6.34,0.355)(6.34,0.28)
\curveto(6.34,0.205)(6.345,0.09)(6.35,0.05)
\curveto(6.355,0.01)(6.36,-0.035)(6.36,-0.05)
}
\pscustom[linewidth=0.04,linecolor=orange]
{
\newpath
\moveto(6.28,0.59)
\lineto(6.29,0.51)
\curveto(6.295,0.47)(6.3,0.355)(6.3,0.28)
\curveto(6.3,0.205)(6.295,0.09)(6.29,0.05)
\curveto(6.285,0.01)(6.28,-0.035)(6.28,-0.05)
}
\pscustom[linewidth=0.04,linecolor=orange]
{
\newpath
\moveto(5.54,0.67)
\lineto(5.55,0.59)
\curveto(5.555,0.55)(5.56,0.44)(5.56,0.37)
\curveto(5.56,0.3)(5.555,0.215)(5.54,0.17)
}
\pscustom[linewidth=0.04,linecolor=blue]
{
\newpath
\moveto(5.98,-0.01)
\lineto(5.99,-0.09)
\curveto(5.995,-0.13)(6.0,-0.245)(6.0,-0.32)
\curveto(6.0,-0.395)(5.995,-0.51)(5.99,-0.55)
\curveto(5.985,-0.59)(5.98,-0.635)(5.98,-0.65)
}
\pscustom[linewidth=0.04,linecolor=blue]
{
\newpath
\moveto(5.28,0.09)
\lineto(5.29,0.01)
\curveto(5.295,-0.03)(5.3,-0.14)(5.3,-0.21)
\curveto(5.3,-0.28)(5.295,-0.365)(5.28,-0.41)
}
\usefont{T1}{ptm}{m}{n}
\rput(3.3314064,1.16){$Interphase$}
\psline[linewidth=0.06cm,arrowsize=0.02cm 2.0,arrowlength=1.4,arrowinset=0.4,doubleline=true,doublesep=0.12,doublecolor=white]{->}(2.72,0.09)(4.16,0.11)
\usefont{T1}{ptm}{m}{n}
\rput{-90.0}(2.635,3.8859375){\rput(3.2414062,0.62){$:=$}}
\end{pspicture}
}
\end{equation*}

In section \ref{Interphase}, we refined our interphase model as
%interphase3
\begin{equation*}\scalebox{0.7} % Change this value to rescale the drawing.
{
\begin{pspicture}(0,-1.9092188)(17.58,1.9492188)
\definecolor{color69}{rgb}{0.0,0.8,0.0}
\definecolor{color73}{rgb}{0.2,0.8,0.0}
\pscustom[linewidth=0.04,linecolor=color69]
{
\newpath
\moveto(9.930552,0.012394104)
\lineto(9.915749,-0.110186465)
\curveto(9.908347,-0.17147675)(9.900945,-0.34768617)(9.900945,-0.4626056)
\curveto(9.900945,-0.577525)(9.908347,-0.75373477)(9.915749,-0.81502503)
\curveto(9.923149,-0.8763153)(9.930552,-0.9452673)(9.930552,-0.968251)
}
\pscustom[linewidth=0.04,linecolor=color69]
{
\newpath
\moveto(7.738268,0.012394104)
\lineto(7.7234645,-0.110186465)
\curveto(7.716063,-0.17147675)(7.7086616,-0.34768617)(7.7086616,-0.4626056)
\curveto(7.7086616,-0.577525)(7.716063,-0.75373477)(7.7234645,-0.81502503)
\curveto(7.730866,-0.8763153)(7.738268,-0.9452673)(7.738268,-0.968251)
}
\pscustom[linewidth=0.04,linecolor=color69]
{
\newpath
\moveto(7.6198425,0.012394104)
\lineto(7.634646,-0.110186465)
\curveto(7.642047,-0.17147675)(7.649449,-0.34768617)(7.649449,-0.4626056)
\curveto(7.649449,-0.577525)(7.642047,-0.75373477)(7.634646,-0.81502503)
\curveto(7.6272445,-0.8763153)(7.6198425,-0.9452673)(7.6198425,-0.968251)
}
\psframe[linewidth=0.04,dimen=outer](10.72,0.41078126)(6.96,-1.4892187)
\pscircle[linewidth=0.04,linecolor=color73,dimen=outer](15.95,-0.51921874){0.59}
\psline[linewidth=0.02cm,linestyle=dashed,dash=0.16cm 0.16cm](15.94,-0.5492188)(16.5,-0.5492188)
\usefont{T1}{ptm}{m}{n}
\rput(15.98,-0.35921875){r}
\pscircle[linewidth=0.04,linecolor=color73,dimen=outer](13.5,-0.50921875){0.68}
\psline[linewidth=0.02cm,linestyle=dashed,dash=0.16cm 0.16cm](13.54,-0.52921873)(14.16,-0.52921873)
\usefont{T1}{ptm}{m}{n}
\rput(13.561406,-0.25921875){$R_1$}
\psframe[linewidth=0.04,dimen=outer](17.04,0.41078126)(12.34,-1.4892187)
\rput{-90.0}(15.325309,14.392398){\pstriangle[linewidth=0.04,dimen=outer](14.858853,-0.747602)(0.64552706,0.56229365)}
\rput{-90.0}(9.305308,8.332398){\pstriangle[linewidth=0.04,dimen=outer](8.818853,-0.767602)(0.64552706,0.56229365)}
\psline[linewidth=0.04cm,arrowsize=0.16cm 8.0,arrowlength=1.4,arrowinset=0.4]{->}(10.94,-0.5492188)(12.12,-0.5492188)
\usefont{T1}{ptm}{m}{n}
\rput(14.151406,0.74078125){$CellGrowth(R_1, r):=$}
\usefont{T1}{ptm}{m}{n}
\rput(8.401406,0.72078127){$DnaReplication:=$}
\psline[linewidth=0.04cm,arrowsize=0.16cm 8.0,arrowlength=1.4,arrowinset=0.4]{->}(5.52,-0.50921875)(6.7,-0.50921875)
\psframe[linewidth=0.04,linestyle=dotted,dotsep=0.16cm,dimen=outer](17.58,1.3107812)(0.0,-1.9092188)
\usefont{T1}{ptm}{m}{n}
\rput(1.4514062,1.7607813){$Interphase_3=$}
\pscircle[linewidth=0.04,linecolor=color73,dimen=outer](4.2,-0.5492188){0.68}
\psline[linewidth=0.02cm,linestyle=dashed,dash=0.16cm 0.16cm](4.24,-0.56921875)(4.86,-0.56921875)
\usefont{T1}{ptm}{m}{n}
\rput(4.2614064,-0.29921874){$R_1$}
\psframe[linewidth=0.04,dimen=outer](5.22,0.39078125)(0.52,-1.5092187)
\rput{-90.0}(3.5253084,2.552398){\pstriangle[linewidth=0.04,dimen=outer](3.0388532,-0.767602)(0.64552706,0.56229365)}
\usefont{T1}{ptm}{m}{n}
\rput(2.4814062,0.72078127){$CellGrowth(R_2,R_1)=$}
\pscircle[linewidth=0.04,linecolor=color73,dimen=outer](1.64,-0.5492188){0.8}
\psline[linewidth=0.02cm,linestyle=dashed,dash=0.16cm 0.16cm](1.74,-0.56921875)(2.38,-0.56921875)
\usefont{T1}{ptm}{m}{n}
\rput(1.7014062,-0.31921875){$R_2$}
\usefont{T1}{ptm}{m}{n}
\rput(11.031406,0.46078125){$\sharp S$}
\end{pspicture}
}
\end{equation*}
where the right-hand side will be defined a $\{Cell\}$. In section \ref{DNA replication} we refined the $DnaReplication$ collection of transformations.

We now give a simple example of an equation where the solution is a process. We take the following equation where $X$ is the only variable.

$$CellGrowth(R_2,R_1)\rightarrow X\rightarrow CellGrowth(R_1,r)\rightarrow \{Cell\}\; \asymp \; Interphase \rightarrow \{Cell\}$$
Here, recall that the symbol $\asymp$ indicates that each side reduces to the same form.

We can also write this as follows with the understanding that the subscript indicates that both are over the initial form $Cell$.
$$CellGrowth(R_2,R_1)\rightarrow X\rightarrow CellGrowth(R_1,r)\; \asymp_{\{Cell\}} \; Interphase.$$

We already know that one solution for this equation is $X=DnaReplication$, but if we were given such an equation, we could see by inspecting $Interphase$ and the $CellGrowth$ that we would only need to copy the chromosomes to satisfy the equation. This is a simple example of solving an equation to find a process. When we are solving equations we can ask for a solution to be coarse or fine. For example, we could ask for $X$ to contain only series of transformations on nucleotides, but this would imply that we have to write the DNA composed of nucleotides and not just represented by orange lines.

\subsection{Solving for antiviral drugs}

A reproducing virus can be represented at the coarse level by
$$\fbox{$virus, virus\rhd virus$}\rightarrow\{virus\}.$$

An antiviral drug $X$ would be a molecule that changes a virus in a way that it stops replicating in the host cell. We can model this by the following system of equations.

This first equation modifies the virus.
$$\fbox{$X\rhd \text{virus}$}\rightarrow\{\text{virus}\}\Rightarrow \{\text{virusModified}\}$$

The next two equations indicates that the modified virus cannot replicate into a virus or modified virus.

$$\begin{array}{c}
\fbox{$\text{virusModified}, \text{virusModified}\rhd \text{virusModified}$}\rightarrow\{\text{virusModified}\}\\
\Downarrow\\
\{\text{virusModified}, \text{virusModified}\}
\end{array}$$

$$\fbox{$\text{virusModified}, \text{virus}\rhd \text{virusModified}$}\rightarrow\{\text{virusModified}\}\nRightarrow \{\text{virusModified}, \text{virus}\}$$

The fourth equation concerns the efficiency of the drug and says that a modified virus should not lose its modification.
$$\fbox{$\text{virus}\rhd \text{virusModified}$}\rightarrow\{\text{virusModified}\}\nRightarrow \{\text{virus}\}$$

The fifth equation indicates that the drug should not interfere with the host cell.
$$\fbox{$X\rhd \text{host}$}\rightarrow\{\text{host}\}\nRightarrow \{\text{unhealthyHost}\}$$

Note that we have used the notation `$\nRightarrow$' to indicate that there was no reduction.

This was a simple example, but the setting would be the same for a much more refined model. With a refined model of a virus at the protein or molecular level including many of its processes, a computer would be able to compute new drugs based on such systems of equations.

\subsection{Understanding cancer}

Let's define a cell along with a marker $\langle Cell, marker\rangle$ where `marker' can take the values $quiet$ and $ready$. The process of cell division follows the sequence of transformations below.

$$\fbox{$\langle Cell,quiet\rangle, \langle Cell,quiet\rangle \rhd \langle Cell,quiet\rangle$} \rightarrow \fbox{$ready \rhd quiet$}\rightarrow\{\langle Cell,quiet\rangle\}$$
The transformation $\fbox{$ready \rhd quiet$}$ is what initiates the cell division cycle when needed. When the cycle is initiated, this will reduce to the two quiet cells $\{\langle Cell,quiet\rangle,\langle Cell,quiet\rangle\}$.

One way to model something that causes cancer and a cancerous cell is with the two following equations where $X$ is an unknown transformation.
$$X\rightarrow\{\langle Cell,quiet\rangle\}\Rightarrow \{\langle CancerCell,ready\rangle\}$$
$$\fbox{$\langle CancerCell,ready\rangle, \langle CancerCell,quiet\rangle \rhd \langle CancerCell,quiet\rangle$}$$
Note that one of the cancerous cell is staying in the ready state without the cycle being initiated. A more aggressive form of cancer would be
$$\fbox{$\langle CancerCell,ready\rangle, \langle CancerCell,ready\rangle \rhd \langle CancerCell,quiet\rangle$}$$

To refine this model further we need define how the immune system reacts to the cancerous cells by defining the transformations $Immune$ which can only apply to cells as follows.
$$Immune\rightarrow\{\langle Cell,quiet\rangle\}\Rightarrow \{\langle Cell,quiet\rangle\}$$
$$Immune\rightarrow\{\langle Cell,ready\rangle\}\Rightarrow \{\langle Cell,ready\rangle\}$$
$$Immune\rightarrow\{\langle UnhealthyCell,quiet\rangle\}\Rightarrow \{\quad\quad\}$$
$$Immune\rightarrow\{\langle UnhealthyCell,ready\rangle\}\Rightarrow \{\quad\quad\}$$

For the cancerous cell not to be removed, the immune system would be leaving many cancerous cells untouched and we would have to include the following in a definition of cancer.
$$Immune\rightarrow\{\langle CancerCell,ready\rangle\}\Rightarrow \{\langle CancerCell,ready\rangle\}$$

\subsection{Calculating mathematical solutions}\label{calculating mathematical solutions}

Now that we can write questions from the domain of medicine as transformation equations, we have to develop ways to solve them.

 An abstract example is the equation $$\fbox{$b\rhd c$} \rightarrow\fbox{$X\rhd a$} \rightarrow\{a\}\quad=\quad ab.$$
 By inspection and trial, a solution is $X=ac$. But there is a computational way to do it. When we are solving equations such as $2+x=5$, we can subtract on both sides to get $-2+2+x=-2+5$ which gives $0+x=-2+5$ and a final solution $x=3$.

 For us, the inverse of $\fbox{$c\rhd b$}^{\,\sharp S}$ will be $\fbox{$b\rhd c$}^{\,\sharp S}$, since we have that $$\fbox{$b\rhd c$}^{\,\sharp S}\rightarrow\fbox{$c\rhd b$}^{\,\sharp S}\rightarrow \{Y\}$$ reduces to $\{Y\}$ when $Y$ is any string of letters. If we want to to solve the following equation, we can use the inverse to isolate $X$ on one side.
\begin{eqnarray*}
\fbox{$b\rhd c$}^{\,\sharp S} \rightarrow\fbox{$X\rhd a$}^{\,\sharp S} \rightarrow\{a\} & \quad=\quad & \{ab\}\\
\fbox{$c\rhd b$}^{\,\sharp S}\rightarrow\fbox{$b\rhd c$}^{\,\sharp S}\rightarrow \fbox{$X\rhd a$}^{\,\sharp S} \rightarrow\{a\} & \quad=\quad & \fbox{$c\rhd b$}^{\,\sharp S}\rightarrow ab\\
\fbox{$X\rhd a$}^{\,\sharp S} \rightarrow\{a\} & \quad=\quad & \{ac\}\\
\{X\} & \quad=\quad & \{ac\}\\
X & \quad =\quad & ac \end{eqnarray*}

\subsection{Producing the remedies}

If we know that a certain molecular structure will have a wanted effect on a system, for example an antiviral drug, we can test this drug in a large and precise model of the body to see where this drug acts and discover the potential side effects.

If we have a satisfying drug, one problem is to be able to synthesize this molecule $Antiviral$. This can be done by solving the equation for $X$ and $Y$ where $X$ is composed of transformations at the molecular level.

$$X\rightarrow Y\Rightarrow Antiviral$$

In this case, $Y$ would be the starting compound and $X$ would the a series of transformations. At first the set of allowable transformations for $X$ would be from a database of understood reactions. If we cannot synthesize the antiviral drug with known reactions, we can focus on a step of the series for which we don't know the reaction. Suppose that we know the reactions $R_1,R_2,R_3$ and $R_4$ and that
$$X\rightarrow R_1\rightarrow R_2 \rightarrow Z\rightarrow R_3\rightarrow R_4\rightarrow Y\Rightarrow Antiviral,$$
then we would need to focus on the unknown reaction $Z$.

A useful tool would be to define a mathematical measure of closeness between molecules. This would help us evaluate which molecule out of two is closer under that measure to another molecule. For example, if we want to reach $X$ from $Y$, we could calculate a value of `molecular distance' $D$ between the molecule resulting from the reaction $R_4\rightarrow Y$ and $X$, and between $R_4'\rightarrow Y$ and $X$. If $D(R_4,X)<D(R_4',X)$, we will choose $R_4$ as the first transformation applied to $Y$.

\section{Computing techniques}

We will now briefly discuss some potential areas of investigation related to computing the solutions of a system of equations.

\subsection{Solving equations}

In mathematics, there are many techniques and algorithms to solve different types of equations. For example, algebra (and other methods) permits us to solve equations involving polynomials. We have seen simple examples of a technique to solve a system of transformations in section \ref{calculating mathematical solutions}, but there are many more techniques that would need to be developed to handle more complex systems involving transformations.

Linear algebra equations and matrices were initially developed to solve systems of linear equations. It would be interesting to develop objects which would themselves be forms that can be applied to systems of equations of transformations. These objects could act in a way that is similar to matrices and help solve certain types of systems.

\subsection{Principle of resilience and propagation}

Most processes and features we observe in the body are resilient. This means that errors, mutations and modified processes are naturally eliminated from the body. We could use this as a way to compute unknown processes. When investigating a hypothetical process, we can evaluate the resilience of a process to determine if it is stable in the body and could have survived. We can probe and test the process by introducing errors in the flow of steps and see if the process destabilizes or not.  If the process destabilizes with small interactions from the body itself or the environment, then this hypothetical process must be discarded. This is a very useful principle from a computational point of view. To solve a problem or find a representation of a biological process, the computer could generate many different processes and eliminate them by testing if they are resilient.

For example, in a certain series of steps, if we know that something is produced at one of the steps, but do not know how it was produced, the computer could generate different ways (using what is present in the cell or not) and then test the resilience of each process. Each process could also be tested for how they propagate in the system. If it implies a buildup somewhere else and it is never observed in reality, this process could be discarded. If the process disturbs the resilience of another process, thus making the other process disappear, then the process could also by discarded. A classical algorithm which might be useful here would be a genetic algorithm.

\subsection{Optimization}

Usually biological mechanisms are very well optimized to accomplish their task. Considering the complexity of the brain and DNA replication, we see they are the result of very efficient processes. Processes that we want to discover can often be close to an optimal solution. This can help us discard many possible solutions by considering measures associated with it such as speed, polyvalence, number of steps, stability and precision.

When we have a general framework of a process, there are many unknown constants that need to be determined in order to obtain an appropriate model. If there is a finite number of unknowns, this model can be optimized by entering different values and evaluated to see how the model will perform. The choice of values can done by using a genetic algorithm or quantum computers using annealing algorithms.

\subsection{Main algorithm}

Below is a sketch of an algorithm to find new treatments and biological processes with a computer. Eventually, this main algorithm should be developed and refined.

\begin{enumerate}
  \item Write the equations representing your question.
  \item Access a database of known processes, transformations and molecules. Including quantities and proportions.
  \item Separate the problem into almost disjoint parts or clusters.
  \item Try solving the equation directly using direct techniques such as inverse transformations.
  \item Use known computing techniques and algorithms. For example, algorithms based on resilience and optimization.
  \item Introduce a molecule or a process that is not usually present in the body to try solving the equations.
  \item Evaluate a solution to see if there are unwanted side-effects and interactions with biological mechanism.
\end{enumerate}

\subsection{Computable and uncomputable functions}

At some point we will meet transformation equations that are uncomputable or out of reach. An idea would be to define an object that is a solution to the transformation equation and build a mathematical theory based on this definition. The generic example behind this is trying to solve the polynomial $x^2=-1$ over the real numbers. Since squaring any real number will result in a positive number we see that this equation has no solution over the real numbers. Euler's idea was to define an object $i$ such that if you square it you get $-1$. This way, we have that $i$ is a solution to the equation $x^2=-1$. Mathematicians did not stop there, using this; they constructed the set of complex numbers along with its algebra and amazing properties. In a way, $i$ and complex numbers could be seen as a `programming oracle' with an algebraic structure. A similar approach for transformations could be beneficial in investigating the world of uncomputable functions, transformations and related structures.

\section{Computer Science}

\subsection{Programming statements}

 Programming statements can be viewed as the atoms or sentences of programming. We will now demonstrate that we can also represent programming statements in our language.

\subsubsection{If-statement}

We can represent an if-statement as follows. Example of conditions are $x=5$, $x<10$ and $b=true$. We will say more about conditions in the section about the for-loop.
\begin{flalign*} \fbox{$B\rhd A$}\rightarrow \fbox{$\text{condition}, A\rhd \text{condition}, A$}&\rightarrow \{\text{condition}, A\}
\end{flalign*}
If the condition is satisfied, then the identity transformation which rewrites $condition, A$ by $condition, A$ is applied and thus permits the next transformation in the sequence to change $A$ or perform an operation depending on what $A$ is.

We can represent an if-else-statement as follows.

\begin{flalign*}[\,\fbox{$B\rhd A$}\rightarrow \fbox{$\text{condition}, A\rhd \text{condition}, A$},& \\
 \fbox{$C\rhd A$}\rightarrow \fbox{$\text{condition}, A\rhd {\color{red}\text{condition}}, A$}\,]&\rightarrow \{\text{condition}, A\}
\end{flalign*}
Thus, since $condition$ is written in red, it means that the transformation applies when the condition is not satisfied, only one of these sequences of two transformations will affect $A$. If we want all the transformations to disappear when one of the sequences of transformation has been applied, we can use the sharp symbol $\sharp$ as follows.
\begin{flalign*}[\fbox{$B\rhd A$}\rightarrow \fbox{$\text{condition}, A\rhd \text{condition}, A$},& \\
 \fbox{$C\rhd A$}\rightarrow \fbox{$\text{condition}, A\rhd {\color{red}\text{condition}}, A$}]^{\,\sharp}&\rightarrow \{\text{condition}, A\}
\end{flalign*}

\subsubsection{Switch-statement}
A switch statement is done in a similar way than an if-then-else statement, but with the use of multiple cases. A general example for three cases is written as
\begin{flalign*} (\fbox{$B\rhd A$}\rightarrow \fbox{$\text{case1}, A\rhd \text{case1}, A$},& \\
 \fbox{$C\rhd A$}\rightarrow \fbox{$\text{case2}, A\rhd \text{case2}, A$},& \\
 \fbox{$D\rhd A$}\rightarrow \fbox{$\text{case3}, A\rhd \text{case3}, A$}),&\rightarrow \{\text{condition}, A\}.
\end{flalign*}

\subsubsection{For-loop}

We present two ways to do a for-loop. The first is the simplest.

$$\fbox{$A'\rhd A$}^{\,n}\rightarrow \{A\}$$
For $n=4$, this will return $A''''$.

In a classical for-loop, we need an increment $i$ and a condition based on the increment, for example $i<4$. The increment will be a natural number. To initiate the loop, we need to first define $i:=1$.
\begin{flalign*} [\,\fbox{$ i:=i+1, A' \rhd i, A$}\rightarrow\fbox{$i\leq 4, A\rhd i\leq 4, A$}\,]^{\,\sharp} & \rightarrow  \{i, A\} \\
\end{flalign*}
Note that in the transformation $\fbox{$ i:=i+1, A' \rhd i, A$}$ we are giving another value to $i$, this can be viewed as a new feature of our language. Concerning the condition $i\leq 4$, we could apply some sequence of transformations to make the symbols $i\leq 4$ appear in the set before applying the transformation $\fbox{$i\leq 4, A\rhd i\leq 4, A$}$, but there is another way. We can define $\fbox{$i\leq 4, A\rhd i\leq 4, A$}$ as a collection of parallel transformations as follows.

$$\fbox{$i\leq 4, A\rhd i\leq 4, A$}:=[\fbox{$1, A\rhd 1, A$}, \fbox{$2, A\rhd 2, A$}, \fbox{$3, A\rhd 3, A$}, \fbox{$4, A\rhd 4, A$}]^{\,\sharp}$$
Thus, when one of the transformations is applied, $\fbox{$i\leq 4, A\rhd i\leq 4, A$}$ will disappear and let the transformation $\fbox{$ i:=i+1, A' \rhd i, A$}$ from the for-loop to be applied.

\subsubsection{While-loop}

The while-loop will check the condition and then execute the content until the condition is not satisfied.

$$\begin{array}{c}
[\,\fbox{$A'\rhd A$}\rightarrow[\,\fbox{$\text{condition}, A\rhd \text{condition}, A$}, \fbox{$\quad\quad\quad\quad, A\rhd {\color{red}\text{condition}}, A$}\,]^{\,\sharp}\,]^{\,\infty} \\
\downarrow\\
\{\text{condition}, A\}
\end{array}$$

\subsubsection{Do-loop}

The do-loop will execute the content and then check the condition to see if it should continue.

$$\begin{array}{c}
[\,[\,\fbox{$\text{condition}, A\rhd \text{condition}, A$}, \fbox{$\quad\quad\quad\quad, A\rhd {\color{red}\text{condition}}, A$}\,]^{\,\sharp}\rightarrow\fbox{$A'\rhd A$}\,]^{\,\infty} \\
\downarrow\\
\{\text{condition}, A\}
\end{array}$$

\subsection{Forms programming}
We have already seen in section \ref{function forms} and \ref{abstraction} that we can name any form with added variables. This permits us to use object programming techniques in our language. Since we can build programming statements with transformations, this means that after building a scientific model in the language of transformations, we can build programs without leaving the language. This points toward a new and powerful programming language based on forms and transformations.

If such a programming language is developed based on modern higher programming languages, one has to make extensive use of arrays and sub-arrays. For example, an inline transformation can be viewed as an array of two components where each component contains an array of $n$ components. For transformations in diagram form, a transformation would be viewed as an array of two components where each component contains a $m$ by $n$ two-dimensional array. Similarly for a 3D representation, one would use transformation components that are three-dimensional arrays.

Interestingly, this array-based approach to transformations points towards what we could understand as $n$-dimensional transformations and forms. This means that the language of transformations could be extended to higher dimensions.

\subsection{Conway's game of life}

We will now give a model of Conway's game of life. Conway's game of life is one example of a wider class of objects called cellular automata.

It is understood that the initial set is a 3 by 3 grid along with the live cell in it. The following transformation defines Conway's game of life. For a larger grid, this collection of transformations is applied to each 3 by 3 grid of the larger grid.

%Conway
\begin{equation*}\scalebox{0.6} % Change this value to rescale the drawing.
{
\begin{pspicture}(0,-8.63)(20.774376,8.63)
\definecolor{color386c}{rgb}{0.5019607843137255,0.5019607843137255,0.5019607843137255}
\rput{-90.0}(5.5884376,15.368438){\pstriangle[linewidth=0.04,dimen=outer,fillstyle=solid](10.478437,4.53)(0.84,0.72)}
\psframe[linewidth=0.04,linecolor=green,dimen=outer,fillstyle=solid,fillcolor=green](13.278438,5.13)(12.818438,4.67)
\rput(12.298437,4.15){\psgrid[gridwidth=0.0482,subgridwidth=0.014111111,gridlabels=0.0pt,subgriddiv=1,unit=0.5cm,subgridcolor=color386c](0,0)(0,0)(3,3)
\psset{unit=1.0cm}}
\usefont{T1}{ptm}{m}{n}
\rput(14.2425,4.91){\Huge ,}
\usefont{T1}{ptm}{m}{n}
\rput(15.9425,4.93){\Huge ,}
\usefont{T1}{ptm}{m}{n}
\rput(11.698281,4.97){\Huge $\{$}
\psframe[linewidth=0.04,dimen=outer,fillstyle=solid,fillcolor=black](15.3984375,5.13)(14.938437,4.67)
\psframe[linewidth=0.04,linecolor=red,dimen=outer,fillstyle=solid,fillcolor=red](17.038437,5.15)(16.578438,4.69)
\usefont{T1}{ptm}{m}{n}
\rput(17.618437,4.97){\Huge $\}$}
\rput{-90.0}(11.468437,9.488438){\pstriangle[linewidth=0.04,dimen=outer,fillstyle=solid](10.478437,-1.35)(0.84,0.72)}
\psframe[linewidth=0.04,linecolor=green,dimen=outer,fillstyle=solid,fillcolor=green](13.278438,-0.75)(12.818438,-1.21)
\rput(12.298437,-1.73){\psgrid[gridwidth=0.0482,subgridwidth=0.014111111,gridlabels=0.0pt,subgriddiv=1,unit=0.5cm,subgridcolor=color386c](0,0)(0,0)(3,3)
\psset{unit=1.0cm}}
\usefont{T1}{ptm}{m}{n}
\rput(14.2425,-0.97){\Huge ,}
\usefont{T1}{ptm}{m}{n}
\rput(15.9425,-0.95){\Huge ,}
\usefont{T1}{ptm}{m}{n}
\rput(11.698281,-0.91){\Huge $\{$}
\psframe[linewidth=0.04,dimen=outer,fillstyle=solid,fillcolor=black](15.3984375,-0.75)(14.938437,-1.21)
\psframe[linewidth=0.04,dimen=outer,fillstyle=solid,fillcolor=black](18.458437,-0.73)(17.998438,-1.19)
\usefont{T1}{ptm}{m}{n}
\rput(17.5425,-0.93){\Huge ,}
\psframe[linewidth=0.04,dimen=outer,fillstyle=solid,fillcolor=black](16.998438,-0.73)(16.538437,-1.19)
\rput{-90.0}(2.6684375,18.288437){\pstriangle[linewidth=0.04,dimen=outer,fillstyle=solid](10.478437,7.45)(0.84,0.72)}
\psframe[linewidth=0.04,linecolor=green,dimen=outer,fillstyle=solid,fillcolor=green](13.298437,8.05)(12.838437,7.59)
\rput(12.318438,7.07){\psgrid[gridwidth=0.0482,subgridwidth=0.014111111,gridlabels=0.0pt,subgriddiv=1,unit=0.5cm,subgridcolor=color386c](0,0)(0,0)(3,3)
\psset{unit=1.0cm}}
\usefont{T1}{ptm}{m}{n}
\rput(14.2625,7.83){\Huge ,}
\usefont{T1}{ptm}{m}{n}
\rput(11.718281,7.89){\Huge $\{$}
\psframe[linewidth=0.04,linecolor=red,dimen=outer,fillstyle=solid,fillcolor=red](15.418438,8.05)(14.958438,7.59)
\usefont{T1}{ptm}{m}{n}
\rput(16.038437,7.93){\Huge $\}$}
\usefont{T1}{ptm}{m}{n}
\rput(18.8625,-0.93){\Huge ,}
\psframe[linewidth=0.04,linecolor=red,dimen=outer,fillstyle=solid,fillcolor=red](19.958437,-0.71)(19.498438,-1.17)
\usefont{T1}{ptm}{m}{n}
\rput(20.538437,-0.89){\Huge $\}$}
\rput{-90.0}(8.408438,12.508437){\pstriangle[linewidth=0.04,dimen=outer,fillstyle=solid](10.458438,1.69)(0.84,0.72)}
\psframe[linewidth=0.04,linecolor=green,dimen=outer,fillstyle=solid,fillcolor=green](13.278438,2.29)(12.818438,1.83)
\rput(12.298437,1.31){\psgrid[gridwidth=0.0482,subgridwidth=0.014111111,gridlabels=0.0pt,subgriddiv=1,unit=0.5cm,subgridcolor=color386c](0,0)(0,0)(3,3)
\psset{unit=1.0cm}}
\usefont{T1}{ptm}{m}{n}
\rput(14.2425,2.07){\Huge ,}
\usefont{T1}{ptm}{m}{n}
\rput(15.9425,2.09){\Huge ,}
\usefont{T1}{ptm}{m}{n}
\rput(11.698281,2.13){\Huge $\{$}
\psframe[linewidth=0.04,dimen=outer,fillstyle=solid,fillcolor=black](15.3984375,2.29)(14.938437,1.83)
\psframe[linewidth=0.04,dimen=outer,fillstyle=solid,fillcolor=black](17.038437,2.31)(16.578438,1.85)
\usefont{T1}{ptm}{m}{n}
\rput(17.5825,2.07){\Huge ,}
\psframe[linewidth=0.04,linecolor=red,dimen=outer,fillstyle=solid,fillcolor=red](18.678438,2.29)(18.218437,1.83)
\usefont{T1}{ptm}{m}{n}
\rput(19.258438,2.11){\Huge $\}$}
\rput{-90.0}(14.488438,6.4284377){\pstriangle[linewidth=0.04,dimen=outer,fillstyle=solid](10.458438,-4.39)(0.84,0.72)}
\psframe[linewidth=0.04,linecolor=green,dimen=outer,fillstyle=solid,fillcolor=green](13.278438,-3.77)(12.818438,-4.23)
\usefont{T1}{ptm}{m}{n}
\rput(14.2225,-4.01){\Huge ,}
\usefont{T1}{ptm}{m}{n}
\rput(15.9225,-3.99){\Huge ,}
\usefont{T1}{ptm}{m}{n}
\rput(11.678281,-3.95){\Huge $\{$}
\psframe[linewidth=0.04,dimen=outer,fillstyle=solid,fillcolor=black](15.378437,-3.79)(14.918438,-4.25)
\psframe[linewidth=0.04,dimen=outer,fillstyle=solid,fillcolor=black](18.438438,-3.77)(17.978437,-4.23)
\usefont{T1}{ptm}{m}{n}
\rput(17.5225,-3.97){\Huge ,}
\psframe[linewidth=0.04,dimen=outer,fillstyle=solid,fillcolor=black](16.978437,-3.77)(16.518438,-4.23)
\usefont{T1}{ptm}{m}{n}
\rput(18.8425,-3.97){\Huge ,}
\psframe[linewidth=0.04,dimen=outer,fillstyle=solid,fillcolor=black](19.938438,-3.75)(19.478437,-4.21)
\usefont{T1}{ptm}{m}{n}
\rput(20.518438,-3.93){\Huge $\}$}
\rput(12.298437,-4.75){\psgrid[gridwidth=0.0482,subgridwidth=0.014111111,gridlabels=0.0pt,subgriddiv=1,unit=0.5cm,subgridcolor=color386c](0,0)(0,0)(3,3)
\psset{unit=1.0cm}}
\rput{-90.0}(18.308437,2.5284376){\pstriangle[linewidth=0.04,dimen=outer,fillstyle=solid](10.418438,-8.25)(0.84,0.72)}
\rput(12.238438,-8.63){\psgrid[gridwidth=0.0482,subgridwidth=0.014111111,gridlabels=0.0pt,subgriddiv=1,unit=0.5cm,subgridcolor=color386c](0,0)(0,0)(3,3)
\psset{unit=1.0cm}}
\usefont{T1}{ptm}{m}{n}
\rput(14.1825,-7.87){\Huge ,}
\usefont{T1}{ptm}{m}{n}
\rput(15.8825,-7.85){\Huge ,}
\usefont{T1}{ptm}{m}{n}
\rput(11.638281,-7.81){\Huge $\{$}
\psframe[linewidth=0.04,dimen=outer,fillstyle=solid,fillcolor=black](15.338437,-7.65)(14.878437,-8.11)
\psframe[linewidth=0.04,dimen=outer,fillstyle=solid,fillcolor=black](18.398438,-7.63)(17.938438,-8.09)
\usefont{T1}{ptm}{m}{n}
\rput(17.4825,-7.83){\Huge ,}
\psframe[linewidth=0.04,dimen=outer,fillstyle=solid,fillcolor=black](16.938438,-7.63)(16.478437,-8.09)
\usefont{T1}{ptm}{m}{n}
\rput(18.8025,-7.83){\Huge ,}
\psframe[linewidth=0.04,linecolor=red,dimen=outer,fillstyle=solid,fillcolor=red](19.898438,-7.61)(19.438438,-8.07)
\usefont{T1}{ptm}{m}{n}
\rput(20.478437,-7.79){\Huge $\}$}
\rput(5.1584377,7.13){\psgrid[gridwidth=0.0482,subgridwidth=0.014111111,gridlabels=0.0pt,subgriddiv=1,unit=0.5cm,subgridcolor=color386c](0,0)(0,0)(3,3)
\psset{unit=1.0cm}}
\usefont{T1}{ptm}{m}{n}
\rput(7.1025,7.89){\Huge ,}
\usefont{T1}{ptm}{m}{n}
\rput(4.5582814,7.95){\Huge $\{$}
\psframe[linewidth=0.04,linecolor=red,dimen=outer,fillstyle=solid,fillcolor=red](8.258437,8.11)(7.7984376,7.65)
\usefont{T1}{ptm}{m}{n}
\rput(8.878437,7.99){\Huge $\}$}
\rput(3.5984375,4.17){\psgrid[gridwidth=0.0482,subgridwidth=0.014111111,gridlabels=0.0pt,subgriddiv=1,unit=0.5cm,subgridcolor=color386c](0,0)(0,0)(3,3)
\psset{unit=1.0cm}}
\usefont{T1}{ptm}{m}{n}
\rput(5.5425,4.93){\Huge ,}
\usefont{T1}{ptm}{m}{n}
\rput(7.2425,4.95){\Huge ,}
\usefont{T1}{ptm}{m}{n}
\rput(2.9982812,4.99){\Huge $\{$}
\psframe[linewidth=0.04,dimen=outer,fillstyle=solid,fillcolor=black](6.6984377,5.15)(6.2384377,4.69)
\psframe[linewidth=0.04,linecolor=red,dimen=outer,fillstyle=solid,fillcolor=red](8.338437,5.17)(7.8784375,4.71)
\usefont{T1}{ptm}{m}{n}
\rput(8.918438,4.99){\Huge $\}$}
\psframe[linewidth=0.04,linecolor=green,dimen=outer,fillstyle=solid,fillcolor=green](2.9384375,2.31)(2.4784374,1.85)
\rput(1.9584374,1.33){\psgrid[gridwidth=0.0482,subgridwidth=0.014111111,gridlabels=0.0pt,subgriddiv=1,unit=0.5cm,subgridcolor=color386c](0,0)(0,0)(3,3)
\psset{unit=1.0cm}}
\usefont{T1}{ptm}{m}{n}
\rput(3.9025,2.09){\Huge ,}
\usefont{T1}{ptm}{m}{n}
\rput(5.6025,2.11){\Huge ,}
\usefont{T1}{ptm}{m}{n}
\rput(1.3582813,2.15){\Huge $\{$}
\psframe[linewidth=0.04,dimen=outer,fillstyle=solid,fillcolor=black](5.0584373,2.31)(4.5984373,1.85)
\psframe[linewidth=0.04,dimen=outer,fillstyle=solid,fillcolor=black](6.6984377,2.33)(6.2384377,1.87)
\usefont{T1}{ptm}{m}{n}
\rput(7.2425,2.09){\Huge ,}
\psframe[linewidth=0.04,linecolor=red,dimen=outer,fillstyle=solid,fillcolor=red](8.338437,2.31)(7.8784375,1.85)
\usefont{T1}{ptm}{m}{n}
\rput(8.918438,2.13){\Huge $\}$}
\psframe[linewidth=0.04,linecolor=green,dimen=outer,fillstyle=solid,fillcolor=green](1.5984375,-0.75)(1.1384375,-1.21)
\rput(0.6184375,-1.73){\psgrid[gridwidth=0.0482,subgridwidth=0.014111111,gridlabels=0.0pt,subgriddiv=1,unit=0.5cm,subgridcolor=color386c](0,0)(0,0)(3,3)
\psset{unit=1.0cm}}
\usefont{T1}{ptm}{m}{n}
\rput(2.5625,-0.97){\Huge ,}
\usefont{T1}{ptm}{m}{n}
\rput(4.2625,-0.95){\Huge ,}
\usefont{T1}{ptm}{m}{n}
\rput(0.01828125,-0.91){\Huge $\{$}
\psframe[linewidth=0.04,dimen=outer,fillstyle=solid,fillcolor=black](3.7184374,-0.75)(3.2584374,-1.21)
\psframe[linewidth=0.04,dimen=outer,fillstyle=solid,fillcolor=black](6.7784376,-0.73)(6.3184376,-1.19)
\usefont{T1}{ptm}{m}{n}
\rput(5.8625,-0.93){\Huge ,}
\psframe[linewidth=0.04,dimen=outer,fillstyle=solid,fillcolor=black](5.3184376,-0.73)(4.8584375,-1.19)
\usefont{T1}{ptm}{m}{n}
\rput(7.1825,-0.93){\Huge ,}
\psframe[linewidth=0.04,linecolor=red,dimen=outer,fillstyle=solid,fillcolor=red](8.278438,-0.71)(7.8184376,-1.17)
\usefont{T1}{ptm}{m}{n}
\rput(8.858438,-0.89){\Huge $\}$}
\usefont{T1}{ptm}{m}{n}
\rput(2.5425,-4.01){\Huge ,}
\usefont{T1}{ptm}{m}{n}
\rput(4.2425,-3.99){\Huge ,}
\usefont{T1}{ptm}{m}{n}
\rput(-0.00171875,-3.95){\Huge $\{$}
\psframe[linewidth=0.04,dimen=outer,fillstyle=solid,fillcolor=black](3.6984375,-3.79)(3.2384374,-4.25)
\psframe[linewidth=0.04,dimen=outer,fillstyle=solid,fillcolor=black](6.7584376,-3.77)(6.2984376,-4.23)
\usefont{T1}{ptm}{m}{n}
\rput(5.8425,-3.97){\Huge ,}
\psframe[linewidth=0.04,dimen=outer,fillstyle=solid,fillcolor=black](5.2984376,-3.77)(4.8384376,-4.23)
\usefont{T1}{ptm}{m}{n}
\rput(7.1625,-3.97){\Huge ,}
\psframe[linewidth=0.04,dimen=outer,fillstyle=solid,fillcolor=black](8.258437,-3.75)(7.7984376,-4.21)
\usefont{T1}{ptm}{m}{n}
\rput(8.838437,-3.93){\Huge $\}$}
\rput(0.6184375,-4.75){\psgrid[gridwidth=0.0482,subgridwidth=0.014111111,gridlabels=0.0pt,subgriddiv=1,unit=0.5cm,subgridcolor=color386c](0,0)(0,0)(3,3)
\psset{unit=1.0cm}}
\rput(0.6384375,-8.59){\psgrid[gridwidth=0.0482,subgridwidth=0.014111111,gridlabels=0.0pt,subgriddiv=1,unit=0.5cm,subgridcolor=color386c](0,0)(0,0)(3,3)
\psset{unit=1.0cm}}
\usefont{T1}{ptm}{m}{n}
\rput(2.5825,-7.83){\Huge ,}
\usefont{T1}{ptm}{m}{n}
\rput(4.2825,-7.81){\Huge ,}
\usefont{T1}{ptm}{m}{n}
\rput(0.03828125,-7.77){\Huge $\{$}
\psframe[linewidth=0.04,dimen=outer,fillstyle=solid,fillcolor=black](3.7384374,-7.61)(3.2784376,-8.07)
\psframe[linewidth=0.04,dimen=outer,fillstyle=solid,fillcolor=black](6.7984376,-7.59)(6.3384376,-8.05)
\usefont{T1}{ptm}{m}{n}
\rput(5.8825,-7.79){\Huge ,}
\psframe[linewidth=0.04,dimen=outer,fillstyle=solid,fillcolor=black](5.3384376,-7.59)(4.8784375,-8.05)
\usefont{T1}{ptm}{m}{n}
\rput(7.2025,-7.79){\Huge ,}
\psframe[linewidth=0.04,linecolor=red,dimen=outer,fillstyle=solid,fillcolor=red](8.298437,-7.57)(7.8384376,-8.03)
\usefont{T1}{ptm}{m}{n}
\rput(8.878437,-7.75){\Huge $\}$}
\end{pspicture}
}\end{equation*}

Interestingly, this opens up a lot of possibilities of cellular automata based on Conway's game of life. For example, the sub-grid does not have to be 3 by 3 but can have an irregular shape, we can have much more complex rules or the grid can take the form of another type of tiling of the plane. We are also not only restricted to the plane, but could define something similar for the surface of a sphere.

\subsection{Turing machines}

We can also define Turing machines by defining transformations similar to the one below.
In state $S_1$ read $1$, write $1$ and move one step to the right and go into state $S_2$.
\begin{center}\fbox{$\begin{array}{ccccc}
     & S_2 &  & S_1 &  \\
     &\mid  & \rhd & \mid &  \\
    1 & 0 &   & 1 & 0 \\
  \end{array}$}\end{center}

In state $S_1$ read $0$, write $1$ and move one step to the right and go into state $S_2$.
  \begin{center}\fbox{$\begin{array}{ccccc}
     & S_2 &  & S_1 &  \\
     & \mid & \rhd & \mid  & \\
    1 & 0 &   & 0 & 0 \\
  \end{array}$}\end{center}

In state $S_2$ read $1$, write $1$ and move one step to the left and go into state $S_3$.
  \begin{center}\fbox{$\begin{array}{ccccc}
    S_3 &  &  &  & S_2 \\
    \mid &  & \rhd &  & \mid \\
    0 & 1 &   & 0 & 1 \\
  \end{array}$}\end{center}

Interestingly, we now have no restriction on what the symbols or the tape can be. The tape could even take the form of a tree or a fractal. Moreover, we could take advantage of the capability for transformations to affect themselves or other transformations to generalize Turing machines.

\section{Mathematics}

%\subsection{Arithmetic}
%
%\textit{Notes: Natural number versus real numbers line (see Mathematical universes).}
%
%We will now expand the model of arithmetic we started in section \ref{naming}.

\subsection{Functions}

Functions are essential objects of modern mathematics and can also be described with transformations. A function $f$ from $X$ to $Y$ associates to each element of the set $X$ an element of the set $Y$. One way to define a function with a transformation is as follows where $y_x$ is the element of $Y$ which is associated to $x$. This is basically the same as defining a collection of ordered pairs to define a function.

\begin{equation*}f(x):=\fbox{$y_x\rhd x$}\end{equation*}

If we want to define the polynomial function $f(x)=x^2$, we can write this as follows based on the multiplication defined in section \ref{abstraction}.

\begin{equation*}f(x):=\fbox{$x\times x\rhd x$}\end{equation*}

Transformations are more flexible than functions or more generally, sets of ordered pairs. Like sets of ordered pairs, transformations can point to two objects and don't need to apply to all elements. Unlike functions, we don't need to define the image set, transformations can also be applied to other transformations and we can have multiple occurrences of an element in a initial set. Usually, when working with functions, the domain and the image set of the function are known beforehand and the elements are of the same type. This is not required for transformations. If we look at an initial set and the reduced set, we can create a set of ordered pairs based on the transformation.

\subsection{Fractals}

By construction, the language of transformations can naturally define fractals. We will now look at and define two well-known fractals, the Koch and Sierpinsky fractals.

At first thought, the Koch snowflake

\begin{center}\begin{tikzpicture}[decoration=Koch snowflake]   \draw decorate{ decorate{ decorate{ decorate{        (0,0) -- ++(60:3)  -- ++(300:3) -- ++(180:3)}}}}; \end{tikzpicture}\end{center}
could be defined by the following transformation.

%Koch
\begin{equation*}\scalebox{0.6} % Change this value to rescale the drawing.
{
\begin{pspicture}(0,-0.774092)(10.24,0.8098076)
\rput{-90.0}(5.473808,5.1061926){\pstriangle[linewidth=0.04,dimen=outer](5.29,-0.7298076)(1.2,1.092)}
\psline[linewidth=0.04cm](6.6,-0.24980763)(10.22,-0.24980763)
\psline[linewidth=0.04cm](2.36,-0.22980762)(3.56,-0.22980762)
\psline[linewidth=0.04cm](1.78,0.7898076)(2.38,-0.24942286)
\psline[linewidth=0.04cm](1.2,-0.24942286)(1.8,0.7898076)
\psline[linewidth=0.04cm](0.0,-0.22980762)(1.22,-0.22980762)
\end{pspicture}
}
\end{equation*}

But, one problem with the above transformation is that there is nothing that is forcing the snowflake to be oriented outside. Orientation can be achieved by using a dot on one side of the line.

%Koch oriented
\begin{equation*}\scalebox{0.6} % Change this value to rescale the drawing.
{
\begin{pspicture}(0,-0.77709204)(10.24,0.80680764)
\rput{-90.0}(5.4768076,5.1031923){\pstriangle[linewidth=0.04,dimen=outer](5.29,-0.73280764)(1.2,1.092)}
\psline[linewidth=0.04cm](6.6,-0.25280762)(10.22,-0.25280762)
\psline[linewidth=0.04cm](2.36,-0.23280762)(3.56,-0.23280762)
\psline[linewidth=0.04cm](1.78,0.7868076)(2.38,-0.25242287)
\psline[linewidth=0.04cm](1.2,-0.25242287)(1.8,0.7868076)
\psline[linewidth=0.04cm](0.0,-0.23280762)(1.22,-0.23280762)
\psdots[dotsize=0.12](2.0,0.18719238)
\psdots[dotsize=0.12](1.58,0.18719238)
\psdots[dotsize=0.36](8.38,-0.5128076)
\psdots[dotsize=0.12](2.9,-0.33280763)
\psdots[dotsize=0.12](0.58,-0.33280763)
\end{pspicture}
}\end{equation*}

Applying this transformation to a triangle will result in the Koch snowflake.

%Koch applied
\begin{equation*}\scalebox{0.6} % Change this value to rescale the drawing.
{
\begin{pspicture}(0,-4.8192186)(12.621875,4.8592186)
\rput{-90.0}(2.7432187,8.796782){\pstriangle[linewidth=0.04,dimen=outer](5.77,2.4807813)(1.2,1.092)}
\psline[linewidth=0.04cm](7.08,2.9607813)(10.7,2.9607813)
\psline[linewidth=0.04cm](2.84,2.9807813)(4.04,2.9807813)
\psline[linewidth=0.04cm](2.26,4.0003967)(2.86,2.961166)
\psline[linewidth=0.04cm](1.68,2.961166)(2.28,4.0003967)
\psline[linewidth=0.04cm](0.48,2.9807813)(1.7,2.9807813)
\psdots[dotsize=0.12](2.48,3.4007812)
\psdots[dotsize=0.12](2.06,3.4007812)
\psdots[dotsize=0.36](8.86,2.7007813)
\psdots[dotsize=0.12](3.38,2.8807812)
\psdots[dotsize=0.12](1.06,2.8807812)
\pstriangle[linewidth=0.04,dimen=outer](5.56,-4.079219)(3.62,3.135)
\psdots[dotsize=0.36](6.26,-2.6792188)
\psdots[dotsize=0.36](4.86,-2.6992188)
\psdots[dotsize=0.36](5.56,-3.7992187)
\psframe[linewidth=0.02,dimen=outer](11.2,4.6207814)(0.0,1.9207813)
\psline[linewidth=0.04cm,arrowsize=0.32cm 8.0,arrowlength=1.4,arrowinset=0.4]{->}(5.54,1.3607812)(5.54,-0.05921875)
\psframe[linewidth=0.02,linestyle=dashed,dash=0.16cm 0.16cm,dimen=outer](8.06,-0.57921875)(2.98,-4.8192186)
\usefont{T1}{ptm}{m}{n}
\rput(12.021406,4.670781){$\infty S$}
\end{pspicture}
}\end{equation*}

\bigskip

For the Sierpinsky carpet, the following transformation (which we will call the Sierpinsky transformation) will be used.

%Sierpinsky
\begin{equation*}\scalebox{0.8} % Change this value to rescale the drawing.
{
\begin{pspicture}(0,-1.52)(9.66,1.52)
\definecolor{color643}{rgb}{0.6,0.6,0.6}
\psframe[linewidth=0.04,dimen=outer,fillstyle=solid,fillcolor=black](2.04,1.5)(1.02,0.48)
\rput{-90.0}(4.92,4.98){\pstriangle[linewidth=0.04,dimen=outer](4.95,-0.58)(1.22,1.22)}
\psframe[linewidth=0.04,dimen=outer,fillstyle=solid,fillcolor=black](9.66,1.5)(6.68,-1.48)
\psframe[linewidth=0.04,dimen=outer,fillstyle=solid,fillcolor=black](1.02,1.52)(0.0,0.5)
\psframe[linewidth=0.04,dimen=outer,fillstyle=solid,fillcolor=black](3.02,1.52)(2.0,0.5)
\psframe[linewidth=0.04,dimen=outer,fillstyle=solid,fillcolor=black](1.02,0.52)(0.0,-0.5)
\psframe[linewidth=0.04,dimen=outer,fillstyle=solid,fillcolor=black](3.02,0.52)(2.0,-0.5)
\psframe[linewidth=0.04,dimen=outer,fillstyle=solid,fillcolor=black](1.02,-0.5)(0.0,-1.52)
\psframe[linewidth=0.04,dimen=outer,fillstyle=solid,fillcolor=black](2.02,-0.5)(1.0,-1.52)
\psframe[linewidth=0.04,dimen=outer,fillstyle=solid,fillcolor=black](3.02,-0.5)(2.0,-1.52)
\rput(0.02,-1.5){\psgrid[gridwidth=0.0582,subgridwidth=0.014111111,gridlabels=0.0pt,subgriddiv=1,gridcolor=color643,subgridcolor=white](0,0)(0,0)(3,3)}
\rput(6.66,-1.5){\psgrid[gridwidth=0.0582,subgridwidth=0.014111111,gridlabels=0.0pt,subgriddiv=1,unit=3.0cm,gridcolor=color643,subgridcolor=white](0,0)(0,0)(1,1)
\psset{unit=1.0cm}}
\end{pspicture}
}
\end{equation*}

If we applied the Sierpinsky transformation to a black square nine times we get the following shape.

%Sierpinsky 2 applications
\begin{equation*}\scalebox{0.8} % Change this value to rescale the drawing.
{
\begin{pspicture}(0,-1.509995)(3.02,1.509995)
\definecolor{color722}{rgb}{0.6,0.6,0.6}
\definecolor{color722c}{rgb}{0.5019607843137255,0.5019607843137255,0.5019607843137255}
\psframe[linewidth=0.04,dimen=outer,fillstyle=solid,fillcolor=black](1.02,-0.82999504)(0.68,-1.169995)
\psframe[linewidth=0.04,dimen=outer,fillstyle=solid,fillcolor=black](0.36,-0.849995)(0.02,-1.189995)
\psframe[linewidth=0.04,dimen=outer,fillstyle=solid,fillcolor=black](0.64,-1.169995)(0.3,-1.509995)
\psframe[linewidth=0.04,dimen=outer,fillstyle=solid,fillcolor=black](1.02,-1.169995)(0.68,-1.509995)
\psframe[linewidth=0.04,dimen=outer,fillstyle=solid,fillcolor=black](0.36,-1.169995)(0.02,-1.509995)
\psframe[linewidth=0.04,dimen=outer,fillstyle=solid,fillcolor=black](0.64,-0.509995)(0.3,-0.849995)
\psframe[linewidth=0.04,dimen=outer,fillstyle=solid,fillcolor=black](1.02,-0.509995)(0.68,-0.849995)
\psframe[linewidth=0.04,dimen=outer,fillstyle=solid,fillcolor=black](0.36,-0.509995)(0.02,-0.849995)
\rput(0.0,-1.489995){\psgrid[gridwidth=0.0582,subgridwidth=0.014111111,gridlabels=0.0pt,subgriddiv=1,unit=0.33333cm,gridcolor=color722,subgridcolor=color722c](0,0)(0,0)(3,3)
\psset{unit=1.0cm}}
\psframe[linewidth=0.04,dimen=outer,fillstyle=solid,fillcolor=black](2.02,-0.82999504)(1.68,-1.169995)
\psframe[linewidth=0.04,dimen=outer,fillstyle=solid,fillcolor=black](1.36,-0.849995)(1.02,-1.189995)
\psframe[linewidth=0.04,dimen=outer,fillstyle=solid,fillcolor=black](1.64,-1.169995)(1.3,-1.509995)
\psframe[linewidth=0.04,dimen=outer,fillstyle=solid,fillcolor=black](2.02,-1.169995)(1.68,-1.509995)
\psframe[linewidth=0.04,dimen=outer,fillstyle=solid,fillcolor=black](1.36,-1.169995)(1.02,-1.509995)
\psframe[linewidth=0.04,dimen=outer,fillstyle=solid,fillcolor=black](1.64,-0.509995)(1.3,-0.849995)
\psframe[linewidth=0.04,dimen=outer,fillstyle=solid,fillcolor=black](2.02,-0.509995)(1.68,-0.849995)
\psframe[linewidth=0.04,dimen=outer,fillstyle=solid,fillcolor=black](1.36,-0.509995)(1.02,-0.849995)
\rput(1.0,-1.489995){\psgrid[gridwidth=0.0582,subgridwidth=0.014111111,gridlabels=0.0pt,subgriddiv=1,unit=0.33333cm,gridcolor=color722,subgridcolor=color722c](0,0)(0,0)(3,3)
\psset{unit=1.0cm}}
\psframe[linewidth=0.04,dimen=outer,fillstyle=solid,fillcolor=black](3.02,-0.82999504)(2.68,-1.169995)
\psframe[linewidth=0.04,dimen=outer,fillstyle=solid,fillcolor=black](2.36,-0.849995)(2.02,-1.189995)
\psframe[linewidth=0.04,dimen=outer,fillstyle=solid,fillcolor=black](2.64,-1.169995)(2.3,-1.509995)
\psframe[linewidth=0.04,dimen=outer,fillstyle=solid,fillcolor=black](3.02,-1.169995)(2.68,-1.509995)
\psframe[linewidth=0.04,dimen=outer,fillstyle=solid,fillcolor=black](2.36,-1.169995)(2.02,-1.509995)
\psframe[linewidth=0.04,dimen=outer,fillstyle=solid,fillcolor=black](2.64,-0.509995)(2.3,-0.849995)
\psframe[linewidth=0.04,dimen=outer,fillstyle=solid,fillcolor=black](3.02,-0.509995)(2.68,-0.849995)
\psframe[linewidth=0.04,dimen=outer,fillstyle=solid,fillcolor=black](2.36,-0.509995)(2.02,-0.849995)
\rput(2.0,-1.489995){\psgrid[gridwidth=0.0582,subgridwidth=0.014111111,gridlabels=0.0pt,subgriddiv=1,unit=0.33333cm,gridcolor=color722,subgridcolor=color722c](0,0)(0,0)(3,3)
\psset{unit=1.0cm}}
\psframe[linewidth=0.04,dimen=outer,fillstyle=solid,fillcolor=black](3.02,0.150005)(2.68,-0.189995)
\psframe[linewidth=0.04,dimen=outer,fillstyle=solid,fillcolor=black](2.36,0.13000499)(2.02,-0.209995)
\psframe[linewidth=0.04,dimen=outer,fillstyle=solid,fillcolor=black](2.64,-0.189995)(2.3,-0.529995)
\psframe[linewidth=0.04,dimen=outer,fillstyle=solid,fillcolor=black](3.02,-0.189995)(2.68,-0.529995)
\psframe[linewidth=0.04,dimen=outer,fillstyle=solid,fillcolor=black](2.36,-0.189995)(2.02,-0.529995)
\psframe[linewidth=0.04,dimen=outer,fillstyle=solid,fillcolor=black](2.64,0.470005)(2.3,0.13000499)
\psframe[linewidth=0.04,dimen=outer,fillstyle=solid,fillcolor=black](3.02,0.470005)(2.68,0.13000499)
\psframe[linewidth=0.04,dimen=outer,fillstyle=solid,fillcolor=black](2.36,0.470005)(2.02,0.13000499)
\rput(2.0,-0.509995){\psgrid[gridwidth=0.0582,subgridwidth=0.014111111,gridlabels=0.0pt,subgriddiv=1,unit=0.33333cm,gridcolor=color722,subgridcolor=color722c](0,0)(0,0)(3,3)
\psset{unit=1.0cm}}
\psframe[linewidth=0.04,dimen=outer,fillstyle=solid,fillcolor=black](3.02,1.170005)(2.68,0.830005)
\psframe[linewidth=0.04,dimen=outer,fillstyle=solid,fillcolor=black](2.36,1.150005)(2.02,0.810005)
\psframe[linewidth=0.04,dimen=outer,fillstyle=solid,fillcolor=black](2.64,0.830005)(2.3,0.490005)
\psframe[linewidth=0.04,dimen=outer,fillstyle=solid,fillcolor=black](3.02,0.830005)(2.68,0.490005)
\psframe[linewidth=0.04,dimen=outer,fillstyle=solid,fillcolor=black](2.36,0.830005)(2.02,0.490005)
\psframe[linewidth=0.04,dimen=outer,fillstyle=solid,fillcolor=black](2.64,1.490005)(2.3,1.150005)
\psframe[linewidth=0.04,dimen=outer,fillstyle=solid,fillcolor=black](3.02,1.490005)(2.68,1.150005)
\psframe[linewidth=0.04,dimen=outer,fillstyle=solid,fillcolor=black](2.36,1.490005)(2.02,1.150005)
\rput(2.0,0.510005){\psgrid[gridwidth=0.0582,subgridwidth=0.014111111,gridlabels=0.0pt,subgriddiv=1,unit=0.33333cm,gridcolor=color722,subgridcolor=color722c](0,0)(0,0)(3,3)
\psset{unit=1.0cm}}
\psframe[linewidth=0.04,dimen=outer,fillstyle=solid,fillcolor=black](2.02,1.170005)(1.68,0.830005)
\psframe[linewidth=0.04,dimen=outer,fillstyle=solid,fillcolor=black](1.36,1.150005)(1.02,0.810005)
\psframe[linewidth=0.04,dimen=outer,fillstyle=solid,fillcolor=black](1.64,0.830005)(1.3,0.490005)
\psframe[linewidth=0.04,dimen=outer,fillstyle=solid,fillcolor=black](2.02,0.830005)(1.68,0.490005)
\psframe[linewidth=0.04,dimen=outer,fillstyle=solid,fillcolor=black](1.36,0.830005)(1.02,0.490005)
\psframe[linewidth=0.04,dimen=outer,fillstyle=solid,fillcolor=black](1.64,1.490005)(1.3,1.150005)
\psframe[linewidth=0.04,dimen=outer,fillstyle=solid,fillcolor=black](2.02,1.490005)(1.68,1.150005)
\psframe[linewidth=0.04,dimen=outer,fillstyle=solid,fillcolor=black](1.36,1.490005)(1.02,1.150005)
\rput(1.0,0.510005){\psgrid[gridwidth=0.0582,subgridwidth=0.014111111,gridlabels=0.0pt,subgriddiv=1,unit=0.33333cm,gridcolor=color722,subgridcolor=color722c](0,0)(0,0)(3,3)
\psset{unit=1.0cm}}
\psframe[linewidth=0.04,dimen=outer,fillstyle=solid,fillcolor=black](1.02,0.170005)(0.68,-0.16999501)
\psframe[linewidth=0.04,dimen=outer,fillstyle=solid,fillcolor=black](0.36,0.150005)(0.02,-0.189995)
\psframe[linewidth=0.04,dimen=outer,fillstyle=solid,fillcolor=black](0.64,-0.16999501)(0.3,-0.509995)
\psframe[linewidth=0.04,dimen=outer,fillstyle=solid,fillcolor=black](1.02,-0.16999501)(0.68,-0.509995)
\psframe[linewidth=0.04,dimen=outer,fillstyle=solid,fillcolor=black](0.36,-0.16999501)(0.02,-0.509995)
\psframe[linewidth=0.04,dimen=outer,fillstyle=solid,fillcolor=black](0.64,0.490005)(0.3,0.150005)
\psframe[linewidth=0.04,dimen=outer,fillstyle=solid,fillcolor=black](1.02,0.490005)(0.68,0.150005)
\psframe[linewidth=0.04,dimen=outer,fillstyle=solid,fillcolor=black](0.36,0.490005)(0.02,0.150005)
\rput(0.0,-0.489995){\psgrid[gridwidth=0.0582,subgridwidth=0.014111111,gridlabels=0.0pt,subgriddiv=1,unit=0.33333cm,gridcolor=color722,subgridcolor=color722c](0,0)(0,0)(3,3)
\psset{unit=1.0cm}}
\psframe[linewidth=0.04,dimen=outer,fillstyle=solid,fillcolor=black](1.02,1.170005)(0.68,0.830005)
\psframe[linewidth=0.04,dimen=outer,fillstyle=solid,fillcolor=black](0.36,1.150005)(0.02,0.810005)
\psframe[linewidth=0.04,dimen=outer,fillstyle=solid,fillcolor=black](0.64,0.830005)(0.3,0.490005)
\psframe[linewidth=0.04,dimen=outer,fillstyle=solid,fillcolor=black](1.02,0.830005)(0.68,0.490005)
\psframe[linewidth=0.04,dimen=outer,fillstyle=solid,fillcolor=black](0.36,0.830005)(0.02,0.490005)
\psframe[linewidth=0.04,dimen=outer,fillstyle=solid,fillcolor=black](0.64,1.490005)(0.3,1.150005)
\psframe[linewidth=0.04,dimen=outer,fillstyle=solid,fillcolor=black](1.02,1.490005)(0.68,1.150005)
\psframe[linewidth=0.04,dimen=outer,fillstyle=solid,fillcolor=black](0.36,1.490005)(0.02,1.150005)
\rput(0.0,0.510005){\psgrid[gridwidth=0.0582,subgridwidth=0.014111111,gridlabels=0.0pt,subgriddiv=1,unit=0.33333cm,gridcolor=color722,subgridcolor=color722c](0,0)(0,0)(3,3)
\psset{unit=1.0cm}}
\psframe[linewidth=0.04,dimen=outer,fillstyle=solid,fillcolor=black](1.02,-0.82999504)(0.68,-1.169995)
\psframe[linewidth=0.04,dimen=outer,fillstyle=solid,fillcolor=black](0.36,-0.849995)(0.02,-1.189995)
\psframe[linewidth=0.04,dimen=outer,fillstyle=solid,fillcolor=black](0.64,-1.169995)(0.3,-1.509995)
\psframe[linewidth=0.04,dimen=outer,fillstyle=solid,fillcolor=black](1.02,-1.169995)(0.68,-1.509995)
\psframe[linewidth=0.04,dimen=outer,fillstyle=solid,fillcolor=black](0.36,-1.169995)(0.02,-1.509995)
\psframe[linewidth=0.04,dimen=outer,fillstyle=solid,fillcolor=black](0.64,-0.509995)(0.3,-0.849995)
\psframe[linewidth=0.04,dimen=outer,fillstyle=solid,fillcolor=black](1.02,-0.509995)(0.68,-0.849995)
\psframe[linewidth=0.04,dimen=outer,fillstyle=solid,fillcolor=black](0.36,-0.509995)(0.02,-0.849995)
\rput(0.0,-1.489995){\psgrid[gridwidth=0.0582,subgridwidth=0.014111111,gridlabels=0.0pt,subgriddiv=1,unit=0.33333cm,gridcolor=color722,subgridcolor=color722c](0,0)(0,0)(3,3)
\psset{unit=1.0cm}}
\psframe[linewidth=0.04,dimen=outer,fillstyle=solid,fillcolor=black](2.02,-0.82999504)(1.68,-1.169995)
\psframe[linewidth=0.04,dimen=outer,fillstyle=solid,fillcolor=black](1.36,-0.849995)(1.02,-1.189995)
\psframe[linewidth=0.04,dimen=outer,fillstyle=solid,fillcolor=black](1.64,-1.169995)(1.3,-1.509995)
\psframe[linewidth=0.04,dimen=outer,fillstyle=solid,fillcolor=black](2.02,-1.169995)(1.68,-1.509995)
\psframe[linewidth=0.04,dimen=outer,fillstyle=solid,fillcolor=black](1.36,-1.169995)(1.02,-1.509995)
\psframe[linewidth=0.04,dimen=outer,fillstyle=solid,fillcolor=black](1.64,-0.509995)(1.3,-0.849995)
\psframe[linewidth=0.04,dimen=outer,fillstyle=solid,fillcolor=black](2.02,-0.509995)(1.68,-0.849995)
\psframe[linewidth=0.04,dimen=outer,fillstyle=solid,fillcolor=black](1.36,-0.509995)(1.02,-0.849995)
\rput(1.0,-1.489995){\psgrid[gridwidth=0.0582,subgridwidth=0.014111111,gridlabels=0.0pt,subgriddiv=1,unit=0.33333cm,gridcolor=color722,subgridcolor=color722c](0,0)(0,0)(3,3)
\psset{unit=1.0cm}}
\psframe[linewidth=0.04,dimen=outer,fillstyle=solid,fillcolor=black](3.02,-0.82999504)(2.68,-1.169995)
\psframe[linewidth=0.04,dimen=outer,fillstyle=solid,fillcolor=black](2.36,-0.849995)(2.02,-1.189995)
\psframe[linewidth=0.04,dimen=outer,fillstyle=solid,fillcolor=black](2.64,-1.169995)(2.3,-1.509995)
\psframe[linewidth=0.04,dimen=outer,fillstyle=solid,fillcolor=black](3.02,-1.169995)(2.68,-1.509995)
\psframe[linewidth=0.04,dimen=outer,fillstyle=solid,fillcolor=black](2.36,-1.169995)(2.02,-1.509995)
\psframe[linewidth=0.04,dimen=outer,fillstyle=solid,fillcolor=black](2.64,-0.509995)(2.3,-0.849995)
\psframe[linewidth=0.04,dimen=outer,fillstyle=solid,fillcolor=black](3.02,-0.509995)(2.68,-0.849995)
\psframe[linewidth=0.04,dimen=outer,fillstyle=solid,fillcolor=black](2.36,-0.509995)(2.02,-0.849995)
\rput(2.0,-1.489995){\psgrid[gridwidth=0.0582,subgridwidth=0.014111111,gridlabels=0.0pt,subgriddiv=1,unit=0.33333cm,gridcolor=color722,subgridcolor=color722c](0,0)(0,0)(3,3)
\psset{unit=1.0cm}}
\psframe[linewidth=0.04,dimen=outer,fillstyle=solid,fillcolor=black](3.02,0.150005)(2.68,-0.189995)
\psframe[linewidth=0.04,dimen=outer,fillstyle=solid,fillcolor=black](2.36,0.13000499)(2.02,-0.209995)
\psframe[linewidth=0.04,dimen=outer,fillstyle=solid,fillcolor=black](2.64,-0.189995)(2.3,-0.529995)
\psframe[linewidth=0.04,dimen=outer,fillstyle=solid,fillcolor=black](3.02,-0.189995)(2.68,-0.529995)
\psframe[linewidth=0.04,dimen=outer,fillstyle=solid,fillcolor=black](2.36,-0.189995)(2.02,-0.529995)
\psframe[linewidth=0.04,dimen=outer,fillstyle=solid,fillcolor=black](2.64,0.470005)(2.3,0.13000499)
\psframe[linewidth=0.04,dimen=outer,fillstyle=solid,fillcolor=black](3.02,0.470005)(2.68,0.13000499)
\psframe[linewidth=0.04,dimen=outer,fillstyle=solid,fillcolor=black](2.36,0.470005)(2.02,0.13000499)
\rput(2.0,-0.509995){\psgrid[gridwidth=0.0582,subgridwidth=0.014111111,gridlabels=0.0pt,subgriddiv=1,unit=0.33333cm,gridcolor=color722,subgridcolor=color722c](0,0)(0,0)(3,3)
\psset{unit=1.0cm}}
\psframe[linewidth=0.04,dimen=outer,fillstyle=solid,fillcolor=black](3.02,1.170005)(2.68,0.830005)
\psframe[linewidth=0.04,dimen=outer,fillstyle=solid,fillcolor=black](2.36,1.150005)(2.02,0.810005)
\psframe[linewidth=0.04,dimen=outer,fillstyle=solid,fillcolor=black](2.64,0.830005)(2.3,0.490005)
\psframe[linewidth=0.04,dimen=outer,fillstyle=solid,fillcolor=black](3.02,0.830005)(2.68,0.490005)
\psframe[linewidth=0.04,dimen=outer,fillstyle=solid,fillcolor=black](2.36,0.830005)(2.02,0.490005)
\psframe[linewidth=0.04,dimen=outer,fillstyle=solid,fillcolor=black](2.64,1.490005)(2.3,1.150005)
\psframe[linewidth=0.04,dimen=outer,fillstyle=solid,fillcolor=black](3.02,1.490005)(2.68,1.150005)
\psframe[linewidth=0.04,dimen=outer,fillstyle=solid,fillcolor=black](2.36,1.490005)(2.02,1.150005)
\rput(2.0,0.510005){\psgrid[gridwidth=0.0582,subgridwidth=0.014111111,gridlabels=0.0pt,subgriddiv=1,unit=0.33333cm,gridcolor=color722,subgridcolor=color722c](0,0)(0,0)(3,3)
\psset{unit=1.0cm}}
\psframe[linewidth=0.04,dimen=outer,fillstyle=solid,fillcolor=black](2.02,1.170005)(1.68,0.830005)
\psframe[linewidth=0.04,dimen=outer,fillstyle=solid,fillcolor=black](1.36,1.150005)(1.02,0.810005)
\psframe[linewidth=0.04,dimen=outer,fillstyle=solid,fillcolor=black](1.64,0.830005)(1.3,0.490005)
\psframe[linewidth=0.04,dimen=outer,fillstyle=solid,fillcolor=black](2.02,0.830005)(1.68,0.490005)
\psframe[linewidth=0.04,dimen=outer,fillstyle=solid,fillcolor=black](1.36,0.830005)(1.02,0.490005)
\psframe[linewidth=0.04,dimen=outer,fillstyle=solid,fillcolor=black](1.64,1.490005)(1.3,1.150005)
\psframe[linewidth=0.04,dimen=outer,fillstyle=solid,fillcolor=black](2.02,1.490005)(1.68,1.150005)
\psframe[linewidth=0.04,dimen=outer,fillstyle=solid,fillcolor=black](1.36,1.490005)(1.02,1.150005)
\rput(1.0,0.510005){\psgrid[gridwidth=0.0582,subgridwidth=0.014111111,gridlabels=0.0pt,subgriddiv=1,unit=0.33333cm,gridcolor=color722,subgridcolor=color722c](0,0)(0,0)(3,3)
\psset{unit=1.0cm}}
\psframe[linewidth=0.04,dimen=outer,fillstyle=solid,fillcolor=black](1.02,0.170005)(0.68,-0.16999501)
\psframe[linewidth=0.04,dimen=outer,fillstyle=solid,fillcolor=black](0.36,0.150005)(0.02,-0.189995)
\psframe[linewidth=0.04,dimen=outer,fillstyle=solid,fillcolor=black](0.64,-0.16999501)(0.3,-0.509995)
\psframe[linewidth=0.04,dimen=outer,fillstyle=solid,fillcolor=black](1.02,-0.16999501)(0.68,-0.509995)
\psframe[linewidth=0.04,dimen=outer,fillstyle=solid,fillcolor=black](0.36,-0.16999501)(0.02,-0.509995)
\psframe[linewidth=0.04,dimen=outer,fillstyle=solid,fillcolor=black](0.64,0.490005)(0.3,0.150005)
\psframe[linewidth=0.04,dimen=outer,fillstyle=solid,fillcolor=black](1.02,0.490005)(0.68,0.150005)
\psframe[linewidth=0.04,dimen=outer,fillstyle=solid,fillcolor=black](0.36,0.490005)(0.02,0.150005)
\rput(0.0,-0.489995){\psgrid[gridwidth=0.0582,subgridwidth=0.014111111,gridlabels=0.0pt,subgriddiv=1,unit=0.33333cm,gridcolor=color722,subgridcolor=color722c](0,0)(0,0)(3,3)
\psset{unit=1.0cm}}
\psframe[linewidth=0.04,dimen=outer,fillstyle=solid,fillcolor=black](1.02,1.170005)(0.68,0.830005)
\psframe[linewidth=0.04,dimen=outer,fillstyle=solid,fillcolor=black](0.36,1.150005)(0.02,0.810005)
\psframe[linewidth=0.04,dimen=outer,fillstyle=solid,fillcolor=black](0.64,0.830005)(0.3,0.490005)
\psframe[linewidth=0.04,dimen=outer,fillstyle=solid,fillcolor=black](1.02,0.830005)(0.68,0.490005)
\psframe[linewidth=0.04,dimen=outer,fillstyle=solid,fillcolor=black](0.36,0.830005)(0.02,0.490005)
\psframe[linewidth=0.04,dimen=outer,fillstyle=solid,fillcolor=black](0.64,1.490005)(0.3,1.150005)
\psframe[linewidth=0.04,dimen=outer,fillstyle=solid,fillcolor=black](1.02,1.490005)(0.68,1.150005)
\psframe[linewidth=0.04,dimen=outer,fillstyle=solid,fillcolor=black](0.36,1.490005)(0.02,1.150005)
\rput(0.0,0.510005){\psgrid[gridwidth=0.0582,subgridwidth=0.014111111,gridlabels=0.0pt,subgriddiv=1,unit=0.33333cm,gridcolor=color722,subgridcolor=color722c](0,0)(0,0)(3,3)
\psset{unit=1.0cm}}
\end{pspicture}
}\end{equation*}
The Sierpinsky carpet is defined by applying the Sierpinsky transformation an infinite number of times to a black square.

Again, this opens many possibilities of modified Koch or Sierpinsky fractals. More interestingly, this indicates that the language of transformation could be a natural mathematical choice to develop a calculus for fractals.

\subsection{Differential calculus}\label{differential calculus}

Although we will not present it here, it is possible to define division and the entire number system with the use of transformations. Assuming that we have division, we can define the derivative of a function as follows.

$$f'(x):=\fbox{${\color{green}\dfrac{h}{2}}\rhd {\color{green}h}$}^{\,\infty S}\rightarrow \frac{f(x+h)-f(x)}{h}$$

Note, that the infinite sequence $\fbox{${\color{green}\dfrac{h}{2}}\rhd {\color{green}h}$}^{\,\infty S}$ acts like the limit when $h$ tends towards $0$.

\subsection{Meta-mathematics}
If we translate two mathematical theories $T_1$ and $T_2$ into the language of transformations, we can investigate how close two such theories are to each other. For this we can write
\begin{equation*}\fbox{$X\rhd T_1$}\rightarrow T_1 \asymp \fbox{$Y\rhd T_2$}\rightarrow T_2.\end{equation*}

This means that under the transformations $X$ and $Y$ the theories will reduce to the same form. Depending on what $X, Y$ are and what the theories reduce to, we could evaluate how close the theories are to one other or how they relate. This is similar to what is done in category theory where functors between categories and comparisons between specific internal structures of categories give a way to compare and classify the categories. One advantage with transformations is that we have a means to relax some restrictions required by categorical structures and thus generalize the tools of category theory.

One of the main activity, and maybe the only activities in Mathematics is to classify abstract objects. In essence, objects are analysed with different tools and then classified as the same or different from other objects. If objects are different, a degree or level of difference can be associated. Here are a few examples spanning basic and advanced mathematics.

A key element of arithmetic is equality. $1+2=3$ can be seen as stating that $1+2$ is the same as $3$. Commutativity $xy=yx$ is also an expression of sameness classifying the structures $xy$ and $yx$. The expression $2x=1$ is asking if there is a way to have $2x$ to be the same as $1$. This is not possible for integers, but defining a new number $\frac{1}{2}$ forces this expression to be valid. Also, $30=n\times15$ can be viewed as asking how far is $15$ is from $30$.

Classification is done at all levels of mathematics, other examples are homology, category theory and topology. An interesting question is to ask if mathematics can solely be viewed as the activity of classification along with the introduction of new definitions and objects.

The advantage of the language of transformation is that all mathematical objects can be constructed with forms and this allows us to compare all concepts and objects with each other. A homogeneous language might the key to build a large map of the mathematical snd scientific  landscape.

\subsection{Continuous transformations}

The language that was presented here (like classical logic and set theory) has a discrete flavor. Since we can represent an infinite number of transformations, as seen in section \ref{differential calculus}, we can approach infinitely small elements. We have seen that we can use probabilities on transformations and we can wonder if there a way to make the application of transformations less discrete. For us, there is before and after the application of a transformation, but we could eventually define a way to continuously apply a transformation where there is a continuum ranging from not applied to applied. We would eventually have to describe the real number with transformations.

\subsection{Formal description of the language}

We have presented the language of transformations in a progressive way with the aim to make it easy to understand. Although we will not be doing it here, it is possible to give a formal description of the language by using axioms and definitions.

\section{Physics}

We will now present simple models in physics.

\subsection{Motion}
For the physics of moving bodies, we will consider that there is a minimal unit of length $u_d$ that an object can move and that there is a minimal unit of time $u_t$. Since they are arbitrary values, they can always be adjusted based on our measuring tools.

We describe an object moving in a straight line along with its associated clock as follows.

\begin{equation*}v_{1/1}=:\fbox{$\dashrightarrow\bullet,\rightsquigarrow \intercal\quad\quad\rhd\quad\quad\bullet,\intercal$}\end{equation*}

When applied to the initial set $\{\bullet, \intercal\}$, this transformation can be interpreted as giving  the speed $1$ unit of distance per unit of time to the particle depicted by `$\bullet$'. Here the dashed arrow $\dashrightarrow$ indicates that one unit was covered and the arrow $\rightsquigarrow$ indicates that one unit of time elapsed.

The following transformation defines a speed of $2$ units of distance per unit of time ($2u_d/u_t$).
\begin{equation*}v_{2/1}:=\fbox{$\dashrightarrow\dashrightarrow\bullet,\rightsquigarrow \intercal\quad\quad\rhd\quad\quad\bullet,\intercal$}\end{equation*}

If we apply this speed $v_{2/1}$ repeatedly to $\{\bullet, \intercal\}$ our particle is seen as having an acceleration of $0$.

We can define uniform acceleration of $1$ by using the following series of five speed transformations.

\begin{equation*}a_{1/1}:=v_{5/1}\rightarrow v_{4/1}\rightarrow v_{3/1}\rightarrow v_{2/1}\rightarrow v_{1/1}\end{equation*}

Although much more work is needed to develop this approach, transformations can bring a new way to understand physics. An interesting property of this approach is that these transformations for speed and acceleration do not need an absolute coordinate system to be expressed, they can be seen as being relative to the moving body.

\subsection{Attractive and repulsive forces}

We now present a model of an attractive and repulsive force. The goal is to show that transformations can provide a new way to look a the fundamental forces.

Let's define an object $(x,E)$ where $x$ is the length of the last distance covered and $E$ can be seen to represent its internal energy or mass. If $x$ is negative, it means that it covered the distance from right to left and if it is positive, it covered the distance from left to right. We will consider the number $1$ written between square brackets as $[1]$ to represent space without the presence of objects, this can be viewed as natural space. If the number is different, it will mean that the space is stretched or contracted. For our model, we put an object $(0,M)$ at rest of mass $M$ where $M$ is larger than $1$. Thus we have the following.
$$[1](0,M)[1][1][1][1][1][1][1]$$
In our model, the mass will change the size of space elements. If the mass is larger than $1$ then the surrounding space will stretch and if the mass is smaller than $1$ then the surrounding space will contract. The closer the space to the object, the larger the stretch or contraction depending on our model. For example, a mass $M$ larger than $1$ will have a stretching effect on its surrounding space.
$$[1+\frac{M-1}{2}](0,M)[1+\frac{M-1}{2}][1+\frac{M-1}{3}][1+\frac{M-1}{4}][1+\frac{M-1}{5}][1+\frac{M-1}{6}].$$
Note that because of the square brackets $[\frac{M-1}{2}]$ must be viewed as a distance and not a mass. Our choice in the amount of space stretching is arbitrary and should be adapted with a more precise sequence to approach a realistic physical model.

If we place an object of mass $m>1$ on the right along with more space on the right of it, such that $m$ is negligible compared to $M$, we get
$$[1+\frac{M-1}{2}](0,M)[1+\frac{M-1}{2}][1+\frac{M-1}{3}][1+\frac{M-1}{4}][1+\frac{M-1}{5}](0,m)[1+\frac{M-1}{6}].$$

The following transformations can be seen as a force acting on the mass $m$. If $u>v$ we apply
$$\fbox{$(-u,m)[u][v]\rhd[u](x,m)[v]$}.$$
If $v>u$, then we apply
$$\fbox{$[u][v](+v,m)\rhd[u](x,m)[v]$}.$$
These transformations can be interpreted to say that a mass will move in the direction where there is the most space available and will record the distance it just covered.

By applying these transformations to
$$[1+\frac{M-1}{2}](0,M)[1+\frac{M-1}{2}][1+\frac{M-1}{3}][1+\frac{M-1}{4}][1+\frac{M-1}{5}][1+\frac{M-1}{6}],$$
we find that the mass $m$ will cover more and more distance, thus accelerating toward the mass $M$. The distance it just covered can be viewed as its speed or kinetic energy.

Now let's take a mass $m_e$ smaller than $1$, this mass could be understood as the mass of an electron. If we construct its surrounding space in a similar manner as the mass $M$, we have that $m_e-1$ is a negative number. This means that the space around the mass $m_e$ is contracting.

$$[1+\frac{m_e-1}{2}](0,m_e)[1+\frac{m_e-1}{2}][1+\frac{m_e-1}{3}][1+\frac{m_e-1}{4}][1+\frac{m_e-1}{5}][1+\frac{m_e-1}{6}].$$

If we put a test particle of mass smaller that $m_e$ and negligible compared to $m_e$ close to the electron,  we have

$$[1+\frac{m_e-1}{2}](0,m_e)[1+\frac{m_e-1}{2}](0,m_t)[1+\frac{m_e-1}{3}][1+\frac{m_e-1}{4}][1+\frac{m_e-1}{5}][1+\frac{m_e-1}{6}].$$
Since there is more space on the right, the test particle will move towards the right after applying the transformation force.

With two electrons we will have repulsion since each will act by repulsing the other. But if we have a mass $M$ larger than $1$ and larger than the difference between $m_e$ and $1$, then the electron will repulse the large mass a very small amount, but the large mass will attract the electron much more.

From this point of view, we have one force which expressed as attractive for a large mass and repulsive for light particles such as electrons. In this simple model, we have to consider that protons have a mass larger than $1$.

A variant of our transformation force is to require for $x<0$ that we apply
$$\fbox{$(-u,m)[u][v]\rhd[u](x,E)[v]$},$$
when $u>v+xE$ and
$$\fbox{$[u][v](+v,m)\rhd[u](x,E)[v]$}.$$
when if $v+xE>u$.

If $x>0$ then we use
$$\fbox{$(-u,m)[u][v]\rhd[u](x,E)[v]$},$$
when $u<v+xE$ and
$$\fbox{$[u][v](+v,m)\rhd[u](x,E)[v]$}.$$
when if $v+xE>u$.

Finally if $x=0$, we use
%If $x>0$ then we use
$$\fbox{$(-u,m)[u][v]\rhd[u](0,E)[v]$},$$
when $u>v$ and
$$\fbox{$[u][v](+v,m)\rhd[u](0,E)[v]$}.$$
when if $v>u$.

With these, the object $(x,E)$ will accelerate towards a larger mass, pass it and continue on its way until it is stopped and goes back towards the mass. This can be viewed as an oscillation or a one dimensional orbit.

\subsection{Future Models}

We could extend this model by adding the components $s$ and $t$ to our object $(x,E)$. Here, $s$ is the amount of space the energy occupies and $t$ is a unit of time $t$ associated to the object. In this extended model, objects could be written as $(x,E,s,t)$. Based on this, our natural space could be written as $(x=0,E=0,s=1,t=1)$, an object with no speed, occupying a space and associated with a unit of time of $1$. A photon might be interpreted as $(x=c/t,E=0,s=0,t=1)$ and a neutrino as $(x=v_n/t,E=m_n,s=0,t=1)$ where $v_n$ is the neutrino speed at some energy $m_n$. Based on these objects, we could try to find a transformation force which account to many observed properties. Our object $(x,E,s,t)$ could be represented as a rectangular prism such that the length is $s$, the width is $t$ and height is $E$. Attached to the prism is also a vector of $x$ pointing in some direction. If we find the right force transformation, we could modify the unit of time as seen in relativity, investigate consequences of an invariant volume and maybe discover inner mechanisms explaining the interaction between space, energy and time.

For example, if we take the sheet of space $(x=0,E=0,s=1,t=1)$ and assume that the area stays the same, stretching $s$ to $2$ would make $t=1/2$. If we interpret $t$ as being the time it took to cover the last distance $x$, then an object passing through this space will get the values $x=s=2$ and $t=1/2$, thus making the speed of the object $x/t=4$ which can be understood as time dilatation.

Take again a sheet of space $(x=0,E=0,s=1,t=1)$ and assume that the area is an invariant. As time spends itself, then $s$ becomes greater which could be interpreted as an expanding space. Time would need to be viewed as decreasing and not increasing, making time more like water evaporating.

Based on the assumption that photons are emitted in quanta, we are forced to conclude that there is a limit to what we can measure.  This is also assuming that there are no other particles or techniques which can be used to have a lower limit. Since what we want to measure now takes the form of a rectangular prism, one way to look at Heisenberg's uncertainty principle is to take a small prism that has the sides of Planck's units and assume the volume is invariant. Thus, stretching a side of the prism will reduce the side of the others in a way which is similar to a reduction of $\triangle x$ implies an augmentation of $\triangle p_x$. From this point of view, saying that our prism has an invariant volume, is another way to express the Heisenberg uncertainty principle.

\section{Metascience}

We now have the possibility to express each domain of science by using the same mathematical notation, this means that each science can directly benefit from the others. Structures found or studied in one science can be applied or studied in another domain. For example, we can try to see if a new mathematical structure also appears in biology, neuroscience or chemistry. The point of view that comes with the language of transformations might be able to provide insights on the concept of emergence.

We are presently at a place where most researchers have to interact with different fields of science to improve their understanding of the world and create beneficial technologies. The language which was presented is only a few steps towards improving the free exchange of ideas and the creation of a large interactive open-access scientific database.

Languages such as the one presented combined with a large scientific database can prove to be fruitful and powerful tools. It is the sincere hope of the author that every person involved will take the responsibility of making ethical choices aimed at benefiting each and every living being.

\pagebreak

\section{Appendix}

We will now give a few more definitions and tools that are useful to manipulate systems of transformations.

\subsection{Open Transformations}

Until now, in our transformations, the cause of the transformation was defined as being the same as the elements in the initial form. We will now introduce a notation saying that the transformation will be applied to any disconnected form it meets. The transformation

\begin{equation}\fbox{$X\rhd \star$}\end{equation}
is understand as replacing any symbol it meet by the form $X$. For example,
\begin{equation}\fbox{$X\rhd \star$}\rightarrow (a)\end{equation}
will reduce to $(X)$ and
\begin{equation}\fbox{$X\rhd \star$}\rightarrow (b)\end{equation}
will also reduce to $(X)$.

We can also have more than one star in the cause of a transformation. For example,
\begin{equation}\fbox{$X,Y\rhd \star,\star$}\rightarrow (c,d)\end{equation}
will also reduce to $(X, Y)$.

We can also restrict the scope of the star symbol to a certain color by writing $\fbox{$X\rhd {\color{blue}\star}$}$. If we apply this transformation three times to $\{{\color{blue}a}, {\color{blue}b}, {\color{blue}a}, c\}$, this will give $\{X, X, X, c \}$.

The star symbol can also be written in the result of a transformation. An example of this is
$$Duplication:=\fbox{${\color{green}\star}, {\color{green}\star}\rhd {\color{green}\star}$}$$
which permits the duplication of any form it meets. Applying this transformation to an initial form containing a rabbit will give us a set of two rabbits.

\begin{equation}\fbox{${\color{green}\star}, {\color{green}\star}\rhd {\color{green}\star}$}\rightarrow \{rabbit\}\Rightarrow \{rabbit, rabbit\}\end{equation}

Applying duplication two times to a set of three rabbits, will give a total of five rabbits and applying duplication three times to a set of three rabbits, will give a total of six rabbits. If we want to apply the duplication to each rabbit in the initial set, we can use the sharp symbol and write the following.

$$\begin{array}{c}
\fbox{${\color{green}\star}, {\color{green}\star}$}^{\,\sharp}\rightarrow \{rabbit, rabbit, rabbit\} \\
\Downarrow\\
 \{rabbit, rabbit,rabbit, rabbit,rabbit, rabbit\}
\end{array}$$

In essence, we have multiplied the number of rabbits of the initial set by $2$. Similarly, multiplication by $5$ of a set of elements can be done with the transformation $\fbox{${\color{green}\star}, {\color{green}\star}, {\color{green}\star}, {\color{green}\star}, {\color{green}\star}\rhd {\color{green}\star}$}$.

Another example from chemistry, a transformation replacing any element having a bond with carbon by an hydrogen can be written as $\fbox{$C-{\color{green}H}\rhd (C-{\color{green}\star})$}$.

\subsection{Negation}

We now introduce the negation symbol $\neg$. When this symbol is written before a form, it means that we are referring to all that is not that form in a set. For example, take the set $\textit{Office}:=\{\textit{desk, computer, person, pen, phone}\}$, we can replace by a beautiful meadow all that is not the person in the office by using the symbol $\rceil_{\textit{Office}}$.

$$\begin{array}{c}
[\textit{beautiful meadow} \rhd \rceil_{\textit{Office}}(\textit{person})]\rightarrow  \{\textit{desk, computer, person, pen, phone}\} \\
\Downarrow\\
 \{\textit{beautiful meadow, person} \}
\end{array}$$

We can also use the negation symbol in the effect of the transformation. Let $\textit{PollutedWater}:=\{Pollutants, water\}$, also denoted by $PW$, be the set containing water mixed with different pollutants. The following transformation can be interpreted as a water filter which filters out all pollutants.

$$\fbox{$\rceil_{\textit{PW}}(\textit{Pollutants})\rhd \textit{PollutedWater}$}$$

\subsection{Initial set displacement}

Until now the position or content of the initial set was fixed and did not change unless transformations were applied to it. We now want to show how we can have an initial form which is displaced after the application of a transformation. Take the string of letters $ababbbaab$, we can define a transformation such that it is applied to the first three terms, then applied to the next three terms and then to the last three terms. We can use the `$\star$' notation to allow us to move the initial form.

If we follow each transformation by the transformation
$$\fbox{$\star\star\star(\star\star\star)\rhd (\star\star\star)\star\star\star$}$$
we will displace the initial form.

An example is given as follows.
$$\begin{array}{c}
\fbox{$\star\star\star(\star\star\star)\rhd (\star\star\star)\star\star\star$} \rightarrow\fbox{$cc\rhd ab$}\rightarrow (aba)bbbaab\\
\Downarrow\\
\fbox{$\star\star\star(\star\star\star)\rhd (\star\star\star)\star\star\star$}\rightarrow (cca)bbbaab\\
\Downarrow\\
cca(bbb)aab
\end{array}$$

\subsection{Series and parallel invariance}

A collection of transformations applied in series to an initial form will often have different effects than if the transformations were applied in parallel. But depending on how the model was built, whether we apply the transformations in series or parallel, will have the same effect. Eventually, it would be interesting to find a precise characterisation of systems which are reduced in the same way whether the transformations are applied in series or in parallel.
Here follows two examples of what we could call \textit{parallel-series invariant} systems.
$$\fbox{$B\rhd b$}\rightarrow\fbox{$A\rhd a$}\rightarrow (ab)\,\,\asymp\,\, [\,\fbox{$B\rhd b$},\fbox{$A\rhd a$}\,]\rightarrow (ab).$$

$$\fbox{$cc\rhd ba$}\rightarrow\fbox{$bba\rhd aaa$}\rightarrow (aaa)\,\,\asymp\,\, [\,\fbox{$bba\rhd aaa$},\fbox{$cc\rhd ba$}\,]\rightarrow (a,a,a).$$

\subsection{Typesetting}

The paper was written in the typesetting language LaTeX. Most figures and diagrams where drawn in LaTeX Draw which generate PSTricks code that can be copied directly in the LaTex document. The LaTeX animate package was used for the animations. Note that for the LaTeX code to compile well, all the images needed to be of the Encapsulated PostScript (\textit{.EPS}). For the animations to display well, the code needed to be compiled with XeLaTeX, but for rapid compilation the sequence LaTeX, DVI2PS and PS2PDF was used.

The transformations are written with the following LaTex code.
\begin{center}\$\textbackslash fbox\{\$B \textbackslash rhd A\$\}\text{^}\{\textbackslash ,\textbackslash infty S\}\$\end{center}
This code will display as follows.
\begin{center}$\fbox{$B \rhd A$}^{\,\infty S}$\end{center}

\pagebreak
\section*{Acknowledgement}

I would like to thank V\'eronique Pag\'e and Nathalie Patenaude for their help and encouragement. I am also infinitely grateful to Anyen Rinpoche and Johanna Okker for their unfailing help and support. They are the ones who made the realization of this project possible.

\pagebreak

\end{document}